\pdfoutput=1
\documentclass[10pt]{article}
\usepackage[utf8]{inputenc}
\usepackage[T1]{fontenc}
\usepackage{lmodern}
\input{glyphtounicode}
\usepackage{amsmath}
\usepackage{booktabs}
\usepackage{array}
\usepackage{graphicx}
\usepackage{natbib}
\usepackage{url}
\usepackage{placeins}
\usepackage{float}
\usepackage{multirow}
\usepackage{adjustbox}

\newcommand{\RsaGapReliableGOne}{0.028}
\newcommand{\RidgeGapGOne}{\ensuremath{-}0.029}
\newcommand{\RidgeGapGFour}{\ensuremath{-}0.051}
\newcommand{\RidgeGapGEight}{\ensuremath{-}0.058}
\newcommand{\POneRidgeRandBncal}{0.002}
\newcommand{\POneRsaRandBncal}{0.047}
\newcommand{\ETwoPerFoldGap}{\ensuremath{-}0.023}
\newcommand{\ETwoHiFracDiff}{0.055}
\newcommand{\EThreeRunFoldGap}{\ensuremath{-}0.027}
\newcommand{\EFourEncGapRandom}{\ensuremath{-}0.004}
\newcommand{\EFourEncGapRun}{\ensuremath{-}0.008}
\newcommand{\EFourEncPosSeedsRandom}{2}
\newcommand{\SOneStdGap}{\ensuremath{-}0.015}
\newcommand{\SOneStdRandBncal}{0.002}
\newcommand{\STwoSubTwoRsa}{0.056}
\newcommand{\STwoSubTwoRidge}{\ensuremath{-}0.023}
\newcommand{\STwoSubThreeRsa}{0.008}
\newcommand{\STwoSubThreeRidge}{\ensuremath{-}0.011}
\newcommand{\VoneGapRawTwoTwoFour}{0.030}
\newcommand{\VoneGapRawTwoTwoFourLo}{0.018}
\newcommand{\VoneGapRawTwoTwoFourHi}{0.042}
\newcommand{\VoneGapRawThreeTwo}{\ensuremath{-}0.001}
\newcommand{\VoneGapRawThreeTwoLo}{\ensuremath{-}0.011}
\newcommand{\VoneGapRawThreeTwoHi}{0.009}
\newcommand{\VoneGapZTwoTwoFour}{\ensuremath{-}0.011}
\newcommand{\VoneGapZTwoTwoFourLo}{\ensuremath{-}0.017}
\newcommand{\VoneGapZTwoTwoFourHi}{\ensuremath{-}0.005}
\newcommand{\VoneGapZThreeTwo}{\ensuremath{-}0.012}
\newcommand{\VoneGapZThreeTwoLo}{\ensuremath{-}0.018}
\newcommand{\VoneGapZThreeTwoHi}{\ensuremath{-}0.007}
\newcommand{\VoneGapRawUp}{0.035}
\newcommand{\VoneGapRawUpLo}{0.023}
\newcommand{\VoneGapRawUpHi}{0.046}
\newcommand{\VoneGapZUp}{0.002}
\newcommand{\VoneGapZUpLo}{\ensuremath{-}0.001}
\newcommand{\VoneGapZUpHi}{0.005}
\newcommand{\SlopeRaw}{0.031}
\newcommand{\SlopeRawLo}{0.024}
\newcommand{\SlopeRawHi}{0.038}
\newcommand{\SlopeZ}{0.002}
\newcommand{\SlopeZLo}{\ensuremath{-}0.004}
\newcommand{\SlopeZHi}{0.007}
\newcommand{\LocGapRaw}{0.011}
\newcommand{\LocGapRawLo}{0.003}
\newcommand{\LocGapRawHi}{0.021}
\newcommand{\LocGapZ}{0.007}
\newcommand{\LocGapZLo}{0.002}
\newcommand{\LocGapZHi}{0.011}
\newcommand{\MOneSpearman}{0.68}
\newcommand{\PRidgeGapThreeTwo}{\ensuremath{-}0.026}
\newcommand{\PRidgeGapTwoTwoFour}{\ensuremath{-}0.026}
\newcommand{\PRidgeSlope}{\ensuremath{-}0.000}
\newcommand{\PMechOneRandRange}{0.068}
\newcommand{\PMechOneBpSeeds}{5}
\newcommand{\PMechTwoNotchSeeds}{5}
\newcommand{\PMechTwoSpearman}{0.27}

\newcommand{\MultiSlopePaper}{0.031}
\newcommand{\MultiSlopePaperLo}{0.024}
\newcommand{\MultiSlopePaperHi}{0.038}
\newcommand{\MultiEuclidGap}{0.023}
\newcommand{\MultiEuclidGapLo}{0.004}
\newcommand{\MultiEuclidGapHi}{0.041}
\newcommand{\MultiOTwoViolations}{0}
\newcommand{\MultiLocPositive}{36}
\newcommand{\MultiLocCiPositive}{20}
\newcommand{\MultiPosCiCells}{13}
\newcommand{\MultiNegCiCells}{9}
\newcommand{\GOneSpearman}{0.97}
\newcommand{\GTwoPartialGap}{0.004}
\newcommand{\GTwoPartialGapLo}{\ensuremath{-}0.004}
\newcommand{\GTwoPartialGapHi}{0.013}
\newcommand{\GTwoPartialSlope}{0.017}
\newcommand{\GTwoPartialSlopeLo}{0.011}
\newcommand{\GTwoPartialSlopeHi}{0.023}
\newcommand{\GenOneSpearman}{0.79}
\newcommand{\GenTwoModels}{4}
\newcommand{\GenThreeModels}{0}
\newcommand{\GenFourQualified}{5}
\newcommand{\GenAlexnetSlopeRawCorr}{0.008}
\newcommand{\GenAlexnetSlopeZCorr}{0.005}

\newcommand{\GenAlexnetLocRawEucl}{0.009}
\newcommand{\GenAlexnetLocRawEuclLo}{\ensuremath{-}0.024}
\newcommand{\GenAlexnetLocRawEuclHi}{0.043}
\newcommand{\GenVggelevenSlopeRawCorr}{0.008}
\newcommand{\GenVggelevenSlopeZCorr}{0.010}
\newcommand{\GenConvnexttinyLocRawEucl}{\ensuremath{-}0.023}
\newcommand{\GenConvnexttinyLocRawEuclLo}{\ensuremath{-}0.061}
\newcommand{\GenConvnexttinyLocRawEuclHi}{0.013}
\newcommand{\SepOwnVoneNormCenter}{0.017}
\newcommand{\SepOwnVoneNormCenterLo}{\ensuremath{-}0.004}
\newcommand{\SepOwnVoneNormCenterHi}{0.036}

\newcommand{\SepSlopeOwnNormCenter}{\ensuremath{-}0.004}
\newcommand{\SepSlopeOwnNormCenterLo}{\ensuremath{-}0.013}
\newcommand{\SepSlopeOwnNormCenterHi}{0.005}
\newcommand{\SepOwnVoneAngleCenter}{\ensuremath{-}0.009}
\newcommand{\SepOwnVoneAngleCenterLo}{\ensuremath{-}0.017}
\newcommand{\SepOwnVoneAngleCenterHi}{\ensuremath{-}0.002}
\newcommand{\SepOwnLocAngleCenter}{0.007}
\newcommand{\SepOwnLocAngleCenterLo}{0.002}
\newcommand{\SepOwnLocAngleCenterHi}{0.012}
\newcommand{\SepSlopeOwnAngleCenter}{0.001}
\newcommand{\SepSlopeOwnAngleCenterLo}{\ensuremath{-}0.004}
\newcommand{\SepSlopeOwnAngleCenterHi}{0.006}
\newcommand{\SepSlopeResnetfiftyAngleCenter}{0.010}
\newcommand{\SepSlopeResnetfiftyAngleCenterLo}{0.007}
\newcommand{\SepSlopeResnetfiftyAngleCenterHi}{0.014}
\newcommand{\SepConvnexttinyVoneAngleCenter}{\ensuremath{-}0.003}
\newcommand{\SepConvnexttinyVoneAngleCenterLo}{\ensuremath{-}0.007}
\newcommand{\SepConvnexttinyVoneAngleCenterHi}{0.002}

\newcommand{\SepSlopeConvnexttinyAngleCenter}{0.001}
\newcommand{\SepSlopeConvnexttinyAngleCenterLo}{\ensuremath{-}0.000}
\newcommand{\SepSlopeConvnexttinyAngleCenterHi}{0.002}
\newcommand{\SepOneMaxRelError}{\ensuremath{3\times 10^{-12}}}
\newcommand{\SepTwoModels}{4}
\newcommand{\SepThreeModels}{5}
\newcommand{\SepFourModels}{6}

\newcommand{\SepSlopeAngleCiPositive}{4}
\newcommand{\PreregRandBpGOne}{\ensuremath{-}0.021}

\newcommand{\PreregRandBpGOnePos}{0}
\newcommand{\PreregRandBpGOneP}{1.000}
\newcommand{\PreregCondRandGOne}{0.043}
\newcommand{\PreregCondBpGOne}{0.064}

\newcommand{\PreregCtlLumcontrastSubOneGOne}{0.118}

\newcommand{\PreregCtlLumcontrastSubTwoGOne}{0.102}

\newcommand{\PreregCtlLumcontrastSubThreeGOne}{0.023}

\newcommand{\PreregRandBpGTwo}{\ensuremath{-}0.035}

\newcommand{\PreregRandBpGTwoPos}{0}
\newcommand{\PreregRandBpGTwoP}{1.000}
\newcommand{\PreregCondBpGTwo}{0.070}

\newcommand{\PreregCtlLumcontrastSubOneGTwo}{0.129}

\newcommand{\PreregCtlLumcontrastSubTwoGTwo}{0.106}

\newcommand{\PreregCtlLumcontrastSubThreeGTwo}{0.027}

\newcommand{\PreregRandBpGFour}{\ensuremath{-}0.043}

\newcommand{\PreregRandBpGFourPos}{0}
\newcommand{\PreregRandBpGFourP}{1.000}
\newcommand{\PreregRandBpGFourDmin}{0.016}
\newcommand{\PreregCondRandGFour}{0.043}
\newcommand{\PreregCondBpGFour}{0.086}

\newcommand{\PreregCtlLumcontrastSubOneGFour}{0.155}

\newcommand{\PreregCtlLumcontrastSubTwoGFour}{0.132}

\newcommand{\PreregCtlLumcontrastSubThreeGFour}{0.030}

\newcommand{\PreregRandBpGEight}{\ensuremath{-}0.045}

\newcommand{\PreregRandBpGEightPos}{0}
\newcommand{\PreregRandBpGEightP}{1.000}
\newcommand{\PreregRandBpGEightDmin}{0.016}
\newcommand{\PreregCondBpGEight}{0.088}

\newcommand{\PreregCtlLumcontrastSubOneGEight}{0.161}

\newcommand{\PreregCtlLumcontrastSubTwoGEight}{0.135}

\newcommand{\PreregCtlLumcontrastSubThreeGEight}{0.035}

\newcommand{\PreregReferenceEffect}{0.044}
\newcommand{\PreregLowEdgeHits}{26}
\newcommand{\PreregLowEdgePlateau}{26}
\newcommand{\GtwoGateMaxDelta}{\ensuremath{1.3\times 10^{-8}}}

\newcommand{\PreregPilotSd}{0.0082}
\newcommand{\PreregDminSim}{0.014}

\newcommand{\PreregShuffleWinSubjects}{3}
\newcommand{\NReliableSubOne}{501}
\newcommand{\NReliableSubTwo}{518}
\newcommand{\NReliableSubThree}{286}
\newcommand{\LumDefPearson}{0.997}
\newcommand{\LumDefSpearman}{0.996}
\newcommand{\PreregReferenceSem}{0.006}
\newcommand{\PreregSecondaryMaxMean}{\ensuremath{-}0.018}
\newcommand{\PreregSecondaryMaxPos}{0}
\newcommand{\MultiOOneMaxDelta}{\ensuremath{4\times 10^{-14}}}
\newcommand{\MultiSignPositive}{21}
\newcommand{\MultiSignNegative}{15}
\newcommand{\VoxGainLumSubOne}{\ensuremath{-}0.28}
\newcommand{\VoxGainConSubOne}{0.13}
\newcommand{\VoxGainLumSubTwo}{\ensuremath{-}0.27}
\newcommand{\VoxGainConSubTwo}{0.20}
\newcommand{\VoxGainLumSubThree}{\ensuremath{-}0.23}
\newcommand{\VoxGainConSubThree}{0.13}
\newcommand{\ModelGainRandLum}{\ensuremath{-}0.44}
\newcommand{\ModelGainRandCon}{0.45}
\newcommand{\ModelGainBpLum}{0.40}
\newcommand{\ModelGainBpCon}{0.00}
\newcommand{\LowEdgePilotHits}{4}
\newcommand{\LowEdgePilotPlateau}{4}
\newcommand{\LowEdgePilotExcluded}{0}
\newcommand{\LowEdgePilotMaxGain}{\ensuremath{1.6\times 10^{-4}}}
\newcommand{\LowEdgeEtwoHits}{3}
\newcommand{\LowEdgeEtwoPlateau}{3}
\newcommand{\LowEdgeEtwoExcluded}{0}
\newcommand{\LowEdgeEtwoMaxGain}{\ensuremath{4.4\times 10^{-8}}}
\newcommand{\LowEdgeEthreeHits}{4}
\newcommand{\LowEdgeEthreePlateau}{4}
\newcommand{\LowEdgeEthreeExcluded}{0}
\newcommand{\LowEdgeEthreeMaxGain}{\ensuremath{4.8\times 10^{-6}}}
\newcommand{\LowEdgeEfourHits}{7}
\newcommand{\LowEdgeEfourPlateau}{7}
\newcommand{\LowEdgeEfourExcluded}{0}
\newcommand{\LowEdgeEfourMaxGain}{\ensuremath{4.8\times 10^{-6}}}
\newcommand{\LowEdgeSubjectsHits}{1}
\newcommand{\LowEdgeSubjectsPlateau}{1}
\newcommand{\LowEdgeSubjectsExcluded}{0}
\newcommand{\LowEdgeSubjectsMaxGain}{\ensuremath{3.0\times 10^{-5}}}
\newcommand{\LowEdgeSubThreeHits}{3}
\newcommand{\LowEdgeSubThreePlateau}{3}
\newcommand{\LowEdgeSubThreeExcluded}{0}
\newcommand{\LowEdgeSubThreeMaxGain}{\ensuremath{1.3\times 10^{-5}}}
\newcommand{\LowEdgeRidgeHits}{20}
\newcommand{\LowEdgeRidgePlateau}{20}
\newcommand{\LowEdgeRidgeExcluded}{0}
\newcommand{\LowEdgeRidgeMaxGain}{\ensuremath{4.6\times 10^{-5}}}
\newcommand{\LowEdgePrimaryHits}{26}
\newcommand{\LowEdgePrimaryPlateau}{26}
\newcommand{\LowEdgePrimaryExcluded}{0}
\newcommand{\LowEdgePrimaryMaxGain}{\ensuremath{5.8\times 10^{-5}}}
\newcommand{\RunFoldPooledGOneMean}{\ensuremath{-}0.027}
\newcommand{\RunFoldPooledGOneSd}{0.010}
\newcommand{\RunFoldPooledGOnePos}{0}
\newcommand{\RunFoldPerfoldGOneMean}{\ensuremath{-}0.026}
\newcommand{\RunFoldPerfoldGOneSd}{0.008}
\newcommand{\RunFoldPerfoldGOnePos}{0}
\newcommand{\RunFoldPooledRandGOneMean}{\ensuremath{-}0.029}
\newcommand{\RunFoldPooledRandGOneSd}{0.008}
\newcommand{\RunFoldPooledRandGOnePos}{0}
\newcommand{\RunFoldLevelGOne}{\ensuremath{-}0.046}
\newcommand{\RunFoldPooledGFourMean}{\ensuremath{-}0.048}
\newcommand{\RunFoldPooledGFourSd}{0.011}
\newcommand{\RunFoldPooledGFourPos}{0}
\newcommand{\RunFoldPerfoldGFourMean}{\ensuremath{-}0.044}
\newcommand{\RunFoldPerfoldGFourSd}{0.006}
\newcommand{\RunFoldPerfoldGFourPos}{0}
\newcommand{\RunFoldPooledRandGFourMean}{\ensuremath{-}0.051}
\newcommand{\RunFoldPooledRandGFourSd}{0.011}
\newcommand{\RunFoldPooledRandGFourPos}{0}
\newcommand{\RunFoldLevelGFour}{\ensuremath{-}0.051}
\newcommand{\RunFoldPooledGEightMean}{\ensuremath{-}0.058}
\newcommand{\RunFoldPooledGEightSd}{0.008}
\newcommand{\RunFoldPooledGEightPos}{0}
\newcommand{\RunFoldPerfoldGEightMean}{\ensuremath{-}0.048}
\newcommand{\RunFoldPerfoldGEightSd}{0.006}
\newcommand{\RunFoldPerfoldGEightPos}{0}
\newcommand{\RunFoldPooledRandGEightMean}{\ensuremath{-}0.058}
\newcommand{\RunFoldPooledRandGEightSd}{0.010}
\newcommand{\RunFoldPooledRandGEightPos}{0}
\newcommand{\RunFoldLevelGEight}{\ensuremath{-}0.050}
\newcommand{\EncRsaRandomCentredMean}{\ensuremath{-}0.004}
\newcommand{\EncRsaRandomCentredSd}{0.009}
\newcommand{\EncRsaRandomCentredPos}{2}
\newcommand{\EncRsaRandomUncentredMean}{\ensuremath{-}0.001}
\newcommand{\EncRsaRandomUncentredSd}{0.009}
\newcommand{\EncRsaRandomUncentredPos}{3}
\newcommand{\EncRsaRunCentredMean}{\ensuremath{-}0.008}
\newcommand{\EncRsaRunCentredSd}{0.008}
\newcommand{\EncRsaRunCentredPos}{0}
\newcommand{\EncRsaRunUncentredMean}{\ensuremath{-}0.007}
\newcommand{\EncRsaRunUncentredSd}{0.009}
\newcommand{\EncRsaRunUncentredPos}{2}
\newcommand{\BandOnlyVoneRaw}{0.033}
\newcommand{\BandOnlyVoneRawLo}{0.021}
\newcommand{\BandOnlyVoneRawHi}{0.044}
\newcommand{\BandOnlyVoneZ}{0.004}
\newcommand{\BandOnlyVoneZLo}{0.001}
\newcommand{\BandOnlyVoneZHi}{0.007}
\newcommand{\BandOnlyLocRaw}{0.002}
\newcommand{\BandOnlyLocRawLo}{\ensuremath{-}0.004}
\newcommand{\BandOnlyLocRawHi}{0.009}
\newcommand{\BandOnlyLocZ}{0.002}
\newcommand{\BandOnlyLocZLo}{\ensuremath{-}0.002}
\newcommand{\BandOnlyLocZHi}{0.006}
\newcommand{\BandNotchVoneRaw}{\ensuremath{-}0.021}
\newcommand{\BandNotchVoneRawLo}{\ensuremath{-}0.035}
\newcommand{\BandNotchVoneRawHi}{\ensuremath{-}0.007}
\newcommand{\BandNotchVoneZ}{\ensuremath{-}0.010}
\newcommand{\BandNotchVoneZLo}{\ensuremath{-}0.018}
\newcommand{\BandNotchVoneZHi}{\ensuremath{-}0.003}
\newcommand{\BandNotchLocRaw}{0.004}
\newcommand{\BandNotchLocRawLo}{\ensuremath{-}0.005}
\newcommand{\BandNotchLocRawHi}{0.013}
\newcommand{\BandNotchLocZ}{0.003}
\newcommand{\BandNotchLocZLo}{\ensuremath{-}0.001}
\newcommand{\BandNotchLocZHi}{0.006}
\newcommand{\BandMOnlyRand}{0.619}
\newcommand{\BandMOnlyBp}{0.484}
\newcommand{\BandMNotchRand}{0.832}
\newcommand{\BandMNotchBp}{0.830}
\newcommand{\BandMNativeRand}{0.707}
\newcommand{\BandMNativeBp}{0.638}
\newcommand{\PerSubDiffSubOneGOneMean}{\ensuremath{-}0.029}
\newcommand{\PerSubDiffSubOneGOneSd}{0.008}
\newcommand{\PerSubDiffSubOneGOnePos}{0}
\newcommand{\PerSubBpSubOneGOne}{0.094}
\newcommand{\PerSubDiffSubOneGTwoMean}{\ensuremath{-}0.049}
\newcommand{\PerSubDiffSubOneGTwoSd}{0.011}
\newcommand{\PerSubDiffSubOneGTwoPos}{0}
\newcommand{\PerSubBpSubOneGTwo}{0.104}
\newcommand{\PerSubDiffSubOneGFourMean}{\ensuremath{-}0.051}
\newcommand{\PerSubDiffSubOneGFourSd}{0.011}
\newcommand{\PerSubDiffSubOneGFourPos}{0}
\newcommand{\PerSubBpSubOneGFour}{0.123}
\newcommand{\PerSubDiffSubOneGEightMean}{\ensuremath{-}0.058}
\newcommand{\PerSubDiffSubOneGEightSd}{0.010}
\newcommand{\PerSubDiffSubOneGEightPos}{0}
\newcommand{\PerSubBpSubOneGEight}{0.126}
\newcommand{\PerSubDiffSubTwoGOneMean}{\ensuremath{-}0.023}
\newcommand{\PerSubDiffSubTwoGOneSd}{0.004}
\newcommand{\PerSubDiffSubTwoGOnePos}{0}
\newcommand{\PerSubBpSubTwoGOne}{0.096}
\newcommand{\PerSubDiffSubTwoGTwoMean}{\ensuremath{-}0.034}
\newcommand{\PerSubDiffSubTwoGTwoSd}{0.006}
\newcommand{\PerSubDiffSubTwoGTwoPos}{0}
\newcommand{\PerSubBpSubTwoGTwo}{0.104}
\newcommand{\PerSubDiffSubTwoGFourMean}{\ensuremath{-}0.045}
\newcommand{\PerSubDiffSubTwoGFourSd}{0.011}
\newcommand{\PerSubDiffSubTwoGFourPos}{0}
\newcommand{\PerSubBpSubTwoGFour}{0.122}
\newcommand{\PerSubDiffSubTwoGEightMean}{\ensuremath{-}0.047}
\newcommand{\PerSubDiffSubTwoGEightSd}{0.011}
\newcommand{\PerSubDiffSubTwoGEightPos}{0}
\newcommand{\PerSubBpSubTwoGEight}{0.122}
\newcommand{\PerSubDiffSubThreeGOneMean}{\ensuremath{-}0.011}
\newcommand{\PerSubDiffSubThreeGOneSd}{0.006}
\newcommand{\PerSubDiffSubThreeGOnePos}{0}
\newcommand{\PerSubBpSubThreeGOne}{0.002}
\newcommand{\PerSubDiffSubThreeGTwoMean}{\ensuremath{-}0.021}
\newcommand{\PerSubDiffSubThreeGTwoSd}{0.006}
\newcommand{\PerSubDiffSubThreeGTwoPos}{0}
\newcommand{\PerSubBpSubThreeGTwo}{0.002}
\newcommand{\PerSubDiffSubThreeGFourMean}{\ensuremath{-}0.032}
\newcommand{\PerSubDiffSubThreeGFourSd}{0.009}
\newcommand{\PerSubDiffSubThreeGFourPos}{0}
\newcommand{\PerSubBpSubThreeGFour}{0.013}
\newcommand{\PerSubDiffSubThreeGEightMean}{\ensuremath{-}0.030}
\newcommand{\PerSubDiffSubThreeGEightSd}{0.009}
\newcommand{\PerSubDiffSubThreeGEightPos}{0}
\newcommand{\PerSubBpSubThreeGEight}{0.014}

\newcommand{\MpFullOwnVoneNatTwoTwoFour}{0.030}
\newcommand{\MpFullOwnVoneNatTwoTwoFourLo}{0.018}
\newcommand{\MpFullOwnVoneNatTwoTwoFourHi}{0.042}

\newcommand{\MpFullOwnLoc}{0.011}
\newcommand{\MpFullOwnLocLo}{0.003}
\newcommand{\MpFullOwnLocHi}{0.021}
\newcommand{\MpOrthoOwnVoneNatTwoTwoFour}{\ensuremath{-}0.002}
\newcommand{\MpOrthoOwnVoneNatTwoTwoFourLo}{\ensuremath{-}0.007}
\newcommand{\MpOrthoOwnVoneNatTwoTwoFourHi}{0.003}

\newcommand{\MpOrthoOwnLoc}{0.005}
\newcommand{\MpOrthoOwnLocLo}{0.001}
\newcommand{\MpOrthoOwnLocHi}{0.010}
\newcommand{\MpOrthoOwnSlope}{0.000}
\newcommand{\MpOrthoOwnSlopeLo}{\ensuremath{-}0.003}
\newcommand{\MpOrthoOwnSlopeHi}{0.003}
\newcommand{\MpParOwnVoneNatTwoTwoFour}{0.012}
\newcommand{\MpParOwnVoneNatTwoTwoFourLo}{\ensuremath{-}0.011}
\newcommand{\MpParOwnVoneNatTwoTwoFourHi}{0.035}

\newcommand{\MpParOwnLoc}{\ensuremath{-}0.002}
\newcommand{\MpParOwnLocLo}{\ensuremath{-}0.020}
\newcommand{\MpParOwnLocHi}{0.017}
\newcommand{\MpParOwnSlope}{0.006}
\newcommand{\MpParOwnSlopeLo}{\ensuremath{-}0.007}
\newcommand{\MpParOwnSlopeHi}{0.020}
\newcommand{\MpMFourModels}{6}
\newcommand{\MpGateMaxDelta}{\ensuremath{9.9\times 10^{-17}}}
\newcommand{\MultiDistinct}{28}
\newcommand{\MultiDistinctPos}{5}
\newcommand{\MultiDistinctNeg}{9}
\newcommand{\MultiDistinctIncl}{14}
\newcommand{\MultiDistinctEucl}{4}
\newcommand{\AbTwoOwnFullFull}{0.030}
\newcommand{\AbTwoOwnFullFullLo}{0.018}
\newcommand{\AbTwoOwnFullFullHi}{0.042}
\newcommand{\AbTwoOwnFullOrtho}{0.006}
\newcommand{\AbTwoOwnFullOrthoLo}{\ensuremath{-}0.001}
\newcommand{\AbTwoOwnFullOrthoHi}{0.012}
\newcommand{\AbTwoAlexnetFullOrtho}{\ensuremath{-}0.001}
\newcommand{\AbTwoAlexnetFullOrthoLo}{\ensuremath{-}0.007}
\newcommand{\AbTwoAlexnetFullOrthoHi}{0.005}
\newcommand{\AbTwoVggelevenFullOrtho}{0.005}
\newcommand{\AbTwoVggelevenFullOrthoLo}{\ensuremath{-}0.001}
\newcommand{\AbTwoVggelevenFullOrthoHi}{0.013}
\newcommand{\AbTwoResneteighteenFullOrtho}{\ensuremath{-}0.003}
\newcommand{\AbTwoResneteighteenFullOrthoLo}{\ensuremath{-}0.008}
\newcommand{\AbTwoResneteighteenFullOrthoHi}{0.003}
\newcommand{\AbTwoResnetfiftyFullOrtho}{\ensuremath{-}0.003}
\newcommand{\AbTwoResnetfiftyFullOrthoLo}{\ensuremath{-}0.008}
\newcommand{\AbTwoResnetfiftyFullOrthoHi}{0.003}
\newcommand{\AbTwoConvnexttinyFullOrtho}{0.003}
\newcommand{\AbTwoConvnexttinyFullOrthoLo}{\ensuremath{-}0.002}
\newcommand{\AbTwoConvnexttinyFullOrthoHi}{0.008}
\newcommand{\AbTwoVitbsixteenFullOrtho}{\ensuremath{-}0.018}
\newcommand{\AbTwoVitbsixteenFullOrthoLo}{\ensuremath{-}0.029}
\newcommand{\AbTwoVitbsixteenFullOrthoHi}{\ensuremath{-}0.008}
\newcommand{\AbTwoImnFullOrthoPos}{0}
\newcommand{\AbTwoImnFullOrthoNeg}{1}
\newcommand{\AbTwoOwnOrthoFull}{\ensuremath{-}0.001}
\newcommand{\AbTwoOwnOrthoFullLo}{\ensuremath{-}0.007}
\newcommand{\AbTwoOwnOrthoFullHi}{0.006}
\newcommand{\AbTwoAlexnetOrthoFull}{\ensuremath{-}0.002}
\newcommand{\AbTwoAlexnetOrthoFullLo}{\ensuremath{-}0.012}
\newcommand{\AbTwoAlexnetOrthoFullHi}{0.008}
\newcommand{\AbTwoVggelevenOrthoFull}{\ensuremath{-}0.003}
\newcommand{\AbTwoVggelevenOrthoFullLo}{\ensuremath{-}0.015}
\newcommand{\AbTwoVggelevenOrthoFullHi}{0.008}
\newcommand{\AbTwoResneteighteenOrthoFull}{\ensuremath{-}0.011}
\newcommand{\AbTwoResneteighteenOrthoFullLo}{\ensuremath{-}0.018}
\newcommand{\AbTwoResneteighteenOrthoFullHi}{\ensuremath{-}0.004}
\newcommand{\AbTwoResnetfiftyOrthoFull}{\ensuremath{-}0.012}
\newcommand{\AbTwoResnetfiftyOrthoFullLo}{\ensuremath{-}0.020}
\newcommand{\AbTwoResnetfiftyOrthoFullHi}{\ensuremath{-}0.005}
\newcommand{\AbTwoConvnexttinyOrthoFull}{0.005}
\newcommand{\AbTwoConvnexttinyOrthoFullLo}{\ensuremath{-}0.002}
\newcommand{\AbTwoConvnexttinyOrthoFullHi}{0.012}
\newcommand{\AbTwoVitbsixteenOrthoFull}{\ensuremath{-}0.024}
\newcommand{\AbTwoVitbsixteenOrthoFullLo}{\ensuremath{-}0.041}
\newcommand{\AbTwoVitbsixteenOrthoFullHi}{\ensuremath{-}0.008}
\newcommand{\AbTwoImnOrthoFullPos}{0}
\newcommand{\AbTwoImnOrthoFullNeg}{3}
\newcommand{\AbTwoOwnOrthoOrtho}{\ensuremath{-}0.002}
\newcommand{\AbTwoOwnOrthoOrthoLo}{\ensuremath{-}0.007}
\newcommand{\AbTwoOwnOrthoOrthoHi}{0.003}
\newcommand{\AbTwoXOne}{0.032}
\newcommand{\AbTwoXOneLo}{0.022}
\newcommand{\AbTwoXOneHi}{0.042}
\newcommand{\WAngOwnVone}{0.013}
\newcommand{\WAngOwnVoneLo}{\ensuremath{-}0.003}
\newcommand{\WAngOwnVoneHi}{0.030}

\newcommand{\WAngVggelevenVone}{0.026}
\newcommand{\WAngVggelevenVoneLo}{0.004}
\newcommand{\WAngVggelevenVoneHi}{0.046}
\newcommand{\WAngImnPosModels}{1}
\newcommand{\WRankNAOwn}{0.09}
\newcommand{\WRankNASubOne}{0.00}
\newcommand{\WRankNASubTwo}{0.01}
\newcommand{\WRankNASubThree}{0.01}

\newcommand{\WRankAWOwn}{0.75}
\newcommand{\WRankAWSubOne}{0.63}
\newcommand{\WRankAWSubTwo}{0.68}
\newcommand{\WRankAWSubThree}{0.67}

\newcommand{\WRtwoNARand}{0.0017}
\newcommand{\WRtwoNABp}{0.0012}

\newcommand{\FSkillMseSkillGOneMean}{\ensuremath{-}0.0041}
\newcommand{\FSkillMseSkillGOneSd}{0.0008}
\newcommand{\FSkillMseSkillGOnePos}{0}
\newcommand{\FSkillMseSkillGTwoMean}{\ensuremath{-}0.0055}
\newcommand{\FSkillMseSkillGTwoSd}{0.0010}
\newcommand{\FSkillMseSkillGTwoPos}{0}
\newcommand{\FSkillMseSkillGFourMean}{\ensuremath{-}0.0069}
\newcommand{\FSkillMseSkillGFourSd}{0.0013}
\newcommand{\FSkillMseSkillGFourPos}{0}
\newcommand{\FSkillMseSkillGEightMean}{\ensuremath{-}0.0074}
\newcommand{\FSkillMseSkillGEightSd}{0.0013}
\newcommand{\FSkillMseSkillGEightPos}{0}
\newcommand{\FSkillPerfoldGOneMean}{\ensuremath{-}0.0166}
\newcommand{\FSkillPerfoldGOneSd}{0.0051}
\newcommand{\FSkillPerfoldGOnePos}{0}
\newcommand{\FSkillPerfoldGTwoMean}{\ensuremath{-}0.0220}
\newcommand{\FSkillPerfoldGTwoSd}{0.0053}
\newcommand{\FSkillPerfoldGTwoPos}{0}
\newcommand{\FSkillPerfoldGFourMean}{\ensuremath{-}0.0281}
\newcommand{\FSkillPerfoldGFourSd}{0.0068}
\newcommand{\FSkillPerfoldGFourPos}{0}
\newcommand{\FSkillPerfoldGEightMean}{\ensuremath{-}0.0311}
\newcommand{\FSkillPerfoldGEightSd}{0.0067}
\newcommand{\FSkillPerfoldGEightPos}{0}
\newcommand{\FSkillAbsBpGFour}{0.0111}

\newcommand{\FSkillAbsRandGFour}{0.0042}

\newcommand{\FFloorSubOne}{\ensuremath{-}0.091}
\newcommand{\FFloorSubTwo}{\ensuremath{-}0.088}
\newcommand{\FFloorSubThree}{\ensuremath{-}0.097}
\newcommand{\FFloorMean}{\ensuremath{-}0.092}
\newcommand{\FSkillGateMaxDelta}{\ensuremath{1.8\times 10^{-13}}}
\newcommand{\FSkillLowEdgeHits}{4}

\newcommand{\LowEdgePrimaryModelHits}{6}
\newcommand{\LowEdgeStepSkillHits}{4}
\newcommand{\LowEdgeStepSkillPlateau}{4}
\newcommand{\LowEdgeStepSkillExcluded}{0}
\newcommand{\LowEdgeStepSkillMaxGain}{\ensuremath{3.0\times 10^{-5}}}
\newcommand{\LowEdgeStepJointHits}{3}
\newcommand{\LowEdgeStepJointPlateau}{3}
\newcommand{\LowEdgeStepJointExcluded}{0}
\newcommand{\LowEdgeStepJointMaxGain}{\ensuremath{3.6\times 10^{-5}}}
\newcommand{\LowEdgeStepPermHits}{3}
\newcommand{\LowEdgeStepPermPlateau}{3}
\newcommand{\LowEdgeStepPermExcluded}{0}
\newcommand{\LowEdgeStepPermMaxGain}{\ensuremath{2.9\times 10^{-5}}}

\newcommand{\FJointRandGOneJoint}{0.060}
\newcommand{\FJointRandGOneBank}{0.081}

\newcommand{\FJointRandGFourJoint}{0.064}
\newcommand{\FJointRandGFourBank}{0.106}

\newcommand{\FJointBpGOneJoint}{0.068}
\newcommand{\FJointBpGOneBank}{0.081}

\newcommand{\FJointBpGFourJoint}{0.090}
\newcommand{\FJointBpGFourBank}{0.106}
\newcommand{\FPermSubOneRandGOneMean}{\ensuremath{-}0.070}
\newcommand{\FPermSubOneRandGOneLo}{\ensuremath{-}0.083}
\newcommand{\FPermSubOneRandGOneHi}{\ensuremath{-}0.040}
\newcommand{\FPermSubOneRandGOneReal}{0.052}
\newcommand{\FPermSubOneBpGOneMean}{\ensuremath{-}0.072}
\newcommand{\FPermSubOneBpGOneLo}{\ensuremath{-}0.083}
\newcommand{\FPermSubOneBpGOneHi}{\ensuremath{-}0.046}
\newcommand{\FPermSubOneBpGOneReal}{0.088}
\newcommand{\FPermSubOneRandGFourMean}{\ensuremath{-}0.073}
\newcommand{\FPermSubOneRandGFourLo}{\ensuremath{-}0.084}
\newcommand{\FPermSubOneRandGFourHi}{\ensuremath{-}0.041}
\newcommand{\FPermSubOneRandGFourReal}{0.060}
\newcommand{\FPermSubOneBpGFourMean}{\ensuremath{-}0.072}
\newcommand{\FPermSubOneBpGFourLo}{\ensuremath{-}0.086}
\newcommand{\FPermSubOneBpGFourHi}{\ensuremath{-}0.046}
\newcommand{\FPermSubOneBpGFourReal}{0.118}
\newcommand{\FPermSubTwoRandGOneMean}{\ensuremath{-}0.068}
\newcommand{\FPermSubTwoRandGOneLo}{\ensuremath{-}0.078}
\newcommand{\FPermSubTwoRandGOneHi}{\ensuremath{-}0.056}
\newcommand{\FPermSubTwoRandGOneReal}{0.073}
\newcommand{\FPermSubTwoBpGOneMean}{\ensuremath{-}0.069}
\newcommand{\FPermSubTwoBpGOneLo}{\ensuremath{-}0.079}
\newcommand{\FPermSubTwoBpGOneHi}{\ensuremath{-}0.062}
\newcommand{\FPermSubTwoBpGOneReal}{0.096}
\newcommand{\FPermSubTwoRandGFourMean}{\ensuremath{-}0.071}
\newcommand{\FPermSubTwoRandGFourLo}{\ensuremath{-}0.081}
\newcommand{\FPermSubTwoRandGFourHi}{\ensuremath{-}0.060}
\newcommand{\FPermSubTwoRandGFourReal}{0.071}
\newcommand{\FPermSubTwoBpGFourMean}{\ensuremath{-}0.069}
\newcommand{\FPermSubTwoBpGFourLo}{\ensuremath{-}0.080}
\newcommand{\FPermSubTwoBpGFourHi}{\ensuremath{-}0.055}
\newcommand{\FPermSubTwoBpGFourReal}{0.120}
\newcommand{\FPermSubThreeRandGOneMean}{\ensuremath{-}0.074}
\newcommand{\FPermSubThreeRandGOneLo}{\ensuremath{-}0.091}
\newcommand{\FPermSubThreeRandGOneHi}{\ensuremath{-}0.038}
\newcommand{\FPermSubThreeRandGOneReal}{\ensuremath{-}0.014}
\newcommand{\FPermSubThreeBpGOneMean}{\ensuremath{-}0.077}
\newcommand{\FPermSubThreeBpGOneLo}{\ensuremath{-}0.093}
\newcommand{\FPermSubThreeBpGOneHi}{\ensuremath{-}0.048}
\newcommand{\FPermSubThreeBpGOneReal}{\ensuremath{-}0.004}
\newcommand{\FPermSubThreeRandGFourMean}{\ensuremath{-}0.075}
\newcommand{\FPermSubThreeRandGFourLo}{\ensuremath{-}0.094}
\newcommand{\FPermSubThreeRandGFourHi}{\ensuremath{-}0.037}
\newcommand{\FPermSubThreeRandGFourReal}{\ensuremath{-}0.021}
\newcommand{\FPermSubThreeBpGFourMean}{\ensuremath{-}0.076}
\newcommand{\FPermSubThreeBpGFourLo}{\ensuremath{-}0.093}
\newcommand{\FPermSubThreeBpGFourHi}{\ensuremath{-}0.048}
\newcommand{\FPermSubThreeBpGFourReal}{0.008}
\newcommand{\FPermPNineSevenFiveMin}{\ensuremath{-}0.062}
\newcommand{\FPermPNineSevenFiveMax}{\ensuremath{-}0.037}
\newcommand{\FPermRealMinMain}{0.052}
\newcommand{\FPermRealMaxMain}{0.120}
\newcommand{\FPermRealMinSubThree}{\ensuremath{-}0.021}
\newcommand{\FPermRealMaxSubThree}{0.008}
\newcommand{\FJointMinDiff}{\ensuremath{-}0.047}
\newcommand{\FJointMaxDiff}{\ensuremath{-}0.013}
\newcommand{\AbsRsaPaperMin}{0.008}
\newcommand{\AbsRsaPaperMax}{0.053}
\newcommand{\LosoBoundVone}{0.051}
\newcommand{\LosoBoundVoneLo}{0.039}
\newcommand{\LosoBoundVoneHi}{0.064}
\newcommand{\BNormZCorrPaper}{\ensuremath{-}0.011}
\newcommand{\BNormZCorrPaperLo}{\ensuremath{-}0.017}
\newcommand{\BNormZCorrPaperHi}{\ensuremath{-}0.005}
\newcommand{\BNormZCorrPaperFixedLo}{\ensuremath{-}0.017}
\newcommand{\BNormZCorrPaperFixedHi}{\ensuremath{-}0.005}
\newcommand{\BNormCentEuclPaper}{0.023}
\newcommand{\BNormCentEuclPaperLo}{0.002}
\newcommand{\BNormCentEuclPaperHi}{0.039}
\newcommand{\BNormCentEuclPaperFixedLo}{0.004}
\newcommand{\BNormCentEuclPaperFixedHi}{0.041}
\newcommand{\BNormCentCorrCent}{0.001}
\newcommand{\BNormCentCorrCentLo}{\ensuremath{-}0.005}
\newcommand{\BNormCentCorrCentHi}{0.007}
\newcommand{\BNormCentCorrCentFixedLo}{\ensuremath{-}0.005}
\newcommand{\BNormCentCorrCentFixedHi}{0.007}

\newcommand{\PreregCtlLumcontrastMeanGOne}{0.081}

\newcommand{\PreregCtlPermutedMeanGOne}{\ensuremath{-}0.078}

\newcommand{\PreregCtlLumcontrastMeanGTwo}{0.087}

\newcommand{\PreregCtlGaborMeanGTwo}{\ensuremath{-}0.006}
\newcommand{\PreregCtlPermutedMeanGTwo}{\ensuremath{-}0.075}

\newcommand{\PreregCtlLumcontrastMeanGFour}{0.106}
\newcommand{\PreregCtlPixelsMeanGFour}{0.021}

\newcommand{\PreregCtlPermutedMeanGFour}{\ensuremath{-}0.074}

\newcommand{\PreregCtlLumcontrastMeanGEight}{0.111}

\newcommand{\PreregCtlPermutedMeanGEight}{\ensuremath{-}0.075}
\newcommand{\FailedPredictions}{24}
\newcommand{\FailedPredictionsVone}{6}
\newcommand{\FailedPredictionsFollowup}{18}
\newcommand{\TotalPredictions}{88}
\newcommand{\TotalPredictionsVone}{33}
\newcommand{\TotalPredictionsFollowup}{55}
\newcommand{\NotInterpretedPredictions}{3}

\newcommand{\AmendmentsFollowup}{12}
\newcommand{\CvTsevTrawVoneRawCorr}{+0.100}
\newcommand{\CvTsevTrawVoneRawCorrPos}{3}

\newcommand{\CvTtestTrawVoneRawCorrKOne}{+0.078}
\newcommand{\CvTtestTrawVoneRawCorrKOnePos}{3}
\newcommand{\CvTtestTrawVoneRawCorrKOneTwo}{+0.165}
\newcommand{\CvTtestTrawVoneRawCorrKOneTwoPos}{3}
\newcommand{\CvTsevTrawVoneZCorr}{+0.002}
\newcommand{\CvTsevTrawVoneZCorrPos}{3}

\newcommand{\CvTtestTrawVoneZCorrKOne}{\ensuremath{-}0.002}
\newcommand{\CvTtestTrawVoneZCorrKOnePos}{2}
\newcommand{\CvTtestTrawVoneZCorrKOneTwo}{\ensuremath{-}0.006}
\newcommand{\CvTtestTrawVoneZCorrKOneTwoPos}{1}
\newcommand{\CvTsevTrawVoneRawEucl}{+0.023}
\newcommand{\CvTsevTrawVoneRawEuclPos}{3}

\newcommand{\CvTtestTrawVoneRawEuclKOne}{0.000}
\newcommand{\CvTtestTrawVoneRawEuclKOnePos}{1}
\newcommand{\CvTtestTrawVoneRawEuclKOneTwo}{\ensuremath{-}0.024}
\newcommand{\CvTtestTrawVoneRawEuclKOneTwoPos}{1}
\newcommand{\CvTsevTrawVoneZEucl}{+0.016}
\newcommand{\CvTsevTrawVoneZEuclPos}{3}

\newcommand{\CvTtestTrawVoneZEuclKOne}{\ensuremath{-}0.001}
\newcommand{\CvTtestTrawVoneZEuclKOnePos}{1}
\newcommand{\CvTtestTrawVoneZEuclKOneTwo}{\ensuremath{-}0.032}
\newcommand{\CvTtestTrawVoneZEuclKOneTwoPos}{0}
\newcommand{\CvTsevTrawLocRawCorr}{\ensuremath{-}0.032}
\newcommand{\CvTsevTrawLocRawCorrPos}{0}

\newcommand{\CvTtestTrawLocRawCorrKOne}{+0.010}
\newcommand{\CvTtestTrawLocRawCorrKOnePos}{2}
\newcommand{\CvTtestTrawLocRawCorrKOneTwo}{+0.001}
\newcommand{\CvTtestTrawLocRawCorrKOneTwoPos}{2}
\newcommand{\CvTsevTrawLocZCorr}{\ensuremath{-}0.004}
\newcommand{\CvTsevTrawLocZCorrPos}{0}

\newcommand{\CvTtestTrawLocZCorrKOne}{\ensuremath{-}0.002}
\newcommand{\CvTtestTrawLocZCorrKOnePos}{1}
\newcommand{\CvTtestTrawLocZCorrKOneTwo}{\ensuremath{-}0.003}
\newcommand{\CvTtestTrawLocZCorrKOneTwoPos}{1}
\newcommand{\CvTsevTrawLocRawEucl}{\ensuremath{-}0.010}
\newcommand{\CvTsevTrawLocRawEuclPos}{1}

\newcommand{\CvTtestTrawLocRawEuclKOne}{+0.005}
\newcommand{\CvTtestTrawLocRawEuclKOnePos}{2}
\newcommand{\CvTtestTrawLocRawEuclKOneTwo}{+0.013}
\newcommand{\CvTtestTrawLocRawEuclKOneTwoPos}{2}
\newcommand{\CvTsevTrawLocZEucl}{\ensuremath{-}0.010}
\newcommand{\CvTsevTrawLocZEuclPos}{1}

\newcommand{\CvTtestTrawLocZEuclKOne}{0.000}
\newcommand{\CvTtestTrawLocZEuclKOnePos}{1}
\newcommand{\CvTtestTrawLocZEuclKOneTwo}{+0.003}
\newcommand{\CvTtestTrawLocZEuclKOneTwoPos}{1}
\newcommand{\CvTsevTglobalVoneRawCorr}{+0.030}
\newcommand{\CvTsevTglobalVoneRawCorrPos}{3}

\newcommand{\CvTtestTglobalVoneRawCorrKOne}{0.000}
\newcommand{\CvTtestTglobalVoneRawCorrKOnePos}{1}
\newcommand{\CvTtestTglobalVoneRawCorrKOneTwo}{\ensuremath{-}0.018}
\newcommand{\CvTtestTglobalVoneRawCorrKOneTwoPos}{1}
\newcommand{\CvTsevTglobalVoneZCorr}{\ensuremath{-}0.011}
\newcommand{\CvTsevTglobalVoneZCorrPos}{0}

\newcommand{\CvTtestTglobalVoneZCorrKOne}{\ensuremath{-}0.016}
\newcommand{\CvTtestTglobalVoneZCorrKOnePos}{0}
\newcommand{\CvTtestTglobalVoneZCorrKOneTwo}{\ensuremath{-}0.073}
\newcommand{\CvTtestTglobalVoneZCorrKOneTwoPos}{0}
\newcommand{\CvTsevTglobalVoneRawEucl}{+0.023}
\newcommand{\CvTsevTglobalVoneRawEuclPos}{3}

\newcommand{\CvTtestTglobalVoneRawEuclKOne}{0.000}
\newcommand{\CvTtestTglobalVoneRawEuclKOnePos}{1}
\newcommand{\CvTtestTglobalVoneRawEuclKOneTwo}{\ensuremath{-}0.022}
\newcommand{\CvTtestTglobalVoneRawEuclKOneTwoPos}{1}
\newcommand{\CvTsevTglobalVoneZEucl}{+0.016}
\newcommand{\CvTsevTglobalVoneZEuclPos}{3}

\newcommand{\CvTtestTglobalVoneZEuclKOne}{\ensuremath{-}0.001}
\newcommand{\CvTtestTglobalVoneZEuclKOnePos}{1}
\newcommand{\CvTtestTglobalVoneZEuclKOneTwo}{\ensuremath{-}0.030}
\newcommand{\CvTtestTglobalVoneZEuclKOneTwoPos}{0}
\newcommand{\CvTsevTglobalLocRawCorr}{\ensuremath{-}0.011}
\newcommand{\CvTsevTglobalLocRawCorrPos}{0}

\newcommand{\CvTtestTglobalLocRawCorrKOne}{0.000}
\newcommand{\CvTtestTglobalLocRawCorrKOnePos}{1}
\newcommand{\CvTtestTglobalLocRawCorrKOneTwo}{\ensuremath{-}0.016}
\newcommand{\CvTtestTglobalLocRawCorrKOneTwoPos}{0}
\newcommand{\CvTsevTglobalLocZCorr}{\ensuremath{-}0.007}
\newcommand{\CvTsevTglobalLocZCorrPos}{0}

\newcommand{\CvTtestTglobalLocZCorrKOne}{\ensuremath{-}0.002}
\newcommand{\CvTtestTglobalLocZCorrKOnePos}{1}
\newcommand{\CvTtestTglobalLocZCorrKOneTwo}{\ensuremath{-}0.007}
\newcommand{\CvTtestTglobalLocZCorrKOneTwoPos}{0}
\newcommand{\CvTsevTglobalLocRawEucl}{\ensuremath{-}0.012}
\newcommand{\CvTsevTglobalLocRawEuclPos}{1}

\newcommand{\CvTtestTglobalLocRawEuclKOne}{+0.003}
\newcommand{\CvTtestTglobalLocRawEuclKOnePos}{2}
\newcommand{\CvTtestTglobalLocRawEuclKOneTwo}{+0.012}
\newcommand{\CvTtestTglobalLocRawEuclKOneTwoPos}{2}
\newcommand{\CvTsevTglobalLocZEucl}{\ensuremath{-}0.011}
\newcommand{\CvTsevTglobalLocZEuclPos}{1}

\newcommand{\CvTtestTglobalLocZEuclKOne}{0.000}
\newcommand{\CvTtestTglobalLocZEuclKOnePos}{2}
\newcommand{\CvTtestTglobalLocZEuclKOneTwo}{+0.004}
\newcommand{\CvTtestTglobalLocZEuclKOneTwoPos}{1}
\newcommand{\CvTsevTsessionVoneRawCorr}{+0.005}
\newcommand{\CvTsevTsessionVoneRawCorrPos}{3}

\newcommand{\CvTtestTsessionVoneRawCorrKOne}{\ensuremath{-}0.007}
\newcommand{\CvTtestTsessionVoneRawCorrKOnePos}{0}
\newcommand{\CvTtestTsessionVoneRawCorrKOneTwo}{\ensuremath{-}0.018}
\newcommand{\CvTtestTsessionVoneRawCorrKOneTwoPos}{1}
\newcommand{\CvTsevTsessionVoneZCorr}{\ensuremath{-}0.014}
\newcommand{\CvTsevTsessionVoneZCorrPos}{0}

\newcommand{\CvTtestTsessionVoneZCorrKOne}{\ensuremath{-}0.017}
\newcommand{\CvTtestTsessionVoneZCorrKOnePos}{0}
\newcommand{\CvTtestTsessionVoneZCorrKOneTwo}{\ensuremath{-}0.073}
\newcommand{\CvTtestTsessionVoneZCorrKOneTwoPos}{0}
\newcommand{\CvTsevTsessionVoneRawEucl}{+0.023}
\newcommand{\CvTsevTsessionVoneRawEuclPos}{3}

\newcommand{\CvTtestTsessionVoneRawEuclKOne}{0.000}
\newcommand{\CvTtestTsessionVoneRawEuclKOnePos}{1}
\newcommand{\CvTtestTsessionVoneRawEuclKOneTwo}{\ensuremath{-}0.022}
\newcommand{\CvTtestTsessionVoneRawEuclKOneTwoPos}{1}
\newcommand{\CvTsevTsessionVoneZEucl}{+0.016}
\newcommand{\CvTsevTsessionVoneZEuclPos}{3}

\newcommand{\CvTtestTsessionVoneZEuclKOne}{\ensuremath{-}0.001}
\newcommand{\CvTtestTsessionVoneZEuclKOnePos}{1}
\newcommand{\CvTtestTsessionVoneZEuclKOneTwo}{\ensuremath{-}0.030}
\newcommand{\CvTtestTsessionVoneZEuclKOneTwoPos}{0}
\newcommand{\CvTsevTsessionLocRawCorr}{\ensuremath{-}0.004}
\newcommand{\CvTsevTsessionLocRawCorrPos}{0}

\newcommand{\CvTtestTsessionLocRawCorrKOne}{\ensuremath{-}0.003}
\newcommand{\CvTtestTsessionLocRawCorrKOnePos}{1}
\newcommand{\CvTtestTsessionLocRawCorrKOneTwo}{\ensuremath{-}0.016}
\newcommand{\CvTtestTsessionLocRawCorrKOneTwoPos}{0}
\newcommand{\CvTsevTsessionLocZCorr}{\ensuremath{-}0.007}
\newcommand{\CvTsevTsessionLocZCorrPos}{0}

\newcommand{\CvTtestTsessionLocZCorrKOne}{\ensuremath{-}0.002}
\newcommand{\CvTtestTsessionLocZCorrKOnePos}{1}
\newcommand{\CvTtestTsessionLocZCorrKOneTwo}{\ensuremath{-}0.007}
\newcommand{\CvTtestTsessionLocZCorrKOneTwoPos}{0}
\newcommand{\CvTsevTsessionLocRawEucl}{\ensuremath{-}0.012}
\newcommand{\CvTsevTsessionLocRawEuclPos}{1}

\newcommand{\CvTtestTsessionLocRawEuclKOne}{+0.003}
\newcommand{\CvTtestTsessionLocRawEuclKOnePos}{2}
\newcommand{\CvTtestTsessionLocRawEuclKOneTwo}{+0.012}
\newcommand{\CvTtestTsessionLocRawEuclKOneTwoPos}{2}
\newcommand{\CvTsevTsessionLocZEucl}{\ensuremath{-}0.011}
\newcommand{\CvTsevTsessionLocZEuclPos}{1}

\newcommand{\CvTtestTsessionLocZEuclKOne}{0.000}
\newcommand{\CvTtestTsessionLocZEuclKOnePos}{2}
\newcommand{\CvTtestTsessionLocZEuclKOneTwo}{+0.003}
\newcommand{\CvTtestTsessionLocZEuclKOneTwoPos}{1}
\newcommand{\CvNsdNrawVoneRawCorrKOne}{+0.116}
\newcommand{\CvNsdNrawVoneRawCorrKOnePos}{8}
\newcommand{\CvNsdNrawVoneRawCorrKOneP}{0.0039}
\newcommand{\CvNsdNrawVoneRawCorrKThree}{+0.166}
\newcommand{\CvNsdNrawVoneRawCorrKThreePos}{8}

\newcommand{\CvNsdNrawVoneZCorrKOne}{\ensuremath{-}0.004}
\newcommand{\CvNsdNrawVoneZCorrKOnePos}{0}

\newcommand{\CvNsdNrawVoneZCorrKThree}{\ensuremath{-}0.008}
\newcommand{\CvNsdNrawVoneZCorrKThreePos}{0}

\newcommand{\CvNsdNrawVoneRawEuclKOne}{\ensuremath{-}0.005}
\newcommand{\CvNsdNrawVoneRawEuclKOnePos}{2}

\newcommand{\CvNsdNrawVoneRawEuclKThree}{\ensuremath{-}0.010}
\newcommand{\CvNsdNrawVoneRawEuclKThreePos}{2}

\newcommand{\CvNsdNrawVoneZEuclKOne}{\ensuremath{-}0.012}
\newcommand{\CvNsdNrawVoneZEuclKOnePos}{1}

\newcommand{\CvNsdNrawVoneZEuclKThree}{\ensuremath{-}0.020}
\newcommand{\CvNsdNrawVoneZEuclKThreePos}{1}

\newcommand{\CvNsdNrawLoobjRawCorrKOne}{+0.007}
\newcommand{\CvNsdNrawLoobjRawCorrKOnePos}{6}

\newcommand{\CvNsdNrawLoobjRawCorrKThree}{+0.013}
\newcommand{\CvNsdNrawLoobjRawCorrKThreePos}{6}

\newcommand{\CvNsdNrawLoobjZCorrKOne}{\ensuremath{-}0.001}
\newcommand{\CvNsdNrawLoobjZCorrKOnePos}{2}

\newcommand{\CvNsdNrawLoobjZCorrKThree}{\ensuremath{-}0.002}
\newcommand{\CvNsdNrawLoobjZCorrKThreePos}{1}

\newcommand{\CvNsdNrawLoobjRawEuclKOne}{\ensuremath{-}0.005}
\newcommand{\CvNsdNrawLoobjRawEuclKOnePos}{1}

\newcommand{\CvNsdNrawLoobjRawEuclKThree}{\ensuremath{-}0.010}
\newcommand{\CvNsdNrawLoobjRawEuclKThreePos}{1}

\newcommand{\CvNsdNrawLoobjZEuclKOne}{\ensuremath{-}0.002}
\newcommand{\CvNsdNrawLoobjZEuclKOnePos}{2}

\newcommand{\CvNsdNrawLoobjZEuclKThree}{\ensuremath{-}0.002}
\newcommand{\CvNsdNrawLoobjZEuclKThreePos}{3}

\newcommand{\CvNsdNrawLatRawCorrKOne}{\ensuremath{-}0.001}
\newcommand{\CvNsdNrawLatRawCorrKOnePos}{5}

\newcommand{\CvNsdNrawLatRawCorrKThree}{0.000}
\newcommand{\CvNsdNrawLatRawCorrKThreePos}{6}

\newcommand{\CvNsdNrawLatZCorrKOne}{\ensuremath{-}0.002}
\newcommand{\CvNsdNrawLatZCorrKOnePos}{0}

\newcommand{\CvNsdNrawLatZCorrKThree}{\ensuremath{-}0.005}
\newcommand{\CvNsdNrawLatZCorrKThreePos}{0}

\newcommand{\CvNsdNrawLatRawEuclKOne}{\ensuremath{-}0.006}
\newcommand{\CvNsdNrawLatRawEuclKOnePos}{1}

\newcommand{\CvNsdNrawLatRawEuclKThree}{\ensuremath{-}0.010}
\newcommand{\CvNsdNrawLatRawEuclKThreePos}{2}

\newcommand{\CvNsdNrawLatZEuclKOne}{\ensuremath{-}0.002}
\newcommand{\CvNsdNrawLatZEuclKOnePos}{1}

\newcommand{\CvNsdNrawLatZEuclKThree}{\ensuremath{-}0.002}
\newcommand{\CvNsdNrawLatZEuclKThreePos}{3}

\newcommand{\CvNsdNglobalVoneRawCorrKOne}{\ensuremath{-}0.012}
\newcommand{\CvNsdNglobalVoneRawCorrKOnePos}{0}

\newcommand{\CvNsdNglobalVoneRawCorrKThree}{\ensuremath{-}0.019}
\newcommand{\CvNsdNglobalVoneRawCorrKThreePos}{0}

\newcommand{\CvNsdNglobalVoneZCorrKOne}{\ensuremath{-}0.015}
\newcommand{\CvNsdNglobalVoneZCorrKOnePos}{0}

\newcommand{\CvNsdNglobalVoneZCorrKThree}{\ensuremath{-}0.025}
\newcommand{\CvNsdNglobalVoneZCorrKThreePos}{0}

\newcommand{\CvNsdNglobalVoneRawEuclKOne}{\ensuremath{-}0.005}
\newcommand{\CvNsdNglobalVoneRawEuclKOnePos}{2}

\newcommand{\CvNsdNglobalVoneRawEuclKThree}{\ensuremath{-}0.008}
\newcommand{\CvNsdNglobalVoneRawEuclKThreePos}{2}

\newcommand{\CvNsdNglobalVoneZEuclKOne}{\ensuremath{-}0.008}
\newcommand{\CvNsdNglobalVoneZEuclKOnePos}{2}

\newcommand{\CvNsdNglobalVoneZEuclKThree}{\ensuremath{-}0.014}
\newcommand{\CvNsdNglobalVoneZEuclKThreePos}{1}

\newcommand{\CvNsdNglobalLoobjRawCorrKOne}{\ensuremath{-}0.007}
\newcommand{\CvNsdNglobalLoobjRawCorrKOnePos}{0}

\newcommand{\CvNsdNglobalLoobjRawCorrKThree}{\ensuremath{-}0.013}
\newcommand{\CvNsdNglobalLoobjRawCorrKThreePos}{0}

\newcommand{\CvNsdNglobalLoobjZCorrKOne}{\ensuremath{-}0.003}
\newcommand{\CvNsdNglobalLoobjZCorrKOnePos}{0}

\newcommand{\CvNsdNglobalLoobjZCorrKThree}{\ensuremath{-}0.006}
\newcommand{\CvNsdNglobalLoobjZCorrKThreePos}{0}

\newcommand{\CvNsdNglobalLoobjRawEuclKOne}{\ensuremath{-}0.004}
\newcommand{\CvNsdNglobalLoobjRawEuclKOnePos}{1}

\newcommand{\CvNsdNglobalLoobjRawEuclKThree}{\ensuremath{-}0.012}
\newcommand{\CvNsdNglobalLoobjRawEuclKThreePos}{1}

\newcommand{\CvNsdNglobalLoobjZEuclKOne}{\ensuremath{-}0.002}
\newcommand{\CvNsdNglobalLoobjZEuclKOnePos}{2}

\newcommand{\CvNsdNglobalLoobjZEuclKThree}{\ensuremath{-}0.004}
\newcommand{\CvNsdNglobalLoobjZEuclKThreePos}{3}

\newcommand{\CvNsdNglobalLatRawCorrKOne}{\ensuremath{-}0.009}
\newcommand{\CvNsdNglobalLatRawCorrKOnePos}{0}

\newcommand{\CvNsdNglobalLatRawCorrKThree}{\ensuremath{-}0.015}
\newcommand{\CvNsdNglobalLatRawCorrKThreePos}{0}

\newcommand{\CvNsdNglobalLatZCorrKOne}{\ensuremath{-}0.004}
\newcommand{\CvNsdNglobalLatZCorrKOnePos}{0}

\newcommand{\CvNsdNglobalLatZCorrKThree}{\ensuremath{-}0.007}
\newcommand{\CvNsdNglobalLatZCorrKThreePos}{0}

\newcommand{\CvNsdNglobalLatRawEuclKOne}{\ensuremath{-}0.006}
\newcommand{\CvNsdNglobalLatRawEuclKOnePos}{1}

\newcommand{\CvNsdNglobalLatRawEuclKThree}{\ensuremath{-}0.011}
\newcommand{\CvNsdNglobalLatRawEuclKThreePos}{2}

\newcommand{\CvNsdNglobalLatZEuclKOne}{\ensuremath{-}0.002}
\newcommand{\CvNsdNglobalLatZEuclKOnePos}{3}

\newcommand{\CvNsdNglobalLatZEuclKThree}{\ensuremath{-}0.004}
\newcommand{\CvNsdNglobalLatZEuclKThreePos}{3}

\newcommand{\CvNsdNsessionVoneRawCorrKOne}{\ensuremath{-}0.012}
\newcommand{\CvNsdNsessionVoneRawCorrKOnePos}{0}

\newcommand{\CvNsdNsessionVoneRawCorrKThree}{\ensuremath{-}0.017}
\newcommand{\CvNsdNsessionVoneRawCorrKThreePos}{0}

\newcommand{\CvNsdNsessionVoneZCorrKOne}{\ensuremath{-}0.016}
\newcommand{\CvNsdNsessionVoneZCorrKOnePos}{0}

\newcommand{\CvNsdNsessionVoneZCorrKThree}{\ensuremath{-}0.026}
\newcommand{\CvNsdNsessionVoneZCorrKThreePos}{0}

\newcommand{\CvNsdNsessionVoneRawEuclKOne}{\ensuremath{-}0.003}
\newcommand{\CvNsdNsessionVoneRawEuclKOnePos}{2}

\newcommand{\CvNsdNsessionVoneRawEuclKThree}{\ensuremath{-}0.008}
\newcommand{\CvNsdNsessionVoneRawEuclKThreePos}{2}

\newcommand{\CvNsdNsessionVoneZEuclKOne}{\ensuremath{-}0.003}
\newcommand{\CvNsdNsessionVoneZEuclKOnePos}{3}

\newcommand{\CvNsdNsessionVoneZEuclKThree}{\ensuremath{-}0.009}
\newcommand{\CvNsdNsessionVoneZEuclKThreePos}{2}

\newcommand{\CvNsdNsessionLoobjRawCorrKOne}{\ensuremath{-}0.008}
\newcommand{\CvNsdNsessionLoobjRawCorrKOnePos}{0}

\newcommand{\CvNsdNsessionLoobjRawCorrKThree}{\ensuremath{-}0.014}
\newcommand{\CvNsdNsessionLoobjRawCorrKThreePos}{0}

\newcommand{\CvNsdNsessionLoobjZCorrKOne}{\ensuremath{-}0.003}
\newcommand{\CvNsdNsessionLoobjZCorrKOnePos}{0}

\newcommand{\CvNsdNsessionLoobjZCorrKThree}{\ensuremath{-}0.006}
\newcommand{\CvNsdNsessionLoobjZCorrKThreePos}{0}

\newcommand{\CvNsdNsessionLoobjRawEuclKOne}{\ensuremath{-}0.005}
\newcommand{\CvNsdNsessionLoobjRawEuclKOnePos}{1}

\newcommand{\CvNsdNsessionLoobjRawEuclKThree}{\ensuremath{-}0.011}
\newcommand{\CvNsdNsessionLoobjRawEuclKThreePos}{1}

\newcommand{\CvNsdNsessionLoobjZEuclKOne}{\ensuremath{-}0.001}
\newcommand{\CvNsdNsessionLoobjZEuclKOnePos}{1}

\newcommand{\CvNsdNsessionLoobjZEuclKThree}{\ensuremath{-}0.002}
\newcommand{\CvNsdNsessionLoobjZEuclKThreePos}{3}

\newcommand{\CvNsdNsessionLatRawCorrKOne}{\ensuremath{-}0.010}
\newcommand{\CvNsdNsessionLatRawCorrKOnePos}{0}

\newcommand{\CvNsdNsessionLatRawCorrKThree}{\ensuremath{-}0.016}
\newcommand{\CvNsdNsessionLatRawCorrKThreePos}{0}

\newcommand{\CvNsdNsessionLatZCorrKOne}{\ensuremath{-}0.004}
\newcommand{\CvNsdNsessionLatZCorrKOnePos}{0}

\newcommand{\CvNsdNsessionLatZCorrKThree}{\ensuremath{-}0.007}
\newcommand{\CvNsdNsessionLatZCorrKThreePos}{0}

\newcommand{\CvNsdNsessionLatRawEuclKOne}{\ensuremath{-}0.006}
\newcommand{\CvNsdNsessionLatRawEuclKOnePos}{1}

\newcommand{\CvNsdNsessionLatRawEuclKThree}{\ensuremath{-}0.010}
\newcommand{\CvNsdNsessionLatRawEuclKThreePos}{3}

\newcommand{\CvNsdNsessionLatZEuclKOne}{\ensuremath{-}0.002}
\newcommand{\CvNsdNsessionLatZEuclKOnePos}{1}

\newcommand{\CvNsdNsessionLatZEuclKThree}{\ensuremath{-}0.001}
\newcommand{\CvNsdNsessionLatZEuclKThreePos}{3}

\newcommand{\MvNrawDistinct}{28}
\newcommand{\MvNsessionDistinct}{28}

\newcommand{\MvNrawAbove}{1}
\newcommand{\MvNsessionAbove}{0}

\newcommand{\MvNrawBelow}{10}
\newcommand{\MvNsessionBelow}{13}

\newcommand{\MvNrawIncl}{17}
\newcommand{\MvNsessionIncl}{15}
\newcommand{\DecVoneVZero}{+0.0304}
\newcommand{\DecVoneVOne}{+0.0050}
\newcommand{\DecVoneVTwo}{+0.0352}
\newcommand{\DecVoneVThree}{+0.0070}
\newcommand{\DecVoneVFour}{+0.0053}
\newcommand{\DecVoneVFive}{+0.0059}
\newcommand{\DecVoneVZeroSubOne}{+0.0330}
\newcommand{\DecVoneVOneSubOne}{+0.0044}
\newcommand{\DecVoneVTwoSubOne}{+0.0325}
\newcommand{\DecVoneVThreeSubOne}{+0.0037}
\newcommand{\DecVoneVZeroSubTwo}{+0.0509}
\newcommand{\DecVoneVOneSubTwo}{+0.0093}
\newcommand{\DecVoneVTwoSubTwo}{+0.0654}
\newcommand{\DecVoneVThreeSubTwo}{+0.0157}
\newcommand{\DecVoneVZeroSubThree}{+0.0074}
\newcommand{\DecVoneVOneSubThree}{+0.0013}
\newcommand{\DecVoneVTwoSubThree}{+0.0077}
\newcommand{\DecVoneVThreeSubThree}{+0.0017}
\newcommand{\DecLocVZero}{\ensuremath{-}0.0113}
\newcommand{\DecLocVOne}{\ensuremath{-}0.0044}
\newcommand{\DecLocVTwo}{\ensuremath{-}0.0130}
\newcommand{\DecLocVThree}{\ensuremath{-}0.0041}
\newcommand{\DecLocVFour}{\ensuremath{-}0.0042}
\newcommand{\DecLocVFive}{\ensuremath{-}0.0042}
\newcommand{\DecThreeVoneVZero}{+0.030}
\newcommand{\DecThreeVoneVOne}{+0.005}

\newcommand{\DecSdSubOne}{2.07}
\newcommand{\DecShareSubOne}{11}

\newcommand{\DecSdSubTwo}{3.94}
\newcommand{\DecShareSubTwo}{25}

\newcommand{\DecSdSubThree}{0.64}
\newcommand{\DecShareSubThree}{2}

\newcommand{\CenterSevenTwentyCell}{+0.0046}
\newcommand{\CenterSevenTwentyCellLo}{\ensuremath{-}0.0034}
\newcommand{\CenterSevenTwentyCellHi}{+0.0125}

\newcommand{\RunCentVoneRc}{+0.0070}
\newcommand{\RunCentLocRc}{\ensuremath{-}0.0041}
\newcommand{\RunCentVoneRcLc}{+0.0014}
\newcommand{\RunCentLocRcLc}{\ensuremath{-}0.0033}

\newcommand{\LumBoundRc}{0.63}
\newcommand{\PartFfVoneNone}{+0.0304}
\newcommand{\PartFfLocNone}{\ensuremath{-}0.0113}
\newcommand{\PartFfVoneLc}{+0.0163}
\newcommand{\PartFfLocLc}{\ensuremath{-}0.0126}
\newcommand{\PartFfVoneT}{+0.0311}
\newcommand{\PartFfLocT}{\ensuremath{-}0.0121}
\newcommand{\PartFfVoneLct}{+0.0171}
\newcommand{\PartFfLocLct}{\ensuremath{-}0.0130}
\newcommand{\PartFffVoneLp}{+0.0266}
\newcommand{\PartFffLocLp}{\ensuremath{-}0.0122}
\newcommand{\PartFffVoneLclp}{+0.0180}
\newcommand{\PartFffLocLclp}{\ensuremath{-}0.0127}
\newcommand{\PartFffVoneLclpt}{+0.0190}
\newcommand{\PartFffLocLclpt}{\ensuremath{-}0.0131}
\newcommand{\PartTestVoneNone}{\ensuremath{-}0.0002}
\newcommand{\PartTestLocNone}{\ensuremath{-}0.0004}
\newcommand{\PartTestVoneLc}{\ensuremath{-}0.0014}
\newcommand{\PartTestLocLc}{+0.0005}
\newcommand{\PartTestVoneLp}{\ensuremath{-}0.0009}
\newcommand{\PartTestLocLp}{\ensuremath{-}0.0006}
\newcommand{\PartTestVoneLclp}{\ensuremath{-}0.0003}
\newcommand{\PartTestLocLclp}{+0.0013}
\newcommand{\PartNsdVoneNone}{\ensuremath{-}0.0120}
\newcommand{\PartNsdLoobjNone}{\ensuremath{-}0.0081}
\newcommand{\PartNsdVoneLc}{\ensuremath{-}0.0155}
\newcommand{\PartNsdLoobjLc}{\ensuremath{-}0.0089}
\newcommand{\PartNsdVoneLp}{\ensuremath{-}0.0154}
\newcommand{\PartNsdLoobjLp}{\ensuremath{-}0.0087}
\newcommand{\PartNsdVoneLclp}{\ensuremath{-}0.0095}
\newcommand{\PartNsdLoobjLclp}{\ensuremath{-}0.0082}
\newcommand{\OrderRhoSubTwo}{+0.139}
\newcommand{\OrderRhoSubOne}{\ensuremath{-}0.007}
\newcommand{\OrderRhoSubThree}{+0.021}
\newcommand{\RuleWOne}{97}
\newcommand{\RuleWTwo}{+0.016}
\newcommand{\RuleWFive}{0.0007}

\newcommand{\LumRatioTsev}{1.04}
\newcommand{\LpRatioTsev}{0.87}
\newcommand{\LumRatioTest}{0.26}
\newcommand{\LpRatioTest}{0.36}
\newcommand{\LumRatioNsd}{0.07}
\newcommand{\LpRatioNsd}{0.17}
\newcommand{\RuleWSix}{+0.018}
\newcommand{\RuleWSixFraction}{41}
\newcommand{\RuleRcOne}{+0.007}
\newcommand{\RuleRcTwo}{0.63}
\newcommand{\RuleRcFour}{4}
\newcommand{\RuleCoOne}{\ensuremath{-6.7\times 10^{-6}}}
\newcommand{\RunLumPctSubOneTwo}{64.8}
\newcommand{\RunLumPctSubThree}{99.6}
\newcommand{\MechDeltaASubOne}{+0.046}
\newcommand{\MechPctSubOne}{62.8}
\newcommand{\MechRunIdxSubOne}{+0.06}
\newcommand{\MechRunPartSubOne}{+0.029}
\newcommand{\MechDeltaASubTwo}{\ensuremath{-}0.074}
\newcommand{\MechPctSubTwo}{34.6}
\newcommand{\MechRunIdxSubTwo}{+0.42}
\newcommand{\MechRunPartSubTwo}{+0.035}
\newcommand{\MechDeltaASubThree}{+0.114}
\newcommand{\MechPctSubThree}{84.0}
\newcommand{\MechRunIdxSubThree}{+0.71}
\newcommand{\MechRunPartSubThree}{+0.006}
\newcommand{\BandedBOneMinGFour}{\ensuremath{-}0.0072}
\newcommand{\BandedBOneMinGOne}{\ensuremath{-}0.0091}
\newcommand{\BandedBOneMinGEight}{\ensuremath{-}0.0071}
\newcommand{\BandedUBpGOne}{\ensuremath{-}0.0024}
\newcommand{\BandedURandGOne}{\ensuremath{-}0.0041}
\newcommand{\BandedUSkillBpGOne}{\ensuremath{-}0.0013}
\newcommand{\BandedUSkillRandGOne}{\ensuremath{-}0.0018}
\newcommand{\BandedBankGOne}{0.082}
\newcommand{\BandedNetBpGOne}{0.059}
\newcommand{\BandedNetRandGOne}{0.041}
\newcommand{\BandedUBpGFour}{+0.0017}
\newcommand{\BandedURandGFour}{\ensuremath{-}0.0023}
\newcommand{\BandedUSkillBpGFour}{\ensuremath{-}0.0009}
\newcommand{\BandedUSkillRandGFour}{\ensuremath{-}0.0012}
\newcommand{\BandedBankGFour}{0.106}
\newcommand{\BandedNetBpGFour}{0.086}
\newcommand{\BandedNetRandGFour}{0.040}
\newcommand{\BandedUBpGEight}{+0.0001}
\newcommand{\BandedURandGEight}{\ensuremath{-}0.0038}
\newcommand{\BandedUSkillBpGEight}{0.0000}
\newcommand{\BandedUSkillRandGEight}{\ensuremath{-}0.0009}
\newcommand{\BandedBankGEight}{0.109}
\newcommand{\BandedNetBpGEight}{0.083}
\newcommand{\BandedNetRandGEight}{0.040}
\newcommand{\CondUBp}{+0.0036}
\newcommand{\CondURand}{\ensuremath{-}0.0002}
\newcommand{\CondUSkillBp}{\ensuremath{-}0.0003}
\newcommand{\CondUSkillRand}{\ensuremath{-}0.0009}
\newcommand{\DAOneDev}{\ensuremath{3.8\times 10^{-9}}}
\newcommand{\DATwoMin}{\ensuremath{-}0.0050}
\newcommand{\NTwoSpearman}{\ensuremath{-}0.94}
\newcommand{\CvRsaVoneRawBOne}{\ensuremath{-}0.018}
\newcommand{\CvRsaVoneRawBOneLo}{\ensuremath{-}0.060}
\newcommand{\CvRsaVoneRawBOneHi}{+0.033}
\newcommand{\CvRsaVoneRawBTwo}{\ensuremath{-}0.022}

\newcommand{\CvRsaVoneRawBThree}{\ensuremath{-}0.023}

\newcommand{\CvRsaVoneZBOne}{\ensuremath{-}0.073}
\newcommand{\CvRsaVoneZBOneLo}{\ensuremath{-}0.121}
\newcommand{\CvRsaVoneZBOneHi}{\ensuremath{-}0.028}
\newcommand{\CvRsaVoneZBTwo}{\ensuremath{-}0.030}

\newcommand{\CvRsaVoneZBThree}{\ensuremath{-}0.037}
\newcommand{\NoiseVoneRawKOne}{0.000}
\newcommand{\NoiseVoneRawKOneFour}{\ensuremath{-}0.0002}
\newcommand{\NoiseVoneRawKOneLo}{\ensuremath{-}0.015}
\newcommand{\NoiseVoneRawKOneHi}{+0.014}
\newcommand{\NoiseVoneRawKTwo}{\ensuremath{-}0.003}

\newcommand{\NoiseVoneRawKThree}{\ensuremath{-}0.007}

\newcommand{\NoiseVoneRawKFour}{\ensuremath{-}0.006}

\newcommand{\NoiseVoneRawKSix}{\ensuremath{-}0.013}

\newcommand{\NoiseVoneRawKOneTwo}{\ensuremath{-}0.018}

\newcommand{\NoiseVoneRawKOneTwoLo}{\ensuremath{-}0.060}
\newcommand{\NoiseVoneRawKOneTwoHi}{+0.033}
\newcommand{\NoiseVoneZKOne}{\ensuremath{-}0.016}

\newcommand{\NoiseVoneZKOneLo}{\ensuremath{-}0.030}
\newcommand{\NoiseVoneZKOneHi}{\ensuremath{-}0.005}
\newcommand{\NoiseVoneZKTwo}{\ensuremath{-}0.025}

\newcommand{\NoiseVoneZKThree}{\ensuremath{-}0.034}

\newcommand{\NoiseVoneZKFour}{\ensuremath{-}0.040}

\newcommand{\NoiseVoneZKSix}{\ensuremath{-}0.054}

\newcommand{\NoiseVoneZKOneTwo}{\ensuremath{-}0.073}

\newcommand{\NoiseVoneZKOneTwoLo}{\ensuremath{-}0.121}
\newcommand{\NoiseVoneZKOneTwoHi}{\ensuremath{-}0.028}
\newcommand{\NoiseLocRawKOne}{0.000}

\newcommand{\NoiseLocRawKOneLo}{\ensuremath{-}0.010}
\newcommand{\NoiseLocRawKOneHi}{+0.008}
\newcommand{\NoiseLocRawKTwo}{+0.001}

\newcommand{\NoiseLocRawKThree}{\ensuremath{-}0.005}

\newcommand{\NoiseLocRawKFour}{\ensuremath{-}0.005}

\newcommand{\NoiseLocRawKSix}{\ensuremath{-}0.010}

\newcommand{\NoiseLocRawKOneTwo}{\ensuremath{-}0.016}

\newcommand{\NoiseLocRawKOneTwoLo}{\ensuremath{-}0.043}
\newcommand{\NoiseLocRawKOneTwoHi}{+0.014}
\newcommand{\NoiseLocZKOne}{\ensuremath{-}0.002}

\newcommand{\NoiseLocZKOneLo}{\ensuremath{-}0.009}
\newcommand{\NoiseLocZKOneHi}{+0.005}
\newcommand{\NoiseLocZKTwo}{\ensuremath{-}0.001}

\newcommand{\NoiseLocZKThree}{\ensuremath{-}0.002}

\newcommand{\NoiseLocZKFour}{\ensuremath{-}0.003}

\newcommand{\NoiseLocZKSix}{\ensuremath{-}0.004}

\newcommand{\NoiseLocZKOneTwo}{\ensuremath{-}0.007}

\newcommand{\NoiseLocZKOneTwoLo}{\ensuremath{-}0.028}
\newcommand{\NoiseLocZKOneTwoHi}{+0.014}
\newcommand{\NsdAllVoneKOneRawCorrelation}{\ensuremath{-}0.012}

\newcommand{\NsdAllVoneKOneRawCorrelationNeg}{8}

\newcommand{\NsdAllVoneKOneZCorrelation}{\ensuremath{-}0.016}

\newcommand{\NsdAllVoneKOneZCorrelationNeg}{8}

\newcommand{\NsdAllVoneKOneRawEuclidean}{\ensuremath{-}0.003}

\newcommand{\NsdAllVoneKOneRawEuclideanNeg}{6}

\newcommand{\NsdAllVoneKOneZEuclidean}{\ensuremath{-}0.003}

\newcommand{\NsdAllVoneKOneZEuclideanNeg}{5}

\newcommand{\NsdAllVoneKThreeRawCorrelation}{\ensuremath{-}0.017}
\newcommand{\NsdAllVoneKThreeRawCorrelationLo}{\ensuremath{-}0.030}
\newcommand{\NsdAllVoneKThreeRawCorrelationHi}{\ensuremath{-}0.002}
\newcommand{\NsdAllVoneKThreeRawCorrelationNeg}{8}

\newcommand{\NsdAllVoneKThreeZCorrelation}{\ensuremath{-}0.026}
\newcommand{\NsdAllVoneKThreeZCorrelationLo}{\ensuremath{-}0.036}
\newcommand{\NsdAllVoneKThreeZCorrelationHi}{\ensuremath{-}0.018}
\newcommand{\NsdAllVoneKThreeZCorrelationNeg}{8}

\newcommand{\NsdAllVoneKThreeZCorrelationPn}{0.0039}

\newcommand{\NsdAllVoneKThreeRawEuclidean}{\ensuremath{-}0.008}

\newcommand{\NsdAllVoneKThreeRawEuclideanNeg}{6}

\newcommand{\NsdAllVoneKThreeZEuclidean}{\ensuremath{-}0.009}

\newcommand{\NsdAllVoneKThreeZEuclideanNeg}{6}

\newcommand{\NsdAllVoneKThreeCenterN}{\ensuremath{-}0.004}

\newcommand{\NsdAllVoneKThreeCenterNNeg}{5}

\newcommand{\NsdAllVoneKThreeCenterA}{\ensuremath{-}0.017}

\newcommand{\NsdAllVoneKThreeCenterANeg}{8}

\newcommand{\NsdAllVoneKThreeCenterW}{\ensuremath{-}0.010}

\newcommand{\NsdAllVoneKThreeCenterWNeg}{7}

\newcommand{\NsdAllLoobjKOneRawCorrelation}{\ensuremath{-}0.008}

\newcommand{\NsdAllLoobjKOneRawCorrelationNeg}{8}

\newcommand{\NsdAllLoobjKOneZCorrelation}{\ensuremath{-}0.003}

\newcommand{\NsdAllLoobjKOneZCorrelationNeg}{8}

\newcommand{\NsdAllLoobjKOneRawEuclidean}{\ensuremath{-}0.005}

\newcommand{\NsdAllLoobjKOneRawEuclideanNeg}{7}

\newcommand{\NsdAllLoobjKOneZEuclidean}{\ensuremath{-}0.001}

\newcommand{\NsdAllLoobjKOneZEuclideanNeg}{7}

\newcommand{\NsdAllLoobjKThreeRawCorrelation}{\ensuremath{-}0.014}

\newcommand{\NsdAllLoobjKThreeRawCorrelationNeg}{8}

\newcommand{\NsdAllLoobjKThreeZCorrelation}{\ensuremath{-}0.006}

\newcommand{\NsdAllLoobjKThreeZCorrelationNeg}{8}

\newcommand{\NsdAllLoobjKThreeRawEuclidean}{\ensuremath{-}0.011}

\newcommand{\NsdAllLoobjKThreeRawEuclideanNeg}{7}

\newcommand{\NsdAllLoobjKThreeZEuclidean}{\ensuremath{-}0.002}

\newcommand{\NsdAllLoobjKThreeZEuclideanNeg}{5}

\newcommand{\NsdAllLoobjKThreeCenterN}{\ensuremath{-}0.012}

\newcommand{\NsdAllLoobjKThreeCenterNNeg}{8}

\newcommand{\NsdAllLoobjKThreeCenterA}{\ensuremath{-}0.009}

\newcommand{\NsdAllLoobjKThreeCenterANeg}{8}

\newcommand{\NsdAllLoobjKThreeCenterW}{\ensuremath{-}0.008}

\newcommand{\NsdAllLoobjKThreeCenterWNeg}{6}

\newcommand{\NsdAllLatKOneRawCorrelation}{\ensuremath{-}0.010}

\newcommand{\NsdAllLatKOneRawCorrelationNeg}{8}

\newcommand{\NsdAllLatKOneZCorrelation}{\ensuremath{-}0.004}

\newcommand{\NsdAllLatKOneZCorrelationNeg}{8}

\newcommand{\NsdAllLatKOneRawEuclidean}{\ensuremath{-}0.006}

\newcommand{\NsdAllLatKOneRawEuclideanNeg}{7}

\newcommand{\NsdAllLatKOneZEuclidean}{\ensuremath{-}0.002}

\newcommand{\NsdAllLatKOneZEuclideanNeg}{7}

\newcommand{\NsdAllLatKThreeRawCorrelation}{\ensuremath{-}0.016}

\newcommand{\NsdAllLatKThreeRawCorrelationNeg}{8}

\newcommand{\NsdAllLatKThreeZCorrelation}{\ensuremath{-}0.007}

\newcommand{\NsdAllLatKThreeZCorrelationNeg}{8}

\newcommand{\NsdAllLatKThreeRawEuclidean}{\ensuremath{-}0.010}

\newcommand{\NsdAllLatKThreeRawEuclideanNeg}{5}

\newcommand{\NsdAllLatKThreeZEuclidean}{\ensuremath{-}0.001}

\newcommand{\NsdAllLatKThreeZEuclideanNeg}{5}

\newcommand{\NsdAllLatKThreeCenterN}{\ensuremath{-}0.013}

\newcommand{\NsdAllLatKThreeCenterNNeg}{8}

\newcommand{\NsdAllLatKThreeCenterA}{\ensuremath{-}0.011}

\newcommand{\NsdAllLatKThreeCenterANeg}{8}

\newcommand{\NsdAllLatKThreeCenterW}{\ensuremath{-}0.005}

\newcommand{\NsdAllLatKThreeCenterWNeg}{6}
\newcommand{\NsdHeldVoneKThreeRawEuclidean}{\ensuremath{-}0.011}

\newcommand{\NsdHeldVoneKThreeRawEuclideanPos}{1}
\newcommand{\NsdHeldVoneKThreeCenterN}{\ensuremath{-}0.008}

\newcommand{\NsdHeldVoneKThreeCenterNPos}{2}

\newcommand{\NsdHeldVoneKThreeCenterA}{\ensuremath{-}0.018}

\newcommand{\NsdHeldVoneKThreeCenterANeg}{7}

\newcommand{\NsdHeldVoneKThreeCenterAPn}{0.0078}
\newcommand{\NsdHeldLoobjKThreeCenterA}{\ensuremath{-}0.009}

\newcommand{\NsdHeldLoobjKThreeCenterANeg}{7}
\newcommand{\PilotVoneRawCorrelation}{\ensuremath{-}0.017}
\newcommand{\PilotVoneZCorrelation}{\ensuremath{-}0.025}
\newcommand{\PilotVoneRawEuclidean}{+0.013}
\newcommand{\PilotVoneZEuclidean}{+0.023}
\newcommand{\PilotLoobjRawCorrelation}{\ensuremath{-}0.027}
\newcommand{\PilotLoobjZCorrelation}{\ensuremath{-}0.010}
\newcommand{\PilotLoobjRawEuclidean}{\ensuremath{-}0.006}
\newcommand{\PilotLoobjZEuclidean}{0.000}
\newcommand{\NsdRidgeVoneGOne}{\ensuremath{-}0.017}
\newcommand{\NsdRidgeVoneGOneNeg}{8}

\newcommand{\NsdRidgeVoneGOneSubjOne}{\ensuremath{-}0.024}

\newcommand{\NsdRidgeVoneGOneSubjTwo}{\ensuremath{-}0.016}

\newcommand{\NsdRidgeVoneGOneSubjThree}{\ensuremath{-}0.015}

\newcommand{\NsdRidgeVoneGOneSubjFour}{\ensuremath{-}0.012}

\newcommand{\NsdRidgeVoneGOneSubjFive}{\ensuremath{-}0.013}

\newcommand{\NsdRidgeVoneGOneSubjSix}{\ensuremath{-}0.019}

\newcommand{\NsdRidgeVoneGOneSubjSeven}{\ensuremath{-}0.022}

\newcommand{\NsdRidgeVoneGOneSubjEight}{\ensuremath{-}0.012}

\newcommand{\NsdRidgeVoneGFour}{\ensuremath{-}0.054}
\newcommand{\NsdRidgeVoneGFourNeg}{8}
\newcommand{\NsdRidgeVoneGFourP}{0.0039}
\newcommand{\NsdRidgeVoneGFourBp}{0.395}
\newcommand{\NsdRidgeVoneGFourRand}{0.342}
\newcommand{\NsdRidgeVoneGFourBank}{0.417}
\newcommand{\NsdRidgeVoneGFourSubjOne}{\ensuremath{-}0.048}
\newcommand{\NsdRidgeVoneGFourSubjOneBank}{0.406}
\newcommand{\NsdRidgeVoneGFourSubjOneBp}{0.384}
\newcommand{\NsdRidgeVoneGFourSubjOneRand}{0.336}
\newcommand{\NsdReliableSubjOne}{1181}

\newcommand{\NsdRidgeVoneGFourSubjTwo}{\ensuremath{-}0.055}
\newcommand{\NsdRidgeVoneGFourSubjTwoBank}{0.436}
\newcommand{\NsdRidgeVoneGFourSubjTwoBp}{0.408}
\newcommand{\NsdRidgeVoneGFourSubjTwoRand}{0.353}
\newcommand{\NsdReliableSubjTwo}{815}

\newcommand{\NsdRidgeVoneGFourSubjThree}{\ensuremath{-}0.057}
\newcommand{\NsdRidgeVoneGFourSubjThreeBank}{0.444}
\newcommand{\NsdRidgeVoneGFourSubjThreeBp}{0.432}
\newcommand{\NsdRidgeVoneGFourSubjThreeRand}{0.375}
\newcommand{\NsdReliableSubjThree}{912}

\newcommand{\NsdRidgeVoneGFourSubjFour}{\ensuremath{-}0.053}
\newcommand{\NsdRidgeVoneGFourSubjFourBank}{0.439}
\newcommand{\NsdRidgeVoneGFourSubjFourBp}{0.408}
\newcommand{\NsdRidgeVoneGFourSubjFourRand}{0.355}
\newcommand{\NsdReliableSubjFour}{600}

\newcommand{\NsdRidgeVoneGFourSubjFive}{\ensuremath{-}0.048}
\newcommand{\NsdRidgeVoneGFourSubjFiveBank}{0.466}
\newcommand{\NsdRidgeVoneGFourSubjFiveBp}{0.449}
\newcommand{\NsdRidgeVoneGFourSubjFiveRand}{0.401}
\newcommand{\NsdReliableSubjFive}{798}

\newcommand{\NsdRidgeVoneGFourSubjSix}{\ensuremath{-}0.057}
\newcommand{\NsdRidgeVoneGFourSubjSixBank}{0.353}
\newcommand{\NsdRidgeVoneGFourSubjSixBp}{0.314}
\newcommand{\NsdRidgeVoneGFourSubjSixRand}{0.257}
\newcommand{\NsdReliableSubjSix}{920}

\newcommand{\NsdRidgeVoneGFourSubjSeven}{\ensuremath{-}0.056}
\newcommand{\NsdRidgeVoneGFourSubjSevenBank}{0.368}
\newcommand{\NsdRidgeVoneGFourSubjSevenBp}{0.356}
\newcommand{\NsdRidgeVoneGFourSubjSevenRand}{0.301}
\newcommand{\NsdReliableSubjSeven}{793}

\newcommand{\NsdRidgeVoneGFourSubjEight}{\ensuremath{-}0.056}
\newcommand{\NsdRidgeVoneGFourSubjEightBank}{0.426}
\newcommand{\NsdRidgeVoneGFourSubjEightBp}{0.413}
\newcommand{\NsdRidgeVoneGFourSubjEightRand}{0.357}
\newcommand{\NsdReliableSubjEight}{550}

\newcommand{\NsdRidgeVoneGEight}{\ensuremath{-}0.070}
\newcommand{\NsdRidgeVoneGEightNeg}{8}

\newcommand{\NsdRidgeVoneGEightSubjOne}{\ensuremath{-}0.070}

\newcommand{\NsdRidgeVoneGEightSubjTwo}{\ensuremath{-}0.080}

\newcommand{\NsdRidgeVoneGEightSubjThree}{\ensuremath{-}0.077}

\newcommand{\NsdRidgeVoneGEightSubjFour}{\ensuremath{-}0.074}

\newcommand{\NsdRidgeVoneGEightSubjFive}{\ensuremath{-}0.058}

\newcommand{\NsdRidgeVoneGEightSubjSix}{\ensuremath{-}0.067}

\newcommand{\NsdRidgeVoneGEightSubjSeven}{\ensuremath{-}0.068}

\newcommand{\NsdRidgeVoneGEightSubjEight}{\ensuremath{-}0.069}
\newcommand{\NsdRidgeLoobjGFour}{\ensuremath{-}0.022}
\newcommand{\NsdRidgeLoobjGFourNeg}{8}

\newcommand{\NsdRidgeLoobjGFourSubjOne}{\ensuremath{-}0.020}

\newcommand{\NsdRidgeLoobjGFourSubjTwo}{\ensuremath{-}0.018}

\newcommand{\NsdRidgeLoobjGFourSubjThree}{\ensuremath{-}0.018}

\newcommand{\NsdRidgeLoobjGFourSubjFour}{\ensuremath{-}0.009}

\newcommand{\NsdRidgeLoobjGFourSubjFive}{\ensuremath{-}0.020}

\newcommand{\NsdRidgeLoobjGFourSubjSix}{\ensuremath{-}0.028}

\newcommand{\NsdRidgeLoobjGFourSubjSeven}{\ensuremath{-}0.031}

\newcommand{\NsdRidgeLoobjGFourSubjEight}{\ensuremath{-}0.031}
\newcommand{\NsdVoxVoneSubjOne}{1350}
\newcommand{\NsdVoxVoneSubjTwo}{1102}
\newcommand{\NsdVoxVoneSubjThree}{1254}
\newcommand{\NsdVoxVoneSubjFour}{877}
\newcommand{\NsdVoxVoneSubjFive}{1113}
\newcommand{\NsdVoxVoneSubjSix}{1121}
\newcommand{\NsdVoxVoneSubjSeven}{1142}
\newcommand{\NsdVoxVoneSubjEight}{1074}
\newcommand{\NsdVoxLoobjSubjOne}{2921}
\newcommand{\NsdVoxLoobjSubjTwo}{2700}
\newcommand{\NsdVoxLoobjSubjThree}{3402}
\newcommand{\NsdVoxLoobjSubjFour}{3078}
\newcommand{\NsdVoxLoobjSubjFive}{2330}
\newcommand{\NsdVoxLoobjSubjSix}{3877}
\newcommand{\NsdVoxLoobjSubjSeven}{3160}
\newcommand{\NsdVoxLoobjSubjEight}{3858}
\newcommand{\NsdVoxLatSubjOne}{7799}
\newcommand{\NsdVoxLatSubjTwo}{6752}
\newcommand{\NsdVoxLatSubjThree}{8572}
\newcommand{\NsdVoxLatSubjFour}{7131}
\newcommand{\NsdVoxLatSubjFive}{7037}
\newcommand{\NsdVoxLatSubjSix}{9030}
\newcommand{\NsdVoxLatSubjSeven}{6118}
\newcommand{\NsdVoxLatSubjEight}{8674}
\newcommand{\YOneThingsRc}{\ensuremath{-}0.040}
\newcommand{\YOneThingsRaw}{\ensuremath{-}0.043}
\newcommand{\YThreeNsdRc}{\ensuremath{-}0.055}
\newcommand{\YFourNsdAll}{\ensuremath{-}0.054}

\newcommand{\ImnAlexnetVoneRawCorrelation}{\ensuremath{-}0.018}
\newcommand{\ImnAlexnetVoneRawCorrelationNeg}{8}
\newcommand{\ImnAlexnetVoneZCorrelation}{\ensuremath{-}0.067}
\newcommand{\ImnAlexnetVoneZCorrelationNeg}{8}
\newcommand{\ImnAlexnetVoneCenterN}{\ensuremath{-}0.010}
\newcommand{\ImnAlexnetVoneCenterNNeg}{6}
\newcommand{\ImnAlexnetVoneCenterA}{\ensuremath{-}0.085}
\newcommand{\ImnAlexnetVoneCenterANeg}{8}
\newcommand{\ImnAlexnetVoneCenterW}{\ensuremath{-}0.039}
\newcommand{\ImnAlexnetVoneCenterWNeg}{7}
\newcommand{\ImnAlexnetLoobjRawCorrelation}{\ensuremath{-}0.092}
\newcommand{\ImnAlexnetLoobjRawCorrelationNeg}{8}
\newcommand{\ImnAlexnetLoobjZCorrelation}{\ensuremath{-}0.096}
\newcommand{\ImnAlexnetLoobjZCorrelationNeg}{8}
\newcommand{\ImnAlexnetLoobjCenterN}{+0.007}
\newcommand{\ImnAlexnetLoobjCenterNNeg}{3}
\newcommand{\ImnAlexnetLoobjCenterA}{\ensuremath{-}0.095}
\newcommand{\ImnAlexnetLoobjCenterANeg}{8}
\newcommand{\ImnAlexnetLoobjCenterW}{\ensuremath{-}0.015}
\newcommand{\ImnAlexnetLoobjCenterWNeg}{7}
\newcommand{\ImnVggOneOneBnVoneRawCorrelation}{\ensuremath{-}0.005}
\newcommand{\ImnVggOneOneBnVoneRawCorrelationNeg}{6}
\newcommand{\ImnVggOneOneBnVoneZCorrelation}{\ensuremath{-}0.057}
\newcommand{\ImnVggOneOneBnVoneZCorrelationNeg}{8}
\newcommand{\ImnVggOneOneBnVoneCenterN}{+0.019}
\newcommand{\ImnVggOneOneBnVoneCenterNNeg}{2}
\newcommand{\ImnVggOneOneBnVoneCenterA}{\ensuremath{-}0.070}
\newcommand{\ImnVggOneOneBnVoneCenterANeg}{8}
\newcommand{\ImnVggOneOneBnVoneCenterW}{\ensuremath{-}0.009}
\newcommand{\ImnVggOneOneBnVoneCenterWNeg}{6}
\newcommand{\ImnVggOneOneBnLoobjRawCorrelation}{\ensuremath{-}0.093}
\newcommand{\ImnVggOneOneBnLoobjRawCorrelationNeg}{8}
\newcommand{\ImnVggOneOneBnLoobjZCorrelation}{\ensuremath{-}0.104}
\newcommand{\ImnVggOneOneBnLoobjZCorrelationNeg}{8}
\newcommand{\ImnVggOneOneBnLoobjCenterN}{+0.008}
\newcommand{\ImnVggOneOneBnLoobjCenterNNeg}{2}
\newcommand{\ImnVggOneOneBnLoobjCenterA}{\ensuremath{-}0.101}
\newcommand{\ImnVggOneOneBnLoobjCenterANeg}{8}
\newcommand{\ImnVggOneOneBnLoobjCenterW}{+0.014}
\newcommand{\ImnVggOneOneBnLoobjCenterWNeg}{3}
\newcommand{\ImnResnetOneEightVoneRawCorrelation}{\ensuremath{-}0.012}
\newcommand{\ImnResnetOneEightVoneRawCorrelationNeg}{8}
\newcommand{\ImnResnetOneEightVoneZCorrelation}{\ensuremath{-}0.043}
\newcommand{\ImnResnetOneEightVoneZCorrelationNeg}{8}
\newcommand{\ImnResnetOneEightVoneCenterN}{\ensuremath{-}0.009}
\newcommand{\ImnResnetOneEightVoneCenterNNeg}{5}
\newcommand{\ImnResnetOneEightVoneCenterA}{\ensuremath{-}0.075}
\newcommand{\ImnResnetOneEightVoneCenterANeg}{8}
\newcommand{\ImnResnetOneEightVoneCenterW}{\ensuremath{-}0.024}
\newcommand{\ImnResnetOneEightVoneCenterWNeg}{7}
\newcommand{\ImnResnetOneEightLoobjRawCorrelation}{\ensuremath{-}0.085}
\newcommand{\ImnResnetOneEightLoobjRawCorrelationNeg}{8}
\newcommand{\ImnResnetOneEightLoobjZCorrelation}{\ensuremath{-}0.087}
\newcommand{\ImnResnetOneEightLoobjZCorrelationNeg}{8}
\newcommand{\ImnResnetOneEightLoobjCenterN}{+0.020}
\newcommand{\ImnResnetOneEightLoobjCenterNNeg}{0}
\newcommand{\ImnResnetOneEightLoobjCenterA}{\ensuremath{-}0.091}
\newcommand{\ImnResnetOneEightLoobjCenterANeg}{8}
\newcommand{\ImnResnetOneEightLoobjCenterW}{\ensuremath{-}0.054}
\newcommand{\ImnResnetOneEightLoobjCenterWNeg}{7}
\newcommand{\ImnResnetFiveZeroVoneRawCorrelation}{\ensuremath{-}0.013}
\newcommand{\ImnResnetFiveZeroVoneRawCorrelationNeg}{8}
\newcommand{\ImnResnetFiveZeroVoneZCorrelation}{\ensuremath{-}0.042}
\newcommand{\ImnResnetFiveZeroVoneZCorrelationNeg}{8}
\newcommand{\ImnResnetFiveZeroVoneCenterN}{\ensuremath{-}0.007}
\newcommand{\ImnResnetFiveZeroVoneCenterNNeg}{5}
\newcommand{\ImnResnetFiveZeroVoneCenterA}{\ensuremath{-}0.066}
\newcommand{\ImnResnetFiveZeroVoneCenterANeg}{8}
\newcommand{\ImnResnetFiveZeroVoneCenterW}{\ensuremath{-}0.023}
\newcommand{\ImnResnetFiveZeroVoneCenterWNeg}{7}
\newcommand{\ImnResnetFiveZeroLoobjRawCorrelation}{\ensuremath{-}0.096}
\newcommand{\ImnResnetFiveZeroLoobjRawCorrelationNeg}{8}
\newcommand{\ImnResnetFiveZeroLoobjZCorrelation}{\ensuremath{-}0.092}
\newcommand{\ImnResnetFiveZeroLoobjZCorrelationNeg}{8}
\newcommand{\ImnResnetFiveZeroLoobjCenterN}{\ensuremath{-}0.008}
\newcommand{\ImnResnetFiveZeroLoobjCenterNNeg}{6}
\newcommand{\ImnResnetFiveZeroLoobjCenterA}{\ensuremath{-}0.092}
\newcommand{\ImnResnetFiveZeroLoobjCenterANeg}{8}
\newcommand{\ImnResnetFiveZeroLoobjCenterW}{\ensuremath{-}0.052}
\newcommand{\ImnResnetFiveZeroLoobjCenterWNeg}{7}
\newcommand{\ImnConvnextTinyVoneRawCorrelation}{\ensuremath{-}0.005}
\newcommand{\ImnConvnextTinyVoneRawCorrelationNeg}{7}
\newcommand{\ImnConvnextTinyVoneZCorrelation}{\ensuremath{-}0.018}
\newcommand{\ImnConvnextTinyVoneZCorrelationNeg}{8}
\newcommand{\ImnConvnextTinyVoneCenterN}{+0.007}
\newcommand{\ImnConvnextTinyVoneCenterNNeg}{2}
\newcommand{\ImnConvnextTinyVoneCenterA}{\ensuremath{-}0.016}
\newcommand{\ImnConvnextTinyVoneCenterANeg}{8}
\newcommand{\ImnConvnextTinyVoneCenterW}{\ensuremath{-}0.028}
\newcommand{\ImnConvnextTinyVoneCenterWNeg}{8}
\newcommand{\ImnConvnextTinyLoobjRawCorrelation}{\ensuremath{-}0.024}
\newcommand{\ImnConvnextTinyLoobjRawCorrelationNeg}{8}
\newcommand{\ImnConvnextTinyLoobjZCorrelation}{\ensuremath{-}0.032}
\newcommand{\ImnConvnextTinyLoobjZCorrelationNeg}{8}
\newcommand{\ImnConvnextTinyLoobjCenterN}{+0.007}
\newcommand{\ImnConvnextTinyLoobjCenterNNeg}{5}
\newcommand{\ImnConvnextTinyLoobjCenterA}{\ensuremath{-}0.016}
\newcommand{\ImnConvnextTinyLoobjCenterANeg}{8}
\newcommand{\ImnConvnextTinyLoobjCenterW}{\ensuremath{-}0.021}
\newcommand{\ImnConvnextTinyLoobjCenterWNeg}{6}
\newcommand{\ImnVitBOneSixVoneRawCorrelation}{\ensuremath{-}0.057}
\newcommand{\ImnVitBOneSixVoneRawCorrelationNeg}{8}
\newcommand{\ImnVitBOneSixVoneZCorrelation}{\ensuremath{-}0.083}
\newcommand{\ImnVitBOneSixVoneZCorrelationNeg}{8}
\newcommand{\ImnVitBOneSixVoneCenterN}{+0.016}
\newcommand{\ImnVitBOneSixVoneCenterNNeg}{4}
\newcommand{\ImnVitBOneSixVoneCenterA}{\ensuremath{-}0.100}
\newcommand{\ImnVitBOneSixVoneCenterANeg}{8}
\newcommand{\ImnVitBOneSixVoneCenterW}{\ensuremath{-}0.021}
\newcommand{\ImnVitBOneSixVoneCenterWNeg}{6}
\newcommand{\ImnVitBOneSixLoobjRawCorrelation}{\ensuremath{-}0.045}
\newcommand{\ImnVitBOneSixLoobjRawCorrelationNeg}{8}
\newcommand{\ImnVitBOneSixLoobjZCorrelation}{\ensuremath{-}0.060}
\newcommand{\ImnVitBOneSixLoobjZCorrelationNeg}{8}
\newcommand{\ImnVitBOneSixLoobjCenterN}{+0.020}
\newcommand{\ImnVitBOneSixLoobjCenterNNeg}{1}
\newcommand{\ImnVitBOneSixLoobjCenterA}{\ensuremath{-}0.052}
\newcommand{\ImnVitBOneSixLoobjCenterANeg}{8}
\newcommand{\ImnVitBOneSixLoobjCenterW}{+0.007}
\newcommand{\ImnVitBOneSixLoobjCenterWNeg}{4}
\newcommand{\ImnSzeroAlexnetVoneRawCorrelation}{\ensuremath{-}0.016}
\newcommand{\ImnSzeroAlexnetVoneRawCorrelationNeg}{8}
\newcommand{\ImnSzeroAlexnetVoneZCorrelation}{\ensuremath{-}0.064}
\newcommand{\ImnSzeroAlexnetVoneZCorrelationNeg}{8}
\newcommand{\ImnSzeroAlexnetVoneCenterN}{\ensuremath{-}0.004}
\newcommand{\ImnSzeroAlexnetVoneCenterNNeg}{5}
\newcommand{\ImnSzeroAlexnetVoneCenterA}{\ensuremath{-}0.081}
\newcommand{\ImnSzeroAlexnetVoneCenterANeg}{8}
\newcommand{\ImnSzeroAlexnetVoneCenterW}{\ensuremath{-}0.029}
\newcommand{\ImnSzeroAlexnetVoneCenterWNeg}{7}
\newcommand{\ImnSzeroAlexnetLoobjRawCorrelation}{\ensuremath{-}0.092}
\newcommand{\ImnSzeroAlexnetLoobjRawCorrelationNeg}{8}
\newcommand{\ImnSzeroAlexnetLoobjZCorrelation}{\ensuremath{-}0.096}
\newcommand{\ImnSzeroAlexnetLoobjZCorrelationNeg}{8}
\newcommand{\ImnSzeroAlexnetLoobjCenterN}{+0.014}
\newcommand{\ImnSzeroAlexnetLoobjCenterNNeg}{2}
\newcommand{\ImnSzeroAlexnetLoobjCenterA}{\ensuremath{-}0.098}
\newcommand{\ImnSzeroAlexnetLoobjCenterANeg}{8}
\newcommand{\ImnSzeroAlexnetLoobjCenterW}{\ensuremath{-}0.015}
\newcommand{\ImnSzeroAlexnetLoobjCenterWNeg}{7}
\newcommand{\ImnSzeroVggOneOneBnVoneRawCorrelation}{\ensuremath{-}0.004}
\newcommand{\ImnSzeroVggOneOneBnVoneRawCorrelationNeg}{5}
\newcommand{\ImnSzeroVggOneOneBnVoneZCorrelation}{\ensuremath{-}0.057}
\newcommand{\ImnSzeroVggOneOneBnVoneZCorrelationNeg}{8}
\newcommand{\ImnSzeroVggOneOneBnVoneCenterN}{+0.021}
\newcommand{\ImnSzeroVggOneOneBnVoneCenterNNeg}{2}
\newcommand{\ImnSzeroVggOneOneBnVoneCenterA}{\ensuremath{-}0.068}
\newcommand{\ImnSzeroVggOneOneBnVoneCenterANeg}{8}
\newcommand{\ImnSzeroVggOneOneBnVoneCenterW}{\ensuremath{-}0.007}
\newcommand{\ImnSzeroVggOneOneBnVoneCenterWNeg}{6}
\newcommand{\ImnSzeroVggOneOneBnLoobjRawCorrelation}{\ensuremath{-}0.093}
\newcommand{\ImnSzeroVggOneOneBnLoobjRawCorrelationNeg}{8}
\newcommand{\ImnSzeroVggOneOneBnLoobjZCorrelation}{\ensuremath{-}0.105}
\newcommand{\ImnSzeroVggOneOneBnLoobjZCorrelationNeg}{8}
\newcommand{\ImnSzeroVggOneOneBnLoobjCenterN}{+0.005}
\newcommand{\ImnSzeroVggOneOneBnLoobjCenterNNeg}{3}
\newcommand{\ImnSzeroVggOneOneBnLoobjCenterA}{\ensuremath{-}0.101}
\newcommand{\ImnSzeroVggOneOneBnLoobjCenterANeg}{8}
\newcommand{\ImnSzeroVggOneOneBnLoobjCenterW}{+0.014}
\newcommand{\ImnSzeroVggOneOneBnLoobjCenterWNeg}{3}
\newcommand{\ImnSzeroResnetOneEightVoneRawCorrelation}{\ensuremath{-}0.011}
\newcommand{\ImnSzeroResnetOneEightVoneRawCorrelationNeg}{8}
\newcommand{\ImnSzeroResnetOneEightVoneZCorrelation}{\ensuremath{-}0.046}
\newcommand{\ImnSzeroResnetOneEightVoneZCorrelationNeg}{8}
\newcommand{\ImnSzeroResnetOneEightVoneCenterN}{\ensuremath{-}0.010}
\newcommand{\ImnSzeroResnetOneEightVoneCenterNNeg}{6}
\newcommand{\ImnSzeroResnetOneEightVoneCenterA}{\ensuremath{-}0.064}
\newcommand{\ImnSzeroResnetOneEightVoneCenterANeg}{8}
\newcommand{\ImnSzeroResnetOneEightVoneCenterW}{\ensuremath{-}0.023}
\newcommand{\ImnSzeroResnetOneEightVoneCenterWNeg}{7}
\newcommand{\ImnSzeroResnetOneEightLoobjRawCorrelation}{\ensuremath{-}0.085}
\newcommand{\ImnSzeroResnetOneEightLoobjRawCorrelationNeg}{8}
\newcommand{\ImnSzeroResnetOneEightLoobjZCorrelation}{\ensuremath{-}0.087}
\newcommand{\ImnSzeroResnetOneEightLoobjZCorrelationNeg}{8}
\newcommand{\ImnSzeroResnetOneEightLoobjCenterN}{+0.012}
\newcommand{\ImnSzeroResnetOneEightLoobjCenterNNeg}{2}
\newcommand{\ImnSzeroResnetOneEightLoobjCenterA}{\ensuremath{-}0.090}
\newcommand{\ImnSzeroResnetOneEightLoobjCenterANeg}{8}
\newcommand{\ImnSzeroResnetOneEightLoobjCenterW}{\ensuremath{-}0.051}
\newcommand{\ImnSzeroResnetOneEightLoobjCenterWNeg}{7}
\newcommand{\ImnSzeroResnetFiveZeroVoneRawCorrelation}{\ensuremath{-}0.011}
\newcommand{\ImnSzeroResnetFiveZeroVoneRawCorrelationNeg}{8}
\newcommand{\ImnSzeroResnetFiveZeroVoneZCorrelation}{\ensuremath{-}0.045}
\newcommand{\ImnSzeroResnetFiveZeroVoneZCorrelationNeg}{8}
\newcommand{\ImnSzeroResnetFiveZeroVoneCenterN}{\ensuremath{-}0.006}
\newcommand{\ImnSzeroResnetFiveZeroVoneCenterNNeg}{4}
\newcommand{\ImnSzeroResnetFiveZeroVoneCenterA}{\ensuremath{-}0.072}
\newcommand{\ImnSzeroResnetFiveZeroVoneCenterANeg}{8}
\newcommand{\ImnSzeroResnetFiveZeroVoneCenterW}{\ensuremath{-}0.024}
\newcommand{\ImnSzeroResnetFiveZeroVoneCenterWNeg}{7}
\newcommand{\ImnSzeroResnetFiveZeroLoobjRawCorrelation}{\ensuremath{-}0.094}
\newcommand{\ImnSzeroResnetFiveZeroLoobjRawCorrelationNeg}{8}
\newcommand{\ImnSzeroResnetFiveZeroLoobjZCorrelation}{\ensuremath{-}0.090}
\newcommand{\ImnSzeroResnetFiveZeroLoobjZCorrelationNeg}{8}
\newcommand{\ImnSzeroResnetFiveZeroLoobjCenterN}{\ensuremath{-}0.010}
\newcommand{\ImnSzeroResnetFiveZeroLoobjCenterNNeg}{7}
\newcommand{\ImnSzeroResnetFiveZeroLoobjCenterA}{\ensuremath{-}0.092}
\newcommand{\ImnSzeroResnetFiveZeroLoobjCenterANeg}{8}
\newcommand{\ImnSzeroResnetFiveZeroLoobjCenterW}{\ensuremath{-}0.048}
\newcommand{\ImnSzeroResnetFiveZeroLoobjCenterWNeg}{7}
\newcommand{\ImnSzeroConvnextTinyVoneRawCorrelation}{\ensuremath{-}0.006}
\newcommand{\ImnSzeroConvnextTinyVoneRawCorrelationNeg}{7}
\newcommand{\ImnSzeroConvnextTinyVoneZCorrelation}{\ensuremath{-}0.018}
\newcommand{\ImnSzeroConvnextTinyVoneZCorrelationNeg}{8}
\newcommand{\ImnSzeroConvnextTinyVoneCenterN}{+0.007}
\newcommand{\ImnSzeroConvnextTinyVoneCenterNNeg}{2}
\newcommand{\ImnSzeroConvnextTinyVoneCenterA}{\ensuremath{-}0.017}
\newcommand{\ImnSzeroConvnextTinyVoneCenterANeg}{8}
\newcommand{\ImnSzeroConvnextTinyVoneCenterW}{\ensuremath{-}0.028}
\newcommand{\ImnSzeroConvnextTinyVoneCenterWNeg}{8}
\newcommand{\ImnSzeroConvnextTinyLoobjRawCorrelation}{\ensuremath{-}0.024}
\newcommand{\ImnSzeroConvnextTinyLoobjRawCorrelationNeg}{8}
\newcommand{\ImnSzeroConvnextTinyLoobjZCorrelation}{\ensuremath{-}0.033}
\newcommand{\ImnSzeroConvnextTinyLoobjZCorrelationNeg}{8}
\newcommand{\ImnSzeroConvnextTinyLoobjCenterN}{+0.007}
\newcommand{\ImnSzeroConvnextTinyLoobjCenterNNeg}{5}
\newcommand{\ImnSzeroConvnextTinyLoobjCenterA}{\ensuremath{-}0.016}
\newcommand{\ImnSzeroConvnextTinyLoobjCenterANeg}{8}
\newcommand{\ImnSzeroConvnextTinyLoobjCenterW}{\ensuremath{-}0.021}
\newcommand{\ImnSzeroConvnextTinyLoobjCenterWNeg}{6}
\newcommand{\ImnSzeroVitBOneSixVoneRawCorrelation}{\ensuremath{-}0.056}
\newcommand{\ImnSzeroVitBOneSixVoneRawCorrelationNeg}{8}
\newcommand{\ImnSzeroVitBOneSixVoneZCorrelation}{\ensuremath{-}0.083}
\newcommand{\ImnSzeroVitBOneSixVoneZCorrelationNeg}{8}
\newcommand{\ImnSzeroVitBOneSixVoneCenterN}{+0.017}
\newcommand{\ImnSzeroVitBOneSixVoneCenterNNeg}{2}
\newcommand{\ImnSzeroVitBOneSixVoneCenterA}{\ensuremath{-}0.099}
\newcommand{\ImnSzeroVitBOneSixVoneCenterANeg}{8}
\newcommand{\ImnSzeroVitBOneSixVoneCenterW}{\ensuremath{-}0.020}
\newcommand{\ImnSzeroVitBOneSixVoneCenterWNeg}{6}
\newcommand{\ImnSzeroVitBOneSixLoobjRawCorrelation}{\ensuremath{-}0.046}
\newcommand{\ImnSzeroVitBOneSixLoobjRawCorrelationNeg}{8}
\newcommand{\ImnSzeroVitBOneSixLoobjZCorrelation}{\ensuremath{-}0.060}
\newcommand{\ImnSzeroVitBOneSixLoobjZCorrelationNeg}{8}
\newcommand{\ImnSzeroVitBOneSixLoobjCenterN}{+0.020}
\newcommand{\ImnSzeroVitBOneSixLoobjCenterNNeg}{1}
\newcommand{\ImnSzeroVitBOneSixLoobjCenterA}{\ensuremath{-}0.052}
\newcommand{\ImnSzeroVitBOneSixLoobjCenterANeg}{8}
\newcommand{\ImnSzeroVitBOneSixLoobjCenterW}{+0.007}
\newcommand{\ImnSzeroVitBOneSixLoobjCenterWNeg}{4}
\newcommand{\ImnVoneAMin}{\ensuremath{-}0.100}
\newcommand{\ImnVoneAMax}{\ensuremath{-}0.016}
\newcommand{\ImnVoneRawCorrMin}{\ensuremath{-}0.057}
\newcommand{\ImnVoneRawCorrMax}{\ensuremath{-}0.005}
\newcommand{\ResRawThreeTwo}{\ensuremath{-}0.021}
\newcommand{\ResRawThreeTwoLo}{\ensuremath{-}0.031}
\newcommand{\ResRawThreeTwoHi}{\ensuremath{-}0.010}
\newcommand{\ResRawSixFour}{\ensuremath{-}0.023}
\newcommand{\ResRawSixFourLo}{\ensuremath{-}0.034}
\newcommand{\ResRawSixFourHi}{\ensuremath{-}0.010}
\newcommand{\ResRawNineSix}{\ensuremath{-}0.022}
\newcommand{\ResRawNineSixLo}{\ensuremath{-}0.034}
\newcommand{\ResRawNineSixHi}{\ensuremath{-}0.008}
\newcommand{\ResRawOneTwoEight}{\ensuremath{-}0.020}
\newcommand{\ResRawOneTwoEightLo}{\ensuremath{-}0.033}
\newcommand{\ResRawOneTwoEightHi}{\ensuremath{-}0.006}
\newcommand{\ResRawOneSixZero}{\ensuremath{-}0.019}
\newcommand{\ResRawOneSixZeroLo}{\ensuremath{-}0.032}
\newcommand{\ResRawOneSixZeroHi}{\ensuremath{-}0.005}
\newcommand{\ResRawTwoTwoFour}{\ensuremath{-}0.017}
\newcommand{\ResRawTwoTwoFourLo}{\ensuremath{-}0.030}
\newcommand{\ResRawTwoTwoFourHi}{\ensuremath{-}0.002}
\newcommand{\ResRawSlope}{+0.004}
\newcommand{\ResRawSlopeLo}{\ensuremath{-}0.004}
\newcommand{\ResRawSlopeHi}{+0.012}
\newcommand{\ResZThreeTwo}{\ensuremath{-}0.037}
\newcommand{\ResZThreeTwoLo}{\ensuremath{-}0.047}
\newcommand{\ResZThreeTwoHi}{\ensuremath{-}0.027}
\newcommand{\ResZSixFour}{\ensuremath{-}0.038}
\newcommand{\ResZSixFourLo}{\ensuremath{-}0.049}
\newcommand{\ResZSixFourHi}{\ensuremath{-}0.029}
\newcommand{\ResZNineSix}{\ensuremath{-}0.035}
\newcommand{\ResZNineSixLo}{\ensuremath{-}0.046}
\newcommand{\ResZNineSixHi}{\ensuremath{-}0.026}
\newcommand{\ResZOneTwoEight}{\ensuremath{-}0.033}
\newcommand{\ResZOneTwoEightLo}{\ensuremath{-}0.043}
\newcommand{\ResZOneTwoEightHi}{\ensuremath{-}0.024}
\newcommand{\ResZOneSixZero}{\ensuremath{-}0.031}
\newcommand{\ResZOneSixZeroLo}{\ensuremath{-}0.041}
\newcommand{\ResZOneSixZeroHi}{\ensuremath{-}0.022}
\newcommand{\ResZTwoTwoFour}{\ensuremath{-}0.026}
\newcommand{\ResZTwoTwoFourLo}{\ensuremath{-}0.036}
\newcommand{\ResZTwoTwoFourHi}{\ensuremath{-}0.018}
\newcommand{\ResZSlope}{+0.011}
\newcommand{\ResZSlopeLo}{+0.003}
\newcommand{\ResZSlopeHi}{+0.017}
\newcommand{\CeilVoneKOne}{0.269}
\newcommand{\CeilVoneKOneLo}{0.248}
\newcommand{\CeilVoneKOneHi}{0.288}
\newcommand{\CeilVoneKOneRand}{0.030}
\newcommand{\CeilVoneKOneBp}{0.042}
\newcommand{\CeilVoneKThree}{0.516}
\newcommand{\CeilVoneKThreeLo}{0.491}
\newcommand{\CeilVoneKThreeHi}{0.541}
\newcommand{\CeilVoneKThreeRand}{0.051}
\newcommand{\CeilVoneKThreeBp}{0.068}
\newcommand{\CeilLoobjKOne}{0.084}
\newcommand{\CeilLoobjKOneLo}{0.076}
\newcommand{\CeilLoobjKOneHi}{0.093}
\newcommand{\CeilLoobjKOneRand}{0.004}
\newcommand{\CeilLoobjKOneBp}{0.013}
\newcommand{\CeilLoobjKThree}{0.207}
\newcommand{\CeilLoobjKThreeLo}{0.190}
\newcommand{\CeilLoobjKThreeHi}{0.224}
\newcommand{\CeilLoobjKThreeRand}{0.011}
\newcommand{\CeilLoobjKThreeBp}{0.025}
\newcommand{\CeilVoneFracMin}{10}
\newcommand{\CeilVoneFracMax}{15}
\newcommand{\CeilThingsLoso}{0.051}
\newcommand{\AmpTsevTrawVoneKOneL}{\ensuremath{-}0.19}
\newcommand{\AmpTsevTrawVoneKOneC}{+0.13}
\newcommand{\AmpTsevTrawVoneKOneCE}{+0.28}
\newcommand{\AmpTsevTrawVoneKOneDD}{+0.19}
\newcommand{\AmpGmTsevTrawVoneKOneRand}{+0.149}
\newcommand{\AmpGmTsevTrawVoneKOneBp}{+0.142}

\newcommand{\AmpTsevTglobalVoneKOneL}{\ensuremath{-}0.18}
\newcommand{\AmpTsevTglobalVoneKOneC}{+0.12}
\newcommand{\AmpTsevTglobalVoneKOneCE}{+0.25}
\newcommand{\AmpTsevTglobalVoneKOneDD}{+0.17}
\newcommand{\AmpGmTsevTglobalVoneKOneRand}{+0.128}
\newcommand{\AmpGmTsevTglobalVoneKOneBp}{+0.124}
\newcommand{\AmpTsevTsessionVoneKOneL}{\ensuremath{-}0.02}
\newcommand{\AmpTsevTsessionVoneKOneC}{+0.07}
\newcommand{\AmpTsevTsessionVoneKOneCE}{+0.08}
\newcommand{\AmpTsevTsessionVoneKOneDD}{+0.02}
\newcommand{\AmpGmTsevTsessionVoneKOneRand}{+0.041}
\newcommand{\AmpGmTsevTsessionVoneKOneBp}{+0.054}
\newcommand{\AmpModelTsevConvOneRandL}{\ensuremath{-}0.21}
\newcommand{\AmpModelTsevConvOneRandC}{+0.46}
\newcommand{\AmpModelTsevConvOneRandCE}{+0.38}
\newcommand{\AmpModelTsevConvOneRandDD}{+0.20}
\newcommand{\AmpModelTsevConvOneBpL}{\ensuremath{-}0.02}
\newcommand{\AmpModelTsevConvOneBpC}{+0.35}
\newcommand{\AmpModelTsevConvOneBpCE}{+0.33}
\newcommand{\AmpModelTsevConvOneBpDD}{+0.01}
\newcommand{\AmpTestTrawVoneKOneTwoL}{\ensuremath{-}0.24}
\newcommand{\AmpTestTrawVoneKOneTwoC}{+0.28}
\newcommand{\AmpTestTrawVoneKOneTwoCE}{+0.47}
\newcommand{\AmpTestTrawVoneKOneTwoDD}{+0.25}
\newcommand{\AmpGmTestTrawVoneKOneTwoRand}{+0.278}
\newcommand{\AmpGmTestTrawVoneKOneTwoBp}{+0.282}
\newcommand{\AmpTestTglobalVoneKOneTwoL}{+0.15}
\newcommand{\AmpTestTglobalVoneKOneTwoC}{\ensuremath{-}0.17}
\newcommand{\AmpTestTglobalVoneKOneTwoCE}{\ensuremath{-}0.30}
\newcommand{\AmpTestTglobalVoneKOneTwoDD}{\ensuremath{-}0.18}
\newcommand{\AmpGmTestTglobalVoneKOneTwoRand}{\ensuremath{-}0.204}
\newcommand{\AmpGmTestTglobalVoneKOneTwoBp}{\ensuremath{-}0.200}
\newcommand{\AmpTestTsessionVoneKOneTwoL}{+0.15}
\newcommand{\AmpTestTsessionVoneKOneTwoC}{\ensuremath{-}0.17}
\newcommand{\AmpTestTsessionVoneKOneTwoCE}{\ensuremath{-}0.30}
\newcommand{\AmpTestTsessionVoneKOneTwoDD}{\ensuremath{-}0.17}
\newcommand{\AmpGmTestTsessionVoneKOneTwoRand}{\ensuremath{-}0.206}
\newcommand{\AmpGmTestTsessionVoneKOneTwoBp}{\ensuremath{-}0.201}
\newcommand{\AmpModelTestConvOneRandL}{\ensuremath{-}0.10}
\newcommand{\AmpModelTestConvOneRandC}{+0.52}
\newcommand{\AmpModelTestConvOneRandCE}{+0.31}
\newcommand{\AmpModelTestConvOneRandDD}{+0.10}
\newcommand{\AmpModelTestConvOneBpL}{+0.07}
\newcommand{\AmpModelTestConvOneBpC}{+0.42}
\newcommand{\AmpModelTestConvOneBpCE}{+0.28}
\newcommand{\AmpModelTestConvOneBpDD}{\ensuremath{-}0.07}
\newcommand{\AmpNsdNrawVoneKOneL}{\ensuremath{-}0.19}
\newcommand{\AmpNsdNrawVoneKOneC}{+0.16}
\newcommand{\AmpNsdNrawVoneKOneCE}{+0.26}
\newcommand{\AmpNsdNrawVoneKOneDD}{+0.18}
\newcommand{\AmpGmNsdNrawVoneKOneRand}{+0.189}
\newcommand{\AmpGmNsdNrawVoneKOneBp}{+0.187}
\newcommand{\AmpNsdNrawVoneKThreeL}{\ensuremath{-}0.27}
\newcommand{\AmpNsdNrawVoneKThreeC}{+0.19}
\newcommand{\AmpNsdNrawVoneKThreeCE}{+0.32}
\newcommand{\AmpNsdNrawVoneKThreeDD}{+0.26}
\newcommand{\AmpGmNsdNrawVoneKThreeRand}{+0.249}
\newcommand{\AmpGmNsdNrawVoneKThreeBp}{+0.243}
\newcommand{\AmpNsdNglobalVoneKOneL}{+0.19}
\newcommand{\AmpNsdNglobalVoneKOneC}{\ensuremath{-}0.08}
\newcommand{\AmpNsdNglobalVoneKOneCE}{\ensuremath{-}0.22}
\newcommand{\AmpNsdNglobalVoneKOneDD}{\ensuremath{-}0.18}
\newcommand{\AmpGmNsdNglobalVoneKOneRand}{\ensuremath{-}0.108}
\newcommand{\AmpGmNsdNglobalVoneKOneBp}{\ensuremath{-}0.083}
\newcommand{\AmpNsdNglobalVoneKThreeL}{+0.32}
\newcommand{\AmpNsdNglobalVoneKThreeC}{\ensuremath{-}0.17}
\newcommand{\AmpNsdNglobalVoneKThreeCE}{\ensuremath{-}0.40}
\newcommand{\AmpNsdNglobalVoneKThreeDD}{\ensuremath{-}0.30}
\newcommand{\AmpGmNsdNglobalVoneKThreeRand}{\ensuremath{-}0.210}
\newcommand{\AmpGmNsdNglobalVoneKThreeBp}{\ensuremath{-}0.181}
\newcommand{\AmpNsdNsessionVoneKOneL}{+0.22}
\newcommand{\AmpNsdNsessionVoneKOneC}{\ensuremath{-}0.10}
\newcommand{\AmpNsdNsessionVoneKOneCE}{\ensuremath{-}0.25}
\newcommand{\AmpNsdNsessionVoneKOneDD}{\ensuremath{-}0.21}
\newcommand{\AmpGmNsdNsessionVoneKOneRand}{\ensuremath{-}0.142}
\newcommand{\AmpGmNsdNsessionVoneKOneBp}{\ensuremath{-}0.110}
\newcommand{\AmpNsdNsessionVoneKThreeL}{+0.32}
\newcommand{\AmpNsdNsessionVoneKThreeC}{\ensuremath{-}0.17}
\newcommand{\AmpNsdNsessionVoneKThreeCE}{\ensuremath{-}0.41}
\newcommand{\AmpNsdNsessionVoneKThreeDD}{\ensuremath{-}0.30}
\newcommand{\AmpGmNsdNsessionVoneKThreeRand}{\ensuremath{-}0.214}
\newcommand{\AmpGmNsdNsessionVoneKThreeBp}{\ensuremath{-}0.183}
\newcommand{\AmpModelNsdConvOneRandL}{\ensuremath{-}0.19}
\newcommand{\AmpModelNsdConvOneRandC}{+0.49}
\newcommand{\AmpModelNsdConvOneRandCE}{+0.38}
\newcommand{\AmpModelNsdConvOneRandDD}{+0.19}
\newcommand{\AmpModelNsdConvOneBpL}{\ensuremath{-}0.01}
\newcommand{\AmpModelNsdConvOneBpC}{+0.47}
\newcommand{\AmpModelNsdConvOneBpCE}{+0.44}
\newcommand{\AmpModelNsdConvOneBpDD}{+0.01}
\newcommand{\AmpTsevTglobalVoneSubOneCE}{+0.25}
\newcommand{\AmpTsevTglobalVoneSubOneL}{\ensuremath{-}0.20}

\newcommand{\AmpTsevTglobalVoneSubTwoCE}{+0.28}
\newcommand{\AmpTsevTglobalVoneSubTwoL}{\ensuremath{-}0.15}

\newcommand{\AmpTsevTglobalVoneSubThreeCE}{+0.22}
\newcommand{\AmpTsevTglobalVoneSubThreeL}{\ensuremath{-}0.18}
\newcommand{\RuleGThree}{5}
\newcommand{\RuleGTwo}{1}

\newcommand{\ImgSpearLDD}{\ensuremath{-}0.98}
\newcommand{\ImgSpearLDDNsd}{\ensuremath{-}0.99}

\newcommand{\DomTsevConvOneSumRand}{+0.466}
\newcommand{\DomTsevConvOneSumRandSeedZero}{+0.588}
\newcommand{\DomTsevConvOneSumRandSeedOne}{+0.333}
\newcommand{\DomTsevConvOneSumRandSeedTwo}{+0.508}
\newcommand{\DomTsevConvOneSumRandSeedThree}{+0.521}
\newcommand{\DomTsevConvOneSumRandSeedFour}{+0.381}
\newcommand{\DomTsevConvOneSumBp}{+0.253}
\newcommand{\DomTsevConvOneSumBpSeedZero}{+0.258}
\newcommand{\DomTsevConvOneSumBpSeedOne}{+0.236}
\newcommand{\DomTsevConvOneSumBpSeedTwo}{+0.320}
\newcommand{\DomTsevConvOneSumBpSeedThree}{+0.242}
\newcommand{\DomTsevConvOneSumBpSeedFour}{+0.207}
\newcommand{\DomMrTsevConvOneSum}{+0.88}
\newcommand{\DomTsevConvOneDiffRand}{+0.393}

\newcommand{\DomTsevConvOneDiffBp}{+0.145}

\newcommand{\DomMrTsevConvOneDiff}{+0.85}
\newcommand{\MratioTsevConvOneRand}{0.71}
\newcommand{\MratioTsevConvOneBp}{0.64}
\newcommand{\BDomTsevTrawVoneKOneSum}{+0.71}
\newcommand{\BDomTsevTrawVoneKOneDiff}{+0.41}
\newcommand{\PbTsevTrawVoneKOneNone}{+0.100}
\newcommand{\PbTsevTrawVoneKOneNonePos}{3}

\newcommand{\PbTsevTrawVoneKOneSum}{+0.021}
\newcommand{\PbTsevTrawVoneKOneSumPos}{3}
\newcommand{\PbTsevTrawVoneKOneSumLe}{1}
\newcommand{\PbTsevTrawVoneKOneDiff}{+0.049}

\newcommand{\PbTsevTrawLocKOneNone}{\ensuremath{-}0.032}
\newcommand{\PbTsevTrawLocKOneNonePos}{0}

\newcommand{\PbTsevTrawLocKOneSum}{\ensuremath{-}0.004}
\newcommand{\PbTsevTrawLocKOneSumPos}{1}
\newcommand{\PbTsevTrawLocKOneSumLe}{3}
\newcommand{\PbTsevTrawLocKOneDiff}{\ensuremath{-}0.019}

\newcommand{\BDomTsevTglobalVoneKOneSum}{+0.22}
\newcommand{\BDomTsevTglobalVoneKOneDiff}{+0.30}
\newcommand{\PbTsevTglobalVoneKOneNone}{+0.030}
\newcommand{\PbTsevTglobalVoneKOneNonePos}{3}

\newcommand{\PbTsevTglobalVoneKOneSum}{+0.007}
\newcommand{\PbTsevTglobalVoneKOneSumPos}{3}
\newcommand{\PbTsevTglobalVoneKOneSumLe}{3}
\newcommand{\PbTsevTglobalVoneKOneDiff}{+0.014}

\newcommand{\PbTsevTglobalLocKOneNone}{\ensuremath{-}0.011}
\newcommand{\PbTsevTglobalLocKOneNonePos}{0}

\newcommand{\PbTsevTglobalLocKOneSum}{\ensuremath{-}0.005}
\newcommand{\PbTsevTglobalLocKOneSumPos}{0}
\newcommand{\PbTsevTglobalLocKOneSumLe}{3}
\newcommand{\PbTsevTglobalLocKOneDiff}{\ensuremath{-}0.014}

\newcommand{\BDomTsevTsessionVoneKOneSum}{+0.02}
\newcommand{\BDomTsevTsessionVoneKOneDiff}{+0.27}
\newcommand{\PbTsevTsessionVoneKOneNone}{+0.005}
\newcommand{\PbTsevTsessionVoneKOneNonePos}{3}

\newcommand{\PbTsevTsessionVoneKOneSum}{+0.004}
\newcommand{\PbTsevTsessionVoneKOneSumPos}{3}
\newcommand{\PbTsevTsessionVoneKOneSumLe}{3}
\newcommand{\PbTsevTsessionVoneKOneDiff}{+0.001}

\newcommand{\PbTsevTsessionLocKOneNone}{\ensuremath{-}0.004}
\newcommand{\PbTsevTsessionLocKOneNonePos}{0}

\newcommand{\PbTsevTsessionLocKOneSum}{\ensuremath{-}0.005}
\newcommand{\PbTsevTsessionLocKOneSumPos}{0}
\newcommand{\PbTsevTsessionLocKOneSumLe}{3}
\newcommand{\PbTsevTsessionLocKOneDiff}{\ensuremath{-}0.008}

\newcommand{\DomTestConvOneSumRand}{+0.448}
\newcommand{\DomTestConvOneSumRandSeedZero}{+0.545}
\newcommand{\DomTestConvOneSumRandSeedOne}{+0.376}
\newcommand{\DomTestConvOneSumRandSeedTwo}{+0.454}
\newcommand{\DomTestConvOneSumRandSeedThree}{+0.463}
\newcommand{\DomTestConvOneSumRandSeedFour}{+0.403}
\newcommand{\DomTestConvOneSumBp}{+0.344}
\newcommand{\DomTestConvOneSumBpSeedZero}{+0.348}
\newcommand{\DomTestConvOneSumBpSeedOne}{+0.355}
\newcommand{\DomTestConvOneSumBpSeedTwo}{+0.404}
\newcommand{\DomTestConvOneSumBpSeedThree}{+0.330}
\newcommand{\DomTestConvOneSumBpSeedFour}{+0.280}
\newcommand{\DomMrTestConvOneSum}{+0.67}
\newcommand{\DomTestConvOneDiffRand}{+0.396}

\newcommand{\DomTestConvOneDiffBp}{+0.202}

\newcommand{\DomMrTestConvOneDiff}{+0.60}

\newcommand{\PbTestTrawVoneKOneNone}{+0.078}
\newcommand{\PbTestTrawVoneKOneNonePos}{3}

\newcommand{\PbTestTrawVoneKOneSum}{+0.014}
\newcommand{\PbTestTrawVoneKOneSumPos}{2}
\newcommand{\PbTestTrawVoneKOneSumLe}{2}
\newcommand{\PbTestTrawVoneKOneDiff}{+0.037}

\newcommand{\BDomTestTrawVoneKOneTwoSum}{+0.91}
\newcommand{\BDomTestTrawVoneKOneTwoDiff}{+0.59}
\newcommand{\PbTestTrawVoneKOneTwoNone}{+0.165}
\newcommand{\PbTestTrawVoneKOneTwoNonePos}{3}

\newcommand{\PbTestTrawVoneKOneTwoSum}{+0.051}
\newcommand{\PbTestTrawVoneKOneTwoSumPos}{2}
\newcommand{\PbTestTrawVoneKOneTwoSumLe}{1}
\newcommand{\PbTestTrawVoneKOneTwoDiff}{+0.071}

\newcommand{\PbTestTrawLocKOneNone}{+0.010}
\newcommand{\PbTestTrawLocKOneNonePos}{2}

\newcommand{\PbTestTrawLocKOneSum}{\ensuremath{-}0.002}
\newcommand{\PbTestTrawLocKOneSumPos}{1}
\newcommand{\PbTestTrawLocKOneSumLe}{3}
\newcommand{\PbTestTrawLocKOneDiff}{+0.011}

\newcommand{\PbTestTrawLocKOneTwoNone}{+0.001}
\newcommand{\PbTestTrawLocKOneTwoNonePos}{2}

\newcommand{\PbTestTrawLocKOneTwoSum}{+0.004}
\newcommand{\PbTestTrawLocKOneTwoSumPos}{2}
\newcommand{\PbTestTrawLocKOneTwoSumLe}{2}
\newcommand{\PbTestTrawLocKOneTwoDiff}{+0.005}

\newcommand{\PbTestTglobalVoneKOneNone}{0.000}
\newcommand{\PbTestTglobalVoneKOneNonePos}{1}

\newcommand{\PbTestTglobalVoneKOneSum}{\ensuremath{-}0.005}
\newcommand{\PbTestTglobalVoneKOneSumPos}{1}
\newcommand{\PbTestTglobalVoneKOneSumLe}{3}
\newcommand{\PbTestTglobalVoneKOneDiff}{\ensuremath{-}0.005}

\newcommand{\BDomTestTglobalVoneKOneTwoSum}{+0.08}
\newcommand{\BDomTestTglobalVoneKOneTwoDiff}{+0.35}
\newcommand{\PbTestTglobalVoneKOneTwoNone}{\ensuremath{-}0.018}
\newcommand{\PbTestTglobalVoneKOneTwoNonePos}{1}

\newcommand{\PbTestTglobalVoneKOneTwoSum}{\ensuremath{-}0.016}
\newcommand{\PbTestTglobalVoneKOneTwoSumPos}{1}
\newcommand{\PbTestTglobalVoneKOneTwoSumLe}{3}
\newcommand{\PbTestTglobalVoneKOneTwoDiff}{\ensuremath{-}0.027}

\newcommand{\PbTestTglobalLocKOneNone}{0.000}
\newcommand{\PbTestTglobalLocKOneNonePos}{1}

\newcommand{\PbTestTglobalLocKOneSum}{\ensuremath{-}0.003}
\newcommand{\PbTestTglobalLocKOneSumPos}{1}
\newcommand{\PbTestTglobalLocKOneSumLe}{3}
\newcommand{\PbTestTglobalLocKOneDiff}{0.000}

\newcommand{\PbTestTglobalLocKOneTwoNone}{\ensuremath{-}0.016}
\newcommand{\PbTestTglobalLocKOneTwoNonePos}{0}

\newcommand{\PbTestTglobalLocKOneTwoSum}{\ensuremath{-}0.011}
\newcommand{\PbTestTglobalLocKOneTwoSumPos}{1}
\newcommand{\PbTestTglobalLocKOneTwoSumLe}{3}
\newcommand{\PbTestTglobalLocKOneTwoDiff}{\ensuremath{-}0.019}

\newcommand{\PbTestTsessionVoneKOneNone}{\ensuremath{-}0.007}
\newcommand{\PbTestTsessionVoneKOneNonePos}{0}

\newcommand{\PbTestTsessionVoneKOneSum}{\ensuremath{-}0.005}
\newcommand{\PbTestTsessionVoneKOneSumPos}{0}
\newcommand{\PbTestTsessionVoneKOneSumLe}{3}
\newcommand{\PbTestTsessionVoneKOneDiff}{\ensuremath{-}0.005}

\newcommand{\BDomTestTsessionVoneKOneTwoSum}{+0.08}
\newcommand{\BDomTestTsessionVoneKOneTwoDiff}{+0.35}
\newcommand{\PbTestTsessionVoneKOneTwoNone}{\ensuremath{-}0.018}
\newcommand{\PbTestTsessionVoneKOneTwoNonePos}{1}

\newcommand{\PbTestTsessionVoneKOneTwoSum}{\ensuremath{-}0.016}
\newcommand{\PbTestTsessionVoneKOneTwoSumPos}{1}
\newcommand{\PbTestTsessionVoneKOneTwoSumLe}{3}
\newcommand{\PbTestTsessionVoneKOneTwoDiff}{\ensuremath{-}0.027}

\newcommand{\PbTestTsessionLocKOneNone}{\ensuremath{-}0.003}
\newcommand{\PbTestTsessionLocKOneNonePos}{1}

\newcommand{\PbTestTsessionLocKOneSum}{\ensuremath{-}0.003}
\newcommand{\PbTestTsessionLocKOneSumPos}{1}
\newcommand{\PbTestTsessionLocKOneSumLe}{3}
\newcommand{\PbTestTsessionLocKOneDiff}{\ensuremath{-}0.002}

\newcommand{\PbTestTsessionLocKOneTwoNone}{\ensuremath{-}0.016}
\newcommand{\PbTestTsessionLocKOneTwoNonePos}{0}

\newcommand{\PbTestTsessionLocKOneTwoSum}{\ensuremath{-}0.011}
\newcommand{\PbTestTsessionLocKOneTwoSumPos}{1}
\newcommand{\PbTestTsessionLocKOneTwoSumLe}{3}
\newcommand{\PbTestTsessionLocKOneTwoDiff}{\ensuremath{-}0.019}

\newcommand{\DomNsdConvOneSumRand}{+0.564}
\newcommand{\DomNsdConvOneSumRandSeedZero}{+0.708}
\newcommand{\DomNsdConvOneSumRandSeedOne}{+0.404}
\newcommand{\DomNsdConvOneSumRandSeedTwo}{+0.616}
\newcommand{\DomNsdConvOneSumRandSeedThree}{+0.634}
\newcommand{\DomNsdConvOneSumRandSeedFour}{+0.459}
\newcommand{\DomNsdConvOneSumBp}{+0.205}
\newcommand{\DomNsdConvOneSumBpSeedZero}{+0.175}
\newcommand{\DomNsdConvOneSumBpSeedOne}{+0.194}
\newcommand{\DomNsdConvOneSumBpSeedTwo}{+0.238}
\newcommand{\DomNsdConvOneSumBpSeedThree}{+0.224}
\newcommand{\DomNsdConvOneSumBpSeedFour}{+0.192}
\newcommand{\DomMrNsdConvOneSum}{+0.84}
\newcommand{\DomNsdConvOneDiffRand}{+0.503}

\newcommand{\DomNsdConvOneDiffBp}{+0.123}

\newcommand{\DomMrNsdConvOneDiff}{+0.84}

\newcommand{\BDomNsdNrawVoneKOneSum}{+0.88}
\newcommand{\BDomNsdNrawVoneKOneDiff}{+0.60}
\newcommand{\PbNsdNrawVoneKOneNone}{+0.116}
\newcommand{\PbNsdNrawVoneKOneNonePos}{8}

\newcommand{\PbNsdNrawVoneKOneSum}{\ensuremath{-}0.009}
\newcommand{\PbNsdNrawVoneKOneSumPos}{0}
\newcommand{\PbNsdNrawVoneKOneSumLe}{8}
\newcommand{\PbNsdNrawVoneKOneDiff}{+0.048}

\newcommand{\BDomNsdNrawVoneKThreeSum}{+0.89}
\newcommand{\BDomNsdNrawVoneKThreeDiff}{+0.58}
\newcommand{\PbNsdNrawVoneKThreeNone}{+0.166}
\newcommand{\PbNsdNrawVoneKThreeNonePos}{8}

\newcommand{\PbNsdNrawVoneKThreeSum}{\ensuremath{-}0.014}
\newcommand{\PbNsdNrawVoneKThreeSumPos}{0}
\newcommand{\PbNsdNrawVoneKThreeSumLe}{8}
\newcommand{\PbNsdNrawVoneKThreeDiff}{+0.068}

\newcommand{\PbNsdNrawLoobjKOneNone}{+0.007}
\newcommand{\PbNsdNrawLoobjKOneNonePos}{6}

\newcommand{\PbNsdNrawLoobjKOneSum}{\ensuremath{-}0.005}
\newcommand{\PbNsdNrawLoobjKOneSumPos}{1}
\newcommand{\PbNsdNrawLoobjKOneSumLe}{8}
\newcommand{\PbNsdNrawLoobjKOneDiff}{+0.004}

\newcommand{\PbNsdNrawLoobjKThreeNone}{+0.013}
\newcommand{\PbNsdNrawLoobjKThreeNonePos}{6}

\newcommand{\PbNsdNrawLoobjKThreeSum}{\ensuremath{-}0.011}
\newcommand{\PbNsdNrawLoobjKThreeSumPos}{0}
\newcommand{\PbNsdNrawLoobjKThreeSumLe}{8}
\newcommand{\PbNsdNrawLoobjKThreeDiff}{+0.005}

\newcommand{\PbNsdNrawLatKOneNone}{\ensuremath{-}0.001}
\newcommand{\PbNsdNrawLatKOneNonePos}{5}

\newcommand{\PbNsdNrawLatKOneSum}{\ensuremath{-}0.007}
\newcommand{\PbNsdNrawLatKOneSumPos}{0}
\newcommand{\PbNsdNrawLatKOneSumLe}{8}
\newcommand{\PbNsdNrawLatKOneDiff}{\ensuremath{-}0.002}

\newcommand{\PbNsdNrawLatKThreeNone}{0.000}
\newcommand{\PbNsdNrawLatKThreeNonePos}{6}

\newcommand{\PbNsdNrawLatKThreeSum}{\ensuremath{-}0.015}
\newcommand{\PbNsdNrawLatKThreeSumPos}{0}
\newcommand{\PbNsdNrawLatKThreeSumLe}{8}
\newcommand{\PbNsdNrawLatKThreeDiff}{\ensuremath{-}0.002}

\newcommand{\BDomNsdNglobalVoneKOneSum}{+0.09}
\newcommand{\BDomNsdNglobalVoneKOneDiff}{+0.22}
\newcommand{\PbNsdNglobalVoneKOneNone}{\ensuremath{-}0.012}
\newcommand{\PbNsdNglobalVoneKOneNonePos}{0}

\newcommand{\PbNsdNglobalVoneKOneSum}{\ensuremath{-}0.001}
\newcommand{\PbNsdNglobalVoneKOneSumPos}{2}
\newcommand{\PbNsdNglobalVoneKOneSumLe}{8}
\newcommand{\PbNsdNglobalVoneKOneDiff}{\ensuremath{-}0.013}

\newcommand{\BDomNsdNglobalVoneKThreeSum}{+0.08}
\newcommand{\BDomNsdNglobalVoneKThreeDiff}{+0.27}
\newcommand{\PbNsdNglobalVoneKThreeNone}{\ensuremath{-}0.019}
\newcommand{\PbNsdNglobalVoneKThreeNonePos}{0}

\newcommand{\PbNsdNglobalVoneKThreeSum}{\ensuremath{-}0.003}
\newcommand{\PbNsdNglobalVoneKThreeSumPos}{2}
\newcommand{\PbNsdNglobalVoneKThreeSumLe}{8}
\newcommand{\PbNsdNglobalVoneKThreeDiff}{\ensuremath{-}0.019}

\newcommand{\PbNsdNglobalLoobjKOneNone}{\ensuremath{-}0.007}
\newcommand{\PbNsdNglobalLoobjKOneNonePos}{0}

\newcommand{\PbNsdNglobalLoobjKOneSum}{\ensuremath{-}0.005}
\newcommand{\PbNsdNglobalLoobjKOneSumPos}{0}
\newcommand{\PbNsdNglobalLoobjKOneSumLe}{8}
\newcommand{\PbNsdNglobalLoobjKOneDiff}{\ensuremath{-}0.008}

\newcommand{\PbNsdNglobalLoobjKThreeNone}{\ensuremath{-}0.013}
\newcommand{\PbNsdNglobalLoobjKThreeNonePos}{0}

\newcommand{\PbNsdNglobalLoobjKThreeSum}{\ensuremath{-}0.010}
\newcommand{\PbNsdNglobalLoobjKThreeSumPos}{0}
\newcommand{\PbNsdNglobalLoobjKThreeSumLe}{8}
\newcommand{\PbNsdNglobalLoobjKThreeDiff}{\ensuremath{-}0.013}

\newcommand{\PbNsdNglobalLatKOneNone}{\ensuremath{-}0.009}
\newcommand{\PbNsdNglobalLatKOneNonePos}{0}

\newcommand{\PbNsdNglobalLatKOneSum}{\ensuremath{-}0.006}
\newcommand{\PbNsdNglobalLatKOneSumPos}{0}
\newcommand{\PbNsdNglobalLatKOneSumLe}{8}
\newcommand{\PbNsdNglobalLatKOneDiff}{\ensuremath{-}0.010}

\newcommand{\PbNsdNglobalLatKThreeNone}{\ensuremath{-}0.015}
\newcommand{\PbNsdNglobalLatKThreeNonePos}{0}

\newcommand{\PbNsdNglobalLatKThreeSum}{\ensuremath{-}0.011}
\newcommand{\PbNsdNglobalLatKThreeSumPos}{0}
\newcommand{\PbNsdNglobalLatKThreeSumLe}{8}
\newcommand{\PbNsdNglobalLatKThreeDiff}{\ensuremath{-}0.015}

\newcommand{\BDomNsdNsessionVoneKOneSum}{+0.07}
\newcommand{\BDomNsdNsessionVoneKOneDiff}{+0.23}
\newcommand{\PbNsdNsessionVoneKOneNone}{\ensuremath{-}0.012}
\newcommand{\PbNsdNsessionVoneKOneNonePos}{0}

\newcommand{\PbNsdNsessionVoneKOneSum}{\ensuremath{-}0.002}
\newcommand{\PbNsdNsessionVoneKOneSumPos}{2}
\newcommand{\PbNsdNsessionVoneKOneSumLe}{8}
\newcommand{\PbNsdNsessionVoneKOneDiff}{\ensuremath{-}0.013}

\newcommand{\BDomNsdNsessionVoneKThreeSum}{+0.08}
\newcommand{\BDomNsdNsessionVoneKThreeDiff}{+0.28}
\newcommand{\PbNsdNsessionVoneKThreeNone}{\ensuremath{-}0.017}
\newcommand{\PbNsdNsessionVoneKThreeNonePos}{0}

\newcommand{\PbNsdNsessionVoneKThreeSum}{\ensuremath{-}0.003}
\newcommand{\PbNsdNsessionVoneKThreeSumPos}{2}
\newcommand{\PbNsdNsessionVoneKThreeSumLe}{8}
\newcommand{\PbNsdNsessionVoneKThreeDiff}{\ensuremath{-}0.018}

\newcommand{\PbNsdNsessionLoobjKOneNone}{\ensuremath{-}0.008}
\newcommand{\PbNsdNsessionLoobjKOneNonePos}{0}

\newcommand{\PbNsdNsessionLoobjKOneSum}{\ensuremath{-}0.005}
\newcommand{\PbNsdNsessionLoobjKOneSumPos}{0}
\newcommand{\PbNsdNsessionLoobjKOneSumLe}{8}
\newcommand{\PbNsdNsessionLoobjKOneDiff}{\ensuremath{-}0.008}

\newcommand{\PbNsdNsessionLoobjKThreeNone}{\ensuremath{-}0.014}
\newcommand{\PbNsdNsessionLoobjKThreeNonePos}{0}

\newcommand{\PbNsdNsessionLoobjKThreeSum}{\ensuremath{-}0.010}
\newcommand{\PbNsdNsessionLoobjKThreeSumPos}{0}
\newcommand{\PbNsdNsessionLoobjKThreeSumLe}{8}
\newcommand{\PbNsdNsessionLoobjKThreeDiff}{\ensuremath{-}0.013}

\newcommand{\PbNsdNsessionLatKOneNone}{\ensuremath{-}0.010}
\newcommand{\PbNsdNsessionLatKOneNonePos}{0}

\newcommand{\PbNsdNsessionLatKOneSum}{\ensuremath{-}0.006}
\newcommand{\PbNsdNsessionLatKOneSumPos}{0}
\newcommand{\PbNsdNsessionLatKOneSumLe}{8}
\newcommand{\PbNsdNsessionLatKOneDiff}{\ensuremath{-}0.010}

\newcommand{\PbNsdNsessionLatKThreeNone}{\ensuremath{-}0.016}
\newcommand{\PbNsdNsessionLatKThreeNonePos}{0}

\newcommand{\PbNsdNsessionLatKThreeSum}{\ensuremath{-}0.011}
\newcommand{\PbNsdNsessionLatKThreeSumPos}{0}
\newcommand{\PbNsdNsessionLatKThreeSumLe}{8}
\newcommand{\PbNsdNsessionLatKThreeDiff}{\ensuremath{-}0.016}
\newcommand{\RuleDThreeLe}{8}
\newcommand{\RuleDTwoNsd}{8}
\newcommand{\MvThingsClsOne}{+0.030}
\newcommand{\MvThingsClsOneLo}{+0.018}
\newcommand{\MvThingsClsOneHi}{+0.042}
\newcommand{\MvNsessionClsOne}{\ensuremath{-}0.017}
\newcommand{\MvNsessionClsOneLo}{\ensuremath{-}0.030}
\newcommand{\MvNsessionClsOneHi}{\ensuremath{-}0.002}
\newcommand{\MvNrawClsOne}{+0.166}
\newcommand{\MvNrawClsOneLo}{+0.125}
\newcommand{\MvNrawClsOneHi}{+0.203}
\newcommand{\MvNrcClsOne}{\ensuremath{-}0.017}
\newcommand{\MvNrcClsOneLo}{\ensuremath{-}0.031}
\newcommand{\MvNrcClsOneHi}{\ensuremath{-}0.002}
\newcommand{\MvThingsClsTwo}{+0.005}
\newcommand{\MvThingsClsTwoLo}{\ensuremath{-}0.003}
\newcommand{\MvThingsClsTwoHi}{+0.012}
\newcommand{\MvNsessionClsTwo}{+0.005}
\newcommand{\MvNsessionClsTwoLo}{\ensuremath{-}0.009}
\newcommand{\MvNsessionClsTwoHi}{+0.021}
\newcommand{\MvNrawClsTwo}{+0.004}
\newcommand{\MvNrawClsTwoLo}{\ensuremath{-}0.011}
\newcommand{\MvNrawClsTwoHi}{+0.021}
\newcommand{\MvNrcClsTwo}{+0.005}
\newcommand{\MvNrcClsTwoLo}{\ensuremath{-}0.009}
\newcommand{\MvNrcClsTwoHi}{+0.021}
\newcommand{\MvThingsClsThree}{+0.005}
\newcommand{\MvThingsClsThreeLo}{\ensuremath{-}0.003}
\newcommand{\MvThingsClsThreeHi}{+0.013}
\newcommand{\MvNsessionClsThree}{+0.003}
\newcommand{\MvNsessionClsThreeLo}{\ensuremath{-}0.009}
\newcommand{\MvNsessionClsThreeHi}{+0.018}
\newcommand{\MvNrawClsThree}{+0.001}
\newcommand{\MvNrawClsThreeLo}{\ensuremath{-}0.011}
\newcommand{\MvNrawClsThreeHi}{+0.015}
\newcommand{\MvNrcClsThree}{+0.004}
\newcommand{\MvNrcClsThreeLo}{\ensuremath{-}0.009}
\newcommand{\MvNrcClsThreeHi}{+0.018}
\newcommand{\MvThingsClsFour}{+0.003}
\newcommand{\MvThingsClsFourLo}{\ensuremath{-}0.004}
\newcommand{\MvThingsClsFourHi}{+0.009}
\newcommand{\MvNsessionClsFour}{\ensuremath{-}0.014}
\newcommand{\MvNsessionClsFourLo}{\ensuremath{-}0.022}
\newcommand{\MvNsessionClsFourHi}{\ensuremath{-}0.004}
\newcommand{\MvNrawClsFour}{+0.021}
\newcommand{\MvNrawClsFourLo}{\ensuremath{-}0.002}
\newcommand{\MvNrawClsFourHi}{+0.044}
\newcommand{\MvNrcClsFour}{\ensuremath{-}0.014}
\newcommand{\MvNrcClsFourLo}{\ensuremath{-}0.023}
\newcommand{\MvNrcClsFourHi}{\ensuremath{-}0.004}
\newcommand{\MvThingsClsFive}{0.000}
\newcommand{\MvThingsClsFiveLo}{\ensuremath{-}0.004}
\newcommand{\MvThingsClsFiveHi}{+0.005}
\newcommand{\MvNsessionClsFive}{\ensuremath{-}0.008}
\newcommand{\MvNsessionClsFiveLo}{\ensuremath{-}0.018}
\newcommand{\MvNsessionClsFiveHi}{+0.003}
\newcommand{\MvNrawClsFive}{\ensuremath{-}0.008}
\newcommand{\MvNrawClsFiveLo}{\ensuremath{-}0.016}
\newcommand{\MvNrawClsFiveHi}{+0.002}
\newcommand{\MvNrcClsFive}{\ensuremath{-}0.009}
\newcommand{\MvNrcClsFiveLo}{\ensuremath{-}0.018}
\newcommand{\MvNrcClsFiveHi}{+0.003}
\newcommand{\MvThingsClsSix}{0.000}
\newcommand{\MvThingsClsSixLo}{\ensuremath{-}0.004}
\newcommand{\MvThingsClsSixHi}{+0.005}
\newcommand{\MvNsessionClsSix}{\ensuremath{-}0.008}
\newcommand{\MvNsessionClsSixLo}{\ensuremath{-}0.017}
\newcommand{\MvNsessionClsSixHi}{+0.002}
\newcommand{\MvNrawClsSix}{\ensuremath{-}0.008}
\newcommand{\MvNrawClsSixLo}{\ensuremath{-}0.016}
\newcommand{\MvNrawClsSixHi}{+0.002}
\newcommand{\MvNrcClsSix}{\ensuremath{-}0.009}
\newcommand{\MvNrcClsSixLo}{\ensuremath{-}0.018}
\newcommand{\MvNrcClsSixHi}{+0.002}
\newcommand{\MvThingsClsSeven}{+0.023}
\newcommand{\MvThingsClsSevenLo}{+0.004}
\newcommand{\MvThingsClsSevenHi}{+0.041}
\newcommand{\MvNsessionClsSeven}{\ensuremath{-}0.008}
\newcommand{\MvNsessionClsSevenLo}{\ensuremath{-}0.032}
\newcommand{\MvNsessionClsSevenHi}{+0.015}
\newcommand{\MvNrawClsSeven}{\ensuremath{-}0.010}
\newcommand{\MvNrawClsSevenLo}{\ensuremath{-}0.032}
\newcommand{\MvNrawClsSevenHi}{+0.010}
\newcommand{\MvNrcClsSeven}{\ensuremath{-}0.007}
\newcommand{\MvNrcClsSevenLo}{\ensuremath{-}0.032}
\newcommand{\MvNrcClsSevenHi}{+0.016}
\newcommand{\MvThingsClsEight}{+0.023}
\newcommand{\MvThingsClsEightLo}{+0.004}
\newcommand{\MvThingsClsEightHi}{+0.041}
\newcommand{\MvNsessionClsEight}{\ensuremath{-}0.009}
\newcommand{\MvNsessionClsEightLo}{\ensuremath{-}0.032}
\newcommand{\MvNsessionClsEightHi}{+0.014}
\newcommand{\MvNrawClsEight}{\ensuremath{-}0.009}
\newcommand{\MvNrawClsEightLo}{\ensuremath{-}0.032}
\newcommand{\MvNrawClsEightHi}{+0.012}
\newcommand{\MvNrcClsEight}{\ensuremath{-}0.008}
\newcommand{\MvNrcClsEightLo}{\ensuremath{-}0.032}
\newcommand{\MvNrcClsEightHi}{+0.015}
\newcommand{\MvThingsClsNine}{+0.004}
\newcommand{\MvThingsClsNineLo}{\ensuremath{-}0.002}
\newcommand{\MvThingsClsNineHi}{+0.011}
\newcommand{\MvNsessionClsNine}{0.000}
\newcommand{\MvNsessionClsNineLo}{\ensuremath{-}0.009}
\newcommand{\MvNsessionClsNineHi}{+0.011}
\newcommand{\MvNrawClsNine}{\ensuremath{-}0.009}
\newcommand{\MvNrawClsNineLo}{\ensuremath{-}0.024}
\newcommand{\MvNrawClsNineHi}{+0.007}
\newcommand{\MvNrcClsNine}{0.000}
\newcommand{\MvNrcClsNineLo}{\ensuremath{-}0.010}
\newcommand{\MvNrcClsNineHi}{+0.011}
\newcommand{\MvThingsClsOneZero}{+0.001}
\newcommand{\MvThingsClsOneZeroLo}{\ensuremath{-}0.005}
\newcommand{\MvThingsClsOneZeroHi}{+0.007}
\newcommand{\MvNsessionClsOneZero}{0.000}
\newcommand{\MvNsessionClsOneZeroLo}{\ensuremath{-}0.010}
\newcommand{\MvNsessionClsOneZeroHi}{+0.011}
\newcommand{\MvNrawClsOneZero}{+0.004}
\newcommand{\MvNrawClsOneZeroLo}{\ensuremath{-}0.006}
\newcommand{\MvNrawClsOneZeroHi}{+0.016}
\newcommand{\MvNrcClsOneZero}{\ensuremath{-}0.001}
\newcommand{\MvNrcClsOneZeroLo}{\ensuremath{-}0.011}
\newcommand{\MvNrcClsOneZeroHi}{+0.011}
\newcommand{\MvThingsClsOneOne}{+0.001}
\newcommand{\MvThingsClsOneOneLo}{\ensuremath{-}0.005}
\newcommand{\MvThingsClsOneOneHi}{+0.007}
\newcommand{\MvNsessionClsOneOne}{\ensuremath{-}0.001}
\newcommand{\MvNsessionClsOneOneLo}{\ensuremath{-}0.010}
\newcommand{\MvNsessionClsOneOneHi}{+0.009}
\newcommand{\MvNrawClsOneOne}{0.000}
\newcommand{\MvNrawClsOneOneLo}{\ensuremath{-}0.009}
\newcommand{\MvNrawClsOneOneHi}{+0.009}
\newcommand{\MvNrcClsOneOne}{\ensuremath{-}0.001}
\newcommand{\MvNrcClsOneOneLo}{\ensuremath{-}0.011}
\newcommand{\MvNrcClsOneOneHi}{+0.009}
\newcommand{\MvThingsClsOneTwo}{\ensuremath{-}0.002}
\newcommand{\MvThingsClsOneTwoLo}{\ensuremath{-}0.008}
\newcommand{\MvThingsClsOneTwoHi}{+0.005}
\newcommand{\MvNsessionClsOneTwo}{\ensuremath{-}0.019}
\newcommand{\MvNsessionClsOneTwoLo}{\ensuremath{-}0.034}
\newcommand{\MvNsessionClsOneTwoHi}{\ensuremath{-}0.002}
\newcommand{\MvNrawClsOneTwo}{0.000}
\newcommand{\MvNrawClsOneTwoLo}{\ensuremath{-}0.018}
\newcommand{\MvNrawClsOneTwoHi}{+0.019}
\newcommand{\MvNrcClsOneTwo}{\ensuremath{-}0.019}
\newcommand{\MvNrcClsOneTwoLo}{\ensuremath{-}0.034}
\newcommand{\MvNrcClsOneTwoHi}{\ensuremath{-}0.001}
\newcommand{\MvThingsClsOneThree}{\ensuremath{-}0.009}
\newcommand{\MvThingsClsOneThreeLo}{\ensuremath{-}0.017}
\newcommand{\MvThingsClsOneThreeHi}{\ensuremath{-}0.002}
\newcommand{\MvNsessionClsOneThree}{\ensuremath{-}0.017}
\newcommand{\MvNsessionClsOneThreeLo}{\ensuremath{-}0.033}
\newcommand{\MvNsessionClsOneThreeHi}{0.000}
\newcommand{\MvNrawClsOneThree}{\ensuremath{-}0.015}
\newcommand{\MvNrawClsOneThreeLo}{\ensuremath{-}0.029}
\newcommand{\MvNrawClsOneThreeHi}{+0.001}
\newcommand{\MvNrcClsOneThree}{\ensuremath{-}0.018}
\newcommand{\MvNrcClsOneThreeLo}{\ensuremath{-}0.034}
\newcommand{\MvNrcClsOneThreeHi}{0.000}
\newcommand{\MvThingsClsOneFour}{\ensuremath{-}0.009}
\newcommand{\MvThingsClsOneFourLo}{\ensuremath{-}0.017}
\newcommand{\MvThingsClsOneFourHi}{\ensuremath{-}0.002}
\newcommand{\MvNsessionClsOneFour}{\ensuremath{-}0.018}
\newcommand{\MvNsessionClsOneFourLo}{\ensuremath{-}0.033}
\newcommand{\MvNsessionClsOneFourHi}{\ensuremath{-}0.001}
\newcommand{\MvNrawClsOneFour}{\ensuremath{-}0.016}
\newcommand{\MvNrawClsOneFourLo}{\ensuremath{-}0.030}
\newcommand{\MvNrawClsOneFourHi}{\ensuremath{-}0.001}
\newcommand{\MvNrcClsOneFour}{\ensuremath{-}0.018}
\newcommand{\MvNrcClsOneFourLo}{\ensuremath{-}0.034}
\newcommand{\MvNrcClsOneFourHi}{0.000}
\newcommand{\MvThingsClsOneFive}{\ensuremath{-}0.008}
\newcommand{\MvThingsClsOneFiveLo}{\ensuremath{-}0.017}
\newcommand{\MvThingsClsOneFiveHi}{+0.001}
\newcommand{\MvNsessionClsOneFive}{\ensuremath{-}0.007}
\newcommand{\MvNsessionClsOneFiveLo}{\ensuremath{-}0.015}
\newcommand{\MvNsessionClsOneFiveHi}{+0.002}
\newcommand{\MvNrawClsOneFive}{\ensuremath{-}0.057}
\newcommand{\MvNrawClsOneFiveLo}{\ensuremath{-}0.083}
\newcommand{\MvNrawClsOneFiveHi}{\ensuremath{-}0.032}
\newcommand{\MvNrcClsOneFive}{\ensuremath{-}0.007}
\newcommand{\MvNrcClsOneFiveLo}{\ensuremath{-}0.016}
\newcommand{\MvNrcClsOneFiveHi}{+0.001}
\newcommand{\MvThingsClsOneSix}{\ensuremath{-}0.004}
\newcommand{\MvThingsClsOneSixLo}{\ensuremath{-}0.009}
\newcommand{\MvThingsClsOneSixHi}{+0.001}
\newcommand{\MvNsessionClsOneSix}{\ensuremath{-}0.005}
\newcommand{\MvNsessionClsOneSixLo}{\ensuremath{-}0.014}
\newcommand{\MvNsessionClsOneSixHi}{+0.003}
\newcommand{\MvNrawClsOneSix}{\ensuremath{-}0.005}
\newcommand{\MvNrawClsOneSixLo}{\ensuremath{-}0.013}
\newcommand{\MvNrawClsOneSixHi}{+0.004}
\newcommand{\MvNrcClsOneSix}{\ensuremath{-}0.005}
\newcommand{\MvNrcClsOneSixLo}{\ensuremath{-}0.014}
\newcommand{\MvNrcClsOneSixHi}{+0.003}
\newcommand{\MvThingsClsOneSeven}{\ensuremath{-}0.004}
\newcommand{\MvThingsClsOneSevenLo}{\ensuremath{-}0.009}
\newcommand{\MvThingsClsOneSevenHi}{+0.001}
\newcommand{\MvNsessionClsOneSeven}{\ensuremath{-}0.005}
\newcommand{\MvNsessionClsOneSevenLo}{\ensuremath{-}0.013}
\newcommand{\MvNsessionClsOneSevenHi}{+0.002}
\newcommand{\MvNrawClsOneSeven}{\ensuremath{-}0.004}
\newcommand{\MvNrawClsOneSevenLo}{\ensuremath{-}0.011}
\newcommand{\MvNrawClsOneSevenHi}{+0.003}
\newcommand{\MvNrcClsOneSeven}{\ensuremath{-}0.006}
\newcommand{\MvNrcClsOneSevenLo}{\ensuremath{-}0.013}
\newcommand{\MvNrcClsOneSevenHi}{+0.003}
\newcommand{\MvThingsClsOneEight}{\ensuremath{-}0.010}
\newcommand{\MvThingsClsOneEightLo}{\ensuremath{-}0.016}
\newcommand{\MvThingsClsOneEightHi}{\ensuremath{-}0.004}
\newcommand{\MvNsessionClsOneEight}{\ensuremath{-}0.008}
\newcommand{\MvNsessionClsOneEightLo}{\ensuremath{-}0.015}
\newcommand{\MvNsessionClsOneEightHi}{\ensuremath{-}0.001}
\newcommand{\MvNrawClsOneEight}{\ensuremath{-}0.043}
\newcommand{\MvNrawClsOneEightLo}{\ensuremath{-}0.061}
\newcommand{\MvNrawClsOneEightHi}{\ensuremath{-}0.024}
\newcommand{\MvNrcClsOneEight}{\ensuremath{-}0.009}
\newcommand{\MvNrcClsOneEightLo}{\ensuremath{-}0.016}
\newcommand{\MvNrcClsOneEightHi}{\ensuremath{-}0.002}
\newcommand{\MvThingsClsOneNine}{\ensuremath{-}0.003}
\newcommand{\MvThingsClsOneNineLo}{\ensuremath{-}0.007}
\newcommand{\MvThingsClsOneNineHi}{+0.001}
\newcommand{\MvNsessionClsOneNine}{\ensuremath{-}0.011}
\newcommand{\MvNsessionClsOneNineLo}{\ensuremath{-}0.019}
\newcommand{\MvNsessionClsOneNineHi}{\ensuremath{-}0.004}
\newcommand{\MvNrawClsOneNine}{\ensuremath{-}0.009}
\newcommand{\MvNrawClsOneNineLo}{\ensuremath{-}0.016}
\newcommand{\MvNrawClsOneNineHi}{\ensuremath{-}0.002}
\newcommand{\MvNrcClsOneNine}{\ensuremath{-}0.012}
\newcommand{\MvNrcClsOneNineLo}{\ensuremath{-}0.019}
\newcommand{\MvNrcClsOneNineHi}{\ensuremath{-}0.004}
\newcommand{\MvThingsClsTwoZero}{\ensuremath{-}0.003}
\newcommand{\MvThingsClsTwoZeroLo}{\ensuremath{-}0.007}
\newcommand{\MvThingsClsTwoZeroHi}{+0.001}
\newcommand{\MvNsessionClsTwoZero}{\ensuremath{-}0.011}
\newcommand{\MvNsessionClsTwoZeroLo}{\ensuremath{-}0.018}
\newcommand{\MvNsessionClsTwoZeroHi}{\ensuremath{-}0.004}
\newcommand{\MvNrawClsTwoZero}{\ensuremath{-}0.010}
\newcommand{\MvNrawClsTwoZeroLo}{\ensuremath{-}0.017}
\newcommand{\MvNrawClsTwoZeroHi}{\ensuremath{-}0.003}
\newcommand{\MvNrcClsTwoZero}{\ensuremath{-}0.011}
\newcommand{\MvNrcClsTwoZeroLo}{\ensuremath{-}0.019}
\newcommand{\MvNrcClsTwoZeroHi}{\ensuremath{-}0.004}
\newcommand{\MvThingsClsTwoOne}{+0.016}
\newcommand{\MvThingsClsTwoOneLo}{+0.001}
\newcommand{\MvThingsClsTwoOneHi}{+0.031}
\newcommand{\MvNsessionClsTwoOne}{\ensuremath{-}0.009}
\newcommand{\MvNsessionClsTwoOneLo}{\ensuremath{-}0.029}
\newcommand{\MvNsessionClsTwoOneHi}{+0.010}
\newcommand{\MvNrawClsTwoOne}{\ensuremath{-}0.020}
\newcommand{\MvNrawClsTwoOneLo}{\ensuremath{-}0.038}
\newcommand{\MvNrawClsTwoOneHi}{\ensuremath{-}0.001}
\newcommand{\MvNrcClsTwoOne}{\ensuremath{-}0.008}
\newcommand{\MvNrcClsTwoOneLo}{\ensuremath{-}0.027}
\newcommand{\MvNrcClsTwoOneHi}{+0.012}
\newcommand{\MvThingsClsTwoTwo}{+0.016}
\newcommand{\MvThingsClsTwoTwoLo}{+0.001}
\newcommand{\MvThingsClsTwoTwoHi}{+0.030}
\newcommand{\MvNsessionClsTwoTwo}{\ensuremath{-}0.010}
\newcommand{\MvNsessionClsTwoTwoLo}{\ensuremath{-}0.029}
\newcommand{\MvNsessionClsTwoTwoHi}{+0.009}
\newcommand{\MvNrawClsTwoTwo}{\ensuremath{-}0.015}
\newcommand{\MvNrawClsTwoTwoLo}{\ensuremath{-}0.034}
\newcommand{\MvNrawClsTwoTwoHi}{+0.005}
\newcommand{\MvNrcClsTwoTwo}{\ensuremath{-}0.008}
\newcommand{\MvNrcClsTwoTwoLo}{\ensuremath{-}0.027}
\newcommand{\MvNrcClsTwoTwoHi}{+0.011}
\newcommand{\MvThingsClsTwoThree}{\ensuremath{-}0.011}
\newcommand{\MvThingsClsTwoThreeLo}{\ensuremath{-}0.017}
\newcommand{\MvThingsClsTwoThreeHi}{\ensuremath{-}0.005}
\newcommand{\MvNsessionClsTwoThree}{\ensuremath{-}0.026}
\newcommand{\MvNsessionClsTwoThreeLo}{\ensuremath{-}0.036}
\newcommand{\MvNsessionClsTwoThreeHi}{\ensuremath{-}0.018}
\newcommand{\MvNrawClsTwoThree}{\ensuremath{-}0.008}
\newcommand{\MvNrawClsTwoThreeLo}{\ensuremath{-}0.024}
\newcommand{\MvNrawClsTwoThreeHi}{+0.006}
\newcommand{\MvNrcClsTwoThree}{\ensuremath{-}0.027}
\newcommand{\MvNrcClsTwoThreeLo}{\ensuremath{-}0.037}
\newcommand{\MvNrcClsTwoThreeHi}{\ensuremath{-}0.019}
\newcommand{\MvThingsClsTwoFour}{\ensuremath{-}0.014}
\newcommand{\MvThingsClsTwoFourLo}{\ensuremath{-}0.020}
\newcommand{\MvThingsClsTwoFourHi}{\ensuremath{-}0.009}
\newcommand{\MvNsessionClsTwoFour}{\ensuremath{-}0.025}
\newcommand{\MvNsessionClsTwoFourLo}{\ensuremath{-}0.035}
\newcommand{\MvNsessionClsTwoFourHi}{\ensuremath{-}0.016}
\newcommand{\MvNrawClsTwoFour}{\ensuremath{-}0.027}
\newcommand{\MvNrawClsTwoFourLo}{\ensuremath{-}0.038}
\newcommand{\MvNrawClsTwoFourHi}{\ensuremath{-}0.018}
\newcommand{\MvNrcClsTwoFour}{\ensuremath{-}0.026}
\newcommand{\MvNrcClsTwoFourLo}{\ensuremath{-}0.036}
\newcommand{\MvNrcClsTwoFourHi}{\ensuremath{-}0.017}
\newcommand{\MvThingsClsTwoFive}{\ensuremath{-}0.014}
\newcommand{\MvThingsClsTwoFiveLo}{\ensuremath{-}0.020}
\newcommand{\MvThingsClsTwoFiveHi}{\ensuremath{-}0.009}
\newcommand{\MvNsessionClsTwoFive}{\ensuremath{-}0.023}
\newcommand{\MvNsessionClsTwoFiveLo}{\ensuremath{-}0.031}
\newcommand{\MvNsessionClsTwoFiveHi}{\ensuremath{-}0.015}
\newcommand{\MvNrawClsTwoFive}{\ensuremath{-}0.022}
\newcommand{\MvNrawClsTwoFiveLo}{\ensuremath{-}0.030}
\newcommand{\MvNrawClsTwoFiveHi}{\ensuremath{-}0.014}
\newcommand{\MvNrcClsTwoFive}{\ensuremath{-}0.024}
\newcommand{\MvNrcClsTwoFiveLo}{\ensuremath{-}0.033}
\newcommand{\MvNrcClsTwoFiveHi}{\ensuremath{-}0.016}
\newcommand{\MvThingsClsTwoSix}{\ensuremath{-}0.013}
\newcommand{\MvThingsClsTwoSixLo}{\ensuremath{-}0.022}
\newcommand{\MvThingsClsTwoSixHi}{\ensuremath{-}0.006}
\newcommand{\MvNsessionClsTwoSix}{\ensuremath{-}0.061}
\newcommand{\MvNsessionClsTwoSixLo}{\ensuremath{-}0.079}
\newcommand{\MvNsessionClsTwoSixHi}{\ensuremath{-}0.043}
\newcommand{\MvNrawClsTwoSix}{+0.007}
\newcommand{\MvNrawClsTwoSixLo}{\ensuremath{-}0.018}
\newcommand{\MvNrawClsTwoSixHi}{+0.030}
\newcommand{\MvNrcClsTwoSix}{\ensuremath{-}0.062}
\newcommand{\MvNrcClsTwoSixLo}{\ensuremath{-}0.079}
\newcommand{\MvNrcClsTwoSixHi}{\ensuremath{-}0.044}
\newcommand{\MvThingsClsTwoSeven}{\ensuremath{-}0.022}
\newcommand{\MvThingsClsTwoSevenLo}{\ensuremath{-}0.030}
\newcommand{\MvThingsClsTwoSevenHi}{\ensuremath{-}0.015}
\newcommand{\MvNsessionClsTwoSeven}{\ensuremath{-}0.058}
\newcommand{\MvNsessionClsTwoSevenLo}{\ensuremath{-}0.076}
\newcommand{\MvNsessionClsTwoSevenHi}{\ensuremath{-}0.040}
\newcommand{\MvNrawClsTwoSeven}{\ensuremath{-}0.052}
\newcommand{\MvNrawClsTwoSevenLo}{\ensuremath{-}0.068}
\newcommand{\MvNrawClsTwoSevenHi}{\ensuremath{-}0.035}
\newcommand{\MvNrcClsTwoSeven}{\ensuremath{-}0.059}
\newcommand{\MvNrcClsTwoSevenLo}{\ensuremath{-}0.077}
\newcommand{\MvNrcClsTwoSevenHi}{\ensuremath{-}0.041}
\newcommand{\MvThingsClsTwoEight}{\ensuremath{-}0.022}
\newcommand{\MvThingsClsTwoEightLo}{\ensuremath{-}0.030}
\newcommand{\MvThingsClsTwoEightHi}{\ensuremath{-}0.014}
\newcommand{\MvNsessionClsTwoEight}{\ensuremath{-}0.057}
\newcommand{\MvNsessionClsTwoEightLo}{\ensuremath{-}0.074}
\newcommand{\MvNsessionClsTwoEightHi}{\ensuremath{-}0.040}
\newcommand{\MvNrawClsTwoEight}{\ensuremath{-}0.053}
\newcommand{\MvNrawClsTwoEightLo}{\ensuremath{-}0.069}
\newcommand{\MvNrawClsTwoEightHi}{\ensuremath{-}0.037}
\newcommand{\MvNrcClsTwoEight}{\ensuremath{-}0.058}
\newcommand{\MvNrcClsTwoEightLo}{\ensuremath{-}0.075}
\newcommand{\MvNrcClsTwoEightHi}{\ensuremath{-}0.040}
\newcommand{\FirstDownloadSubjTwo}{2026-11-17 08:21:07 UTC}
\newcommand{\FirstDownloadRawFortyEight}{2026-11-22 10:46:30 UTC}

\title{Whether untrained networks beat trained ones at V1 depends on a normalisation convention: evidence from two fMRI datasets}
\author{Nils Leutenegger\\Independent researcher, Zug, Switzerland\\\texttt{nils.leutenegger@gmail.com}}
\date{}

\begin{document}
\maketitle

\begin{abstract}
Untrained convolutional networks have been reported to rival or beat trained networks as models of
primary visual cortex (V1)~\citep{truzzi2020, kazemian2025}, and we had reported such an advantage
in THINGS fMRI~\citep{evalres}. Under a pre-specified ridge readout, the trained network was ahead at
every pooling grid in THINGS and in NSD (\NsdRidgeVoneGFourNeg{} of 8 subjects), also when voxel
responses were centred within runs. Representational similarity analysis (RSA) in THINGS depended on
analysis choices (\MultiDistinct{} analyses: intervals above 0 in \MultiDistinctPos, below 0 in
\MultiDistinctNeg, including 0 in \MultiDistinctIncl); in NSD with its standard normalisation, none
of \MvNsessionDistinct{} analyses favoured the untrained network. The untrained advantage appears
when a response profile common to all trials is left in the voxel patterns and model units are not
centred, under correlation distance: removing the session-mean profile in THINGS reduced it from
$\DecThreeVoneVZero$ to $\DecThreeVoneVOne$; with raw betas it reached $\CvTsevTrawVoneRawCorr$ in
THINGS, appeared in the THINGS test images, and appeared in NSD ($\CvNsdNrawVoneRawCorrKOne$,
\CvNsdNrawVoneRawCorrKOnePos{} of 8 subjects, a pre-specified prediction), where it was confined to
this one of \MvNrawDistinct{} analyses. With the profile left in, both representational
dissimilarity matrices are dominated by per-image response amplitude, which in V1 tracks contrast
energy, and the untrained network's matrix is more amplitude-dominated; the amplitude itself is
matched equally by both networks; removing the amplitude structure from both matrices removes the
advantage in NSD ($\PbNsdNrawVoneKOneNone$ to $\PbNsdNrawVoneKOneSum$, \PbNsdNrawVoneKOneSumLe{} of 8
subjects). A luminance-and-contrast baseline outperformed every network under ridge in both datasets
(point estimates); the networks reached \CeilVoneFracMin--\CeilVoneFracMax\% of the NSD noise-ceiling
lower bound. \FailedPredictions{} of \TotalPredictions{} pre-specified predictions were not met. We
recommend centring or normalising voxel responses within session, reporting the sensitivity of RSA
rankings to this choice and to model-unit centring, and replicating on a second dataset.

\end{abstract}

\section{Introduction}
Comparisons between artificial networks and neural data have become a standard way to ask which
models capture the computations of visual cortex. Representational similarity analysis (RSA) is one
of the most widely used tools for this purpose: it summarises each system by the dissimilarities
between its responses to a set of stimuli and compares these summaries directly, without fitting a
mapping \citep{kriegeskorte2008}. A recurring and surprising result is that networks with random,
untrained weights can rival trained networks as models of visual cortex, in some reports exceeding
them \citep{truzzi2020, kazemian2025}. If true, such results bear on how much of early visual
representation follows from architecture alone and how much from learning.

We reported a result of this kind ourselves. In THINGS fMRI \citep{hebart2023}, an untrained network
with uncalibrated batch-normalisation statistics exceeded a backpropagation-trained network at V1
when images were evaluated at 224\,px, but not at 32\,px, the resolution at which the networks had
been trained \citep{evalres}; a related claim appears in \citet{leutenegger2026}. Both analyses used
a single convention: globally pooled features, correlation distance between units, and voxel
responses z-scored over all trials of the experiment. This paper traces that claimed property of
untrained networks to a normalisation convention.

Correlation distance across units is not invariant to constant offsets of individual units over the
stimulus set: when units have different mean responses, the dissimilarity between two stimuli depends
in part on how strongly each pattern is aligned with the shared profile of unit means. The same holds
for voxels. Methodological work has shown that removing or keeping the mean of the measured patterns
changes what RSA measures: demeaning can produce ``representational confusion''
\citep{ramirez2017}, the regional-mean response profile carries information that standard RSA
discards \citep{taylor2025}, and averaging over neurons within voxels amplifies activity shared by
all sampled neurons, a distortion that mean removal largely corrects \citep{bossio2025}. Centring is
also explicit in related similarity measures \citep{kornblith2019}, and the choice of dissimilarity
measure affects the reliability of RSA \citep{walther2016, diedrichsen2017}. Whether such choices
reverse the ranking of candidate models has rarely been tested.

We make six contributions. (i)~A pre-specified ridge readout on the full feature map, in THINGS and,
with the same plan, in the Natural Scenes Dataset \citep[NSD;][]{allen2022}, places the trained network
ahead at every pooling grid. (ii)~A multiverse of 36 combinations of unit normalisation, distance
measure and voxel normalisation in THINGS \citep{steegen2016} shows that the sign of the V1 contrast
depends on the analysis. (iii)~A replication in NSD (8 subjects) finds no analysis in which the
untrained network is ahead under the NSD standard normalisation. (iv)~We identify the convention that
creates the advantage: a voxel response profile common to all analysed trials, which is present in unnormalised betas and, in
THINGS, also after normalising over all trials of the experiment, because all 720 images were shown in
one session; a pre-specified prediction that the advantage appears in NSD when this profile is left in
(raw betas) was confirmed. (v)~We describe the
mechanism as far as it is identified: with the profile left in, both representational dissimilarity
matrices (RDMs) are dominated by per-image response amplitude, and the untrained network's RDM more so.
(vi)~We give recommendations for model--brain comparisons. Every pre-specified prediction is reported,
including the \FailedPredictions{} that were not met, and we revise the corresponding claims of our
earlier work.

\section{Methods}
\subsection{THINGS data}
\label{sec:data}

We used the THINGS-fMRI data set \citep{hebart2023} of 3 subjects. Single-trial
betas were taken from the ROI masks of the voxel metadata (V1; LOC as the union of lLOC
and rLOC) and z-scored per voxel across all 9{,}840 trials of the subject before stimulus
selection. The 720 evaluation stimuli are one exemplar each of 720 distinct concepts,
the same images for all subjects (presentation order and run assignment identical in sub-01 and
sub-02, different in sub-03); all are single-presentation \texttt{train} trials
from one session per subject.

Voxel reliability was measured on the 100 \texttt{test} images, shown 12 times each,
identical across subjects and disjoint from the 720. Per voxel, the Pearson
correlation between the means of two halves of 6 repetitions was averaged over all 462
balanced splits and Spearman--Brown corrected, $r_{\mathrm{SB}} = 2r/(1+r)$. A voxel is
reliable if $r_{\mathrm{SB}} > 0.2$, a threshold fixed in advance for all conditions;
measured on other images, selection cannot leak into fits on the 720. V1 contained \NReliableSubOne,
\NReliableSubTwo{} and \NReliableSubThree{} reliable voxels in sub-01, sub-02 and
sub-03. Ridge scores use reliable V1 voxels; RSA uses all voxels of the ROI.

RDMs were compared on cross-run pairs only: pairs of distinct stimuli shown in different
runs in every subject, 210{,}205 of the 258{,}840 pairs of the full set. In every subject
the 720 fall into 10 runs of 72; sub-01 and sub-02 saw them in identical run order, so
same-run pairs share a run component, and the restriction removes it \citep{evalres}.

\subsection{Models}

\paragraph{CIFAR-trained CNNs.} We evaluated the 30 checkpoints of \citet{evalres}:
6 conditions $\times$ 5 seeds (42, 123, 456, 789, 1337) of one convolutional
architecture, trained on CIFAR-10 \citep{krizhevsky2009} at 32\,px, the training resolution. The
conditions are Backprop, Feedback Alignment (FA) \citep{lillicrap2016}, Predictive Coding (PC)
\citep{rao1999,whittington2017}, STDP \citep{bi1998,masquelier2007} and two untrained networks: untrained (uncalibrated BN
statistics), with BatchNorm running statistics left at initialisation, and untrained
(BN-calibrated), with running statistics set by 50 forward-only passes of CIFAR batches
at 32\,px. ``Untrained'' alone means the uncalibrated variant, as in the published
headline.

The V1 layer is Conv1, the output of the whole first block (convolution, BatchNorm, ReLU,
$2 \times 2$ max pooling; 32 channels, map side half the input size); LOC is mapped to
Conv3. Features were extracted in eval mode with the original code, in two
input arms. NATIVE: the image is resized and centre-cropped
to $p \in \{32, 64, 96, 128, 160, 224\}$\,px. UPSAMPLED: the image is resized and
cropped to 32\,px and then resized to $p$, so its content is limited to 32\,px; we used
it at $p = 224$. Both arms apply the CIFAR channel normalisation.

\paragraph{ImageNet models.} Six torchvision models \citep{paszke2019, marcel2010}: alexnet,
vgg11\_bn, resnet18, resnet50, convnext\_tiny and vit\_b\_16, in eval mode. Trained
models use the IMAGENET1K\_V1 weights. Untrained models were built without weights after
\texttt{torch.manual\_seed($s$)}, $s \in \{0, 1, 2\}$, with BatchNorm running statistics
at their defaults (mean 0, variance 1); convnext\_tiny and vit\_b\_16 use LayerNorm. RGB images
were resized (shorter side) and centre-cropped to $R \in \{112, 224, 448\}$ (training
resolution 224) and ImageNet-normalised; vit\_b\_16 accepts only 224. Layers (V1 / LOC),
fixed before the run: alexnet
\texttt{features[1]} / \texttt{features[12]}; vgg11\_bn \texttt{features[2]} / last max
pooling of \texttt{features}; resnet18 and resnet50 \texttt{maxpool} / \texttt{layer4};
convnext\_tiny \texttt{features[0]} / \texttt{features[7]}; vit\_b\_16
\texttt{encoder.layers[0]} / \texttt{encoder.layers[-1]}. Outputs were averaged over
space, or over tokens without the class token.

\subsection{RSA convention}
\label{sec:rsa}

Unless stated otherwise, RSA follows \citet{evalres}. The model RDM is the
correlation distance ($1 -$ Pearson $r$ across units) between the globally average-pooled
(GAP) features of two stimuli; the fMRI RDM is the correlation distance between their
beta patterns over all ROI voxels. RSA is the Spearman correlation between the two RDMs
over the cross-run pairs, per subject, then averaged. A
contrast is the mean over seeds of the per-seed difference between two conditions; for
ImageNet models it is untrained (mean over the 3 initialisation seeds) minus trained.

Uncertainty is over stimuli: the 1000 bootstrap resamples of \citet{evalres} (720
stimuli drawn with replacement, seed 20261002). Each draw recomputes the cross-run pair
set on the resampled stimuli, dropping pairs of a stimulus with itself, and every
difference, slope and ratio is formed within the draw (paired). Intervals are 95\%
percentile intervals. They describe stimulus variation in 3 subjects, not the population. Unit means and SDs, the feature transformations $F$, the voxel normalisations $B$, the mean profile and the norms are estimated once on the 720 stimuli and held fixed across resamples; the bootstrap resamples pairs of fixed RDMs. Re-estimating them within 200 resamples left the classification of three cells unchanged (Appendix~\ref{app:bootnorm}).

\subsection{Analysis multiverse}
\label{sec:mvmethods}

Three factors were crossed. Feature normalisation $F$, per unit over the 720 stimuli:
raw; center (mean subtracted); scale (divided by the SD, not centred); z (both). Units
with SD below $10^{-8}$ are set to 0. Distance $D$: correlation, cosine or euclidean,
applied to features and betas alike. fMRI normalisation $B$: paper (betas
as in Section~\ref{sec:data}), center720 (centred per voxel over the 720 stimuli) and
z720 (z-scored per voxel over the 720). The cell (raw, correlation, paper) is the
convention of \citet{evalres}. All $4 \times 3 \times 3 = 36$ cells were computed for
the V1 contrast (Conv1, untrained $-$ Backprop) and the LOC contrast (Conv3, Backprop
$-$ untrained). The resolution slope is
$S = \mathrm{Gap}(\text{NATIVE } 224) - \mathrm{Gap}(\text{NATIVE } 32)$, with Gap the V1
contrast. As a check, euclidean RDMs under $F$ = raw and center must coincide. For
ImageNet models, all 36 cells were computed as point estimates and four were
bootstrapped: (raw, correlation, paper), (z, correlation, paper), (raw, euclidean,
paper) and (center, correlation, center720); their slope is
$S_{\mathrm{gen}} = \mathrm{Gap}(448) - \mathrm{Gap}(224)$.

\subsection{Ridge readout}
\label{sec:ridge}

Encoding models used dual ridge regression \citep{hoerl1970}, with the kernel
$K = XX^{\top}$ eigendecomposed once per split. The Conv1 maps were
pooled to a $g \times g$ grid by adaptive average pooling, $g \in \{1, 2, 4, 8\}$, giving
$32g^2$ features independent of resolution; $g = 1$ is GAP, and larger $g$ retains
spatial structure. Features were standardised with statistics of the training rows of
each split only.

Cross-validation is nested: 8 outer folds and 5 inner folds within each outer training
set, from a fixed partition (seed 20260818); with one exemplar per concept, folds are
concept-grouped. Per voxel and outer fold, $\alpha$ was
chosen from 29 log-spaced values between $10^{-6}$ and $10^{8}$ by the minimum mean
squared error over the inner folds, and the model was refit on the outer training set.
The regularisation is selected by mean squared error, not by Pearson $r$, the reported score. In simulations with the dimensions of the primary data set, selection by $r$ placed a substantial share of the selections at reliable voxels on the lowest grid value in every simulated population that contained reliable voxels, irrespective of the strength of the signal, whereas selection by mean squared error placed none there (Phase~2b report in the repository); because $r$ does not depend on the scale of the predictions, which $\alpha$ mainly sets, its inner-fold optimum was not localised on the grid. The selected $\alpha$ therefore need not maximise $r$.
Out-of-fold predictions were pooled over the outer folds and scored by one Pearson $r$
per voxel over all 720 stimuli. When the selected regularisation shrinks the predictions
of a voxel towards the training mean, this pooled correlation is negative in
expectation, so features without signal yield negative rather than zero scores. A model that predicts the mean of the training rows of each fold scored $\FFloorMean$ (mean over subjects); the 97.5th percentile of a permutation null for the seed-0 networks lay between $\FPermPNineSevenFiveMin$ and $\FPermPNineSevenFiveMax$, and observed scores were above it in sub-01 and sub-02 (Appendix~\ref{app:ridgerobust}). The score is the median $r$ over reliable V1 voxels,
averaged over subjects, with contrasts formed within seed.

A low-edge selection at a reliable voxel is logged, not fatal (amendment of 2026-10-03).
With diagnostic values $10^{-7}$, $10^{-8}$ and $10^{-9}$
added, it is a valid plateau if the inner-fold MSE improves by less than a relative
$10^{-3}$ on its value at $10^{-6}$; otherwise the cell is excluded. The gate never
changes a score.

The pre-specified analysis used the 720 single-trial betas at NATIVE 224\,px, all
checkpoints and subjects, and all four $g$; the $g = 2$ features were extracted on a CPU,
subject to a downstream equivalence gate (amendment of 2026-10-13). An exploratory fit
at $g = 1$ covered all NATIVE resolutions and UPSAMPLED 224.

Two controls used the identical readout. The control band consists of low-level banks
pooled to the same grid: pixels; the luminance-and-contrast bank, two feature maps (luminance, Rec.\,709, $Y = 0.2126R + 0.7152G + 0.0722B$, and local RMS contrast in a $9 \times 9$ window) pooled to the $g \times g$ grid, i.e.\ $2g^2$ features (2 at $g = 1$, 128 at $g = 8$); absolute responses of a Gabor bank (4 orientations $\times$
3 spatial frequencies); and, as negative test, Backprop seed-0 features with permuted
stimulus assignment. The spatial-correspondence control permutes the $g \times g$ cells
differently for every image (seed 20260820), keeping dimensionality and per-channel
distributions; it was run for the seed-0 untrained and Backprop checkpoints at
$g \in \{2, 4, 8\}$.

\subsection{Mean-profile components and decomposition}
\label{sec:gainmethods}

For a stimulus $\times$ unit matrix $X$ (720 rows), let $\mu$ be the unit means over the 720
stimuli, $\mu_c = \mu - \overline{\mu}$ the mean profile with its mean over units removed,
$u = \mu_c / \lVert \mu_c \rVert$, and $\tilde{x}_i$ row $i$ centred across units. The
mean-profile alignment of stimulus $i$ is the cosine between its centred pattern and the
mean profile,
\begin{equation}
  a_i = \frac{\langle \tilde{x}_i, u \rangle}{\lVert \tilde{x}_i \rVert}.
\end{equation}
Model alignment was computed on raw features, voxel alignment on all ROI voxels under
$B$ = paper. $c_{\mathrm{align}}$ is the Pearson correlation over the 720 stimuli between
model and V1 voxel alignment, averaged over subjects. The alignment RDM is
$\lvert a_i - a_j \rvert$. Partial RSA regresses the ranks of the model and of the fMRI
RDM, per subject and draw, on an intercept and the ranks of their own alignment RDM and
correlates the residuals (Pearson). Alignments were described against stimulus mean
luminance and RMS contrast, the per-image mean and SD of Rec.\,601 luminance
($Y = 0.299R + 0.587G + 0.114B$); over the 720 stimuli at 224\,px, the image means of the
two luminance definitions correlate at Pearson \LumDefPearson{} (Spearman
\LumDefSpearman).

The projection coefficient $p_i = \langle \tilde{x}_i, u \rangle = a_i \lVert \tilde{x}_i \rVert$
and the remainder $r_i = \tilde{x}_i - p_i u$ split each centred pattern into a part along
the mean profile and a part orthogonal to it. We compare three variants: ``full'', the
correlation distance on $x_i$ (the convention of \citet{evalres}); ``orthogonal'', the
correlation distance on $r_i$; and ``parallel'', the RDM $\lvert p_i - p_j \rvert$. The beta
patterns are decomposed in the same way (voxel means over the 720 stimuli, $B$ = paper),
and each model component is compared with the same component of the brain.

The norm and angular components start from $X_c$, $X$ with every unit centred over the
720 stimuli ($F$ = center; for $F$ = z, units are also scaled to SD 1). With
$n_i = \lVert X_{c,i} \rVert$ and $\cos\theta_{ij}$ the cosine between rows $i$ and $j$,
\begin{equation}
  \lVert X_{c,i} - X_{c,j} \rVert^2 = (n_i - n_j)^2 + W_{ij}, \qquad
  W_{ij} = 2\, n_i n_j \,(1 - \cos\theta_{ij}).
\end{equation}
We report the norm RDM $N_{ij} = \lvert n_i - n_j \rvert$, the angular component
$A_{ij} = 1 - \cos\theta_{ij}$, the unweighted angular dissimilarity of the patterns
centred over stimuli, and the norm-weighted angular term $W_{ij}$. The identity holds for
$W$, not for $A$ alone; we therefore report $N$, $A$ and $W$ separately, and the RSA of the
euclidean RDM is not the sum of the RSA of its components. The literature uses
``representational geometry'' more broadly, for the distance structure including the
norms; ``angular component'' here denotes only $A$. The betas are decomposed in the same
way, and the RSA of a component compares the model and fMRI RDMs of that component. The
identity was checked numerically. These names label quantities defined here; they are
not claims about physiological gain.

\subsection{Analysis plan, amendments and inference}
\label{sec:inference}

\paragraph{Pre-specified primary analysis.}
The ridge-readout analysis (Section~\ref{sec:ridge}) followed a written analysis plan
with a frozen configuration (schema version~2, configuration hash \texttt{206fa875cca5}),
dated 20~August~2026. The plan was not publicly registered; its dating rests on in-text
date stamps and the configuration hash, and the files were placed under version control
on 3~October~2026. We therefore describe its criteria as fixed before the analysis.
The plan asked whether the difference between the untrained and the backpropagation-trained network at V1 reported in \citet{evalres} survives a fitted readout on the full Conv1
feature map. It fixed the layer, the voxel selection (split-half reliability above~0.2),
the grid ladder $g \in \{1, 2, 4, 8\}$, the cross-validation scheme and seed, the
regularisation grid and the aggregation (median over voxels, mean over subjects). The
unit of inference was the training seed ($n = 5$), tested with a one-sided exact
sign-flip test over all 32 sign assignments. The effect was to count as surviving only
if it held at $g \geq 4$. The minimal detectable effect $\Delta_{\min}$ was derived from
the pilot's seed-level standard deviation by simulating this test. The reference effect
was the value reported in version~1 of \citet{evalres}
($+\PreregReferenceEffect \pm \PreregReferenceSem$ at 224\,px); the current value is
$+\VoneGapRawTwoTwoFour$ [$\VoneGapRawTwoTwoFourLo$, $\VoneGapRawTwoTwoFourHi$], and the verdict does not depend on this choice. The pre-specified rule labels the outcome ``The effect disappears'' when $\Delta_{\min}$ is below the reference effect; this is a decision label under the plan, not an equivalence test. Every seed contrast was negative (Section~\ref{sec:signflip}).

\paragraph{Amendments.}
Three amendments were adopted before any fit to THINGS data, on the basis of
simulations with the dimensions of the primary data set: a hard assertion against
selecting the lowest regularisation value for reliable voxels, selection of the
regularisation by mean squared error with Pearson~$r$ reported, and a widening of the
regularisation grid. Four further decisions were taken after data had been seen, and we
list them as such. (i) The regularisation was kept at minimum-MSE selection rather than
a tolerance rule, after a tolerance measurement was known; this adds no degree of
freedom. (ii) After the pilot aborted on the assertion, its remedy clause was amended
to require a diagnosis before any change. (iii) The abort was then replaced by a post-hoc
gate that classifies each lower-boundary selection as a plateau or a true optimum below
the grid; only the latter would invalidate a cell. This was decided after a diagnosis
showed that all observed boundary selections were plateaus, and it leaves all scores
unchanged. No cell was excluded in any analysis. (iv) Features for $g = 2$ were
extracted on CPU, whereas all other features came from a GPU extraction. The
pre-specified reproduction gate (absolute tolerance $10^{-6}$) failed because of
floating-point differences; we replaced it by a downstream-equivalence gate, which
passed (maximum absolute difference in median $r$ of \GtwoGateMaxDelta{} at $g = 4$).
Four operational details not stated in the plan (significance level $0.05$; ``holds at
$g \geq 4$'' as $p \leq 0.05$ at both $g = 4$ and $g = 8$; the decision based on the
largest $\Delta_{\min}$ at $g \geq 4$; averaging over subjects only for complete cells)
were committed before the primary run. An implementation error initially derived
$\Delta_{\min}$ from the primary analysis's own seed variability instead of the pilot's;
we report the pre-specified value and the original value as supplementary. The verdict
is the same under both.

\paragraph{Blinding and deviations.}
The primary analysis was not blind. Results for subject~1 at $g \in \{1, 4, 8\}$ and for
subjects~2 and~3 at $g = 1$ were known from earlier steps; only $g = 2$ and subjects~2
and~3 at $g \geq 4$ were new. Several pre-specified instruments were not implemented:
a capacity sweep, a PLS readout with 25 components, a ridge fit on the repeated-image
test set, and a leave-one-subject-out bound for the ridge scores. The image-wise cell
shuffle was run for one seed at 224\,px only, and the grid ladder only at 224\,px.

\paragraph{Exploratory analyses.}
All analyses in Sections~\ref{sec:standardization}--\ref{sec:gaingeometry} lie outside
the pre-specified plan, which excluded, among others, other layers and band-pass
decompositions. They were developed sequentially. For each step, predictions and
decision rules were written down and committed to version control before the step's
scripts were run; the commit timestamps, not the dates in file names, are authoritative
and are listed in the repository. We report every pre-stated prediction, including the
\FailedPredictionsVone{} of version~1 that were not met (Appendix~\ref{app:failed}). Run-grouped cross-validation
folds were an additional sensitivity analysis and were not part of the plan.

\paragraph{Inference and its limits.}
We distinguish four kinds of statements. (a)~The pre-specified primary analysis, whose decision rule was fixed before its run. (b)~Exploratory analyses with pre-stated rules: each rule was committed before its step was run, and we report it as met or not met. The rules of later steps were written after the results of earlier steps were known, and we chose the thresholds ourselves, so ``met'' means consistent with a pre-stated rule, not independently confirmed. The analyses of Section~\ref{sec:ablation} and Appendices~\ref{app:ridgerobust} and~\ref{app:bootnorm} (ablation of the mean-profile component, ridge robustness and the bootstrap with re-estimated normalisation) were added in response to criticism of an earlier draft; their rules were committed before their scripts (amendment dated 2026-10-16, committed on 2026-10-18 as \texttt{74853dc}; the commit timestamp is authoritative). The count of distinct multiverse analyses and the ablation on each side (Appendix~\ref{app:ablation2x2}) followed further criticism of the revised draft; their rules were committed before their scripts (amendment dated 2026-10-18, committed as \texttt{ac59694}). (c)~Descriptive observations without a rule, which we label as such. (d)~The follow-up analyses of version~2 (Section~\ref{sec:followup}), each again with predictions and decision rules committed before its scripts and, for new data, before the data were loaded; their criteria were written after the results of version~1 and of earlier follow-up steps were known.

The intervals have different sources in different parts of the paper. For RSA, we resampled stimuli (1000 bootstrap resamples, seed 20261002) and computed differences between conditions from the same resamples; intervals are 95\% percentile intervals. They describe variation over the 720 stimuli, for these three subjects and these networks; training seeds are averaged within each resample and are not part of the intervals. For ridge scores, the unit of inference is the training seed ($n = 5$). With 32 sign assignments, the smallest attainable one-sided $p$ is $1/32 \approx 0.031$ and the smallest two-sided $p$ is $2/32 = 0.0625$, so the test cannot support a two-sided claim at the 5\% level; we use it, as pre-specified, for the positive direction only. Seed standard deviations do not include variation over stimuli, voxels or subjects. For each ImageNet model we used one set of trained weights, and counts such as ``5 of 6 models'' are descriptive. The 36 cells of the multiverse share the same stimuli and resamples and are strongly dependent; intervals are not corrected for multiple comparisons, and counts of cells are descriptive. With three THINGS and eight NSD subjects, the results describe these subjects and do not generalise to a population of observers (Table~\ref{tab:evidence}). Code, configuration, amendments and the full commit history are available in the accompanying repository.\footnote{\url{https://github.com/nilsleut/ridge-readout}}

\subsection{NSD}
\label{sec:nsdmethods}

\paragraph{Data.} The Natural Scenes Dataset \citep{allen2022} contains 7T fMRI of 8 subjects who
viewed natural scenes. We used the 515 images that every subject saw three times; all their trials lie
in sessions 1--30. Single-trial betas (\texttt{betas\_fithrf\_GLMdenoise\_RR}, 1.8\,mm) were divided by
300, which gives percent signal change, and z-scored per voxel within session, the convention of the
NSD documentation (N-session, Section~\ref{sec:convmethods}). Voxels invalid in any session were
excluded (in one subject). The image order of the features equals that of the betas for
every subject.

\paragraph{ROIs.} V1 is the union of V1v and V1d of the population-receptive-field ROIs. NSD provides no
object-selective ROI; we constructed LO-obj, fixed before any RSA, as the intersection of the
\texttt{lateral} stream ROI with voxels whose \texttt{floc\_objectstval} exceeds 0, the threshold
convention of the manual for the floc ROIs. The underlying contrast is the objects domain (car,
instrument) against all other categories of the functional localiser (characters, bodies, faces,
places). The \texttt{lateral} stream ROI alone was reported as a sensitivity variant. RSA used all
voxels of the ROI (Appendix~\ref{app:nsd}, Table~\ref{tab:nsdsubjects}).

\paragraph{Patterns and pairs.} For $k = 3$ each image is represented by the mean of its three
presentations (primary); for $k = 1$ each presentation is used separately and the three RSA values
are averaged, excluding the pairs whose chosen presentations fall in the same session and run.
Model RDMs are computed as for THINGS (Section~\ref{sec:rsa}) from the own CNN at Conv1, 224\,px, and
from the ImageNet models with the layers of Section~\ref{sec:data}; RSA is the Spearman correlation
over all pairs.

\paragraph{Ridge.} The pre-specified ridge analysis of Section~\ref{sec:ridge} was applied unchanged
(nested cross-validation 8 $\times$ 5, the same $\alpha$ grid and selection, pooled out-of-fold $r$,
median over reliable voxels, the lower-edge gate), with the $k = 3$ responses as targets and
$g \in \{1, 4, 8\}$. A voxel is reliable if the mean $r$ of the three pairwise split correlations over
the 515 images, Spearman--Brown corrected as $3r/(1+2r)$, exceeds 0.2.

\paragraph{Inference.} Per subject, the contrast is averaged over seeds. Across subjects we used a
one-sided sign-flip test over the 8 subject means (256 assignments) in the predicted direction, and
a stimulus bootstrap of the 8-subject mean (1000 resamples, seed 20261002, the same image draws for
every subject). With 8 subjects the smallest attainable one-sided $p$ is $1/256$. The noise ceiling is
the leave-one-subject-out lower bound on the same pairs.

\subsection{Voxel normalisation conventions}
\label{sec:convmethods}

The THINGS betas of version~1 were z-scored per voxel over all 9{,}840 trials of a subject; the NSD
betas are z-scored per voxel within session. We compared three conventions in each data set
(paper cell unless stated). THINGS: T-raw, the single-trial betas as stored in the THINGS-fMRI files
(not normalised by the data set: voxel means differ from 0 and SDs from 1, also within sessions;
the absolute unit is not documented); T-global, z-scored per voxel over all trials of the
subject, the convention of version~1 (reproduced exactly); T-session, z-scored per voxel over the
trials of one session. NSD: N-raw, the betas in percent signal change; N-global, z-scored per voxel
over all trials of sessions 1--30 (pooled from the session means and SDs); N-session, z-scored within
session (the NSD standard, reproduced exactly from the stored raw betas in every session).

For THINGS-720, whose 720 images lie in one session of 820 trials per subject, let $m$ be the voxel
mean over the 820 trials (the session-mean profile), $\mu_r$ the mean over the 82 trials of run
$r$, and $d_r = \mu_r - m$ the run deviation, split into its part along $\hat m = m / \lVert m \rVert$,
$d_{r\parallel} = (d_r \cdot \hat m)\, \hat m$, and the remainder $d_{r\perp} = d_r - d_{r\parallel}$.
Run centring subtracts $\mu_r$ from every trial of run $r$, i.e.\ it removes $m$ and $d_r$. The
decomposition compares V0 (nothing removed, the paper cell), V1 ($m$ removed), V2 ($d_r$ removed),
V3 (both, run centring), V4 (V3 with $d_{r\parallel}$ re-added) and V5 (V3 with $d_{r\perp}$
re-added). Removing $m$ is equivalent to centring each voxel over the trials of the session.

Partial RSA with image-statistic, position and predecessor covariates (steps 44, 45, 47) is defined in Appendix~\ref{app:why}.

\subsection{Image statistics and response amplitude}
\label{sec:ampmethods}

\paragraph{Image statistics.} Images were resized and centre-cropped to 224\,px and converted to grey
(Rec.\,709). $L$ is the mean grey value and $C$ its SD (the definitions of Section~\ref{sec:gainmethods}
use Rec.\,601; both correlate almost perfectly). Contrast energy CE is the mean over pixels and 16
filters of the squared response of a zero-mean Gabor bank (4 orientations $\times$ 4 spatial
frequencies, 0.05--0.40 cycles/px, cosine phase, $\sigma = 1/(3 f)$, $41 \times 41$ kernels, reflect
padding). OFF and ON are the mean of $\max(0, \mathrm{ref} - x)$ and of $\max(0, x - \mathrm{ref})$
over pixels, and DD $=$ OFF$/($OFF$+$ON$)$. With the image's own mean as reference OFF equals ON for
every image, so the reference is the mean grey over all images of the data set; DD is then close to a
mirror of $L$ (Spearman \ImgSpearLDD{} in THINGS-720, \ImgSpearLDDNsd{} in NSD).

\paragraph{Response amplitude.} The brain gain $g_i$ is the cosine between pattern $i$ (all voxels of
the ROI, under the given convention) and the mean over the analysed patterns; the model gain $g^M_i$
is the cosine between the raw GAP feature vector of image $i$ and the mean feature vector. With a
common profile left in, two patterns are similar when both have a high amplitude along the profile.
The amplitude RDMs therefore have the sum form $A_{ij} = 2 - g_i - g_j$ and
$A^M_{ij} = 2 - g^M_i - g^M_j$. This form follows from a simulation of the preceding step (step 48,
test (ii): with a common profile $c\,p$ added to amplitude-modulated stimulus patterns, the Spearman
correlation between the correlation-distance RDM and the sum-form RDM rose with $c$). A further
simulation with synthetic patterns, run before the amendment of step 48b and without real data,
showed that the difference form $|g_i - g_j|$ does not capture this structure and that partialling it
does not remove a pure amplitude effect, whereas the sum form does; the difference form is reported
for comparison only (Appendix~\ref{app:conv}). Amplitude dominance is Spearman(RDM, $A$) for the brain
and Spearman(RDM, $A^M$) for a model. The two-sided partial residualises the ranks of the brain RDM on
the ranks of $A$ and the ranks of the model RDM on the ranks of $A^M$ (each with intercept) and
correlates the residuals; the amendment calls this the Spearman of the residuals, and, as in the
partial RSA of step 44 to which it refers, it is computed as the Pearson correlation of the rank
residuals. The moment ratio $m = \lVert \mu \rVert^2 / (\lVert \mu \rVert^2 + \sum_k \sigma_k^2)$ of the
GAP features \citep{evalres} was computed per checkpoint.

\subsection{Follow-up analyses}
\label{sec:followup}

After version~1 we ran \AmendmentsFollowup{} further steps (Appendix~\ref{app:chronology}, Table~\ref{tab:chronology}). As before,
each step's predictions, decision rules and interpretation rules were committed before its scripts were
written and run; where new data were needed, also before the data were loaded. Two orderings concern
the NSD data. The predictions Q1--Q4 (step 42b) were written after the subj01 pilot had shown a
positive Euclidean contrast and before any data of subj02--08 were loaded (first beta of subj02
requested at \FirstDownloadSubjTwo, after the amendment commit), and they are evaluated on subj02--08
only. The amendment of step 48, which predicted the effect of the normalisation convention in NSD
(NB1, NB2), was committed before the first raw NSD beta was downloaded (\FirstDownloadRawFortyEight).
The definition of the amplitude RDM in step 48b was chosen after a simulation on synthetic patterns,
before the amendment and before any dominance value was computed on real data
(Section~\ref{sec:ampmethods}).

Each interpretation rule is reported with its literal outcome. The rule of step 48 has no case for the
pattern that occurred (NB1 and NB2 met, G2 or G3 not met); its outcome is reported as ``no statement of
the rule applies'', and the amendment was not changed. All predictions, with wording, commit, result
and source file, are listed in \texttt{paper/PREDICTIONS\_LEDGER.csv}; the predictions that were not
met are in Appendix~\ref{app:failed}.

\section{Results: THINGS}
\subsection{Sign flip between RSA and ridge (including the pre-specified primary analysis)}
\label{sec:signflip}
\begin{figure}[t]
\centering
\includegraphics[width=3.5in]{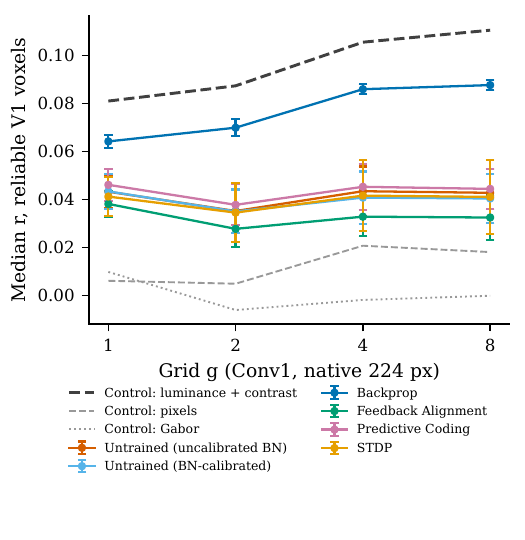}
\caption{Pre-specified primary analysis. Median $r$ over reliable V1 voxels (Conv1, NATIVE 224\,px), averaged over the three subjects, against grid size $g$ (logarithmic axis). Coloured lines: the six conditions, mean over 5 seeds, error bars SD over seeds. Grey dashed and dotted lines: control banks (luminance and contrast, darker; pixels; Gabor). The permuted-assignment control is not shown (Section~\ref{sec:signflip}).}
\label{fig:prereg_ladder}
\end{figure}

In the pre-specified primary analysis (Section~\ref{sec:inference}), the untrained
(uncalibrated BN statistics) network scored below Backprop at every rung of the grid
ladder (Fig.~\ref{fig:prereg_ladder}). The untrained-minus-Backprop contrast in median
$r$ over reliable V1 voxels, averaged over subjects, was $\PreregRandBpGOne$,
$\PreregRandBpGTwo$, $\PreregRandBpGFour$ and $\PreregRandBpGEight$ at $g = 1, 2, 4, 8$.
The number of positive seeds was \PreregRandBpGOnePos, \PreregRandBpGTwoPos,
\PreregRandBpGFourPos{} and \PreregRandBpGEightPos{} of 5, and the one-sided sign-flip
$p$ was \PreregRandBpGOneP, \PreregRandBpGTwoP, \PreregRandBpGFourP{} and
\PreregRandBpGEightP. The pre-specified minimal detectable effect, derived from the
pilot's seed-level standard deviation (\PreregPilotSd), was
$\Delta_{\min} = \PreregDminSim$ (supplementary value from the primary analysis's own
seed variability: \PreregRandBpGFourDmin{} at $g = 4$ and \PreregRandBpGEightDmin{} at
$g = 8$), below the reference effect of $+\PreregReferenceEffect$. The criterion for
survival at $g \geq 4$ was not met. Because $\Delta_{\min}$ was smaller than the
reference effect, absence is informative under the plan, and the pre-specified rule
classifies this outcome as ``The effect disappears'' rather than ``The method cannot
resolve it''. In every seed the contrast was negative. With five seeds the smallest attainable one-sided $p$ is $1/32$ and the test addresses only the positive direction, so the negative direction is an observation, not a tested effect. Per-subject values are given in Appendix~\ref{app:persubject}; the mean over subjects includes sub-03, where ridge scores were close to zero. The four
secondary contrasts against Backprop (untrained BN-calibrated, FA, PC, STDP) were also
negative at every $g$; the largest was $\PreregSecondaryMaxMean$, and at most
\PreregSecondaryMaxPos{} of 5 seeds were positive in any of them.

In an exploratory check of this result (three subjects, Conv1, NATIVE 224\,px;
Appendix~\ref{app:ridgerobust}), the untrained (uncalibrated BN statistics) minus Backprop
contrast was negative in two further scores. In the MSE skill (one minus the ratio of the
squared error to that of the training-fold mean) it was $\FSkillMseSkillGOneMean$,
$\FSkillMseSkillGTwoMean$, $\FSkillMseSkillGFourMean$ and $\FSkillMseSkillGEightMean$ at
$g = 1, 2, 4, 8$, and in the per-fold $r$ it was $\FSkillPerfoldGOneMean$,
$\FSkillPerfoldGTwoMean$, $\FSkillPerfoldGFourMean$ and $\FSkillPerfoldGEightMean$ (mean over
subjects and seeds); no seed was positive in either score at any $g$ (F-R1 and F-R2: met). A
model that predicts the training-fold mean scored a pooled $r$ of $\FFloorMean$ (mean over
subjects); per-fold $r$ is undefined for predictions that are constant within each fold. Under
50 permutations of the stimulus assignment of the features (seed-0 untrained and Backprop,
$g = 1$ and $g = 4$), the 97.5th percentile of the null lay between $\FPermPNineSevenFiveMin$
and $\FPermPNineSevenFiveMax$; the observed scores in sub-01 and sub-02 (from $\FPermRealMinMain$
to $\FPermRealMaxMain$) lay above it (F-R3: met). In sub-03, described without a rule, the
observed scores (from $\FPermRealMinSubThree$ to $\FPermRealMaxSubThree$) also lay above the
97.5th percentiles of their nulls. A readout of each network together with the
luminance-and-contrast bank scored below the bank alone in all six conditions at $g = 1$ and
$g = 4$ (differences from $\FJointMinDiff$ to $\FJointMaxDiff$): for Backprop the joint readout
reached $\FJointBpGOneJoint$ and $\FJointBpGFourJoint$, against $\FJointBpGOneBank$ and
$\FJointBpGFourBank$ for the bank alone, and for the untrained network $\FJointRandGOneJoint$
and $\FJointRandGFourJoint$ (F-R4, joint minus bank at most 0.010 for Backprop: met). Concatenating network features with the bank under a shared ridge penalty did not improve prediction. With 32 bank and 512 network features at $g = 4$, the two groups may need different penalties (banded ridge), which we did not fit; the result does not show how much variance the networks and the bank share. Absolute levels were low: the median $r$ of the best network
condition, Backprop at $g = 4$, was $\PreregCondBpGFour$ (mean over subjects); this correlation score should not be read as variance explained. The absolute MSE skill of Backprop at $g = 4$ was \FSkillAbsBpGFour{} and that of the untrained network \FSkillAbsRandGFour{} (median over reliable voxels, mean over subjects and seeds, relative to the training-fold mean).

Among the control banks, the luminance and contrast bank reached
\PreregCtlLumcontrastMeanGOne, \PreregCtlLumcontrastMeanGTwo,
\PreregCtlLumcontrastMeanGFour{} and \PreregCtlLumcontrastMeanGEight{} at
$g = 1, 2, 4, 8$, above the mean of every network condition at every $g$, including
Backprop (\PreregCondBpGOne, \PreregCondBpGTwo, \PreregCondBpGFour, \PreregCondBpGEight).
The pixel and Gabor banks ranged from \PreregCtlGaborMeanGTwo{} to
\PreregCtlPixelsMeanGFour. At $g = 4$ and $g = 8$ the intact grid exceeded its
image-wise cell shuffle for both the untrained and the Backprop seed-0 network in
\PreregShuffleWinSubjects{} of 3 subjects; this instrument covers one seed per condition
and is descriptive. The permuted-assignment control, the pipeline's negative test, for
which the plan expected a value near zero, gave \PreregCtlPermutedMeanGOne,
\PreregCtlPermutedMeanGTwo, \PreregCtlPermutedMeanGFour{} and
\PreregCtlPermutedMeanGEight; negative values are expected for features without
signal under the pooled out-of-fold correlation (Section~\ref{sec:ridge}), and the
control is not shown in Fig.~\ref{fig:prereg_ladder}. All
\PreregLowEdgeHits{} lower-boundary selections at reliable voxels (Appendix~\ref{app:lowedge}) were classified as
plateaus (\PreregLowEdgePlateau), and no cell was excluded.

\begin{figure}[t]
\centering
\includegraphics[width=\textwidth]{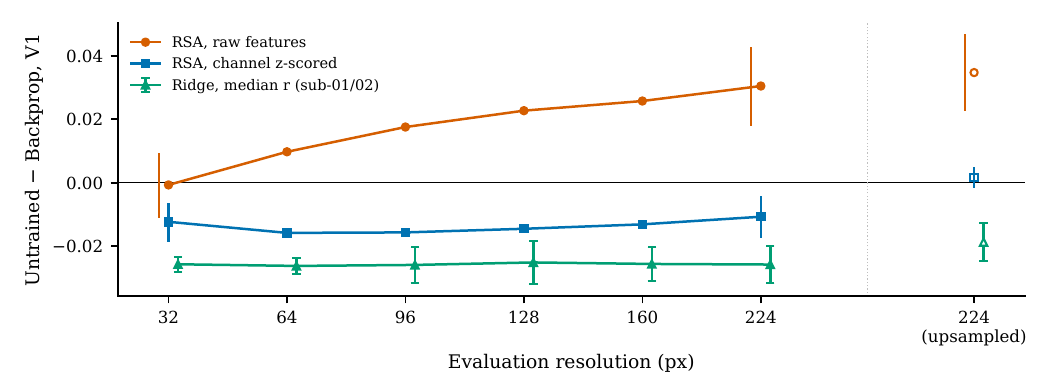}
\caption{Untrained (uncalibrated BN statistics) minus Backprop at V1 (Conv1) against evaluation resolution; NATIVE inputs left of the dotted line, UPSAMPLED 224\,px (open markers) right. RSA with raw features (circles) and channel-z-scored features (squares): per subject on cross-run pairs, averaged over subjects and 5 seeds; bars are 95\% stimulus-bootstrap CIs where computed. Ridge (triangles): median $r$ over reliable voxels, mean over sub-01 and sub-02, error bars SD over seeds.}
\label{fig:sign_flip}
\end{figure}

In an exploratory analysis of sub-01 (\NReliableSubOne{} reliable V1 voxels, $g = 1$,
NATIVE 224\,px), RSA and ridge disagreed in sign on the same voxels. RSA on the reliable
voxels gave an untrained-minus-Backprop difference of $+\RsaGapReliableGOne$; ridge gave
$\RidgeGapGOne$, and $\RidgeGapGFour$ and $\RidgeGapGEight$ at $g = 4$ and $g = 8$
(Fig.~\ref{fig:sign_flip}). Four candidate sources of the disagreement were tested
against rules fixed before each run, and none was supported. The voxel set: the RSA
difference stays positive on the voxels the ridge readout uses. The pooled out-of-fold
score: with per-fold $r$ the ridge contrast was $\ETwoPerFoldGap$ (a positive value would
have implicated pooling), and the two networks differed by \ETwoHiFracDiff{} in the share
of selections at the largest $\alpha$, below the threshold of 0.10 for a collapse
asymmetry. The fold convention: with run-grouped folds (Appendix~\ref{app:runfolds}) the contrast was
$\EThreeRunFoldGap$ (a positive value would have implicated leakage within runs). The
score: correlating RDMs built from the ridge predictions with the fMRI RDM gave
$\EFourEncGapRandom$ with random folds (\EFourEncPosSeedsRandom{} of 5 seeds
positive) and $\EFourEncGapRun$ with run-grouped folds. This meets the pre-stated rule
($\leq 0$), but the random-fold value is close to zero: the analysis shows that the RSA
advantage of the untrained network does not carry over to RDMs of ridge predictions, not
that it reverses there. In
the same subject, ridge did not separate the two untrained networks
($\POneRidgeRandBncal$; rule $\leq 0.010$: met), whereas RSA separated them by
$\POneRsaRandBncal$. In sub-02 and sub-03 the signs of the predicted pattern held: RSA
$+\STwoSubTwoRsa$ and $+\STwoSubThreeRsa$, ridge $\STwoSubTwoRidge$ and
$\STwoSubThreeRidge$. Sub-03 has \NReliableSubThree{} reliable voxels, ridge scores near zero in every condition and RSA and ridge differences close to zero; it contributes little to the pattern.

\subsection{Channel standardization removes gap and slope}
\label{sec:standardization}
In an exploratory analysis, each channel of the GAP features was z-scored over the 720
stimuli before RSA. In sub-01 on the reliable voxels (S1), the untrained (uncalibrated
BN statistics) minus Backprop difference became $\SOneStdGap$, below the threshold of
0.010 for the scale hypothesis, which was therefore supported; the two untrained
networks differed by \SOneStdRandBncal{} (control criterion $\leq 0.010$: met).

With the RSA convention of Section~\ref{sec:rsa} (all ROI voxels, three subjects,
stimulus bootstrap), raw features reproduced the published values (R1, met):
$+\VoneGapRawTwoTwoFour$ [$\VoneGapRawTwoTwoFourLo$, $\VoneGapRawTwoTwoFourHi$] at NATIVE
224\,px and $\VoneGapRawThreeTwo$ [$\VoneGapRawThreeTwoLo$, $\VoneGapRawThreeTwoHi$] at 32\,px.
Absolute RSA values at V1 were low ($\rho$ between \AbsRsaPaperMin{} and \AbsRsaPaperMax{}
in the paper cell at NATIVE 224\,px, means of the six conditions). The leave-one-subject-out
lower bound of the noise ceiling was \LosoBoundVone{} [\LosoBoundVoneLo, \LosoBoundVoneHi] on
cross-run pairs \citep{leutenegger2026}; the upper bound is not informative with three
subjects, so we cannot say how close the models are to the ceiling. With z-scored channels the V1 gap was $\VoneGapZTwoTwoFour$
[$\VoneGapZTwoTwoFourLo$, $\VoneGapZTwoTwoFourHi$] at 224\,px (R2, upper bound
$\leq 0.010$: met) and $\VoneGapZThreeTwo$ [$\VoneGapZThreeTwoLo$, $\VoneGapZThreeTwoHi$]
at 32\,px. The resolution slope fell from $S_{\mathrm{raw}} = +\SlopeRaw$
[$\SlopeRawLo$, $\SlopeRawHi$] to $S_z = +\SlopeZ$ [$\SlopeZLo$, $\SlopeZHi$]
(R3, $S_z \leq 0.5\,S_{\mathrm{raw}}$: met; Fig.~\ref{fig:sign_flip}). At LOC the
Backprop-minus-untrained gap at 224\,px was $+\LocGapRaw$ [$\LocGapRawLo$, $\LocGapRawHi$]
with raw and $+\LocGapZ$ [$\LocGapZLo$, $\LocGapZHi$] with z-scored channels (R4, CI above
0 under z: met). With content limited to 32\,px (UPSAMPLED 224), the V1 gap was
$+\VoneGapRawUp$ [$\VoneGapRawUpLo$, $\VoneGapRawUpHi$] raw and $+\VoneGapZUp$
[$\VoneGapZUpLo$, $\VoneGapZUpHi$] z-scored (R5, upper bound $\leq 0.010$: met).

The moment ratio $m = \lVert\mu\rVert^2 / (\lVert\mu\rVert^2 + \sum_k \sigma_k^2)$ of the
channel means $\mu$ and SDs $\sigma_k$ tracks the difference $d$ between raw and
z-scored RSA: over 60 checkpoint-by-resolution cells of the untrained network and
Backprop, Spearman$(m, d) = \MOneSpearman$ (MR1, called M1 in the step 5 amendment; threshold 0.6: met). For Backprop, $m$ was
ordered UPSAMPLED 224 $<$ NATIVE 224 $<$ NATIVE 32 in \PMechOneBpSeeds{} of 5 seeds, and
the untrained network's range of $m$ (\PMechOneRandRange) was below half of Backprop's
(P\_mech1, met). P\_mech2 was not met (Appendix~\ref{app:bandpass}): Backprop's $m$ was higher for band-pass inputs
without the coarse band than for the coarse band alone in \PMechTwoNotchSeeds{} of 5
seeds, but Spearman$(m, d)$ over all 50 cells including the band-pass inputs was
\PMechTwoSpearman, below 0.6.

Ridge on the same GAP features ($g = 1$) at every NATIVE resolution (P\_ridge; sub-01 and
sub-02) gave untrained-minus-Backprop contrasts of $\PRidgeGapThreeTwo$ at 32\,px and
$\PRidgeGapTwoTwoFour$ at 224\,px, non-positive at all six resolutions, with a slope of
$\PRidgeSlope$, below half of $S_{\mathrm{raw}}$ (met; Fig.~\ref{fig:sign_flip}).

\subsection{Analysis multiverse}
\label{sec:multiverse}
\begin{figure}[t]
\centering
\includegraphics[width=\textwidth]{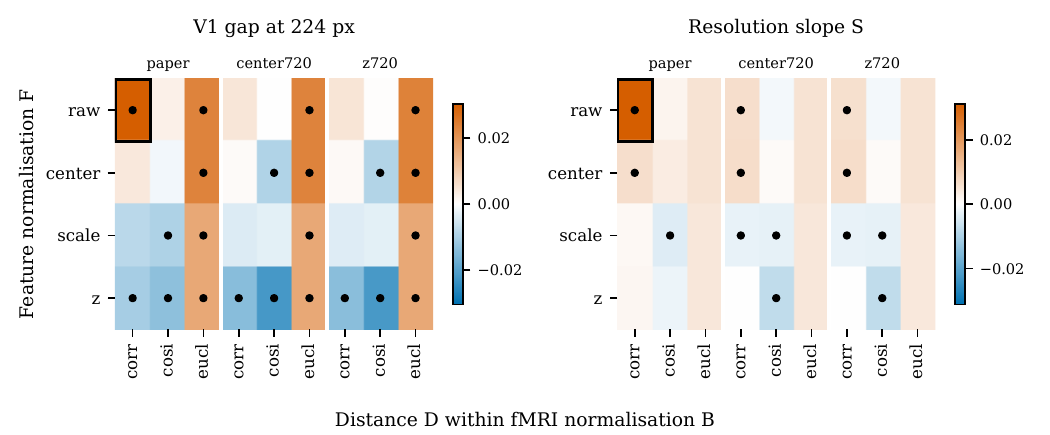}
\caption{Untrained minus Backprop at V1 in the 36 cells of the analysis multiverse. Left: gap at NATIVE 224\,px; right: resolution slope $S$ (224 minus 32\,px). Rows: feature normalisation $F$; columns: distance $D$ within fMRI normalisation $B$. Colour: value on a diverging scale centred at 0. Dots: 95\% stimulus-bootstrap CI excludes 0. Black frame: the convention of \citet{evalres}. Mean over 5 seeds and 3 subjects. Euclidean cells agree in pairs of rows (raw and center; scale and z) and between the paper and center720 columns (centring invariance).}
\label{fig:multiverse_v1}
\end{figure}

In an exploratory analysis, the V1 contrast between the untrained (uncalibrated BN
statistics) network and Backprop was computed in all 36 cells of the multiverse
(Section~\ref{sec:mvmethods}; Fig.~\ref{fig:multiverse_v1}). The mathematical control O1 was met:
euclidean RDMs under raw and centred features, and under paper and center720 betas,
agreed to within \MultiOOneMaxDelta. O2 was met: the only cell with a strong resolution
slope (CI above 0 and $S \geq 0.5\,S_{\mathrm{paper}}$) was the convention of
\citet{evalres}, raw/correlation/paper, with $S = +\MultiSlopePaper$
[$\MultiSlopePaperLo$, $\MultiSlopePaperHi$]; \MultiOTwoViolations{} of 36 cells violated
the prediction (at most 2 allowed). At NATIVE 224\,px the V1 gap had
\MultiSignPositive{} positive and \MultiSignNegative{} negative point estimates. Its CI
excluded 0 in \MultiPosCiCells{} positive cells, the paper's cell and the 12 euclidean
cells, and in \MultiNegCiCells{} negative cells. Euclidean distances are invariant to centring of units and of voxels, so the 12 euclidean cells comprise \MultiDistinctEucl{} distinct analyses and the 36 cells \MultiDistinct{}; counted over distinct analyses, the CI excluded 0 above in \MultiDistinctPos, below in \MultiDistinctNeg{} and included 0 in \MultiDistinctIncl{} analyses. Cell counts are descriptive. Examples are $+\VoneGapRawTwoTwoFour$
[$\VoneGapRawTwoTwoFourLo$, $\VoneGapRawTwoTwoFourHi$] in the paper's cell,
$\VoneGapZTwoTwoFour$ [$\VoneGapZTwoTwoFourLo$, $\VoneGapZTwoTwoFourHi$] with z-scored
features and correlation distance, and $+\MultiEuclidGap$ [$\MultiEuclidGapLo$,
$\MultiEuclidGapHi$] with raw features and euclidean distance. At LOC
(Fig.~\ref{fig:multiverse_loc}), Backprop minus untrained was positive in
\MultiLocPositive{} of 36 cells and had a CI above 0 in \MultiLocCiPositive{} (O3,
thresholds 30 and 18: met); the CI included 0 in all euclidean cells.

For the six ImageNet models (contrast: untrained minus trained at V1, trained minus
untrained at LOC), GEN2 was met: in \GenTwoModels{} of 6 models two of the four bootstrap
cells gave V1 gaps at 224\,px whose CIs excluded 0 in opposite directions. GEN3 was not
met: the LOC contrast had a CI above 0 in all four cells for \GenThreeModels{} models
(at least 5 required). The raw/euclidean cell included 0 for every model, for example
$+\GenAlexnetLocRawEucl$ [$\GenAlexnetLocRawEuclLo$, $\GenAlexnetLocRawEuclHi$] for
alexnet and $\GenConvnexttinyLocRawEucl$ [$\GenConvnexttinyLocRawEuclLo$,
$\GenConvnexttinyLocRawEuclHi$] for convnext\_tiny. GEN4 was not met: \GenFourQualified{}
models had a raw/correlation slope $S_{\mathrm{gen}}$ (448 minus 224\,px) with a CI
excluding 0, and in none of them did channel z-scoring halve its magnitude. A slope
remained under z-scoring, for example $+\GenAlexnetSlopeRawCorr$ raw versus
$+\GenAlexnetSlopeZCorr$ z-scored for alexnet and $+\GenVggelevenSlopeRawCorr$ versus
$+\GenVggelevenSlopeZCorr$ for vgg11\_bn.

\subsection{Mean-profile alignment}
\label{sec:gainaxis}
\begin{figure}[t]
\centering
\includegraphics[width=\textwidth]{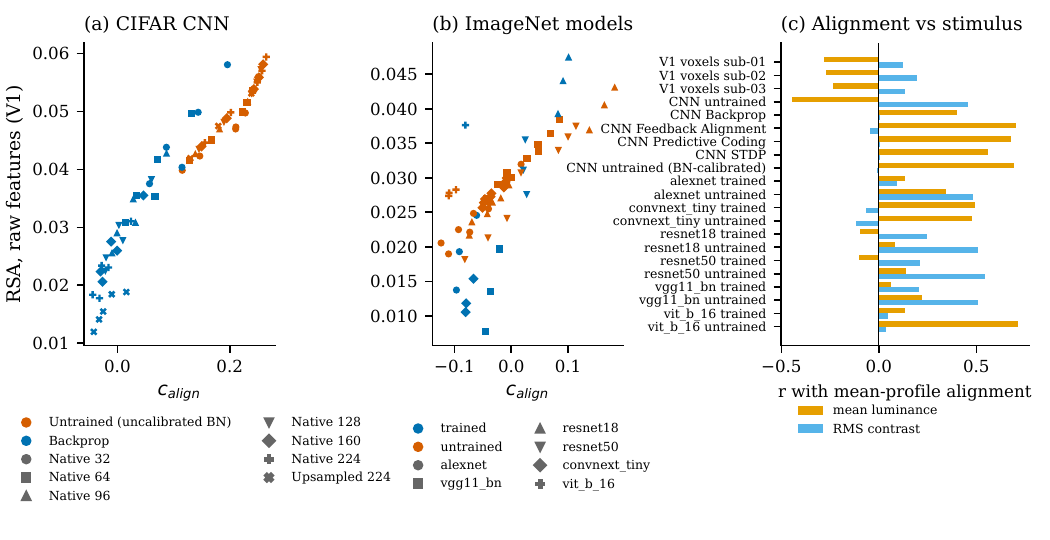}
\caption{(a)~Own CNN: raw-feature RSA at V1 against $c_{\mathrm{align}}$ for the untrained (vermillion) and Backprop (blue) checkpoints; markers code the input condition. (b)~ImageNet models: the same for trained (blue) and untrained (vermillion) networks; markers code the model. (c)~Pearson correlation over stimuli between mean-profile alignment and stimulus mean luminance (orange) or RMS contrast (light blue), for V1 voxels per subject, the own CNN's conditions and the ImageNet models.}
\label{fig:gain}
\end{figure}

In an exploratory analysis, the mean-profile alignment of Section~\ref{sec:gainmethods} was computed for every model
and for the V1 voxels of each subject (Fig.~\ref{fig:gain}). Over the 70 cells of the
untrained (uncalibrated BN statistics) network and Backprop (5 seeds, six NATIVE
resolutions and UPSAMPLED 224\,px), $c_{\mathrm{align}}$ predicts the raw-feature RSA at
V1 with Spearman \GOneSpearman{} (G1, threshold 0.6: met). Over 64 cells of the ImageNet
models the same correlation was \GenOneSpearman{} (GEN1, met). G2 was not met: after the
ranks of both RDMs were regressed on the ranks of their own alignment RDMs, the V1 gap at
NATIVE 224\,px was $+\GTwoPartialGap$ [$\GTwoPartialGapLo$, $\GTwoPartialGapHi$], whose
upper bound exceeds 0.010, and the slope was $S_{\mathrm{partial}} = +\GTwoPartialSlope$
[$\GTwoPartialSlopeLo$, $\GTwoPartialSlopeHi$], above half of $S_{\mathrm{paper}}$.

Descriptively, V1 voxel alignment correlated with stimulus mean luminance at
\VoxGainLumSubOne, \VoxGainLumSubTwo{} and \VoxGainLumSubThree, and with RMS contrast at
\VoxGainConSubOne, \VoxGainConSubTwo{} and \VoxGainConSubThree{} (sub-01, sub-02,
sub-03). At 224\,px the untrained network's alignment had the same signs
(\ModelGainRandLum{} with luminance, \ModelGainRandCon{} with contrast); Backprop's alignment correlated positively with luminance (\ModelGainBpLum), and its correlation with contrast
was close to zero (\ModelGainBpCon).

\subsection{Removing the mean-profile component}
\label{sec:ablation}
In an exploratory analysis prompted by criticism of an earlier draft (Section~\ref{sec:inference}),
each raw GAP pattern and each beta pattern was split into its projection on the mean profile
and the remainder orthogonal to it (Section~\ref{sec:gainmethods}); each model component was
compared with the same component of the brain, with the RSA convention and bootstrap of
Section~\ref{sec:rsa}. The full variant reproduced the stored paper cell of the own CNN and
of the ImageNet models to within \MpGateMaxDelta{} (reproduction gate: met).

M1 was met: in the orthogonal remainder, the V1 contrast of the untrained (uncalibrated BN
statistics) network minus Backprop at NATIVE 224\,px was $\MpOrthoOwnVoneNatTwoTwoFour$
[$\MpOrthoOwnVoneNatTwoTwoFourLo$, $\MpOrthoOwnVoneNatTwoTwoFourHi$], with an upper bound of
at most 0.010, against $+\MpFullOwnVoneNatTwoTwoFour$ [$\MpFullOwnVoneNatTwoTwoFourLo$,
$\MpFullOwnVoneNatTwoTwoFourHi$] in the full variant. M2 was met: the resolution slope of
the orthogonal remainder was $\MpOrthoOwnSlope$ [$\MpOrthoOwnSlopeLo$, $\MpOrthoOwnSlopeHi$],
below half of $S_{\mathrm{paper}}$. M4 was met: for the ImageNet models at 224\,px, the upper
bound of the orthogonal V1 contrast was at most 0.010 in \MpMFourModels{} of 6 models. M3 was
not met: the parallel component gave $+\MpParOwnVoneNatTwoTwoFour$
[$\MpParOwnVoneNatTwoTwoFourLo$, $\MpParOwnVoneNatTwoTwoFourHi$] at NATIVE 224\,px and a
slope of $+\MpParOwnSlope$ [$\MpParOwnSlopeLo$, $\MpParOwnSlopeHi$]. The interval of the
parallel gap includes both 0 and the raw gap, so these data do not decide whether the
parallel component alone suffices. At LOC, described without a rule, Backprop minus
untrained was $+\MpFullOwnLoc$ [$\MpFullOwnLocLo$, $\MpFullOwnLocHi$] in the full variant,
$+\MpOrthoOwnLoc$ [$\MpOrthoOwnLocLo$, $\MpOrthoOwnLocHi$] in the orthogonal remainder and
$\MpParOwnLoc$ [$\MpParOwnLocLo$, $\MpParOwnLocHi$] in the parallel component.

M3 was not met, and sufficiency of the parallel component was not established. The wording
of the pre-stated rule (``M3 not met: it is not sufficient'') claimed more than a missed
threshold with this interval can show, and we state only what the data resolve: the
advantage is absent from the orthogonal remainder, and the parallel component alone did not
resolve it. No physiological claim follows.

On the same resamples, the paired contrast of the V1 gap with the full variant on both
sides minus the gap with the orthogonal remainder on both sides was $+\AbTwoXOne$ [$\AbTwoXOneLo$, $\AbTwoXOneHi$] (X1: met), which restates
that the drop between the full and the orthogonal variant is resolved. With the orthogonal
remainder on one side only (Appendix~\ref{app:ablation2x2}), the V1 contrast was
$\AbTwoOwnOrthoFull$ [$\AbTwoOwnOrthoFullLo$, $\AbTwoOwnOrthoFullHi$] for the orthogonal model
against the full brain RDM and $+\AbTwoOwnFullOrtho$ [$\AbTwoOwnFullOrthoLo$,
$\AbTwoOwnFullOrthoHi$] for the full model against the orthogonal brain RDM; both intervals
include 0, so under the pre-stated rule no attribution to a side is claimed. Removing the component from only one side brought the point estimate from $+\MpFullOwnVoneNatTwoTwoFour$ in the full variant to values close to that of the orthogonal variant, and the intervals of both mixed cells lay below that of the full variant. This is consistent with an advantage that depends on the mean-profile component on both sides, as the alignment correlation $c_{\mathrm{align}}$ of Section~\ref{sec:gainaxis} suggests; it is a descriptive observation, not covered by a rule, and we did not test a paired contrast.

\subsection{Norm and angular components}
\label{sec:gaingeometry}
In an exploratory analysis, the euclidean RDM on unit-centred features was decomposed
into its norm component and its angular component (Section~\ref{sec:gainmethods};
Fig.~\ref{fig:separation}). Contrasts are untrained minus trained at V1 and trained minus
untrained at LOC; for the own CNN they are untrained (uncalibrated BN statistics) and
Backprop. The identity held to a maximum relative error of
\SepOneMaxRelError{} (SEP1, threshold $10^{-8}$: met).

SEP2 was not met: the V1 norm contrast ($F$ = center) was $+\SepOwnVoneNormCenter$
[$\SepOwnVoneNormCenterLo$, $\SepOwnVoneNormCenterHi$] in the own CNN, with a CI
including 0, and had a CI above 0 in \SepTwoModels{} of 6 ImageNet models. SEP3 was met:
the V1 angle contrast was $\SepOwnVoneAngleCenter$ [$\SepOwnVoneAngleCenterLo$,
$\SepOwnVoneAngleCenterHi$] in the own CNN and had a CI below 0 in \SepThreeModels{} of 6
ImageNet models. SEP4 was met: the LOC angle contrast was $+\SepOwnLocAngleCenter$
[$\SepOwnLocAngleCenterLo$, $\SepOwnLocAngleCenterHi$] in the own CNN and had a CI above 0
in \SepFourModels{} of 6 ImageNet models. Because SEP2 was not met, the pre-stated interpretation rule applies: the V1 ranking is not attributed to the norm or the angular component. On the angular component alone, the trained
networks were ahead at V1 in the own CNN and in \SepThreeModels{} ImageNet models, and
at LOC in the own CNN and in \SepFourModels{} ImageNet models. The norm-weighted angular
term $W$ gave a V1 contrast of $+\WAngOwnVone$ [$\WAngOwnVoneLo$, $\WAngOwnVoneHi$] in the own
CNN, and its interval excluded 0 in \WAngImnPosModels{} of 6 ImageNet models (vgg11\_bn,
$+\WAngVggelevenVone$ [$\WAngVggelevenVoneLo$, $\WAngVggelevenVoneHi$]). The angular result
thus concerns the unweighted angular dissimilarity $A$; the norm-weighted term $W$ pointed in
the direction of the norm component and was not resolved. Over the cross-run pairs, $N$ and $A$
were nearly uncorrelated (Spearman \WRankNAOwn{} in the own CNN, mean over checkpoints;
\WRankNASubOne, \WRankNASubTwo{} and \WRankNASubThree{} in the fMRI of the three subjects),
whereas $A$ and $W$ were correlated (\WRankAWOwn; \WRankAWSubOne, \WRankAWSubTwo{} and
\WRankAWSubThree). In a regression of the ranks of the fMRI RDM on the ranks of $N$, $A$ or
both (own CNN, V1, NATIVE 224\,px), each component accounted for about 0.1\% of the rank
variance of the fMRI RDM ($R^2$ for $N$ and $A$ together: \WRtwoNARand{} for the untrained
network, \WRtwoNABp{} for Backprop), and no contrast of $R^2$ was resolved.

The resolution slope of the contrasts was a descriptive point fixed in advance without a
decision rule. For the ImageNet models, the V1 angle contrast grew from 224 to 448\,px
with a CI above 0 in \SepSlopeAngleCiPositive{} of 5 models, for example
$+\SepSlopeResnetfiftyAngleCenter$ [$\SepSlopeResnetfiftyAngleCenterLo$,
$\SepSlopeResnetfiftyAngleCenterHi$] for resnet50; no norm slope had a CI excluding 0. For
the own CNN neither component had a slope from 32 to 224\,px under $F$ = center (norm
$\SepSlopeOwnNormCenter$ [$\SepSlopeOwnNormCenterLo$, $\SepSlopeOwnNormCenterHi$], angle
$+\SepSlopeOwnAngleCenter$ [$\SepSlopeOwnAngleCenterLo$, $\SepSlopeOwnAngleCenterHi$]).
convnext\_tiny was the exception among the ImageNet models: its V1 angle contrast
($\SepConvnexttinyVoneAngleCenter$ [$\SepConvnexttinyVoneAngleCenterLo$,
$\SepConvnexttinyVoneAngleCenterHi$]) and its angle slope ($+\SepSlopeConvnexttinyAngleCenter$
[$\SepSlopeConvnexttinyAngleCenterLo$, $\SepSlopeConvnexttinyAngleCenterHi$]) both had CIs
including 0. We did not test an explanation for this exception.

\subsection{Banded ridge and repeated test images}
\label{sec:thingsfollowup}
\paragraph{Banded ridge.} A banded readout with separate penalties for the luminance-and-contrast bank
and the network (step 40) was meant to estimate the network's unique contribution beyond the bank. Its
sanity rule B1 was not met: the joint readout lay below the bank alone by more than 0.002 in some fits
(minimum $\BandedBOneMinGFour$ at $g = 4$), so B2 and B3 were not interpreted. A diagnosis (step 41)
showed no implementation error at the limit of an infinite network penalty (D-A1 met, maximum
deviation \DAOneDev), but a one-dimensional conditional search did not remove the deficit (D-A2 not met,
minimum $\DATwoMin$); the violation was therefore not attributed to selection variance, and the
conditional estimator was not interpreted. Descriptively, neither penalty scheme improved on the bank
alone in MSE skill (joint minus bank at $g = 4$: $\BandedUSkillBpGFour$ and $\CondUSkillBp$ for
Backprop, $\BandedUSkillRandGFour$ and $\CondUSkillRand$ for the untrained network), and in median $r$
the joint readout exceeded the bank by at most $\CondUBp$ (Appendix~\ref{app:banded},
Table~\ref{tab:banded}).

\paragraph{Repeated test images.} On the 100 test images, each shown 12 times, the paper cell did not
show the untrained advantage on presentation-averaged patterns (C1 not met: $\CvRsaVoneRawBOne$
[$\CvRsaVoneRawBOneLo$, $\CvRsaVoneRawBOneHi$]), whereas the standardised cell resolved a Backprop
advantage (C2 met: $\CvRsaVoneZBOne$ [$\CvRsaVoneZBOneLo$, $\CvRsaVoneZBOneHi$]); cross-validated and
plain Euclidean distances gave the same sign (C3 met). A noise curve over the number $k$ of averaged
presentations (step 41) showed that the contrast fell with $k$ (N1 and N2 met; Spearman of $k$ and
$\Delta(k)$ \NTwoSpearman), but the single-presentation value was not positive either (N3 not met:
$\Delta(1) = \NoiseVoneRawKOneFour$ [$\NoiseVoneRawKOneLo$, $\NoiseVoneRawKOneHi$]); under standardisation
$\Delta(k) \leq 0$ at every $k$ (N4 met; Appendix~\ref{app:banded}, Table~\ref{tab:noisecurve}). The
test images are z-scored with all trials like the 720 (T-global); their contrast under the other conventions
is reported in Section~\ref{sec:conventions}.

\section{Results: NSD replication}
\label{sec:nsd}
\subsection{Pre-specified ridge readout}
\label{sec:nsdridge}
Applied unchanged to NSD (step 47), the pre-specified ridge readout placed the untrained network below
Backprop at V1 in every subject and at every grid: $\Delta = \NsdRidgeVoneGOne$, $\NsdRidgeVoneGFour$ and
$\NsdRidgeVoneGEight$ at $g = 1$, 4 and 8 (\NsdRidgeVoneGOneNeg, \NsdRidgeVoneGFourNeg{} and
\NsdRidgeVoneGEightNeg{} of 8 subjects negative, sign-flip $p = \NsdRidgeVoneGFourP$ each; RD1 and RD2
met). The same held without voxel selection (Y4 met, $\YFourNsdAll$ at $g = 4$), after run centring in NSD
(Y3 met, $\YThreeNsdRc$) and in THINGS (Y1 met, $\YOneThingsRc$ against $\YOneThingsRaw$ without), and at
LO-obj ($\NsdRidgeLoobjGFour$ at $g = 4$, \NsdRidgeLoobjGFourNeg{} of 8). The luminance-and-contrast bank
alone exceeded both networks at V1 at every grid (mean over subjects at $g = 4$: bank
\NsdRidgeVoneGFourBank, Backprop \NsdRidgeVoneGFourBp, untrained \NsdRidgeVoneGFourRand;
Appendix~\ref{app:nsd}, Table~\ref{tab:nsdsubjects}), as in THINGS; at LO-obj it was below both.

\subsection{RSA under the NSD standard normalisation}
\label{sec:nsdrsa}
\begin{figure}[t]
\centering
\includegraphics[width=\textwidth]{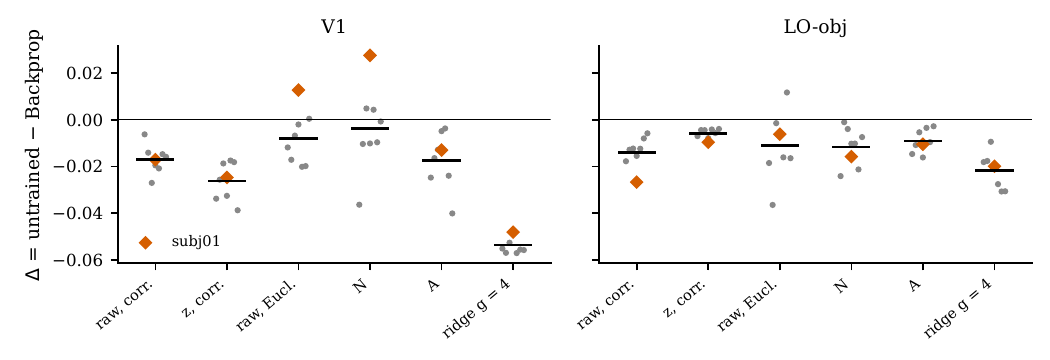}
\caption{NSD under its standard normalisation (N-session), per subject: untrained minus Backprop at V1 and LO-obj for raw and standardised units with correlation distance, raw units with Euclidean distance, the norm ($N$) and angular ($A$) components ($k = 3$), and the pre-specified ridge readout at $g = 4$ (median $r$). Grey: subj02--08; diamond: subj01 (the pilot); bar: mean over the 8 subjects.}
\label{fig:nsd}
\end{figure}

In the paper cell the untrained network was behind Backprop (Fig.~\ref{fig:nsd}; P1 met: $\Delta = \NsdAllVoneKThreeRawCorrelation$
[$\NsdAllVoneKThreeRawCorrelationLo$, $\NsdAllVoneKThreeRawCorrelationHi$], \NsdAllVoneKThreeRawCorrelationNeg{}
of 8 subjects negative), and more so with standardised units (P2 met: $\NsdAllVoneKThreeZCorrelation$
[$\NsdAllVoneKThreeZCorrelationLo$, $\NsdAllVoneKThreeZCorrelationHi$], \NsdAllVoneKThreeZCorrelationNeg{} of 8,
$p = \NsdAllVoneKThreeZCorrelationPn$) and at LO-obj (P3 met: $\NsdAllLoobjKThreeZCorrelation$,
\NsdAllLoobjKThreeZCorrelationNeg{} of 8). Averaging the three presentations did not move the paper cell
towards the untrained network (P4 met: $\NsdAllVoneKThreeRawCorrelation$ at $k = 3$ against
$\NsdAllVoneKOneRawCorrelation$ at $k = 1$).

\subsection{Multiverse}
\label{sec:nsdmultiverse}
The 36 cells of the multiverse formed \MvNsessionDistinct{} distinct analyses in NSD, as in THINGS. None
had an interval above 0 (M1 met); \MvNsessionBelow{} had an interval below 0 and \MvNsessionIncl{}
included 0, so M2 (at least half below 0) was not met by one analysis
(Fig.~\ref{fig:multi}, Appendix~\ref{app:nsd}, Table~\ref{tab:multi28}). Positive point estimates
occurred for raw units with correlation distance on voxels centred or z-scored over the images and, close
to 0, for centred units with correlation distance.

\subsection{Norm and angular components}
\label{sec:nsdcomponents}
In the subj01 pilot the untrained network was ahead under Euclidean distance ($\PilotVoneRawEuclidean$,
Table~\ref{tab:nsdpilot}). The predictions written on this basis and tested on subj02--08 were not met for
the Euclidean cell (Q1: $\NsdHeldVoneKThreeRawEuclidean$, \NsdHeldVoneKThreeRawEuclideanPos{} of 7
positive) or for the norm component (Q2: $\NsdHeldVoneKThreeCenterN$, \NsdHeldVoneKThreeCenterNPos{} of 7),
and were met for the angular component at V1 (Q3: $\NsdHeldVoneKThreeCenterA$,
\NsdHeldVoneKThreeCenterANeg{} of 7 negative, $p = \NsdHeldVoneKThreeCenterAPn$) and at LO-obj (Q4:
$\NsdHeldLoobjKThreeCenterA$, \NsdHeldLoobjKThreeCenterANeg{} of 7). The norm hypothesis motivated by subj01
was thus not replicated.

\subsection{ImageNet models, resolution and noise ceiling}
\label{sec:nsdceiling}
All six ImageNet architectures had trained ahead of untrained at V1 in the angular component (I1 met;
$\Delta$ from $\ImnVoneAMin$ to $\ImnVoneAMax$) and in the paper cell (I2 met; $\ImnVoneRawCorrMin$ to
$\ImnVoneRawCorrMax$; Table~\ref{tab:nsdimagenet}). Reported without prediction, the own CNN's paper-cell
contrast was negative at every evaluation resolution from 32 to 224\,px (slope
$S = \ResRawSlope$ [$\ResRawSlopeLo$, $\ResRawSlopeHi$]; Table~\ref{tab:nsdres}). The leave-one-subject-out
lower bound of the noise ceiling at V1 was \CeilVoneKThree{} at $k = 3$ and \CeilVoneKOne{} at $k = 1$,
against \CeilThingsLoso{} in THINGS-720; both networks reached \CeilVoneFracMin--\CeilVoneFracMax\% of the
NSD bound (Table~\ref{tab:nsdceil}).

\begin{figure}[t]
\centering
\includegraphics[width=\textwidth]{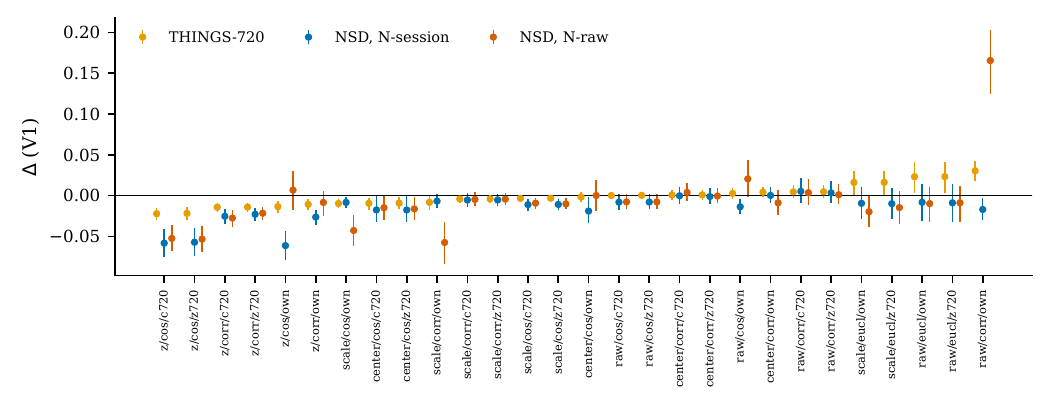}
\caption{The 28 distinct analyses of the V1 multiverse (one representative cell $F$/$D$/$B$ per class; ``own'' denotes the data set's own voxel normalisation, c720 centring over the images) in THINGS-720 (orange), NSD under its standard normalisation N-session (blue) and NSD with raw betas N-raw (vermillion), sorted by the THINGS value. Bars: 95\% stimulus-bootstrap intervals.}
\label{fig:multi}
\end{figure}

\section{A normalisation convention creates the untrained advantage}
\label{sec:convention}
\subsection{Decomposition in THINGS}
\label{sec:decomposition}
Several candidate explanations of the THINGS-720 advantage were tested and excluded before the
convention was identified (Appendix~\ref{app:why}). Sampling: subsets of 100 of the 720 images kept a
contrast above the test-image value in \RuleWOne\% of 1000 subsets (W1 met), so the absence of the
advantage in the test images is not sampling variation of 100 images. Measurement noise: the
single-presentation test-image contrast was not positive either (N3 not met). Presentation order:
partialling the position RDM changed $\Delta$ by \RuleWFive{} (W5 met). Carry-over: partialling the
luminance of the preceding image changed it by $\RuleCoOne$ (CO1 not met). Image statistics explained
only part: after partialling global luminance and contrast $\Delta$ was $\RuleWTwo$ (W2 not met), and
after adding coarse luminance $\RuleWSix$ (W6 not met; \RuleWSixFraction\% of the contrast explained;
interpretation of both steps: ``not identified''). Run centring reduced $\Delta$ to $\RuleRcOne$ (RC1
met), but the further predictions of that step were not met (RC2, RC3, RC4), its interpretation rule
had no case for this pattern (``no statement of the rule applies''), and an alignment of run-level drift
with the untrained network's features was not supported (X1 not met; interpretation: ``not identified'').

The decomposition of run centring (step 47c; Table~\ref{tab:decomp}) located the effect in the
session-mean profile $m$. Removing $m$ alone reduced $\Delta$ from $\DecVoneVZero$ to $\DecVoneVOne$
(V1), whereas removing the run deviations alone left $\DecVoneVTwo$ (V2; Z1 not met), and re-adding
either part of the run deviations to the run-centred patterns left it small ($\DecVoneVFour$ and
$\DecVoneVFive$; Z2 not met). The interpretation rule gave: ``mainly removal of the session mean profile
(equivalent to voxel centring), not between-run structure; the description of step 47 as run structure is
to be corrected in the manuscript.'' This agrees with the multiverse of version~1, in which centring the
voxels over the 720 images (raw / correlation / center720) gave $\CenterSevenTwentyCell$
[$\CenterSevenTwentyCellLo$, $\CenterSevenTwentyCellHi$]: the 720 images lie in one session, and
centring over them removes nearly the same profile.

\subsection{Conventions across data sets}
\label{sec:conventions}
\begin{table}[t]
\centering
\small
\begin{adjustbox}{max width=\textwidth, max totalheight=0.82\textheight}
\begin{tabular}{@{}llrrr@{}}
\toprule
Data & Convention & raw / correlation & z / correlation & raw / Euclidean \\
\midrule
THINGS-720 & T-raw & $\CvTsevTrawVoneRawCorr$ (\CvTsevTrawVoneRawCorrPos/3) & $\CvTsevTrawVoneZCorr$ (\CvTsevTrawVoneZCorrPos/3) & $\CvTsevTrawVoneRawEucl$ (\CvTsevTrawVoneRawEuclPos/3) \\
THINGS-720 & T-global & $\CvTsevTglobalVoneRawCorr$ (\CvTsevTglobalVoneRawCorrPos/3) & $\CvTsevTglobalVoneZCorr$ (\CvTsevTglobalVoneZCorrPos/3) & $\CvTsevTglobalVoneRawEucl$ (\CvTsevTglobalVoneRawEuclPos/3) \\
THINGS-720 & T-session & $\CvTsevTsessionVoneRawCorr$ (\CvTsevTsessionVoneRawCorrPos/3) & $\CvTsevTsessionVoneZCorr$ (\CvTsevTsessionVoneZCorrPos/3) & $\CvTsevTsessionVoneRawEucl$ (\CvTsevTsessionVoneRawEuclPos/3) \\
\addlinespace
THINGS test, $k = 1$ & T-raw & $\CvTtestTrawVoneRawCorrKOne$ (\CvTtestTrawVoneRawCorrKOnePos/3) & $\CvTtestTrawVoneZCorrKOne$ (\CvTtestTrawVoneZCorrKOnePos/3) & $\CvTtestTrawVoneRawEuclKOne$ (\CvTtestTrawVoneRawEuclKOnePos/3) \\
THINGS test, $k = 1$ & T-global & $\CvTtestTglobalVoneRawCorrKOne$ (\CvTtestTglobalVoneRawCorrKOnePos/3) & $\CvTtestTglobalVoneZCorrKOne$ (\CvTtestTglobalVoneZCorrKOnePos/3) & $\CvTtestTglobalVoneRawEuclKOne$ (\CvTtestTglobalVoneRawEuclKOnePos/3) \\
THINGS test, $k = 1$ & T-session & $\CvTtestTsessionVoneRawCorrKOne$ (\CvTtestTsessionVoneRawCorrKOnePos/3) & $\CvTtestTsessionVoneZCorrKOne$ (\CvTtestTsessionVoneZCorrKOnePos/3) & $\CvTtestTsessionVoneRawEuclKOne$ (\CvTtestTsessionVoneRawEuclKOnePos/3) \\
\addlinespace
THINGS test, $k = 12$ & T-raw & $\CvTtestTrawVoneRawCorrKOneTwo$ (\CvTtestTrawVoneRawCorrKOneTwoPos/3) & $\CvTtestTrawVoneZCorrKOneTwo$ (\CvTtestTrawVoneZCorrKOneTwoPos/3) & $\CvTtestTrawVoneRawEuclKOneTwo$ (\CvTtestTrawVoneRawEuclKOneTwoPos/3) \\
THINGS test, $k = 12$ & T-global & $\CvTtestTglobalVoneRawCorrKOneTwo$ (\CvTtestTglobalVoneRawCorrKOneTwoPos/3) & $\CvTtestTglobalVoneZCorrKOneTwo$ (\CvTtestTglobalVoneZCorrKOneTwoPos/3) & $\CvTtestTglobalVoneRawEuclKOneTwo$ (\CvTtestTglobalVoneRawEuclKOneTwoPos/3) \\
THINGS test, $k = 12$ & T-session & $\CvTtestTsessionVoneRawCorrKOneTwo$ (\CvTtestTsessionVoneRawCorrKOneTwoPos/3) & $\CvTtestTsessionVoneZCorrKOneTwo$ (\CvTtestTsessionVoneZCorrKOneTwoPos/3) & $\CvTtestTsessionVoneRawEuclKOneTwo$ (\CvTtestTsessionVoneRawEuclKOneTwoPos/3) \\
\addlinespace
NSD, $k = 1$ & N-raw & $\CvNsdNrawVoneRawCorrKOne$ (\CvNsdNrawVoneRawCorrKOnePos/8) & $\CvNsdNrawVoneZCorrKOne$ (\CvNsdNrawVoneZCorrKOnePos/8) & $\CvNsdNrawVoneRawEuclKOne$ (\CvNsdNrawVoneRawEuclKOnePos/8) \\
NSD, $k = 1$ & N-global & $\CvNsdNglobalVoneRawCorrKOne$ (\CvNsdNglobalVoneRawCorrKOnePos/8) & $\CvNsdNglobalVoneZCorrKOne$ (\CvNsdNglobalVoneZCorrKOnePos/8) & $\CvNsdNglobalVoneRawEuclKOne$ (\CvNsdNglobalVoneRawEuclKOnePos/8) \\
NSD, $k = 1$ & N-session & $\CvNsdNsessionVoneRawCorrKOne$ (\CvNsdNsessionVoneRawCorrKOnePos/8) & $\CvNsdNsessionVoneZCorrKOne$ (\CvNsdNsessionVoneZCorrKOnePos/8) & $\CvNsdNsessionVoneRawEuclKOne$ (\CvNsdNsessionVoneRawEuclKOnePos/8) \\
\addlinespace
NSD, $k = 3$ & N-raw & $\CvNsdNrawVoneRawCorrKThree$ (\CvNsdNrawVoneRawCorrKThreePos/8) & $\CvNsdNrawVoneZCorrKThree$ (\CvNsdNrawVoneZCorrKThreePos/8) & $\CvNsdNrawVoneRawEuclKThree$ (\CvNsdNrawVoneRawEuclKThreePos/8) \\
NSD, $k = 3$ & N-global & $\CvNsdNglobalVoneRawCorrKThree$ (\CvNsdNglobalVoneRawCorrKThreePos/8) & $\CvNsdNglobalVoneZCorrKThree$ (\CvNsdNglobalVoneZCorrKThreePos/8) & $\CvNsdNglobalVoneRawEuclKThree$ (\CvNsdNglobalVoneRawEuclKThreePos/8) \\
NSD, $k = 3$ & N-session & $\CvNsdNsessionVoneRawCorrKThree$ (\CvNsdNsessionVoneRawCorrKThreePos/8) & $\CvNsdNsessionVoneZCorrKThree$ (\CvNsdNsessionVoneZCorrKThreePos/8) & $\CvNsdNsessionVoneRawEuclKThree$ (\CvNsdNsessionVoneRawEuclKThreePos/8) \\
\bottomrule
\end{tabular}
\end{adjustbox}
\caption{Untrained minus Backprop ($\Delta$) at V1 under the three voxel normalisation conventions
(Section~\ref{sec:convmethods}), for the paper cell (raw model units, correlation distance), the
standardised cell (model units z-scored over the images) and raw units with Euclidean distance.
Mean over seeds 0--4, then subjects; in brackets the number of subjects with $\Delta > 0$.
T-global and N-session are the conventions of the respective data sets (THINGS: the paper; NSD:
the NSD standard).}
\label{tab:conventions}
\end{table}

\begin{figure}[t]
\centering
\includegraphics[width=\textwidth]{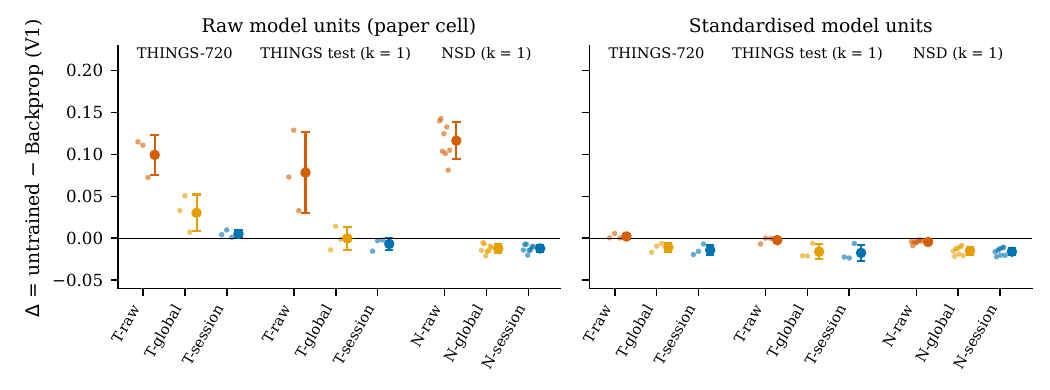}
\caption{Untrained minus Backprop at V1 under the three voxel normalisation conventions of each data set (step 48). Left: paper cell (raw model units, correlation distance); right: model units standardised over the images. Small dots: subjects (THINGS: 3; NSD: 8); large dots: mean with $\pm 1$ SD over subjects. Conventions: raw betas (vermillion), z-scored over all trials (orange), z-scored within session (blue).}
\label{fig:conv}
\end{figure}

Table~\ref{tab:conventions} and Fig.~\ref{fig:conv} show the paper cell under the three conventions of
each data set (step 48). In THINGS-720 the advantage grew when the betas were not normalised
($\CvTsevTrawVoneRawCorr$ under T-raw) and nearly vanished when they were normalised within session
($\CvTsevTsessionVoneRawCorr$ under T-session; NB3 met). The THINGS test images, whose presentations are
spread over 12 sessions, showed no advantage under T-global ($\CvTtestTglobalVoneRawCorrKOne$ at $k = 1$)
but a clear one under T-raw ($\CvTtestTrawVoneRawCorrKOne$ at $k = 1$, $\CvTtestTrawVoneRawCorrKOneTwo$ at
$k = 12$). Normalising over all trials leaves a profile common to the analysed trials only when they come
from one session: under N-global, which normalises NSD over all sessions as T-global does for THINGS, the
paper cell was $\CvNsdNglobalVoneRawCorrKOne$ at $k = 1$ ($\CvNsdNglobalVoneRawCorrKThree$ at $k = 3$;
\CvNsdNglobalVoneRawCorrKOnePos{} of 8 subjects above 0 in both),
because the presentations of the 515 images are spread over 30 sessions; the THINGS test images, spread
over 12 sessions, likewise showed no advantage under T-global. In NSD, the pre-specified prediction that raw betas create the advantage was met: under N-raw
the paper cell was $\CvNsdNrawVoneRawCorrKOne$ at $k = 1$, positive in \CvNsdNrawVoneRawCorrKOnePos{} of
8 subjects (sign-flip $p = \CvNsdNrawVoneRawCorrKOneP$), against $\CvNsdNsessionVoneRawCorrKOne$ under the
NSD standard (NB1 and NB2 met). Under N-raw the multiverse had \MvNrawAbove{} distinct analysis with an
interval above 0, the paper cell itself; \MvNrawBelow{} were below 0 and \MvNrawIncl{} included 0
(Fig.~\ref{fig:multi}). Standardising the model units removed most of the effect
($\CvTsevTrawVoneZCorr$ under T-raw, $\CvNsdNrawVoneZCorrKOne$ under N-raw), and Euclidean distance was
almost unaffected by the convention. The untrained advantage of version~1 is therefore a consequence of
the normalisation convention: it appears when a response profile common to all trials is left in the
voxel patterns and the model units are not centred, under correlation distance.

\subsection{Response amplitude}
\label{sec:amplitude}
\begin{figure}[t]
\centering
\includegraphics[width=\textwidth]{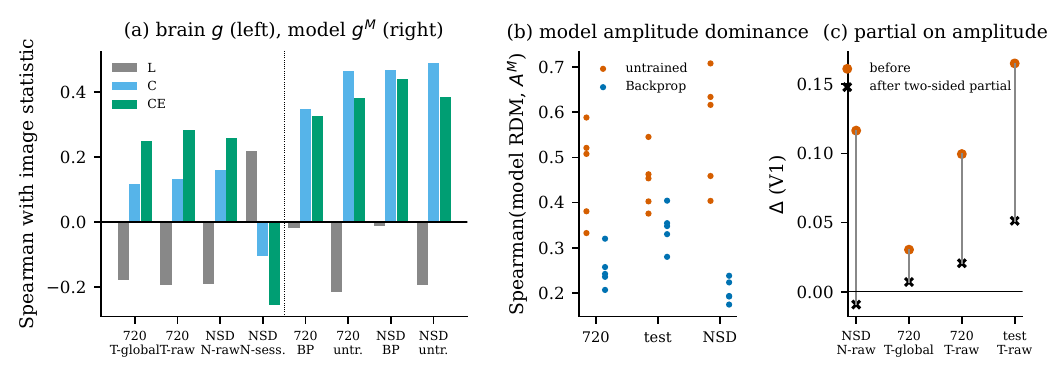}
\caption{Response amplitude. (a)~Spearman correlation of the brain amplitude $g$ at V1 (mean over subjects) and of the model amplitude $g^M$ (Conv1, mean over seeds) with luminance $L$, RMS contrast $C$ and contrast energy CE. (b)~Amplitude dominance of the model RDMs (sum form), one dot per seed (D1). (c)~$\Delta$ at V1 before and after the two-sided partial on the amplitude RDMs (sum form; D3 for NSD).}
\label{fig:amp}
\end{figure}

With a common profile left in, the correlation distance between two patterns depends on how strongly
each is aligned with the profile. In THINGS-720 under T-global, this brain amplitude $g$ correlated more
strongly with contrast energy than with luminance in every subject (G1 met; Spearman with CE
$\AmpTsevTglobalVoneSubOneCE$, $\AmpTsevTglobalVoneSubTwoCE$, $\AmpTsevTglobalVoneSubThreeCE$; with $L$
$\AmpTsevTglobalVoneSubOneL$, $\AmpTsevTglobalVoneSubTwoL$, $\AmpTsevTglobalVoneSubThreeL$), consistent
with the dependence of V1 responses (Fig.~\ref{fig:amp}a) to natural images on contrast energy \citep{kay2013}. Both networks
matched this amplitude about equally: the Spearman correlation between $g$ and the model amplitude $g^M$
was larger for the untrained network in only \RuleGTwo{} of 3 THINGS subjects (G2 not met; mean
$\AmpGmTsevTglobalVoneKOneRand$ against $\AmpGmTsevTglobalVoneKOneBp$) and \RuleGThree{} of 8 NSD
subjects under N-raw (G3 not met; $\AmpGmNsdNrawVoneKOneRand$ against $\AmpGmNsdNrawVoneKOneBp$). The
interpretation rule of step 48 has no case for this pattern (NB1 and NB2 met, G2 or G3 not met), and its
outcome is: ``no statement of the rule applies''.

Step 48b asked instead how much each RDM is dominated by amplitude structure, using the sum form
$A_{ij} = 2 - g_i - g_j$ (Section~\ref{sec:ampmethods}). The untrained network's RDM was more
amplitude-dominated than Backprop's in every seed in each data set (D1 met; Fig.~\ref{fig:amp}b; mean Spearman(RDM, $A^M$)
$\DomTsevConvOneSumRand$ against $\DomTsevConvOneSumBp$ in THINGS-720, $\DomTestConvOneSumRand$ against
$\DomTestConvOneSumBp$ in the test images, $\DomNsdConvOneSumRand$ against $\DomNsdConvOneSumBp$ in NSD).
The brain RDM was amplitude-dominated when the profile was left in and not when it was removed (D2 met;
THINGS-720 $\BDomTsevTrawVoneKOneSum$ under T-raw against $\BDomTsevTsessionVoneKOneSum$ under T-session;
NSD $\BDomNsdNrawVoneKOneSum$ under N-raw against $\BDomNsdNsessionVoneKOneSum$ under N-session, larger in
\RuleDTwoNsd{} of 8 subjects). Removing the amplitude structure from both RDMs by the two-sided partial
reduced the NSD advantage under N-raw from $\PbNsdNrawVoneKOneNone$ to $\PbNsdNrawVoneKOneSum$, with
$\Delta \leq +0.020$ in \RuleDThreeLe{} of 8 subjects (D3 met; Table~\ref{tab:partial48b}). The
interpretation rule gave: ``the untrained advantage under conventions that leave a common profile in both
patterns is attributed to amplitude structure dominating both RDMs, with the untrained network's RDM more
amplitude-dominated; not to a better match of per-image amplitude.'' Descriptively, the model's amplitude
dominance followed the moment ratio $m$ of version~1 across the 10 checkpoints (Spearman
$\DomMrNsdConvOneSum$ in NSD and $\DomMrTsevConvOneSum$ in THINGS-720; $m$ $\MratioTsevConvOneRand$ for the
untrained network against $\MratioTsevConvOneBp$ for Backprop in THINGS-720), which links the result to the
mean-profile alignment of Section~\ref{sec:gainaxis}. With the difference form $|g_i - g_j|$ the partial
reduced the NSD advantage only to $\PbNsdNrawVoneKOneDiff$ (Appendix~\ref{app:conv}).

\subsection{Why ridge is unaffected}
\label{sec:ridgeunaffected}
A response profile common to all trials is a constant per voxel. A voxel-wise readout fits an intercept
per voxel and is scored by a correlation over images per voxel, so the constant is absorbed and the score
does not change. This is consistent with the ridge results under every normalisation that was tested
(RD1, RD2 in NSD; Y1 to Y4 with run centring and without voxel selection). The readout and the
correlation-distance RDM thus differ in exactly the quantity that the convention controls.

\subsection{Open points}
\label{sec:openres}
Three observations are not explained. (1)~In THINGS-720 the untrained network was ahead under raw units
with Euclidean distance under every convention ($\CvTsevTrawVoneRawEucl$, $\CvTsevTglobalVoneRawEucl$,
$\CvTsevTsessionVoneRawEucl$, each in \CvTsevTglobalVoneRawEuclPos{} of 3 subjects). Euclidean distance is
unaffected by subtracting a profile common to all patterns, so this contrast is not created by the
convention; it is absent in NSD (Table~\ref{tab:convfull}). (2)~After removing the amplitude structure
from both RDMs, a residual untrained advantage remains in THINGS, unlike NSD: $\PbTsevTrawVoneKOneSum$
under T-raw (\PbTsevTrawVoneKOneSumPos{} of 3 subjects positive), $\PbTsevTglobalVoneKOneSum$ under
T-global, and $\PbTestTrawVoneKOneTwoSum$ in the test images under T-raw at $k = 12$. (3)~The sign of the
relation between the brain amplitude and the image statistics depends on the convention: in NSD the
Spearman correlation of $g$ with contrast energy was $\AmpNsdNrawVoneKOneCE$ under N-raw and
$\AmpNsdNsessionVoneKOneCE$ under N-session, and in the test images $\AmpTestTrawVoneKOneTwoCE$ under T-raw
and $\AmpTestTglobalVoneKOneTwoCE$ under T-global (Appendix~\ref{app:conv}, Table~\ref{tab:amplitude}).

\subsection{A side finding: presentation order}
\label{sec:sidefinding}
The V1 RDM of sub-02 correlated with the presentation-position RDM (Spearman $\OrderRhoSubTwo$; sub-01
$\OrderRhoSubOne$, sub-03 $\OrderRhoSubThree$). sub-01 and sub-02 saw the 720 images in the same order and
run assignment, so their RDMs share any order-related component, and the agreement between these two
subjects in THINGS is a limited test of cross-subject consistency. Partialling the position RDM did not
change the group contrast (W5).

\section{Discussion}
\paragraph{Strength of the evidence.}
Table~\ref{tab:evidence} collects the V1 contrast under each readout and convention in the three image
sets. Only the ridge rows stem from pre-specified primary analyses (THINGS, and its unchanged application
to NSD). All RSA rows come from exploratory steps with rules committed before each step; the rules of
later steps were written after earlier results were known, so agreement among them is not independent
confirmation. The NSD rows under N-raw, however, test a prediction (NB1, NB2) that was committed before
the raw NSD betas were loaded, on a normalisation of the NSD data (raw betas) that had not been analysed before.

\begin{table}[t]
\centering
\footnotesize
\setlength{\tabcolsep}{4pt}
\begin{tabular}{@{}>{\raggedright\arraybackslash}p{0.33\textwidth}>{\raggedright\arraybackslash}p{0.17\textwidth}>{\raggedright\arraybackslash}p{0.17\textwidth}>{\raggedright\arraybackslash}p{0.25\textwidth}@{}}
\toprule
Readout and convention (V1) & THINGS-720 & THINGS test images ($k = 1$) & NSD ($k = 1$; ridge $k = 3$) \\
\midrule
Ridge, $g = 4$ (pre-specified; RD1) & $\PreregRandBpGFour$ (\PreregRandBpGFourPos/5 seeds $> 0$) & -- & $\NsdRidgeVoneGFour$ (\NsdRidgeVoneGFourNeg/8 $< 0$) \\
RSA, paper cell, standard convention (T-global, N-session) & $\CvTsevTglobalVoneRawCorr$ (\CvTsevTglobalVoneRawCorrPos/3) & $\CvTtestTglobalVoneRawCorrKOne$ (\CvTtestTglobalVoneRawCorrKOnePos/3) & $\CvNsdNsessionVoneRawCorrKOne$ (\CvNsdNsessionVoneRawCorrKOnePos/8) \\
RSA, paper cell, raw betas (T-raw, N-raw) & $\CvTsevTrawVoneRawCorr$ (\CvTsevTrawVoneRawCorrPos/3) & $\CvTtestTrawVoneRawCorrKOne$ (\CvTtestTrawVoneRawCorrKOnePos/3) & $\CvNsdNrawVoneRawCorrKOne$ (\CvNsdNrawVoneRawCorrKOnePos/8) \\
RSA, paper cell, session-normalised (T-session, N-session) & $\CvTsevTsessionVoneRawCorr$ (\CvTsevTsessionVoneRawCorrPos/3) & $\CvTtestTsessionVoneRawCorrKOne$ (\CvTtestTsessionVoneRawCorrKOnePos/3) & $\CvNsdNsessionVoneRawCorrKOne$ (\CvNsdNsessionVoneRawCorrKOnePos/8) \\
RSA, standardised units, raw betas & $\CvTsevTrawVoneZCorr$ (\CvTsevTrawVoneZCorrPos/3) & $\CvTtestTrawVoneZCorrKOne$ (\CvTtestTrawVoneZCorrKOnePos/3) & $\CvNsdNrawVoneZCorrKOne$ (\CvNsdNrawVoneZCorrKOnePos/8) \\
RSA, raw units, Euclidean, standard convention & $\CvTsevTglobalVoneRawEucl$ (\CvTsevTglobalVoneRawEuclPos/3) & $\CvTtestTglobalVoneRawEuclKOne$ (\CvTtestTglobalVoneRawEuclKOnePos/3) & $\CvNsdNsessionVoneRawEuclKOne$ (\CvNsdNsessionVoneRawEuclKOnePos/8) \\
RSA, paper cell, raw betas, amplitude structure removed on both sides (D3) & $\PbTsevTrawVoneKOneSum$ (\PbTsevTrawVoneKOneSumPos/3) & $\PbTestTrawVoneKOneSum$ (\PbTestTrawVoneKOneSumPos/3) & $\PbNsdNrawVoneKOneSum$ (\PbNsdNrawVoneKOneSumPos/8) \\
Multiverse: distinct analyses above / below / including 0 & \MultiDistinctPos{} / \MultiDistinctNeg{} / \MultiDistinctIncl & -- & N-session \MvNsessionAbove{} / \MvNsessionBelow{} / \MvNsessionIncl; N-raw \MvNrawAbove{} / \MvNrawBelow{} / \MvNrawIncl \\
\bottomrule
\end{tabular}
\caption{Evidence table: untrained minus Backprop at V1 under each readout and voxel normalisation convention, mean over seeds and subjects; in brackets the number of subjects with $\Delta > 0$ (ridge rows: the number of seeds or subjects with the sign stated in the cell). THINGS ridge is the pre-specified primary analysis (mean over the three subjects). The rows are not independent (same images, subjects and networks).}
\label{tab:evidence}
\end{table}

\paragraph{What the V1 advantage measured.}
The untrained advantage at V1 that we reported earlier \citep{evalres} was a consequence of the
normalisation convention. It appeared, in both data sets, when a voxel response profile common to all
trials was left in the patterns and model units were not centred, under correlation distance, it nearly
vanished when voxels were normalised within session or model units were standardised, and it was reversed
under a fitted readout. NSD, normalised within session as its documentation recommends, shows no analysis in which
the untrained network is ahead; with raw betas it shows the advantage in every subject, confined to the
paper cell. With the profile left in, both RDMs are dominated by per-image response amplitude, the
untrained network's more so, and removing this structure from both RDMs removed the advantage in NSD;
both networks matched the amplitude itself about equally. In V1 this amplitude tracked contrast energy
more than luminance, as expected from the contrast-energy dependence of V1 BOLD responses to natural
images \citep{kay2013}. The asymmetry between dark and light responses \citep{kremkow2014} is a possible
point of contact, but our dark-dominance index is close to a mirror of luminance and adds no separate
evidence. The amplitude structure is a property of the comparison under one convention, and we do not
claim that it is a physiological gain.

\paragraph{Relation to methodological work on demeaning.}
That removing or keeping the mean of measured patterns changes what RSA measures has been argued with
simulations \citep{ramirez2017}, by proposing to retain the regional-mean profile explicitly
\citep{taylor2025}, and by showing that pooling within voxels amplifies activity shared by all sampled
neurons, which mean removal largely corrects \citep{bossio2025}. Our data give an empirical case in which
this choice reverses a model ranking, and in which the mechanism follows the shared-component account:
the component common to all patterns dominates both RDMs, and the network whose RDM is more dominated by
its own common component is favoured.

\paragraph{The norm hypothesis.}
In the subj01 pilot the untrained network was ahead under Euclidean distance and on the norm component,
which suggested that the advantage sits in pattern norms. On the seven held-out subjects this was not
replicated (Q1, Q2 not met), whereas the trained network was ahead on the angular component at V1 and
LO-obj (Q3, Q4 met), as in THINGS.

\paragraph{Ridge and RSA address different questions.}
Raw RSA asks which unaltered distance structure resembles the measured one, centred angular RSA asks which
direction-based structure resembles it after centring, and ridge asks which features predict the voxel
responses under a regularised linear readout. A different ranking under ridge therefore does not by itself
show that an RSA ranking is invalid; here the two readouts differ in one quantity, a profile common to all
trials, which the ridge intercept absorbs and correlation-distance RSA does not. That rankings change across
defensible analyses shows that the metric matters, not that a single analysis is invalid: a benchmark may have
reasons to keep the regional-mean profile \citep{taylor2025}, and which properties a comparison should
preserve depends on what it is meant to explain \citep{cao2024}. A brain-alignment score is a measurement
defined by the evaluation procedure, including the normalisation of the neural data, not an intrinsic
property of a model.

\paragraph{A luminance-and-contrast bank exceeds every network at V1.}
Under the ridge readout, the two-feature-map luminance-and-contrast bank predicted V1 better than every
network condition at every pooling grid, in THINGS and in NSD (point estimates, mean over subjects), and
in NSD the networks reached only \CeilVoneFracMin--\CeilVoneFracMax\% of the lower bound of the noise
ceiling under RSA. This limits what V1 comparisons in these data can reveal about learning rules.

\paragraph{Recommendations.}
(1)~Centre or normalise voxel responses within session (or over the evaluated stimuli), and report the
sensitivity of RSA rankings to this choice and to the centring of model units; normalising voxels over all
trials of an experiment is not equivalent: it leaves a profile common to the analysed trials whenever
these come from one session, as in THINGS-720. (2)~Replicate model rankings on a second data set with a
different acquisition and normalisation. (3)~Report noise ceilings. (4)~Check simple image-statistic
baselines, such as a luminance-and-contrast bank. (5)~Avoid presentation orders shared across subjects, or
model them. (6)~Report norm and angular components separately and a small multiverse of distance measure
and normalisation for ranking claims; for benchmark platforms, a fixed primary procedure with supplementary
diagnostics for the leading models (the ranking under within-session normalisation and under unit centring)
would make the dependence visible without changing the reference score. (7)~State how the normalisation
statistics of untrained networks were set.

\paragraph{Evaluation resolution.}
The resolution dependence of the V1 gap in our CIFAR-trained networks vanished under unit centring and
standardisation and was absent under ridge; in NSD the own CNN was behind at every resolution. In the ImageNet
models a small dependence above the training resolution remained in THINGS (GEN4 not met).

\paragraph{Consequences for our earlier work.}
The headline of \citet{evalres}, that untrained networks exceed backpropagation at V1 at 224\,px and that
this gap depends on evaluation resolution, and the corresponding claim in \citet{leutenegger2026} are a
consequence of the normalisation convention of those analyses. Revised versions of the earlier preprints
refer to this work.

\paragraph{Limitations.}
The networks are one own CNN evaluated at one layer per ROI and six ImageNet architectures with one set of
trained weights each. NSD shows natural scenes and THINGS isolated objects, so the data sets differ in more
than their normalisation. LO-obj is a constructed ROI. The NSD multiverse missed M2 by one analysis. All
models are far below the noise ceiling. In THINGS, presentation positions were taken from the row order of
the stimulus metadata, the unit of the stored betas is not documented, and sub-01 and sub-02 share order
and run assignment; THINGS has three subjects. The intervals describe stimulus variation (RSA) and seeds
(THINGS ridge) or subjects (NSD sign-flip tests); intervals within the multiverse are not corrected for
multiple comparisons. The THINGS Euclidean contrast and the residual THINGS advantage after the two-sided
partial are not explained. The banded ridge readout failed its sanity rule and was not interpreted. The
primary analysis in THINGS was not blind, several of its pre-specified instruments were not implemented
(Section~\ref{sec:inference}), and \FailedPredictions{} of \TotalPredictions{} evaluated pre-specified
predictions were not met (Appendix~\ref{app:failed}).

\section{Conclusion}
Whether untrained networks beat trained ones at V1 depends on a normalisation convention. In two fMRI data
sets, a pre-specified ridge readout placed the trained network ahead, and correlation-distance RSA favoured
the untrained network only when a voxel response profile common to all trials was left in the patterns and
the model units were not centred. A prediction committed before the NSD data were reloaded confirmed that
this convention creates the advantage in a second data set. Under that convention both RDMs are dominated by
per-image response amplitude, the untrained network's more so, and removing this amplitude structure from
both RDMs removed the advantage in NSD. Normalising voxel responses within session, reporting the
sensitivity of rankings to this choice and to model-unit centring, and replicating on a second data set make
such comparisons interpretable.

\section*{Acknowledgements}
We thank Kartik Pradeepan for detailed feedback on a draft of this manuscript, in particular on the distinction between sensitivity and validity and on the recommendations for benchmarking platforms.

Collection of the NSD dataset was supported by NSF IIS-1822683 and NSF IIS-1822929.

Claude (Anthropic), a large language model, was used as an assistant throughout this work: to suggest and refine analyses and decision rules, to write and run analysis code, and to draft most of the manuscript text. The author defined the research question, decided which analyses to run and which claims to make, checked the results with reproduction tests and gates, and takes full responsibility for the content, including any errors.

\bibliographystyle{plainnat}
\bibliography{refs}

\appendix
\section{Lower-edge gate}
\label{app:lowedge}
The ridge fits select the regularisation $\alpha$ per voxel and outer fold from a grid whose
lowest value is $10^{-6}$. Following the amendment of 2026-10-03 (Section~\ref{sec:inference}),
the selection of the lowest value at a reliable voxel does not abort the run. It is logged
and, after the run, classified on the mean inner-fold MSE curve of that voxel in that outer
fold. The curve is extended by the diagnostic values $10^{-7}$, $10^{-8}$ and $10^{-9}$, and
the relative gain is $(\mathrm{MSE}(10^{-6}) - \mathrm{MSE}_{\min}) / \mathrm{MSE}(10^{-6})$,
with $\mathrm{MSE}_{\min}$ the minimum over the extended grid. A relative gain below $10^{-3}$
is a plateau and the selection is valid; otherwise the cell would be excluded from every
result. Smaller values of $\alpha$ are not used, because there the solution is determined mainly by
eigenvalues of the dual kernel at rounding level ($|\lambda| \leq 1.9 \times 10^{-12}$ in the
diagnosed cases). The gate decides whether a cell is kept and never changes a score.

The first cell with such selections was a condition of the pilot (untrained network with
calibrated BN statistics, one seed, $g = 1$, sub-01). There the standardised training design had full column rank
and no dead features (step report in the repository); the pilot diagnosis used a wider
diagnostic grid than the gate. Table~\ref{tab:lowedge} lists every analysis with such
selections. Steps~7 to~9 involve no ridge fits. The number of selections at the lowest value changed
between runs of identical fits (\FSkillLowEdgeHits{} versus \LowEdgePrimaryModelHits{} on the
model fits of the primary analysis, \LowEdgeEtwoHits{} versus \LowEdgePilotHits{} in the
pilot cell) because of near-ties under single-thread BLAS; scores agreed to
\FSkillGateMaxDelta{} and the gate classification was unchanged.

\begin{table}[H]
\centering
\footnotesize
\setlength{\tabcolsep}{3pt}
\begin{tabular}{@{}p{0.40\textwidth}rrrr@{}}
\toprule
Analysis & Hits & Plateau & Excluded & Max.\ gain \\
\midrule
Pilot diagnosis (sub-01, $g = 1$) & \LowEdgePilotHits & \LowEdgePilotPlateau & \LowEdgePilotExcluded & \LowEdgePilotMaxGain \\
RSA versus ridge: collapse, random folds (sub-01) & \LowEdgeEtwoHits & \LowEdgeEtwoPlateau & \LowEdgeEtwoExcluded & \LowEdgeEtwoMaxGain \\
RSA versus ridge: run-grouped folds (sub-01) & \LowEdgeEthreeHits & \LowEdgeEthreePlateau & \LowEdgeEthreeExcluded & \LowEdgeEthreeMaxGain \\
RSA versus ridge: encoding-RSA (sub-01; repeats the hits above) & \LowEdgeEfourHits & \LowEdgeEfourPlateau & \LowEdgeEfourExcluded & \LowEdgeEfourMaxGain \\
Other subjects, $g = 1$ (sub-02, sub-03) & \LowEdgeSubjectsHits & \LowEdgeSubjectsPlateau & \LowEdgeSubjectsExcluded & \LowEdgeSubjectsMaxGain \\
Diagnosis of sub-03 (controls, $g \in \{1, 4, 8\}$) & \LowEdgeSubThreeHits & \LowEdgeSubThreePlateau & \LowEdgeSubThreeExcluded & \LowEdgeSubThreeMaxGain \\
Ridge across resolutions (sub-01, sub-02, $g = 1$) & \LowEdgeRidgeHits & \LowEdgeRidgePlateau & \LowEdgeRidgeExcluded & \LowEdgeRidgeMaxGain \\
Pre-specified primary analysis (3 subjects) & \LowEdgePrimaryHits & \LowEdgePrimaryPlateau & \LowEdgePrimaryExcluded & \LowEdgePrimaryMaxGain \\
Ridge robustness: model fits (3 subjects) & \LowEdgeStepSkillHits & \LowEdgeStepSkillPlateau & \LowEdgeStepSkillExcluded & \LowEdgeStepSkillMaxGain \\
Ridge robustness: joint readout & \LowEdgeStepJointHits & \LowEdgeStepJointPlateau & \LowEdgeStepJointExcluded & \LowEdgeStepJointMaxGain \\
Ridge robustness: permutation null & \LowEdgeStepPermHits & \LowEdgeStepPermPlateau & \LowEdgeStepPermExcluded & \LowEdgeStepPermMaxGain \\
\bottomrule
\end{tabular}
\caption{Selections of the lowest regularisation value at reliable voxels, classified by the
plateau gate. In the pilot diagnosis the plateau criterion was applied to the relative gain
of the wider diagnostic grid. All hits were plateaus, and no cell was excluded.}
\label{tab:lowedge}
\end{table}

\section{Run-grouped folds}
\label{app:runfolds}
Run-grouped cross-validation was an exploratory sensitivity analysis of sub-01
(\NReliableSubOne{} reliable V1 voxels). It was not part of the pre-specified plan, which fixes
random folds. The 720 stimuli come from 10 runs of 72. The 10 runs were permuted (seed 20260818)
and consecutive pairs of runs formed the 5 outer test folds. Within each outer training set, the
8 remaining runs were permuted (seed $20260818 + 1000\,(k + 1)$ for outer fold $k$) and
consecutive pairs formed the 4 inner folds. Standardisation on training rows, the regularisation
grid, the selection rule and the gate were unchanged. All 30 checkpoints were fitted at
$g \in \{1, 4, 8\}$.

Table~\ref{tab:runfolds} gives the untrained (uncalibrated BN statistics) minus Backprop
contrast, mean $\pm$ SD over the 5 seeds with the number of positive seeds. The contrast was
negative in every seed under both fold schemes. Run-grouped folds lowered the pooled $r$ of all 30
checkpoints by $\RunFoldLevelGOne$, $\RunFoldLevelGFour$ and $\RunFoldLevelGEight$ at
$g = 1, 4, 8$. Table~\ref{tab:encrsa} gives the contrast of encoding-RSA, in which the
out-of-fold predictions of the 501 voxels form a model RDM that is compared with the fMRI RDM by
the RSA statistic of Section~\ref{sec:rsa}; ``centred'' denotes predictions centred within each
outer fold.

\begin{table}[H]
\centering
\footnotesize
\setlength{\tabcolsep}{3pt}
\begin{tabular}{@{}cccc@{}}
\toprule
$g$ & Pooled $r$, random folds & Pooled $r$, run folds & Per-fold $r$, run folds \\
\midrule
1 & $\RunFoldPooledRandGOneMean \pm \RunFoldPooledRandGOneSd$ (\RunFoldPooledRandGOnePos/5) & $\RunFoldPooledGOneMean \pm \RunFoldPooledGOneSd$ (\RunFoldPooledGOnePos/5) & $\RunFoldPerfoldGOneMean \pm \RunFoldPerfoldGOneSd$ (\RunFoldPerfoldGOnePos/5) \\
4 & $\RunFoldPooledRandGFourMean \pm \RunFoldPooledRandGFourSd$ (\RunFoldPooledRandGFourPos/5) & $\RunFoldPooledGFourMean \pm \RunFoldPooledGFourSd$ (\RunFoldPooledGFourPos/5) & $\RunFoldPerfoldGFourMean \pm \RunFoldPerfoldGFourSd$ (\RunFoldPerfoldGFourPos/5) \\
8 & $\RunFoldPooledRandGEightMean \pm \RunFoldPooledRandGEightSd$ (\RunFoldPooledRandGEightPos/5) & $\RunFoldPooledGEightMean \pm \RunFoldPooledGEightSd$ (\RunFoldPooledGEightPos/5) & $\RunFoldPerfoldGEightMean \pm \RunFoldPerfoldGEightSd$ (\RunFoldPerfoldGEightPos/5) \\
\bottomrule
\end{tabular}
\caption{Untrained minus Backprop, median $r$ over reliable voxels of sub-01 (mean $\pm$ SD
over 5 seeds; positive seeds of 5).}
\label{tab:runfolds}
\end{table}

\begin{table}[H]
\centering
\footnotesize
\setlength{\tabcolsep}{3pt}
\begin{tabular}{@{}lcc@{}}
\toprule
Folds ($g = 1$) & Centred & Uncentred \\
\midrule
Random folds & $\EncRsaRandomCentredMean \pm \EncRsaRandomCentredSd$ (\EncRsaRandomCentredPos/5) & $\EncRsaRandomUncentredMean \pm \EncRsaRandomUncentredSd$ (\EncRsaRandomUncentredPos/5) \\
Run-grouped folds & $\EncRsaRunCentredMean \pm \EncRsaRunCentredSd$ (\EncRsaRunCentredPos/5) & $\EncRsaRunUncentredMean \pm \EncRsaRunUncentredSd$ (\EncRsaRunUncentredPos/5) \\
\bottomrule
\end{tabular}
\caption{Untrained minus Backprop, encoding-RSA of sub-01 (mean $\pm$ SD over 5 seeds;
positive seeds of 5).}
\label{tab:encrsa}
\end{table}

\section{Band-pass inputs}
\label{app:bandpass}
The band-pass inputs were built from the image after resizing and centre-cropping to
224\,px and before the channel normalisation. Gaussian low-pass filters $L_r$ with
$r \in \{112, 56, 28, 14\}$ have an amplitude response of 0.5 at $r/2$ cycles per image; the
octave bands are differences of consecutive filters. Band 0--14 is the coarse residual $L_{14}$ and
contains the image mean. Condition ``only 0--14'' keeps this band alone, and condition ``notch
0--14'' is the native image minus this band, shifted by the CIFAR-10 channel-mean grey. Both were
evaluated at NATIVE 224\,px with all 30 checkpoints.

Table~\ref{tab:bandgap} gives the V1 contrast (untrained minus Backprop) and the LOC contrast
(Backprop minus untrained) with raw and channel-z-scored features, with 95\% stimulus-bootstrap
intervals. The V1 contrast with the coarse band alone was positive under raw features, as for
native input ($+\VoneGapRawTwoTwoFour$); without the coarse band it was negative.
Table~\ref{tab:bandm} gives the moment ratio $m$ of Section~\ref{sec:standardization}, mean over
seeds. The prediction P\_mech2 that relates $m$ to the difference between raw and z-scored RSA
was not met (Appendix~\ref{app:failed}).

\begin{table}[H]
\centering
\footnotesize
\setlength{\tabcolsep}{3pt}
\begin{tabular}{@{}lcc@{}}
\toprule
 & Only 0--14 & Notch 0--14 \\
\midrule
V1, raw & $\BandOnlyVoneRaw\ [\BandOnlyVoneRawLo, \BandOnlyVoneRawHi]$ & $\BandNotchVoneRaw\ [\BandNotchVoneRawLo, \BandNotchVoneRawHi]$ \\
V1, z & $\BandOnlyVoneZ\ [\BandOnlyVoneZLo, \BandOnlyVoneZHi]$ & $\BandNotchVoneZ\ [\BandNotchVoneZLo, \BandNotchVoneZHi]$ \\
LOC, raw & $\BandOnlyLocRaw\ [\BandOnlyLocRawLo, \BandOnlyLocRawHi]$ & $\BandNotchLocRaw\ [\BandNotchLocRawLo, \BandNotchLocRawHi]$ \\
LOC, z & $\BandOnlyLocZ\ [\BandOnlyLocZLo, \BandOnlyLocZHi]$ & $\BandNotchLocZ\ [\BandNotchLocZLo, \BandNotchLocZHi]$ \\
\bottomrule
\end{tabular}
\caption{Contrasts for band-pass inputs (NATIVE 224\,px; V1: untrained minus Backprop, LOC:
Backprop minus untrained; 95\% stimulus-bootstrap intervals).}
\label{tab:bandgap}
\end{table}

\begin{table}[H]
\centering
\footnotesize
\setlength{\tabcolsep}{3pt}
\begin{tabular}{@{}lcc@{}}
\toprule
Input & $m$, untrained & $m$, Backprop \\
\midrule
Native 224\,px & \BandMNativeRand & \BandMNativeBp \\
Only 0--14 & \BandMOnlyRand & \BandMOnlyBp \\
Notch 0--14 & \BandMNotchRand & \BandMNotchBp \\
\bottomrule
\end{tabular}
\caption{Moment ratio $m$ of the Conv1 features (mean over 5 seeds).}
\label{tab:bandm}
\end{table}

\section{Failed predictions}
\label{app:failed}
Tables~\ref{tab:failed} and~\ref{tab:failedfollowup} list the \FailedPredictions{} pre-specified
predictions that were not met: \FailedPredictionsVone{} of the \TotalPredictionsVone{} evaluated
predictions of version~1 and \FailedPredictionsFollowup{} of the \TotalPredictionsFollowup{}
evaluated predictions of the follow-up analyses. Each rule was committed before the scripts of its
step were run (Section~\ref{sec:inference}). All predictions, decision rules and interpretation
rules, with wording, amendment commit, result and source file, are in
\texttt{paper/PREDICTIONS\_LEDGER.csv}.

\begin{table}[H]
\centering
\small
\begin{tabular}{@{}p{0.11\textwidth}p{0.37\textwidth}p{0.36\textwidth}p{0.07\textwidth}@{}}
\toprule
Prediction & Rule and threshold & Result & Section \\
\midrule
P\_mech2 & $m_{\mathrm{BP}}(\text{notch}) > m_{\mathrm{BP}}(\text{only})$ in 5 of 5 seeds,
  and Spearman$(m, d) \geq 0.6$ over 50 cells
  & \PMechTwoNotchSeeds{} of 5 seeds; Spearman \PMechTwoSpearman
  & \ref{sec:standardization} \\
G2 & partial V1 gap at NATIVE 224\,px with upper CI bound $\leq 0.010$, and
  $S_{\mathrm{partial}} \leq 0.5\,S_{\mathrm{paper}}$
  & gap $+\GTwoPartialGap$ [$\GTwoPartialGapLo$, $\GTwoPartialGapHi$];
  $S_{\mathrm{partial}} = +\GTwoPartialSlope$ [$\GTwoPartialSlopeLo$, $\GTwoPartialSlopeHi$]
  & \ref{sec:gainaxis} \\
GEN3 & LOC trained $-$ untrained at 224\,px with CI above 0 in all 4 bootstrap cells,
  for $\geq 5$ of 6 models
  & \GenThreeModels{} models; the raw/euclidean cell includes 0 for every model
  & \ref{sec:multiverse} \\
GEN4 & for every model whose $S_{\mathrm{gen}}$ (raw, correlation) has a CI excluding 0:
  $|S_{\mathrm{gen}}(z)| \leq 0.5\,|S_{\mathrm{gen}}(\text{raw})|$
  & \GenFourQualified{} qualifying models, none within the bound
  & \ref{sec:multiverse} \\
SEP2 & V1 norm contrast ($F$ = center) $> 0$ with CI above 0 in the own CNN and in
  $\geq 4$ of 6 ImageNet models
  & own CNN $+\SepOwnVoneNormCenter$ [$\SepOwnVoneNormCenterLo$, $\SepOwnVoneNormCenterHi$];
  ImageNet \SepTwoModels{} of 6
  & \ref{sec:gaingeometry} \\
M3 & parallel component: V1 gap at NATIVE 224\,px $> 0$ with CI above 0, and resolution
  slope with CI above 0
  & gap $+\MpParOwnVoneNatTwoTwoFour$ [$\MpParOwnVoneNatTwoTwoFourLo$, $\MpParOwnVoneNatTwoTwoFourHi$];
  slope $+\MpParOwnSlope$ [$\MpParOwnSlopeLo$, $\MpParOwnSlopeHi$]
  & \ref{sec:ablation} \\
\bottomrule
\end{tabular}
\caption{Pre-specified predictions of version~1 that were not met. Intervals are 95\% stimulus-bootstrap
percentile intervals.}
\label{tab:failed}
\end{table}

\begin{table}[H]
\centering
\footnotesize
\begin{tabular}{@{}p{0.11\textwidth}p{0.44\textwidth}p{0.30\textwidth}p{0.05\textwidth}@{}}
\toprule
Prediction (step) & Rule and threshold & Result & Section \\
\midrule
B1 (40) & sanity: banded joint readout not below the bank alone by more than 0.002 in any fit
  & minimum $\BandedBOneMinGFour$ at $g = 4$ & \ref{sec:thingsfollowup} \\
C1 (40) & test images, raw units, correlation vs presentation-averaged brain RDM: $\Delta > 0$
  & $\CvRsaVoneRawBOne$ [$\CvRsaVoneRawBOneLo$, $\CvRsaVoneRawBOneHi$] & \ref{sec:thingsfollowup} \\
D-A2 (41) & conditional banded search: no fit below the bank alone by more than 0.002
  & minimum $\DATwoMin$ & \ref{sec:thingsfollowup} \\
N3 (41) & test images, paper cell, single presentations: $\Delta(1) > 0$
  & $\NoiseVoneRawKOneFour$ & \ref{sec:thingsfollowup} \\
Q1 (42b) & NSD V1, raw units, Euclidean: $\Delta > 0$ in $\geq 5$ of 7 held-out subjects, $p \leq 0.05$
  & $\NsdHeldVoneKThreeRawEuclidean$, \NsdHeldVoneKThreeRawEuclideanPos{} of 7 positive & \ref{sec:nsdcomponents} \\
Q2 (42b) & NSD V1, norm component $N$: $\Delta > 0$ in $\geq 5$ of 7, $p \leq 0.05$
  & $\NsdHeldVoneKThreeCenterN$, \NsdHeldVoneKThreeCenterNPos{} of 7 positive & \ref{sec:nsdcomponents} \\
M2 (43) & NSD multiverse: interval below 0 in at least half of the distinct analyses
  & \MvNsessionBelow{} of \MvNsessionDistinct{} & \ref{sec:nsdmultiverse} \\
W2 (44) & THINGS-720: partialling luminance and contrast RDMs reduces $\Delta$ to $\leq +0.010$
  & $\RuleWTwo$ & \ref{sec:decomposition} \\
W6 (45) & THINGS-720: partialling luminance, contrast and coarse-luminance RDMs reduces $\Delta$ to $\leq +0.010$
  & $\RuleWSix$ & \ref{sec:decomposition} \\
RC2 (47) & with run centring, luminance correlation of the V1 RDM / LOSO bound $< 0.5$
  & \RuleRcTwo & \ref{sec:decomposition} \\
RC3 (47) & run-level variance of run-mean luminance larger for the sub-01/02 than for the sub-03 assignment
  & percentile \RunLumPctSubOneTwo{} vs \RunLumPctSubThree{} (reversed) & \ref{sec:decomposition} \\
RC4 (47) & with run centring, at most 2 of 28 distinct analyses with interval above 0
  & \RuleRcFour & \ref{sec:decomposition} \\
CO1 (47) & partialling the predecessor-luminance RDM changes $\Delta$ by $\geq 0.005$
  & change $\RuleCoOne$ & \ref{sec:decomposition} \\
X1 (47b) & alignment of the run RDMs, untrained minus Backprop, $> 0$ in sub-01 and sub-02
  & $\MechDeltaASubOne$, $\MechDeltaASubTwo$ & \ref{sec:decomposition} \\
Z1 (47c) & removing the run deviations alone gives $\Delta \leq +0.015$
  & $\DecVoneVTwo$ & \ref{sec:decomposition} \\
Z2 (47c) & $\Delta(\text{V4}) \geq +0.020$ and $\Delta(\text{V5}) \leq +0.012$
  & $\DecVoneVFour$, $\DecVoneVFive$ & \ref{sec:decomposition} \\
G2 (48) & THINGS-720, T-global: $\rho(g, g^M)$ larger for the untrained network in 3 of 3 subjects
  & \RuleGTwo{} of 3 & \ref{sec:amplitude} \\
G3 (48) & NSD, N-raw, $k = 1$: $\rho(g, g^M)$ larger for the untrained network in $\geq 6$ of 8 subjects
  & \RuleGThree{} of 8 & \ref{sec:amplitude} \\
\bottomrule
\end{tabular}
\caption{Pre-specified predictions of the follow-up analyses (steps 40--48b) that were not met. In
addition, \NotInterpretedPredictions{} predictions were not interpreted because a sanity rule
failed (B2, B3 after B1; the conditional estimator after D-A2).}
\label{tab:failedfollowup}
\end{table}

\section{Per-subject results of the primary analysis}
\label{app:persubject}
Table~\ref{tab:persubdiff} gives the untrained (uncalibrated BN statistics) minus Backprop contrast
of the pre-specified primary analysis for each subject (median $r$ over reliable V1 voxels; mean
$\pm$ SD over the 5 seeds; positive seeds of 5). The main text reports the mean over the three
subjects. Table~\ref{tab:persubbank} gives the median $r$ of the luminance and contrast bank and
of the Backprop condition (mean over seeds) for each subject. In each subject and at each $g$ the
bank scored above the Backprop condition. Sub-03 has \NReliableSubThree{} reliable voxels. All values are point estimates per
subject, without an interval over stimuli or voxels.

\begin{table}[H]
\centering
\footnotesize
\setlength{\tabcolsep}{3pt}
\begin{tabular}{@{}lccc@{}}
\toprule
$g$ & sub-01 & sub-02 & sub-03 \\
\midrule
1 & $\PerSubDiffSubOneGOneMean \pm \PerSubDiffSubOneGOneSd$ (\PerSubDiffSubOneGOnePos/5) & $\PerSubDiffSubTwoGOneMean \pm \PerSubDiffSubTwoGOneSd$ (\PerSubDiffSubTwoGOnePos/5) & $\PerSubDiffSubThreeGOneMean \pm \PerSubDiffSubThreeGOneSd$ (\PerSubDiffSubThreeGOnePos/5) \\
2 & $\PerSubDiffSubOneGTwoMean \pm \PerSubDiffSubOneGTwoSd$ (\PerSubDiffSubOneGTwoPos/5) & $\PerSubDiffSubTwoGTwoMean \pm \PerSubDiffSubTwoGTwoSd$ (\PerSubDiffSubTwoGTwoPos/5) & $\PerSubDiffSubThreeGTwoMean \pm \PerSubDiffSubThreeGTwoSd$ (\PerSubDiffSubThreeGTwoPos/5) \\
4 & $\PerSubDiffSubOneGFourMean \pm \PerSubDiffSubOneGFourSd$ (\PerSubDiffSubOneGFourPos/5) & $\PerSubDiffSubTwoGFourMean \pm \PerSubDiffSubTwoGFourSd$ (\PerSubDiffSubTwoGFourPos/5) & $\PerSubDiffSubThreeGFourMean \pm \PerSubDiffSubThreeGFourSd$ (\PerSubDiffSubThreeGFourPos/5) \\
8 & $\PerSubDiffSubOneGEightMean \pm \PerSubDiffSubOneGEightSd$ (\PerSubDiffSubOneGEightPos/5) & $\PerSubDiffSubTwoGEightMean \pm \PerSubDiffSubTwoGEightSd$ (\PerSubDiffSubTwoGEightPos/5) & $\PerSubDiffSubThreeGEightMean \pm \PerSubDiffSubThreeGEightSd$ (\PerSubDiffSubThreeGEightPos/5) \\
\bottomrule
\end{tabular}
\caption{Untrained minus Backprop per subject.}
\label{tab:persubdiff}
\end{table}

\begin{table}[H]
\centering
\footnotesize
\setlength{\tabcolsep}{3pt}
\begin{tabular}{@{}lccc@{}}
\toprule
$g$ & sub-01 & sub-02 & sub-03 \\
\midrule
1 & \PreregCtlLumcontrastSubOneGOne / \PerSubBpSubOneGOne & \PreregCtlLumcontrastSubTwoGOne / \PerSubBpSubTwoGOne & \PreregCtlLumcontrastSubThreeGOne / \PerSubBpSubThreeGOne \\
2 & \PreregCtlLumcontrastSubOneGTwo / \PerSubBpSubOneGTwo & \PreregCtlLumcontrastSubTwoGTwo / \PerSubBpSubTwoGTwo & \PreregCtlLumcontrastSubThreeGTwo / \PerSubBpSubThreeGTwo \\
4 & \PreregCtlLumcontrastSubOneGFour / \PerSubBpSubOneGFour & \PreregCtlLumcontrastSubTwoGFour / \PerSubBpSubTwoGFour & \PreregCtlLumcontrastSubThreeGFour / \PerSubBpSubThreeGFour \\
8 & \PreregCtlLumcontrastSubOneGEight / \PerSubBpSubOneGEight & \PreregCtlLumcontrastSubTwoGEight / \PerSubBpSubTwoGEight & \PreregCtlLumcontrastSubThreeGEight / \PerSubBpSubThreeGEight \\
\bottomrule
\end{tabular}
\caption{Median $r$ of the luminance and contrast bank / of Backprop, per subject.}
\label{tab:persubbank}
\end{table}

\section{Ridge robustness}
\label{app:ridgerobust}
These checks of the pre-specified ridge result are exploratory (Section~\ref{sec:inference}).
They used the pre-specified configuration (three subjects, Conv1, NATIVE 224\,px, reliable V1
voxels, the same nested cross-validation and regularisation grid) with the fitting code
unchanged; the pooled median $r$ of every model fit reproduced the primary analysis to
\FSkillGateMaxDelta. Table~\ref{tab:skill} gives the untrained (uncalibrated BN statistics)
minus Backprop contrast in two further scores, the MSE skill (one minus the squared error
relative to that of the mean of the training rows of each outer fold, median over reliable
voxels) and the per-fold $r$ (Pearson $r$ within each outer fold, averaged over folds, median
over reliable voxels), mean over subjects.

\begin{table}[H]
\centering
\footnotesize
\setlength{\tabcolsep}{3pt}
\begin{tabular}{@{}ccc@{}}
\toprule
$g$ & MSE skill & per-fold $r$ \\
\midrule
1 & $\FSkillMseSkillGOneMean \pm \FSkillMseSkillGOneSd$ (\FSkillMseSkillGOnePos/5) & $\FSkillPerfoldGOneMean \pm \FSkillPerfoldGOneSd$ (\FSkillPerfoldGOnePos/5) \\
2 & $\FSkillMseSkillGTwoMean \pm \FSkillMseSkillGTwoSd$ (\FSkillMseSkillGTwoPos/5) & $\FSkillPerfoldGTwoMean \pm \FSkillPerfoldGTwoSd$ (\FSkillPerfoldGTwoPos/5) \\
4 & $\FSkillMseSkillGFourMean \pm \FSkillMseSkillGFourSd$ (\FSkillMseSkillGFourPos/5) & $\FSkillPerfoldGFourMean \pm \FSkillPerfoldGFourSd$ (\FSkillPerfoldGFourPos/5) \\
8 & $\FSkillMseSkillGEightMean \pm \FSkillMseSkillGEightSd$ (\FSkillMseSkillGEightPos/5) & $\FSkillPerfoldGEightMean \pm \FSkillPerfoldGEightSd$ (\FSkillPerfoldGEightPos/5) \\
\bottomrule
\end{tabular}
\caption{Untrained minus Backprop, mean $\pm$ SD over 5 seeds (positive seeds of 5), mean
over subjects.}
\label{tab:skill}
\end{table}

A model that predicts the mean of the training rows of each fold scored a pooled $r$ of $\FFloorSubOne$, $\FFloorSubTwo$, $\FFloorSubThree$
in sub-01, sub-02 and sub-03; its per-fold $r$ is undefined, because the prediction is constant
within each fold. Table~\ref{tab:permnull} gives the null distribution of the pooled median $r$
under 50 permutations of the stimulus assignment of the feature rows (the same permutation for
all voxels; seed-0 checkpoints) next to the observed value.

\begin{table}[H]
\centering
\footnotesize
\setlength{\tabcolsep}{3pt}
\begin{tabular}{@{}lclcc@{}}
\toprule
Subject & $g$ & Condition & Null mean [2.5\%, 97.5\%] & Observed \\
\midrule
sub-01 & 1 & untrained & $\FPermSubOneRandGOneMean$ [$\FPermSubOneRandGOneLo$, $\FPermSubOneRandGOneHi$] & $\FPermSubOneRandGOneReal$ \\
sub-01 & 1 & Backprop & $\FPermSubOneBpGOneMean$ [$\FPermSubOneBpGOneLo$, $\FPermSubOneBpGOneHi$] & $\FPermSubOneBpGOneReal$ \\
sub-01 & 4 & untrained & $\FPermSubOneRandGFourMean$ [$\FPermSubOneRandGFourLo$, $\FPermSubOneRandGFourHi$] & $\FPermSubOneRandGFourReal$ \\
sub-01 & 4 & Backprop & $\FPermSubOneBpGFourMean$ [$\FPermSubOneBpGFourLo$, $\FPermSubOneBpGFourHi$] & $\FPermSubOneBpGFourReal$ \\
sub-02 & 1 & untrained & $\FPermSubTwoRandGOneMean$ [$\FPermSubTwoRandGOneLo$, $\FPermSubTwoRandGOneHi$] & $\FPermSubTwoRandGOneReal$ \\
sub-02 & 1 & Backprop & $\FPermSubTwoBpGOneMean$ [$\FPermSubTwoBpGOneLo$, $\FPermSubTwoBpGOneHi$] & $\FPermSubTwoBpGOneReal$ \\
sub-02 & 4 & untrained & $\FPermSubTwoRandGFourMean$ [$\FPermSubTwoRandGFourLo$, $\FPermSubTwoRandGFourHi$] & $\FPermSubTwoRandGFourReal$ \\
sub-02 & 4 & Backprop & $\FPermSubTwoBpGFourMean$ [$\FPermSubTwoBpGFourLo$, $\FPermSubTwoBpGFourHi$] & $\FPermSubTwoBpGFourReal$ \\
sub-03 & 1 & untrained & $\FPermSubThreeRandGOneMean$ [$\FPermSubThreeRandGOneLo$, $\FPermSubThreeRandGOneHi$] & $\FPermSubThreeRandGOneReal$ \\
sub-03 & 1 & Backprop & $\FPermSubThreeBpGOneMean$ [$\FPermSubThreeBpGOneLo$, $\FPermSubThreeBpGOneHi$] & $\FPermSubThreeBpGOneReal$ \\
sub-03 & 4 & untrained & $\FPermSubThreeRandGFourMean$ [$\FPermSubThreeRandGFourLo$, $\FPermSubThreeRandGFourHi$] & $\FPermSubThreeRandGFourReal$ \\
sub-03 & 4 & Backprop & $\FPermSubThreeBpGFourMean$ [$\FPermSubThreeBpGFourLo$, $\FPermSubThreeBpGFourHi$] & $\FPermSubThreeBpGFourReal$ \\
\bottomrule
\end{tabular}
\caption{Permutation null of the pooled median $r$ (seed-0 checkpoints).}
\label{tab:permnull}
\end{table}

For the joint readout, the network features and the luminance-and-contrast bank of the same
$g$ were concatenated into one feature matrix. Each feature was standardised with the
statistics of the training rows of each split, as in every other fit, and the cross-validation
scheme and regularisation grid were unchanged. Table~\ref{tab:joint} compares the bank alone,
the network alone and the joint readout.

\begin{table}[H]
\centering
\footnotesize
\setlength{\tabcolsep}{3pt}
\begin{tabular}{@{}clccc@{}}
\toprule
$g$ & Network & Bank alone & Network alone & Network and bank \\
\midrule
1 & untrained & $\FJointRandGOneBank$ & $\PreregCondRandGOne$ & $\FJointRandGOneJoint$ \\
1 & Backprop & $\FJointBpGOneBank$ & $\PreregCondBpGOne$ & $\FJointBpGOneJoint$ \\
4 & untrained & $\FJointRandGFourBank$ & $\PreregCondRandGFour$ & $\FJointRandGFourJoint$ \\
4 & Backprop & $\FJointBpGFourBank$ & $\PreregCondBpGFour$ & $\FJointBpGFourJoint$ \\
\bottomrule
\end{tabular}
\caption{Median $r$ over reliable V1 voxels, mean over seeds and subjects.}
\label{tab:joint}
\end{table}

\section{Bootstrap with re-estimated normalisation}
\label{app:bootnorm}
In the main analyses, unit means and SDs, the feature transformations $F$, the voxel
normalisations $B$, the mean profile and the norms are estimated once on the 720 stimuli,
and the bootstrap resamples pairs of the resulting fixed RDMs (Section~\ref{sec:rsa}). In this
exploratory check, the statistics were re-estimated from the resampled rows of each of the
first 200 resamples (duplicates kept, self-pairs dropped), and the RDMs were rebuilt with
them. Table~\ref{tab:bootnorm} compares the V1 contrast of the untrained (uncalibrated BN
statistics) network minus Backprop at NATIVE 224\,px in three cells. The classification of
the interval relative to 0 was the same in all three cells (BN1: met). With euclidean
distance, centring does not change the distances, so the two intervals agree by construction
over the same resamples.

\begin{table}[H]
\centering
\footnotesize
\setlength{\tabcolsep}{3pt}
\begin{tabular}{@{}lcccc@{}}
\toprule
Cell ($F$, $D$, $B$) & Value & Fixed (1000) & Re-estimated (200) & Interval \\
\midrule
$z$, correlation, paper & $\BNormZCorrPaper$ & [$\BNormZCorrPaperFixedLo$, $\BNormZCorrPaperFixedHi$] & [$\BNormZCorrPaperLo$, $\BNormZCorrPaperHi$] & below 0 \\
center, euclidean, paper & $\BNormCentEuclPaper$ & [$\BNormCentEuclPaperFixedLo$, $\BNormCentEuclPaperFixedHi$] & [$\BNormCentEuclPaperLo$, $\BNormCentEuclPaperHi$] & above 0 \\
center, correlation, center720 & $\BNormCentCorrCent$ & [$\BNormCentCorrCentFixedLo$, $\BNormCentCorrCentFixedHi$] & [$\BNormCentCorrCentLo$, $\BNormCentCorrCentHi$] & includes 0 \\
\bottomrule
\end{tabular}
\caption{V1 contrast with 95\% stimulus-bootstrap intervals, normalisation fixed (1000
resamples) or re-estimated within each resample (first 200 resamples).}
\label{tab:bootnorm}
\end{table}

\section{Ablation on each side}
\label{app:ablation2x2}
This exploratory analysis was added after further criticism of a revised draft; its
rule was committed before its script (Section~\ref{sec:inference}). The mean-profile
decomposition of Section~\ref{sec:gainmethods} was applied to the model side, the brain
side, both or neither. Table~\ref{tab:ablation2x2} gives the V1 contrast at NATIVE 224\,px
(raw GAP features, correlation distance, $B$ = paper) for the untrained (uncalibrated BN
statistics) network minus Backprop in all four combinations, and for the six ImageNet models
(untrained, mean of three initialisations, minus trained, 224\,px) in the two mixed
combinations. All values use the same 1000 stimulus-bootstrap resamples. The full/full and
orthogonal/orthogonal cells reproduce the full and orthogonal variants of
Section~\ref{sec:ablation}. In the ImageNet models, the CI lay above 0 in
\AbTwoImnOrthoFullPos{} and below 0 in \AbTwoImnOrthoFullNeg{} models with the orthogonal
model against the full brain RDM, and above 0 in \AbTwoImnFullOrthoPos{} and below 0 in
\AbTwoImnFullOrthoNeg{} models with the full model against the orthogonal brain RDM
(descriptive, no rule).

\begin{table}[H]
\centering
\footnotesize
\setlength{\tabcolsep}{4pt}
\begin{tabular}{@{}lllc@{}}
\toprule
Model & Model RDM & Brain RDM & V1 contrast [95\% CI] \\
\midrule
own CNN & full & full & $\AbTwoOwnFullFull$ [$\AbTwoOwnFullFullLo$, $\AbTwoOwnFullFullHi$] \\
own CNN & full & orthogonal & $\AbTwoOwnFullOrtho$ [$\AbTwoOwnFullOrthoLo$, $\AbTwoOwnFullOrthoHi$] \\
own CNN & orthogonal & full & $\AbTwoOwnOrthoFull$ [$\AbTwoOwnOrthoFullLo$, $\AbTwoOwnOrthoFullHi$] \\
own CNN & orthogonal & orthogonal & $\AbTwoOwnOrthoOrtho$ [$\AbTwoOwnOrthoOrthoLo$, $\AbTwoOwnOrthoOrthoHi$] \\
\midrule
alexnet & orthogonal & full & $\AbTwoAlexnetOrthoFull$ [$\AbTwoAlexnetOrthoFullLo$, $\AbTwoAlexnetOrthoFullHi$] \\
vgg11\_bn & orthogonal & full & $\AbTwoVggelevenOrthoFull$ [$\AbTwoVggelevenOrthoFullLo$, $\AbTwoVggelevenOrthoFullHi$] \\
resnet18 & orthogonal & full & $\AbTwoResneteighteenOrthoFull$ [$\AbTwoResneteighteenOrthoFullLo$, $\AbTwoResneteighteenOrthoFullHi$] \\
resnet50 & orthogonal & full & $\AbTwoResnetfiftyOrthoFull$ [$\AbTwoResnetfiftyOrthoFullLo$, $\AbTwoResnetfiftyOrthoFullHi$] \\
convnext\_tiny & orthogonal & full & $\AbTwoConvnexttinyOrthoFull$ [$\AbTwoConvnexttinyOrthoFullLo$, $\AbTwoConvnexttinyOrthoFullHi$] \\
vit\_b\_16 & orthogonal & full & $\AbTwoVitbsixteenOrthoFull$ [$\AbTwoVitbsixteenOrthoFullLo$, $\AbTwoVitbsixteenOrthoFullHi$] \\
alexnet & full & orthogonal & $\AbTwoAlexnetFullOrtho$ [$\AbTwoAlexnetFullOrthoLo$, $\AbTwoAlexnetFullOrthoHi$] \\
vgg11\_bn & full & orthogonal & $\AbTwoVggelevenFullOrtho$ [$\AbTwoVggelevenFullOrthoLo$, $\AbTwoVggelevenFullOrthoHi$] \\
resnet18 & full & orthogonal & $\AbTwoResneteighteenFullOrtho$ [$\AbTwoResneteighteenFullOrthoLo$, $\AbTwoResneteighteenFullOrthoHi$] \\
resnet50 & full & orthogonal & $\AbTwoResnetfiftyFullOrtho$ [$\AbTwoResnetfiftyFullOrthoLo$, $\AbTwoResnetfiftyFullOrthoHi$] \\
convnext\_tiny & full & orthogonal & $\AbTwoConvnexttinyFullOrtho$ [$\AbTwoConvnexttinyFullOrthoLo$, $\AbTwoConvnexttinyFullOrthoHi$] \\
vit\_b\_16 & full & orthogonal & $\AbTwoVitbsixteenFullOrtho$ [$\AbTwoVitbsixteenFullOrthoLo$, $\AbTwoVitbsixteenFullOrthoHi$] \\
\midrule
\multicolumn{4}{@{}l}{X1, own CNN: gap[full, full] $-$ gap[orthogonal, orthogonal]: $+\AbTwoXOne$ [$\AbTwoXOneLo$, $\AbTwoXOneHi$] (met)} \\
\bottomrule
\end{tabular}
\caption{V1 contrast with 95\% stimulus-bootstrap CIs (own CNN: untrained minus Backprop,
mean over 5 seeds; ImageNet: untrained minus trained); mean over three subjects.}
\label{tab:ablation2x2}
\end{table}

\section{Additional figures}
\label{app:figs}

\begin{figure}[H]
\centering
\includegraphics[width=3.5in]{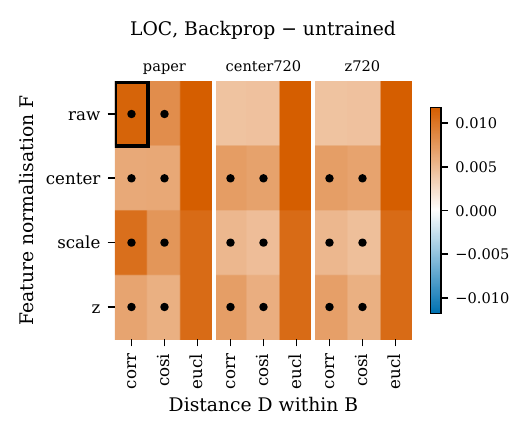}
\caption{Backprop minus untrained at LOC (Conv3), NATIVE 224\,px, in the 36 cells of the analysis multiverse. Layout, colour scale, dots and frame as in Fig.~\ref{fig:multiverse_v1}.}
\label{fig:multiverse_loc}
\end{figure}

\begin{figure}[H]
\centering
\includegraphics[width=\textwidth]{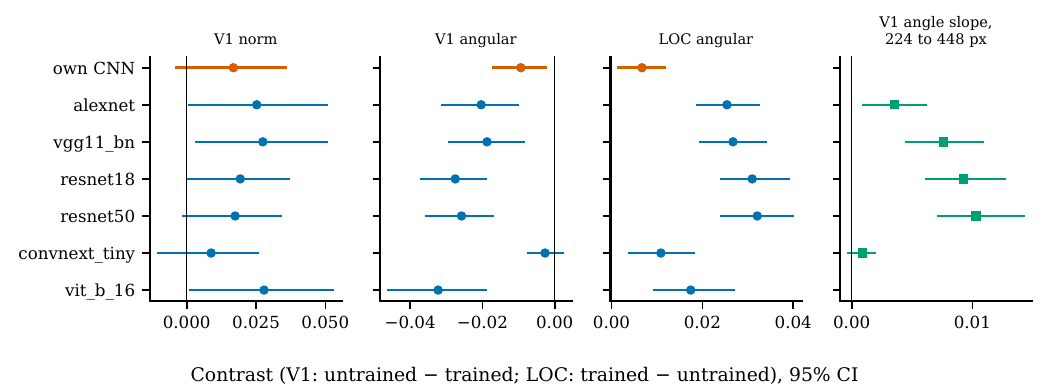}
\caption{Component RSA contrasts with $F$ = center and 95\% stimulus-bootstrap CIs, for the own CNN (vermillion) and six ImageNet models (blue). Panels from left: V1 norm component, V1 angular component, LOC angular component, and the slope of the V1 angular contrast from 224 to 448\,px for the ImageNet models (green; none for vit\_b\_16). V1: untrained minus trained; LOC: trained minus untrained.}
\label{fig:separation}
\end{figure}
\FloatBarrier
\section{Chronology of the follow-up analyses}
\label{app:chronology}
\begin{table}[t]
\centering
\footnotesize
\begin{adjustbox}{max width=\textwidth, max totalheight=0.82\textheight}
\begin{tabular}{@{}llll@{}}
\toprule
Step & Commit & Committed (local time) & Content \\
\midrule
40 & \texttt{02e82da} & 2026-11-16 20:28:31 & banded ridge; cross-validated RSA on the repeated test images \\
41 & \texttt{ef18cfd} & 2026-11-16 22:32:58 & diagnosis of B1; noise curve \\
42 & \texttt{11fac04} & 2026-11-17 08:26:39 & NSD replication (P1--P4) \\
42b & \texttt{366a9eb} & 2026-11-17 09:16:02 & norm and angle components, subj02--08 (Q1--Q4) \\
43 & \texttt{c4db2e2} & 2026-11-17 13:22:10 & NSD multiverse, ImageNet models, resolution, noise ceiling \\
44 & \texttt{d958a05} & 2026-11-21 09:15:50 & candidate explanations in THINGS (W1--W5) \\
45 & \texttt{9648934} & 2026-11-21 09:41:21 & coarse spatial luminance (W6, W7) \\
47 & \texttt{4f16e42} & 2026-11-21 09:57:43 & run centring in THINGS; ridge in NSD \\
47b & \texttt{edbb11d} & 2026-11-21 18:22:05 & run-level mechanism; robustness of ridge and NSD RSA to run centring \\
47c & \texttt{71585be} & 2026-11-22 11:24:32 & decomposition of run centring \\
48 & \texttt{01e3790} & 2026-11-22 11:45:46 & voxel normalisation conventions; response amplitude \\
48b & \texttt{56e49d1} & 2026-11-22 16:20:57 & amplitude dominance of the RDMs \\
\bottomrule
\end{tabular}
\end{adjustbox}
\caption{Amendments of the follow-up analyses, in commit order. Each amendment was committed before the scripts of its step; the commit timestamps, not the dates in the file names, are authoritative.}
\label{tab:chronology}
\end{table}

\FloatBarrier
\section{Banded ridge and repeated test images}
\label{app:banded}
Banded ridge (steps 40, 41) and the repeated test images (steps 40, 41). Table~\ref{tab:banded} lists the
banded readouts; Table~\ref{tab:noisecurve} the noise curve. In the cross-validated RSA of step 40, $b_1$
denotes correlation distance on presentation-averaged patterns, $b_2$ Euclidean distance on them and $b_3$
the cross-validated squared Euclidean distance; at V1, untrained minus Backprop was
$\CvRsaVoneRawBOne$ ($b_1$), $\CvRsaVoneRawBTwo$ ($b_2$) and $\CvRsaVoneRawBThree$ ($b_3$) for raw units, and
$\CvRsaVoneZBOne$, $\CvRsaVoneZBTwo$ and $\CvRsaVoneZBThree$ for standardised units.

\begin{table}[H]
\centering
\small
\begin{adjustbox}{max width=\textwidth, max totalheight=0.82\textheight}
\begin{tabular}{@{}lrrrrrr@{}}
\toprule
Readout & $g$ & bank & network & $U$ (median $r$) & $U$ (MSE skill) & min.\ joint $-$ bank \\
\midrule
banded, 2D search, Backprop & 1 & $\BandedBankGOne$ & $\BandedNetBpGOne$ & $\BandedUBpGOne$ & $\BandedUSkillBpGOne$ & $\BandedBOneMinGOne$ \\
banded, 2D search, untrained & 1 & $\BandedBankGOne$ & $\BandedNetRandGOne$ & $\BandedURandGOne$ & $\BandedUSkillRandGOne$ &  \\
banded, 2D search, Backprop & 4 & $\BandedBankGFour$ & $\BandedNetBpGFour$ & $\BandedUBpGFour$ & $\BandedUSkillBpGFour$ & $\BandedBOneMinGFour$ \\
banded, 2D search, untrained & 4 & $\BandedBankGFour$ & $\BandedNetRandGFour$ & $\BandedURandGFour$ & $\BandedUSkillRandGFour$ &  \\
banded, 2D search, Backprop & 8 & $\BandedBankGEight$ & $\BandedNetBpGEight$ & $\BandedUBpGEight$ & $\BandedUSkillBpGEight$ & $\BandedBOneMinGEight$ \\
banded, 2D search, untrained & 8 & $\BandedBankGEight$ & $\BandedNetRandGEight$ & $\BandedURandGEight$ & $\BandedUSkillRandGEight$ &  \\
banded, conditional search, Backprop & 4 & & & $\CondUBp$ & $\CondUSkillBp$ & $\DATwoMin$ \\
banded, conditional search, untrained & 4 & & & $\CondURand$ & $\CondUSkillRand$ & \\
\bottomrule
\end{tabular}
\end{adjustbox}
\caption{Banded ridge in THINGS-720 (steps 40, 41; V1, reliable voxels; mean over subjects and seeds). $U$: joint readout minus bank alone. Minimum joint minus bank: over all fits (B1 and D-A2 required $\geq -0.002$). B2 and B3 are not interpreted because B1 was not met; the conditional estimator is not interpreted because D-A2 was not met.}
\label{tab:banded}
\end{table}

\begin{table}[H]
\centering
\small
\begin{adjustbox}{max width=\textwidth, max totalheight=0.82\textheight}
\begin{tabular}{@{}llrrrrrr@{}}
\toprule
ROI & units & $k = 1$ & 2 & 3 & 4 & 6 & 12 \\
\midrule
V1 & raw & $\NoiseVoneRawKOne$ {\scriptsize [$\NoiseVoneRawKOneLo$, $\NoiseVoneRawKOneHi$]} & $\NoiseVoneRawKTwo$ & $\NoiseVoneRawKThree$ & $\NoiseVoneRawKFour$ & $\NoiseVoneRawKSix$ & $\NoiseVoneRawKOneTwo$ {\scriptsize [$\NoiseVoneRawKOneTwoLo$, $\NoiseVoneRawKOneTwoHi$]} \\
V1 & z & $\NoiseVoneZKOne$ {\scriptsize [$\NoiseVoneZKOneLo$, $\NoiseVoneZKOneHi$]} & $\NoiseVoneZKTwo$ & $\NoiseVoneZKThree$ & $\NoiseVoneZKFour$ & $\NoiseVoneZKSix$ & $\NoiseVoneZKOneTwo$ {\scriptsize [$\NoiseVoneZKOneTwoLo$, $\NoiseVoneZKOneTwoHi$]} \\
LOC & raw & $\NoiseLocRawKOne$ {\scriptsize [$\NoiseLocRawKOneLo$, $\NoiseLocRawKOneHi$]} & $\NoiseLocRawKTwo$ & $\NoiseLocRawKThree$ & $\NoiseLocRawKFour$ & $\NoiseLocRawKSix$ & $\NoiseLocRawKOneTwo$ {\scriptsize [$\NoiseLocRawKOneTwoLo$, $\NoiseLocRawKOneTwoHi$]} \\
LOC & z & $\NoiseLocZKOne$ {\scriptsize [$\NoiseLocZKOneLo$, $\NoiseLocZKOneHi$]} & $\NoiseLocZKTwo$ & $\NoiseLocZKThree$ & $\NoiseLocZKFour$ & $\NoiseLocZKSix$ & $\NoiseLocZKOneTwo$ {\scriptsize [$\NoiseLocZKOneTwoLo$, $\NoiseLocZKOneTwoHi$]} \\
\bottomrule
\end{tabular}
\end{adjustbox}
\caption{Noise curve on the 100 THINGS test images (step 41): $\Delta(k)$, untrained minus Backprop, correlation distance on the mean of $k$ of the 12 presentations (20 draws per $k < 12$). Intervals: 95\% stimulus bootstrap.}
\label{tab:noisecurve}
\end{table}

\section{NSD details}
\label{app:nsd}
Gates of the NSD pipeline (step 42): every subject has 515 images with three presentations; the z-scored
betas reproduce zero mean and unit SD per voxel and session; the feature extraction passed a
downstream-equivalence gate. Table~\ref{tab:nsdsubjects} gives voxel counts and per-subject ridge results,
Table~\ref{tab:nsdcells} all cells under the NSD standard normalisation (including $k = 1$ and the
\texttt{lateral} ROI), Table~\ref{tab:nsdimagenet} the ImageNet models (also LO-obj and seed 0 only),
Table~\ref{tab:nsdres} the resolution analysis, Table~\ref{tab:nsdceil} the noise ceiling,
Table~\ref{tab:nsdpilot} the subj01 pilot and Table~\ref{tab:multi28} the 28 distinct analyses of the
multiverse under N-session, N-raw and run centring, next to THINGS-720.

\begin{table}[H]
\centering
\footnotesize
\setlength{\tabcolsep}{3pt}
\begin{adjustbox}{max width=\textwidth, max totalheight=0.82\textheight}
\begin{tabular}{@{}lrrrrrrrr@{}}
\toprule
& subj01 & subj02 & subj03 & subj04 & subj05 & subj06 & subj07 & subj08 \\
\midrule
voxels V1 & $\NsdVoxVoneSubjOne$ & $\NsdVoxVoneSubjTwo$ & $\NsdVoxVoneSubjThree$ & $\NsdVoxVoneSubjFour$ & $\NsdVoxVoneSubjFive$ & $\NsdVoxVoneSubjSix$ & $\NsdVoxVoneSubjSeven$ & $\NsdVoxVoneSubjEight$ \\
voxels LO-obj & $\NsdVoxLoobjSubjOne$ & $\NsdVoxLoobjSubjTwo$ & $\NsdVoxLoobjSubjThree$ & $\NsdVoxLoobjSubjFour$ & $\NsdVoxLoobjSubjFive$ & $\NsdVoxLoobjSubjSix$ & $\NsdVoxLoobjSubjSeven$ & $\NsdVoxLoobjSubjEight$ \\
voxels lateral & $\NsdVoxLatSubjOne$ & $\NsdVoxLatSubjTwo$ & $\NsdVoxLatSubjThree$ & $\NsdVoxLatSubjFour$ & $\NsdVoxLatSubjFive$ & $\NsdVoxLatSubjSix$ & $\NsdVoxLatSubjSeven$ & $\NsdVoxLatSubjEight$ \\
reliable V1 voxels (ridge) & $\NsdReliableSubjOne$ & $\NsdReliableSubjTwo$ & $\NsdReliableSubjThree$ & $\NsdReliableSubjFour$ & $\NsdReliableSubjFive$ & $\NsdReliableSubjSix$ & $\NsdReliableSubjSeven$ & $\NsdReliableSubjEight$ \\
ridge $g = 1$, $\Delta$ & $\NsdRidgeVoneGOneSubjOne$ & $\NsdRidgeVoneGOneSubjTwo$ & $\NsdRidgeVoneGOneSubjThree$ & $\NsdRidgeVoneGOneSubjFour$ & $\NsdRidgeVoneGOneSubjFive$ & $\NsdRidgeVoneGOneSubjSix$ & $\NsdRidgeVoneGOneSubjSeven$ & $\NsdRidgeVoneGOneSubjEight$ \\
ridge $g = 4$, $\Delta$ & $\NsdRidgeVoneGFourSubjOne$ & $\NsdRidgeVoneGFourSubjTwo$ & $\NsdRidgeVoneGFourSubjThree$ & $\NsdRidgeVoneGFourSubjFour$ & $\NsdRidgeVoneGFourSubjFive$ & $\NsdRidgeVoneGFourSubjSix$ & $\NsdRidgeVoneGFourSubjSeven$ & $\NsdRidgeVoneGFourSubjEight$ \\
ridge $g = 8$, $\Delta$ & $\NsdRidgeVoneGEightSubjOne$ & $\NsdRidgeVoneGEightSubjTwo$ & $\NsdRidgeVoneGEightSubjThree$ & $\NsdRidgeVoneGEightSubjFour$ & $\NsdRidgeVoneGEightSubjFive$ & $\NsdRidgeVoneGEightSubjSix$ & $\NsdRidgeVoneGEightSubjSeven$ & $\NsdRidgeVoneGEightSubjEight$ \\
ridge $g = 4$, bank & $\NsdRidgeVoneGFourSubjOneBank$ & $\NsdRidgeVoneGFourSubjTwoBank$ & $\NsdRidgeVoneGFourSubjThreeBank$ & $\NsdRidgeVoneGFourSubjFourBank$ & $\NsdRidgeVoneGFourSubjFiveBank$ & $\NsdRidgeVoneGFourSubjSixBank$ & $\NsdRidgeVoneGFourSubjSevenBank$ & $\NsdRidgeVoneGFourSubjEightBank$ \\
ridge $g = 4$, Backprop & $\NsdRidgeVoneGFourSubjOneBp$ & $\NsdRidgeVoneGFourSubjTwoBp$ & $\NsdRidgeVoneGFourSubjThreeBp$ & $\NsdRidgeVoneGFourSubjFourBp$ & $\NsdRidgeVoneGFourSubjFiveBp$ & $\NsdRidgeVoneGFourSubjSixBp$ & $\NsdRidgeVoneGFourSubjSevenBp$ & $\NsdRidgeVoneGFourSubjEightBp$ \\
ridge $g = 4$, untrained & $\NsdRidgeVoneGFourSubjOneRand$ & $\NsdRidgeVoneGFourSubjTwoRand$ & $\NsdRidgeVoneGFourSubjThreeRand$ & $\NsdRidgeVoneGFourSubjFourRand$ & $\NsdRidgeVoneGFourSubjFiveRand$ & $\NsdRidgeVoneGFourSubjSixRand$ & $\NsdRidgeVoneGFourSubjSevenRand$ & $\NsdRidgeVoneGFourSubjEightRand$ \\
ridge LO-obj $g = 4$, $\Delta$ & $\NsdRidgeLoobjGFourSubjOne$ & $\NsdRidgeLoobjGFourSubjTwo$ & $\NsdRidgeLoobjGFourSubjThree$ & $\NsdRidgeLoobjGFourSubjFour$ & $\NsdRidgeLoobjGFourSubjFive$ & $\NsdRidgeLoobjGFourSubjSix$ & $\NsdRidgeLoobjGFourSubjSeven$ & $\NsdRidgeLoobjGFourSubjEight$ \\
\bottomrule
\end{tabular}
\end{adjustbox}
\caption{NSD per subject: voxel counts (all voxels of each ROI that were valid in every session), reliable V1 voxels used by the ridge readout (split-half reliability over the three presentations, Spearman--Brown corrected, $> 0.2$), and ridge median $r$ (untrained minus Backprop, mean over seeds; bank: luminance-and-contrast bank).}
\label{tab:nsdsubjects}
\end{table}

\begin{table}[H]
\centering
\footnotesize
\setlength{\tabcolsep}{3pt}
\begin{adjustbox}{max width=\textwidth, max totalheight=0.82\textheight}
\begin{tabular}{@{}llrrrrrrr@{}}
\toprule
ROI & $k$ & raw, corr. & z, corr. & raw, Eucl. & z, Eucl. & $N$ & $A$ & $W$ \\
\midrule
V1 & 3 & $\NsdAllVoneKThreeRawCorrelation$ (\NsdAllVoneKThreeRawCorrelationNeg) & $\NsdAllVoneKThreeZCorrelation$ (\NsdAllVoneKThreeZCorrelationNeg) & $\NsdAllVoneKThreeRawEuclidean$ (\NsdAllVoneKThreeRawEuclideanNeg) & $\NsdAllVoneKThreeZEuclidean$ (\NsdAllVoneKThreeZEuclideanNeg) & $\NsdAllVoneKThreeCenterN$ (\NsdAllVoneKThreeCenterNNeg) & $\NsdAllVoneKThreeCenterA$ (\NsdAllVoneKThreeCenterANeg) & $\NsdAllVoneKThreeCenterW$ (\NsdAllVoneKThreeCenterWNeg) \\
V1 & 1 & $\NsdAllVoneKOneRawCorrelation$ (\NsdAllVoneKOneRawCorrelationNeg) & $\NsdAllVoneKOneZCorrelation$ (\NsdAllVoneKOneZCorrelationNeg) & $\NsdAllVoneKOneRawEuclidean$ (\NsdAllVoneKOneRawEuclideanNeg) & $\NsdAllVoneKOneZEuclidean$ (\NsdAllVoneKOneZEuclideanNeg) &  &  &  \\
LO-obj & 3 & $\NsdAllLoobjKThreeRawCorrelation$ (\NsdAllLoobjKThreeRawCorrelationNeg) & $\NsdAllLoobjKThreeZCorrelation$ (\NsdAllLoobjKThreeZCorrelationNeg) & $\NsdAllLoobjKThreeRawEuclidean$ (\NsdAllLoobjKThreeRawEuclideanNeg) & $\NsdAllLoobjKThreeZEuclidean$ (\NsdAllLoobjKThreeZEuclideanNeg) & $\NsdAllLoobjKThreeCenterN$ (\NsdAllLoobjKThreeCenterNNeg) & $\NsdAllLoobjKThreeCenterA$ (\NsdAllLoobjKThreeCenterANeg) & $\NsdAllLoobjKThreeCenterW$ (\NsdAllLoobjKThreeCenterWNeg) \\
LO-obj & 1 & $\NsdAllLoobjKOneRawCorrelation$ (\NsdAllLoobjKOneRawCorrelationNeg) & $\NsdAllLoobjKOneZCorrelation$ (\NsdAllLoobjKOneZCorrelationNeg) & $\NsdAllLoobjKOneRawEuclidean$ (\NsdAllLoobjKOneRawEuclideanNeg) & $\NsdAllLoobjKOneZEuclidean$ (\NsdAllLoobjKOneZEuclideanNeg) &  &  &  \\
lateral & 3 & $\NsdAllLatKThreeRawCorrelation$ (\NsdAllLatKThreeRawCorrelationNeg) & $\NsdAllLatKThreeZCorrelation$ (\NsdAllLatKThreeZCorrelationNeg) & $\NsdAllLatKThreeRawEuclidean$ (\NsdAllLatKThreeRawEuclideanNeg) & $\NsdAllLatKThreeZEuclidean$ (\NsdAllLatKThreeZEuclideanNeg) & $\NsdAllLatKThreeCenterN$ (\NsdAllLatKThreeCenterNNeg) & $\NsdAllLatKThreeCenterA$ (\NsdAllLatKThreeCenterANeg) & $\NsdAllLatKThreeCenterW$ (\NsdAllLatKThreeCenterWNeg) \\
lateral & 1 & $\NsdAllLatKOneRawCorrelation$ (\NsdAllLatKOneRawCorrelationNeg) & $\NsdAllLatKOneZCorrelation$ (\NsdAllLatKOneZCorrelationNeg) & $\NsdAllLatKOneRawEuclidean$ (\NsdAllLatKOneRawEuclideanNeg) & $\NsdAllLatKOneZEuclidean$ (\NsdAllLatKOneZEuclideanNeg) &  &  &  \\
\bottomrule
\end{tabular}
\end{adjustbox}
\caption{NSD, all 8 subjects, NSD standard normalisation (N-session): $\Delta$ (untrained minus Backprop, mean over seeds and subjects); in brackets the number of subjects with $\Delta < 0$. $N$, $A$, $W$: norm, angular and norm-weighted angular components on centred units and voxels ($k = 3$).}
\label{tab:nsdcells}
\end{table}

\begin{table}[H]
\centering
\footnotesize
\setlength{\tabcolsep}{3pt}
\begin{adjustbox}{max width=\textwidth, max totalheight=0.82\textheight}
\begin{tabular}{@{}llrrrrr@{}}
\toprule
Model & ROI & raw, corr. & z, corr. & $N$ & $A$ & $W$ \\
\midrule
\texttt{alexnet} & V1 & $\ImnAlexnetVoneRawCorrelation$ (\ImnAlexnetVoneRawCorrelationNeg) & $\ImnAlexnetVoneZCorrelation$ (\ImnAlexnetVoneZCorrelationNeg) & $\ImnAlexnetVoneCenterN$ (\ImnAlexnetVoneCenterNNeg) & $\ImnAlexnetVoneCenterA$ (\ImnAlexnetVoneCenterANeg) & $\ImnAlexnetVoneCenterW$ (\ImnAlexnetVoneCenterWNeg) \\
\texttt{alexnet} & LO-obj & $\ImnAlexnetLoobjRawCorrelation$ (\ImnAlexnetLoobjRawCorrelationNeg) & $\ImnAlexnetLoobjZCorrelation$ (\ImnAlexnetLoobjZCorrelationNeg) & $\ImnAlexnetLoobjCenterN$ (\ImnAlexnetLoobjCenterNNeg) & $\ImnAlexnetLoobjCenterA$ (\ImnAlexnetLoobjCenterANeg) & $\ImnAlexnetLoobjCenterW$ (\ImnAlexnetLoobjCenterWNeg) \\
\texttt{vgg11\_bn} & V1 & $\ImnVggOneOneBnVoneRawCorrelation$ (\ImnVggOneOneBnVoneRawCorrelationNeg) & $\ImnVggOneOneBnVoneZCorrelation$ (\ImnVggOneOneBnVoneZCorrelationNeg) & $\ImnVggOneOneBnVoneCenterN$ (\ImnVggOneOneBnVoneCenterNNeg) & $\ImnVggOneOneBnVoneCenterA$ (\ImnVggOneOneBnVoneCenterANeg) & $\ImnVggOneOneBnVoneCenterW$ (\ImnVggOneOneBnVoneCenterWNeg) \\
\texttt{vgg11\_bn} & LO-obj & $\ImnVggOneOneBnLoobjRawCorrelation$ (\ImnVggOneOneBnLoobjRawCorrelationNeg) & $\ImnVggOneOneBnLoobjZCorrelation$ (\ImnVggOneOneBnLoobjZCorrelationNeg) & $\ImnVggOneOneBnLoobjCenterN$ (\ImnVggOneOneBnLoobjCenterNNeg) & $\ImnVggOneOneBnLoobjCenterA$ (\ImnVggOneOneBnLoobjCenterANeg) & $\ImnVggOneOneBnLoobjCenterW$ (\ImnVggOneOneBnLoobjCenterWNeg) \\
\texttt{resnet18} & V1 & $\ImnResnetOneEightVoneRawCorrelation$ (\ImnResnetOneEightVoneRawCorrelationNeg) & $\ImnResnetOneEightVoneZCorrelation$ (\ImnResnetOneEightVoneZCorrelationNeg) & $\ImnResnetOneEightVoneCenterN$ (\ImnResnetOneEightVoneCenterNNeg) & $\ImnResnetOneEightVoneCenterA$ (\ImnResnetOneEightVoneCenterANeg) & $\ImnResnetOneEightVoneCenterW$ (\ImnResnetOneEightVoneCenterWNeg) \\
\texttt{resnet18} & LO-obj & $\ImnResnetOneEightLoobjRawCorrelation$ (\ImnResnetOneEightLoobjRawCorrelationNeg) & $\ImnResnetOneEightLoobjZCorrelation$ (\ImnResnetOneEightLoobjZCorrelationNeg) & $\ImnResnetOneEightLoobjCenterN$ (\ImnResnetOneEightLoobjCenterNNeg) & $\ImnResnetOneEightLoobjCenterA$ (\ImnResnetOneEightLoobjCenterANeg) & $\ImnResnetOneEightLoobjCenterW$ (\ImnResnetOneEightLoobjCenterWNeg) \\
\texttt{resnet50} & V1 & $\ImnResnetFiveZeroVoneRawCorrelation$ (\ImnResnetFiveZeroVoneRawCorrelationNeg) & $\ImnResnetFiveZeroVoneZCorrelation$ (\ImnResnetFiveZeroVoneZCorrelationNeg) & $\ImnResnetFiveZeroVoneCenterN$ (\ImnResnetFiveZeroVoneCenterNNeg) & $\ImnResnetFiveZeroVoneCenterA$ (\ImnResnetFiveZeroVoneCenterANeg) & $\ImnResnetFiveZeroVoneCenterW$ (\ImnResnetFiveZeroVoneCenterWNeg) \\
\texttt{resnet50} & LO-obj & $\ImnResnetFiveZeroLoobjRawCorrelation$ (\ImnResnetFiveZeroLoobjRawCorrelationNeg) & $\ImnResnetFiveZeroLoobjZCorrelation$ (\ImnResnetFiveZeroLoobjZCorrelationNeg) & $\ImnResnetFiveZeroLoobjCenterN$ (\ImnResnetFiveZeroLoobjCenterNNeg) & $\ImnResnetFiveZeroLoobjCenterA$ (\ImnResnetFiveZeroLoobjCenterANeg) & $\ImnResnetFiveZeroLoobjCenterW$ (\ImnResnetFiveZeroLoobjCenterWNeg) \\
\texttt{convnext\_tiny} & V1 & $\ImnConvnextTinyVoneRawCorrelation$ (\ImnConvnextTinyVoneRawCorrelationNeg) & $\ImnConvnextTinyVoneZCorrelation$ (\ImnConvnextTinyVoneZCorrelationNeg) & $\ImnConvnextTinyVoneCenterN$ (\ImnConvnextTinyVoneCenterNNeg) & $\ImnConvnextTinyVoneCenterA$ (\ImnConvnextTinyVoneCenterANeg) & $\ImnConvnextTinyVoneCenterW$ (\ImnConvnextTinyVoneCenterWNeg) \\
\texttt{convnext\_tiny} & LO-obj & $\ImnConvnextTinyLoobjRawCorrelation$ (\ImnConvnextTinyLoobjRawCorrelationNeg) & $\ImnConvnextTinyLoobjZCorrelation$ (\ImnConvnextTinyLoobjZCorrelationNeg) & $\ImnConvnextTinyLoobjCenterN$ (\ImnConvnextTinyLoobjCenterNNeg) & $\ImnConvnextTinyLoobjCenterA$ (\ImnConvnextTinyLoobjCenterANeg) & $\ImnConvnextTinyLoobjCenterW$ (\ImnConvnextTinyLoobjCenterWNeg) \\
\texttt{vit\_b\_16} & V1 & $\ImnVitBOneSixVoneRawCorrelation$ (\ImnVitBOneSixVoneRawCorrelationNeg) & $\ImnVitBOneSixVoneZCorrelation$ (\ImnVitBOneSixVoneZCorrelationNeg) & $\ImnVitBOneSixVoneCenterN$ (\ImnVitBOneSixVoneCenterNNeg) & $\ImnVitBOneSixVoneCenterA$ (\ImnVitBOneSixVoneCenterANeg) & $\ImnVitBOneSixVoneCenterW$ (\ImnVitBOneSixVoneCenterWNeg) \\
\texttt{vit\_b\_16} & LO-obj & $\ImnVitBOneSixLoobjRawCorrelation$ (\ImnVitBOneSixLoobjRawCorrelationNeg) & $\ImnVitBOneSixLoobjZCorrelation$ (\ImnVitBOneSixLoobjZCorrelationNeg) & $\ImnVitBOneSixLoobjCenterN$ (\ImnVitBOneSixLoobjCenterNNeg) & $\ImnVitBOneSixLoobjCenterA$ (\ImnVitBOneSixLoobjCenterANeg) & $\ImnVitBOneSixLoobjCenterW$ (\ImnVitBOneSixLoobjCenterWNeg) \\
\midrule
\texttt{alexnet} (s0) & V1 & $\ImnSzeroAlexnetVoneRawCorrelation$ (\ImnSzeroAlexnetVoneRawCorrelationNeg) & $\ImnSzeroAlexnetVoneZCorrelation$ (\ImnSzeroAlexnetVoneZCorrelationNeg) & $\ImnSzeroAlexnetVoneCenterN$ (\ImnSzeroAlexnetVoneCenterNNeg) & $\ImnSzeroAlexnetVoneCenterA$ (\ImnSzeroAlexnetVoneCenterANeg) & $\ImnSzeroAlexnetVoneCenterW$ (\ImnSzeroAlexnetVoneCenterWNeg) \\
\texttt{alexnet} (s0) & LO-obj & $\ImnSzeroAlexnetLoobjRawCorrelation$ (\ImnSzeroAlexnetLoobjRawCorrelationNeg) & $\ImnSzeroAlexnetLoobjZCorrelation$ (\ImnSzeroAlexnetLoobjZCorrelationNeg) & $\ImnSzeroAlexnetLoobjCenterN$ (\ImnSzeroAlexnetLoobjCenterNNeg) & $\ImnSzeroAlexnetLoobjCenterA$ (\ImnSzeroAlexnetLoobjCenterANeg) & $\ImnSzeroAlexnetLoobjCenterW$ (\ImnSzeroAlexnetLoobjCenterWNeg) \\
\texttt{vgg11\_bn} (s0) & V1 & $\ImnSzeroVggOneOneBnVoneRawCorrelation$ (\ImnSzeroVggOneOneBnVoneRawCorrelationNeg) & $\ImnSzeroVggOneOneBnVoneZCorrelation$ (\ImnSzeroVggOneOneBnVoneZCorrelationNeg) & $\ImnSzeroVggOneOneBnVoneCenterN$ (\ImnSzeroVggOneOneBnVoneCenterNNeg) & $\ImnSzeroVggOneOneBnVoneCenterA$ (\ImnSzeroVggOneOneBnVoneCenterANeg) & $\ImnSzeroVggOneOneBnVoneCenterW$ (\ImnSzeroVggOneOneBnVoneCenterWNeg) \\
\texttt{vgg11\_bn} (s0) & LO-obj & $\ImnSzeroVggOneOneBnLoobjRawCorrelation$ (\ImnSzeroVggOneOneBnLoobjRawCorrelationNeg) & $\ImnSzeroVggOneOneBnLoobjZCorrelation$ (\ImnSzeroVggOneOneBnLoobjZCorrelationNeg) & $\ImnSzeroVggOneOneBnLoobjCenterN$ (\ImnSzeroVggOneOneBnLoobjCenterNNeg) & $\ImnSzeroVggOneOneBnLoobjCenterA$ (\ImnSzeroVggOneOneBnLoobjCenterANeg) & $\ImnSzeroVggOneOneBnLoobjCenterW$ (\ImnSzeroVggOneOneBnLoobjCenterWNeg) \\
\texttt{resnet18} (s0) & V1 & $\ImnSzeroResnetOneEightVoneRawCorrelation$ (\ImnSzeroResnetOneEightVoneRawCorrelationNeg) & $\ImnSzeroResnetOneEightVoneZCorrelation$ (\ImnSzeroResnetOneEightVoneZCorrelationNeg) & $\ImnSzeroResnetOneEightVoneCenterN$ (\ImnSzeroResnetOneEightVoneCenterNNeg) & $\ImnSzeroResnetOneEightVoneCenterA$ (\ImnSzeroResnetOneEightVoneCenterANeg) & $\ImnSzeroResnetOneEightVoneCenterW$ (\ImnSzeroResnetOneEightVoneCenterWNeg) \\
\texttt{resnet18} (s0) & LO-obj & $\ImnSzeroResnetOneEightLoobjRawCorrelation$ (\ImnSzeroResnetOneEightLoobjRawCorrelationNeg) & $\ImnSzeroResnetOneEightLoobjZCorrelation$ (\ImnSzeroResnetOneEightLoobjZCorrelationNeg) & $\ImnSzeroResnetOneEightLoobjCenterN$ (\ImnSzeroResnetOneEightLoobjCenterNNeg) & $\ImnSzeroResnetOneEightLoobjCenterA$ (\ImnSzeroResnetOneEightLoobjCenterANeg) & $\ImnSzeroResnetOneEightLoobjCenterW$ (\ImnSzeroResnetOneEightLoobjCenterWNeg) \\
\texttt{resnet50} (s0) & V1 & $\ImnSzeroResnetFiveZeroVoneRawCorrelation$ (\ImnSzeroResnetFiveZeroVoneRawCorrelationNeg) & $\ImnSzeroResnetFiveZeroVoneZCorrelation$ (\ImnSzeroResnetFiveZeroVoneZCorrelationNeg) & $\ImnSzeroResnetFiveZeroVoneCenterN$ (\ImnSzeroResnetFiveZeroVoneCenterNNeg) & $\ImnSzeroResnetFiveZeroVoneCenterA$ (\ImnSzeroResnetFiveZeroVoneCenterANeg) & $\ImnSzeroResnetFiveZeroVoneCenterW$ (\ImnSzeroResnetFiveZeroVoneCenterWNeg) \\
\texttt{resnet50} (s0) & LO-obj & $\ImnSzeroResnetFiveZeroLoobjRawCorrelation$ (\ImnSzeroResnetFiveZeroLoobjRawCorrelationNeg) & $\ImnSzeroResnetFiveZeroLoobjZCorrelation$ (\ImnSzeroResnetFiveZeroLoobjZCorrelationNeg) & $\ImnSzeroResnetFiveZeroLoobjCenterN$ (\ImnSzeroResnetFiveZeroLoobjCenterNNeg) & $\ImnSzeroResnetFiveZeroLoobjCenterA$ (\ImnSzeroResnetFiveZeroLoobjCenterANeg) & $\ImnSzeroResnetFiveZeroLoobjCenterW$ (\ImnSzeroResnetFiveZeroLoobjCenterWNeg) \\
\texttt{convnext\_tiny} (s0) & V1 & $\ImnSzeroConvnextTinyVoneRawCorrelation$ (\ImnSzeroConvnextTinyVoneRawCorrelationNeg) & $\ImnSzeroConvnextTinyVoneZCorrelation$ (\ImnSzeroConvnextTinyVoneZCorrelationNeg) & $\ImnSzeroConvnextTinyVoneCenterN$ (\ImnSzeroConvnextTinyVoneCenterNNeg) & $\ImnSzeroConvnextTinyVoneCenterA$ (\ImnSzeroConvnextTinyVoneCenterANeg) & $\ImnSzeroConvnextTinyVoneCenterW$ (\ImnSzeroConvnextTinyVoneCenterWNeg) \\
\texttt{convnext\_tiny} (s0) & LO-obj & $\ImnSzeroConvnextTinyLoobjRawCorrelation$ (\ImnSzeroConvnextTinyLoobjRawCorrelationNeg) & $\ImnSzeroConvnextTinyLoobjZCorrelation$ (\ImnSzeroConvnextTinyLoobjZCorrelationNeg) & $\ImnSzeroConvnextTinyLoobjCenterN$ (\ImnSzeroConvnextTinyLoobjCenterNNeg) & $\ImnSzeroConvnextTinyLoobjCenterA$ (\ImnSzeroConvnextTinyLoobjCenterANeg) & $\ImnSzeroConvnextTinyLoobjCenterW$ (\ImnSzeroConvnextTinyLoobjCenterWNeg) \\
\texttt{vit\_b\_16} (s0) & V1 & $\ImnSzeroVitBOneSixVoneRawCorrelation$ (\ImnSzeroVitBOneSixVoneRawCorrelationNeg) & $\ImnSzeroVitBOneSixVoneZCorrelation$ (\ImnSzeroVitBOneSixVoneZCorrelationNeg) & $\ImnSzeroVitBOneSixVoneCenterN$ (\ImnSzeroVitBOneSixVoneCenterNNeg) & $\ImnSzeroVitBOneSixVoneCenterA$ (\ImnSzeroVitBOneSixVoneCenterANeg) & $\ImnSzeroVitBOneSixVoneCenterW$ (\ImnSzeroVitBOneSixVoneCenterWNeg) \\
\texttt{vit\_b\_16} (s0) & LO-obj & $\ImnSzeroVitBOneSixLoobjRawCorrelation$ (\ImnSzeroVitBOneSixLoobjRawCorrelationNeg) & $\ImnSzeroVitBOneSixLoobjZCorrelation$ (\ImnSzeroVitBOneSixLoobjZCorrelationNeg) & $\ImnSzeroVitBOneSixLoobjCenterN$ (\ImnSzeroVitBOneSixLoobjCenterNNeg) & $\ImnSzeroVitBOneSixLoobjCenterA$ (\ImnSzeroVitBOneSixLoobjCenterANeg) & $\ImnSzeroVitBOneSixLoobjCenterW$ (\ImnSzeroVitBOneSixLoobjCenterWNeg) \\
\bottomrule
\end{tabular}
\end{adjustbox}
\caption{ImageNet models on NSD ($k = 3$, 224\,px): untrained (mean over initialisation seeds 0--2, or seed 0 only) minus trained; in brackets the number of subjects with $\Delta < 0$ of 8.}
\label{tab:nsdimagenet}
\end{table}

\begin{table}[H]
\centering
\small
\begin{adjustbox}{max width=\textwidth, max totalheight=0.82\textheight}
\begin{tabular}{@{}lrr@{}}
\toprule
px & raw, correlation & z, correlation \\
\midrule
32 & $\ResRawThreeTwo$ [$\ResRawThreeTwoLo$, $\ResRawThreeTwoHi$] & $\ResZThreeTwo$ [$\ResZThreeTwoLo$, $\ResZThreeTwoHi$] \\
64 & $\ResRawSixFour$ [$\ResRawSixFourLo$, $\ResRawSixFourHi$] & $\ResZSixFour$ [$\ResZSixFourLo$, $\ResZSixFourHi$] \\
96 & $\ResRawNineSix$ [$\ResRawNineSixLo$, $\ResRawNineSixHi$] & $\ResZNineSix$ [$\ResZNineSixLo$, $\ResZNineSixHi$] \\
128 & $\ResRawOneTwoEight$ [$\ResRawOneTwoEightLo$, $\ResRawOneTwoEightHi$] & $\ResZOneTwoEight$ [$\ResZOneTwoEightLo$, $\ResZOneTwoEightHi$] \\
160 & $\ResRawOneSixZero$ [$\ResRawOneSixZeroLo$, $\ResRawOneSixZeroHi$] & $\ResZOneSixZero$ [$\ResZOneSixZeroLo$, $\ResZOneSixZeroHi$] \\
224 & $\ResRawTwoTwoFour$ [$\ResRawTwoTwoFourLo$, $\ResRawTwoTwoFourHi$] & $\ResZTwoTwoFour$ [$\ResZTwoTwoFourLo$, $\ResZTwoTwoFourHi$] \\
$S = \Delta(224) - \Delta(32)$ & $\ResRawSlope$ [$\ResRawSlopeLo$, $\ResRawSlopeHi$] & $\ResZSlope$ [$\ResZSlopeLo$, $\ResZSlopeHi$] \\
\bottomrule
\end{tabular}
\end{adjustbox}
\caption{Evaluation resolution on NSD (step 43, reported without prediction): $\Delta$ at V1, own CNN, NATIVE inputs; 95\% stimulus-bootstrap intervals of the 8-subject mean.}
\label{tab:nsdres}
\end{table}

\begin{table}[H]
\centering
\small
\begin{adjustbox}{max width=\textwidth, max totalheight=0.82\textheight}
\begin{tabular}{@{}llrrr@{}}
\toprule
ROI & $k$ & lower bound [95\% interval] & $\rho$ untrained & $\rho$ Backprop \\
\midrule
V1 & 3 & $\CeilVoneKThree$ [$\CeilVoneKThreeLo$, $\CeilVoneKThreeHi$] & $\CeilVoneKThreeRand$ & $\CeilVoneKThreeBp$ \\
V1 & 1 & $\CeilVoneKOne$ [$\CeilVoneKOneLo$, $\CeilVoneKOneHi$] & $\CeilVoneKOneRand$ & $\CeilVoneKOneBp$ \\
LO-obj & 3 & $\CeilLoobjKThree$ [$\CeilLoobjKThreeLo$, $\CeilLoobjKThreeHi$] & $\CeilLoobjKThreeRand$ & $\CeilLoobjKThreeBp$ \\
LO-obj & 1 & $\CeilLoobjKOne$ [$\CeilLoobjKOneLo$, $\CeilLoobjKOneHi$] & $\CeilLoobjKOneRand$ & $\CeilLoobjKOneBp$ \\
\bottomrule
\end{tabular}
\end{adjustbox}
\caption{Leave-one-subject-out lower bound of the noise ceiling in NSD (correlation distance) and the RSA of the own CNN (raw units, mean over seeds and subjects).}
\label{tab:nsdceil}
\end{table}

\begin{table}[H]
\centering
\small
\begin{adjustbox}{max width=\textwidth, max totalheight=0.82\textheight}
\begin{tabular}{@{}lrrrr@{}}
\toprule
ROI, cell (subj01, $k = 3$) & raw, corr. & z, corr. & raw, Eucl. & z, Eucl. \\
\midrule
V1 & $\PilotVoneRawCorrelation$ & $\PilotVoneZCorrelation$ & $\PilotVoneRawEuclidean$ & $\PilotVoneZEuclidean$ \\
LO-obj & $\PilotLoobjRawCorrelation$ & $\PilotLoobjZCorrelation$ & $\PilotLoobjRawEuclidean$ & $\PilotLoobjZEuclidean$ \\
\bottomrule
\end{tabular}
\end{adjustbox}
\caption{The subj01 pilot (step 42): $\Delta$, untrained minus Backprop, mean over seeds. The positive Euclidean values motivated Q1--Q4 (step 42b), which were tested on subj02--08.}
\label{tab:nsdpilot}
\end{table}

\begin{table}[H]
\centering
\scriptsize
\setlength{\tabcolsep}{2.5pt}
\begin{adjustbox}{max width=\textwidth, max totalheight=0.82\textheight}
\begin{tabular}{@{}rlrrrr@{}}
\toprule
Class & $F$ / $D$ / $B$ (representative) & THINGS-720 & NSD, N-session & NSD, N-raw & NSD, run-centred \\
\midrule
1 & raw / correlation / paper & $\MvThingsClsOne$ {\scriptsize [$\MvThingsClsOneLo$, $\MvThingsClsOneHi$]} & $\MvNsessionClsOne$ {\scriptsize [$\MvNsessionClsOneLo$, $\MvNsessionClsOneHi$]} & $\MvNrawClsOne$ {\scriptsize [$\MvNrawClsOneLo$, $\MvNrawClsOneHi$]} & $\MvNrcClsOne$ {\scriptsize [$\MvNrcClsOneLo$, $\MvNrcClsOneHi$]} \\
2 & raw / correlation / center720 & $\MvThingsClsTwo$ {\scriptsize [$\MvThingsClsTwoLo$, $\MvThingsClsTwoHi$]} & $\MvNsessionClsTwo$ {\scriptsize [$\MvNsessionClsTwoLo$, $\MvNsessionClsTwoHi$]} & $\MvNrawClsTwo$ {\scriptsize [$\MvNrawClsTwoLo$, $\MvNrawClsTwoHi$]} & $\MvNrcClsTwo$ {\scriptsize [$\MvNrcClsTwoLo$, $\MvNrcClsTwoHi$]} \\
3 & raw / correlation / z720 & $\MvThingsClsThree$ {\scriptsize [$\MvThingsClsThreeLo$, $\MvThingsClsThreeHi$]} & $\MvNsessionClsThree$ {\scriptsize [$\MvNsessionClsThreeLo$, $\MvNsessionClsThreeHi$]} & $\MvNrawClsThree$ {\scriptsize [$\MvNrawClsThreeLo$, $\MvNrawClsThreeHi$]} & $\MvNrcClsThree$ {\scriptsize [$\MvNrcClsThreeLo$, $\MvNrcClsThreeHi$]} \\
4 & raw / cosine / paper & $\MvThingsClsFour$ {\scriptsize [$\MvThingsClsFourLo$, $\MvThingsClsFourHi$]} & $\MvNsessionClsFour$ {\scriptsize [$\MvNsessionClsFourLo$, $\MvNsessionClsFourHi$]} & $\MvNrawClsFour$ {\scriptsize [$\MvNrawClsFourLo$, $\MvNrawClsFourHi$]} & $\MvNrcClsFour$ {\scriptsize [$\MvNrcClsFourLo$, $\MvNrcClsFourHi$]} \\
5 & raw / cosine / center720 & $\MvThingsClsFive$ {\scriptsize [$\MvThingsClsFiveLo$, $\MvThingsClsFiveHi$]} & $\MvNsessionClsFive$ {\scriptsize [$\MvNsessionClsFiveLo$, $\MvNsessionClsFiveHi$]} & $\MvNrawClsFive$ {\scriptsize [$\MvNrawClsFiveLo$, $\MvNrawClsFiveHi$]} & $\MvNrcClsFive$ {\scriptsize [$\MvNrcClsFiveLo$, $\MvNrcClsFiveHi$]} \\
6 & raw / cosine / z720 & $\MvThingsClsSix$ {\scriptsize [$\MvThingsClsSixLo$, $\MvThingsClsSixHi$]} & $\MvNsessionClsSix$ {\scriptsize [$\MvNsessionClsSixLo$, $\MvNsessionClsSixHi$]} & $\MvNrawClsSix$ {\scriptsize [$\MvNrawClsSixLo$, $\MvNrawClsSixHi$]} & $\MvNrcClsSix$ {\scriptsize [$\MvNrcClsSixLo$, $\MvNrcClsSixHi$]} \\
7 & raw / euclidean / paper & $\MvThingsClsSeven$ {\scriptsize [$\MvThingsClsSevenLo$, $\MvThingsClsSevenHi$]} & $\MvNsessionClsSeven$ {\scriptsize [$\MvNsessionClsSevenLo$, $\MvNsessionClsSevenHi$]} & $\MvNrawClsSeven$ {\scriptsize [$\MvNrawClsSevenLo$, $\MvNrawClsSevenHi$]} & $\MvNrcClsSeven$ {\scriptsize [$\MvNrcClsSevenLo$, $\MvNrcClsSevenHi$]} \\
8 & raw / euclidean / z720 & $\MvThingsClsEight$ {\scriptsize [$\MvThingsClsEightLo$, $\MvThingsClsEightHi$]} & $\MvNsessionClsEight$ {\scriptsize [$\MvNsessionClsEightLo$, $\MvNsessionClsEightHi$]} & $\MvNrawClsEight$ {\scriptsize [$\MvNrawClsEightLo$, $\MvNrawClsEightHi$]} & $\MvNrcClsEight$ {\scriptsize [$\MvNrcClsEightLo$, $\MvNrcClsEightHi$]} \\
9 & center / correlation / paper & $\MvThingsClsNine$ {\scriptsize [$\MvThingsClsNineLo$, $\MvThingsClsNineHi$]} & $\MvNsessionClsNine$ {\scriptsize [$\MvNsessionClsNineLo$, $\MvNsessionClsNineHi$]} & $\MvNrawClsNine$ {\scriptsize [$\MvNrawClsNineLo$, $\MvNrawClsNineHi$]} & $\MvNrcClsNine$ {\scriptsize [$\MvNrcClsNineLo$, $\MvNrcClsNineHi$]} \\
10 & center / correlation / center720 & $\MvThingsClsOneZero$ {\scriptsize [$\MvThingsClsOneZeroLo$, $\MvThingsClsOneZeroHi$]} & $\MvNsessionClsOneZero$ {\scriptsize [$\MvNsessionClsOneZeroLo$, $\MvNsessionClsOneZeroHi$]} & $\MvNrawClsOneZero$ {\scriptsize [$\MvNrawClsOneZeroLo$, $\MvNrawClsOneZeroHi$]} & $\MvNrcClsOneZero$ {\scriptsize [$\MvNrcClsOneZeroLo$, $\MvNrcClsOneZeroHi$]} \\
11 & center / correlation / z720 & $\MvThingsClsOneOne$ {\scriptsize [$\MvThingsClsOneOneLo$, $\MvThingsClsOneOneHi$]} & $\MvNsessionClsOneOne$ {\scriptsize [$\MvNsessionClsOneOneLo$, $\MvNsessionClsOneOneHi$]} & $\MvNrawClsOneOne$ {\scriptsize [$\MvNrawClsOneOneLo$, $\MvNrawClsOneOneHi$]} & $\MvNrcClsOneOne$ {\scriptsize [$\MvNrcClsOneOneLo$, $\MvNrcClsOneOneHi$]} \\
12 & center / cosine / paper & $\MvThingsClsOneTwo$ {\scriptsize [$\MvThingsClsOneTwoLo$, $\MvThingsClsOneTwoHi$]} & $\MvNsessionClsOneTwo$ {\scriptsize [$\MvNsessionClsOneTwoLo$, $\MvNsessionClsOneTwoHi$]} & $\MvNrawClsOneTwo$ {\scriptsize [$\MvNrawClsOneTwoLo$, $\MvNrawClsOneTwoHi$]} & $\MvNrcClsOneTwo$ {\scriptsize [$\MvNrcClsOneTwoLo$, $\MvNrcClsOneTwoHi$]} \\
13 & center / cosine / center720 & $\MvThingsClsOneThree$ {\scriptsize [$\MvThingsClsOneThreeLo$, $\MvThingsClsOneThreeHi$]} & $\MvNsessionClsOneThree$ {\scriptsize [$\MvNsessionClsOneThreeLo$, $\MvNsessionClsOneThreeHi$]} & $\MvNrawClsOneThree$ {\scriptsize [$\MvNrawClsOneThreeLo$, $\MvNrawClsOneThreeHi$]} & $\MvNrcClsOneThree$ {\scriptsize [$\MvNrcClsOneThreeLo$, $\MvNrcClsOneThreeHi$]} \\
14 & center / cosine / z720 & $\MvThingsClsOneFour$ {\scriptsize [$\MvThingsClsOneFourLo$, $\MvThingsClsOneFourHi$]} & $\MvNsessionClsOneFour$ {\scriptsize [$\MvNsessionClsOneFourLo$, $\MvNsessionClsOneFourHi$]} & $\MvNrawClsOneFour$ {\scriptsize [$\MvNrawClsOneFourLo$, $\MvNrawClsOneFourHi$]} & $\MvNrcClsOneFour$ {\scriptsize [$\MvNrcClsOneFourLo$, $\MvNrcClsOneFourHi$]} \\
15 & scale / correlation / paper & $\MvThingsClsOneFive$ {\scriptsize [$\MvThingsClsOneFiveLo$, $\MvThingsClsOneFiveHi$]} & $\MvNsessionClsOneFive$ {\scriptsize [$\MvNsessionClsOneFiveLo$, $\MvNsessionClsOneFiveHi$]} & $\MvNrawClsOneFive$ {\scriptsize [$\MvNrawClsOneFiveLo$, $\MvNrawClsOneFiveHi$]} & $\MvNrcClsOneFive$ {\scriptsize [$\MvNrcClsOneFiveLo$, $\MvNrcClsOneFiveHi$]} \\
16 & scale / correlation / center720 & $\MvThingsClsOneSix$ {\scriptsize [$\MvThingsClsOneSixLo$, $\MvThingsClsOneSixHi$]} & $\MvNsessionClsOneSix$ {\scriptsize [$\MvNsessionClsOneSixLo$, $\MvNsessionClsOneSixHi$]} & $\MvNrawClsOneSix$ {\scriptsize [$\MvNrawClsOneSixLo$, $\MvNrawClsOneSixHi$]} & $\MvNrcClsOneSix$ {\scriptsize [$\MvNrcClsOneSixLo$, $\MvNrcClsOneSixHi$]} \\
17 & scale / correlation / z720 & $\MvThingsClsOneSeven$ {\scriptsize [$\MvThingsClsOneSevenLo$, $\MvThingsClsOneSevenHi$]} & $\MvNsessionClsOneSeven$ {\scriptsize [$\MvNsessionClsOneSevenLo$, $\MvNsessionClsOneSevenHi$]} & $\MvNrawClsOneSeven$ {\scriptsize [$\MvNrawClsOneSevenLo$, $\MvNrawClsOneSevenHi$]} & $\MvNrcClsOneSeven$ {\scriptsize [$\MvNrcClsOneSevenLo$, $\MvNrcClsOneSevenHi$]} \\
18 & scale / cosine / paper & $\MvThingsClsOneEight$ {\scriptsize [$\MvThingsClsOneEightLo$, $\MvThingsClsOneEightHi$]} & $\MvNsessionClsOneEight$ {\scriptsize [$\MvNsessionClsOneEightLo$, $\MvNsessionClsOneEightHi$]} & $\MvNrawClsOneEight$ {\scriptsize [$\MvNrawClsOneEightLo$, $\MvNrawClsOneEightHi$]} & $\MvNrcClsOneEight$ {\scriptsize [$\MvNrcClsOneEightLo$, $\MvNrcClsOneEightHi$]} \\
19 & scale / cosine / center720 & $\MvThingsClsOneNine$ {\scriptsize [$\MvThingsClsOneNineLo$, $\MvThingsClsOneNineHi$]} & $\MvNsessionClsOneNine$ {\scriptsize [$\MvNsessionClsOneNineLo$, $\MvNsessionClsOneNineHi$]} & $\MvNrawClsOneNine$ {\scriptsize [$\MvNrawClsOneNineLo$, $\MvNrawClsOneNineHi$]} & $\MvNrcClsOneNine$ {\scriptsize [$\MvNrcClsOneNineLo$, $\MvNrcClsOneNineHi$]} \\
20 & scale / cosine / z720 & $\MvThingsClsTwoZero$ {\scriptsize [$\MvThingsClsTwoZeroLo$, $\MvThingsClsTwoZeroHi$]} & $\MvNsessionClsTwoZero$ {\scriptsize [$\MvNsessionClsTwoZeroLo$, $\MvNsessionClsTwoZeroHi$]} & $\MvNrawClsTwoZero$ {\scriptsize [$\MvNrawClsTwoZeroLo$, $\MvNrawClsTwoZeroHi$]} & $\MvNrcClsTwoZero$ {\scriptsize [$\MvNrcClsTwoZeroLo$, $\MvNrcClsTwoZeroHi$]} \\
21 & scale / euclidean / paper & $\MvThingsClsTwoOne$ {\scriptsize [$\MvThingsClsTwoOneLo$, $\MvThingsClsTwoOneHi$]} & $\MvNsessionClsTwoOne$ {\scriptsize [$\MvNsessionClsTwoOneLo$, $\MvNsessionClsTwoOneHi$]} & $\MvNrawClsTwoOne$ {\scriptsize [$\MvNrawClsTwoOneLo$, $\MvNrawClsTwoOneHi$]} & $\MvNrcClsTwoOne$ {\scriptsize [$\MvNrcClsTwoOneLo$, $\MvNrcClsTwoOneHi$]} \\
22 & scale / euclidean / z720 & $\MvThingsClsTwoTwo$ {\scriptsize [$\MvThingsClsTwoTwoLo$, $\MvThingsClsTwoTwoHi$]} & $\MvNsessionClsTwoTwo$ {\scriptsize [$\MvNsessionClsTwoTwoLo$, $\MvNsessionClsTwoTwoHi$]} & $\MvNrawClsTwoTwo$ {\scriptsize [$\MvNrawClsTwoTwoLo$, $\MvNrawClsTwoTwoHi$]} & $\MvNrcClsTwoTwo$ {\scriptsize [$\MvNrcClsTwoTwoLo$, $\MvNrcClsTwoTwoHi$]} \\
23 & z / correlation / paper & $\MvThingsClsTwoThree$ {\scriptsize [$\MvThingsClsTwoThreeLo$, $\MvThingsClsTwoThreeHi$]} & $\MvNsessionClsTwoThree$ {\scriptsize [$\MvNsessionClsTwoThreeLo$, $\MvNsessionClsTwoThreeHi$]} & $\MvNrawClsTwoThree$ {\scriptsize [$\MvNrawClsTwoThreeLo$, $\MvNrawClsTwoThreeHi$]} & $\MvNrcClsTwoThree$ {\scriptsize [$\MvNrcClsTwoThreeLo$, $\MvNrcClsTwoThreeHi$]} \\
24 & z / correlation / center720 & $\MvThingsClsTwoFour$ {\scriptsize [$\MvThingsClsTwoFourLo$, $\MvThingsClsTwoFourHi$]} & $\MvNsessionClsTwoFour$ {\scriptsize [$\MvNsessionClsTwoFourLo$, $\MvNsessionClsTwoFourHi$]} & $\MvNrawClsTwoFour$ {\scriptsize [$\MvNrawClsTwoFourLo$, $\MvNrawClsTwoFourHi$]} & $\MvNrcClsTwoFour$ {\scriptsize [$\MvNrcClsTwoFourLo$, $\MvNrcClsTwoFourHi$]} \\
25 & z / correlation / z720 & $\MvThingsClsTwoFive$ {\scriptsize [$\MvThingsClsTwoFiveLo$, $\MvThingsClsTwoFiveHi$]} & $\MvNsessionClsTwoFive$ {\scriptsize [$\MvNsessionClsTwoFiveLo$, $\MvNsessionClsTwoFiveHi$]} & $\MvNrawClsTwoFive$ {\scriptsize [$\MvNrawClsTwoFiveLo$, $\MvNrawClsTwoFiveHi$]} & $\MvNrcClsTwoFive$ {\scriptsize [$\MvNrcClsTwoFiveLo$, $\MvNrcClsTwoFiveHi$]} \\
26 & z / cosine / paper & $\MvThingsClsTwoSix$ {\scriptsize [$\MvThingsClsTwoSixLo$, $\MvThingsClsTwoSixHi$]} & $\MvNsessionClsTwoSix$ {\scriptsize [$\MvNsessionClsTwoSixLo$, $\MvNsessionClsTwoSixHi$]} & $\MvNrawClsTwoSix$ {\scriptsize [$\MvNrawClsTwoSixLo$, $\MvNrawClsTwoSixHi$]} & $\MvNrcClsTwoSix$ {\scriptsize [$\MvNrcClsTwoSixLo$, $\MvNrcClsTwoSixHi$]} \\
27 & z / cosine / center720 & $\MvThingsClsTwoSeven$ {\scriptsize [$\MvThingsClsTwoSevenLo$, $\MvThingsClsTwoSevenHi$]} & $\MvNsessionClsTwoSeven$ {\scriptsize [$\MvNsessionClsTwoSevenLo$, $\MvNsessionClsTwoSevenHi$]} & $\MvNrawClsTwoSeven$ {\scriptsize [$\MvNrawClsTwoSevenLo$, $\MvNrawClsTwoSevenHi$]} & $\MvNrcClsTwoSeven$ {\scriptsize [$\MvNrcClsTwoSevenLo$, $\MvNrcClsTwoSevenHi$]} \\
28 & z / cosine / z720 & $\MvThingsClsTwoEight$ {\scriptsize [$\MvThingsClsTwoEightLo$, $\MvThingsClsTwoEightHi$]} & $\MvNsessionClsTwoEight$ {\scriptsize [$\MvNsessionClsTwoEightLo$, $\MvNsessionClsTwoEightHi$]} & $\MvNrawClsTwoEight$ {\scriptsize [$\MvNrawClsTwoEightLo$, $\MvNrawClsTwoEightHi$]} & $\MvNrcClsTwoEight$ {\scriptsize [$\MvNrcClsTwoEightLo$, $\MvNrcClsTwoEightHi$]} \\
\bottomrule
\end{tabular}
\end{adjustbox}
\caption{The 28 distinct analyses of the V1 multiverse (class numbering of step 43; one representative cell per class, $B$ = paper denotes the data set's own normalisation). THINGS-720: the paper's multiverse (NATIVE 224\,px); NSD under N-session (the NSD standard), N-raw and run centring. $\Delta$ with 95\% stimulus-bootstrap interval of the subject mean.}
\label{tab:multi28}
\end{table}

\section{Candidate explanations in THINGS}
\label{app:why}
Partial RSA (Section~\ref{sec:gainmethods}) was used with covariate RDMs built from image statistics:
$|L_i - L_j|$ (mean luminance), $|C_i - C_j|$ (RMS contrast), the Euclidean distance between
coarse-luminance images (the grey image filtered to the lowest spatial-frequency band of
\citet{evalres}, including the mean, at $28 \times 28$; LP), the predecessor-luminance RDM
$|L_{\mathrm{prev}(i)} - L_{\mathrm{prev}(j)}|$, and the presentation-position RDM $|p_i - p_j|$
($T$). Ranks of the model and the brain RDM are regressed on the ranks of the covariates (with
intercept), and the residuals are correlated (Pearson).

The analyses of steps 44, 45, 47, 47b and 47c in THINGS: the decomposition of run centring
(Table~\ref{tab:decomp}), partial RSA with image-statistic, position and predecessor covariates
(Tables~\ref{tab:partial} and~\ref{tab:partialother}), and run-level quantities per subject
(Table~\ref{tab:runlevel}). The luminance correlation of the V1 RDM, divided by the leave-one-subject-out
lower bound, was \LumRatioTsev{} in THINGS-720, \LumRatioTest{} in the test images and \LumRatioNsd{} in NSD
(W4 met); for the coarse-luminance RDM the ratios were \LpRatioTsev, \LpRatioTest{} and \LpRatioNsd{} (W7
met). With run centring, the THINGS-720 ratio fell to \LumBoundRc{} (RC2 not met), and the multiverse had
\RuleRcFour{} distinct analyses with an interval above 0 (RC4 not met). The run-level variance of the
run-mean luminance was at the \RunLumPctSubOneTwo{}th percentile of random assignments for sub-01/02 and at
the \RunLumPctSubThree{}th for sub-03 (RC3 not met, reversed).

\begin{table}[H]
\centering
\small
\begin{adjustbox}{max width=\textwidth, max totalheight=0.82\textheight}
\begin{tabular}{@{}lrrrrr@{}}
\toprule
Variant (Section~\ref{sec:convmethods}) & sub-01 & sub-02 & sub-03 & mean V1 & mean LOC \\
\midrule
V0 nothing removed (paper cell) & $\DecVoneVZeroSubOne$ & $\DecVoneVZeroSubTwo$ & $\DecVoneVZeroSubThree$ & $\DecVoneVZero$ & $\DecLocVZero$ \\
V1 session-mean profile $m$ removed & $\DecVoneVOneSubOne$ & $\DecVoneVOneSubTwo$ & $\DecVoneVOneSubThree$ & $\DecVoneVOne$ & $\DecLocVOne$ \\
V2 run deviations $d_r$ removed & $\DecVoneVTwoSubOne$ & $\DecVoneVTwoSubTwo$ & $\DecVoneVTwoSubThree$ & $\DecVoneVTwo$ & $\DecLocVTwo$ \\
V3 both removed (run centring) & $\DecVoneVThreeSubOne$ & $\DecVoneVThreeSubTwo$ & $\DecVoneVThreeSubThree$ & $\DecVoneVThree$ & $\DecLocVThree$ \\
V4 = V3 + $d_{r\parallel}$ re-added &  &  &  & $\DecVoneVFour$ & $\DecLocVFour$ \\
V5 = V3 + $d_{r\perp}$ re-added &  &  &  & $\DecVoneVFive$ & $\DecLocVFive$ \\
\bottomrule
\end{tabular}
\end{adjustbox}
\caption{Decomposition of run centring in THINGS-720 (step 47c): $\Delta$ (untrained minus Backprop, paper cell, cross-run pairs) after removing parts of the voxel means. Per subject for V0--V3, V1 ROI.}
\label{tab:decomp}
\end{table}

\begin{table}[H]
\centering
\small
\begin{adjustbox}{max width=\textwidth, max totalheight=0.82\textheight}
\begin{tabular}{@{}lrr@{}}
\toprule
Covariates partialled (THINGS-720, paper cell) & V1 & LOC \\
\midrule
none & $\PartFfVoneNone$ & $\PartFfLocNone$ \\
$L$, $C$ (global luminance, RMS contrast) & $\PartFfVoneLc$ & $\PartFfLocLc$ \\
$T$ (presentation position) & $\PartFfVoneT$ & $\PartFfLocT$ \\
$L$, $C$, $T$ & $\PartFfVoneLct$ & $\PartFfLocLct$ \\
LP (coarse luminance image) & $\PartFffVoneLp$ & $\PartFffLocLp$ \\
$L$, $C$, LP & $\PartFffVoneLclp$ & $\PartFffLocLclp$ \\
$L$, $C$, LP, $T$ & $\PartFffVoneLclpt$ & $\PartFffLocLclpt$ \\
\midrule
run centring (step 47) & $\RunCentVoneRc$ & $\RunCentLocRc$ \\
run centring, then $L$, $C$ & $\RunCentVoneRcLc$ & $\RunCentLocRcLc$ \\
\bottomrule
\end{tabular}
\end{adjustbox}
\caption{Partial RSA in THINGS-720 (steps 44, 45, 47): $\Delta$ after rank-regressing both RDMs on covariate RDMs. THINGS test images and NSD: Table~\ref{tab:partialother}.}
\label{tab:partial}
\end{table}

\begin{table}[H]
\centering
\small
\begin{adjustbox}{max width=\textwidth, max totalheight=0.82\textheight}
\begin{tabular}{@{}lrrrr@{}}
\toprule
Covariates & test V1 ($k = 1$) & test LOC ($k = 1$) & NSD V1 ($k = 1$) & NSD LO-obj ($k = 1$) \\
\midrule
none & $\PartTestVoneNone$ & $\PartTestLocNone$ & $\PartNsdVoneNone$ & $\PartNsdLoobjNone$ \\
$L$, $C$ & $\PartTestVoneLc$ & $\PartTestLocLc$ & $\PartNsdVoneLc$ & $\PartNsdLoobjLc$ \\
LP & $\PartTestVoneLp$ & $\PartTestLocLp$ & $\PartNsdVoneLp$ & $\PartNsdLoobjLp$ \\
$L$, $C$, LP & $\PartTestVoneLclp$ & $\PartTestLocLclp$ & $\PartNsdVoneLclp$ & $\PartNsdLoobjLclp$ \\
\bottomrule
\end{tabular}
\end{adjustbox}
\caption{Partial RSA in the THINGS test images and in NSD (step 45), paper cell, standard normalisation of each data set.}
\label{tab:partialother}
\end{table}

\begin{table}[H]
\centering
\small
\begin{adjustbox}{max width=\textwidth, max totalheight=0.82\textheight}
\begin{tabular}{@{}lrrrrrrr@{}}
\toprule
Subject & $\Delta A$ & percentile & $\rho$(run RDM, run index) & run-level part of $\Delta$ & SD of $d_r\!\cdot\!\hat m$ & share along $\hat m$ & order $\rho$ \\
\midrule
sub-01 & $\MechDeltaASubOne$ & $\MechPctSubOne$ & $\MechRunIdxSubOne$ & $\MechRunPartSubOne$ & $\DecSdSubOne$ & $\DecShareSubOne\%$ & $\OrderRhoSubOne$ \\
sub-02 & $\MechDeltaASubTwo$ & $\MechPctSubTwo$ & $\MechRunIdxSubTwo$ & $\MechRunPartSubTwo$ & $\DecSdSubTwo$ & $\DecShareSubTwo\%$ & $\OrderRhoSubTwo$ \\
sub-03 & $\MechDeltaASubThree$ & $\MechPctSubThree$ & $\MechRunIdxSubThree$ & $\MechRunPartSubThree$ & $\DecSdSubThree$ & $\DecShareSubThree\%$ & $\OrderRhoSubThree$ \\
\bottomrule
\end{tabular}
\end{adjustbox}
\caption{Run-level quantities per subject (steps 44, 47b, 47c). $\Delta A$: alignment of the brain run RDM with the untrained minus the Backprop run RDM, and its percentile in the null of random run assignments; run-level part: $\Delta$ without minus with run centring; SD and share: the projection of the run deviations on the session-mean profile; order $\rho$: Spearman between the V1 RDM and the presentation-position RDM.}
\label{tab:runlevel}
\end{table}

\section{Conventions and response amplitude}
\label{app:conv}
All cells of step 48 (Table~\ref{tab:convfull}), the response-amplitude statistics of steps 48 and 48b
(Table~\ref{tab:amplitude}), model amplitude dominance per seed (Table~\ref{tab:dom}) and $\Delta$ before
and after the two-sided partial in both forms of the amplitude RDM (Table~\ref{tab:partial48b}). In
\texttt{results/step48/amplitude.csv} the seed column is empty for the THINGS rows (a merge step of the
step-48 scripts dropped it); the values are complete and in generation order, and the macro and figure
scripts reconstruct the seeds from that order, as the gate of step 48b did.

\begin{table}[H]
\centering
\scriptsize
\setlength{\tabcolsep}{3pt}
\begin{adjustbox}{max width=\textwidth, max totalheight=0.82\textheight}
\begin{tabular}{@{}llllrrrr@{}}
\toprule
Data & ROI & $k$ & Convention & raw, corr. & z, corr. & raw, Eucl. & z, Eucl. \\
\midrule
THINGS-720 & V1 &  & T-raw & $\CvTsevTrawVoneRawCorr$ (\CvTsevTrawVoneRawCorrPos/3) & $\CvTsevTrawVoneZCorr$ (\CvTsevTrawVoneZCorrPos/3) & $\CvTsevTrawVoneRawEucl$ (\CvTsevTrawVoneRawEuclPos/3) & $\CvTsevTrawVoneZEucl$ (\CvTsevTrawVoneZEuclPos/3) \\
THINGS-720 & V1 &  & T-global & $\CvTsevTglobalVoneRawCorr$ (\CvTsevTglobalVoneRawCorrPos/3) & $\CvTsevTglobalVoneZCorr$ (\CvTsevTglobalVoneZCorrPos/3) & $\CvTsevTglobalVoneRawEucl$ (\CvTsevTglobalVoneRawEuclPos/3) & $\CvTsevTglobalVoneZEucl$ (\CvTsevTglobalVoneZEuclPos/3) \\
THINGS-720 & V1 &  & T-session & $\CvTsevTsessionVoneRawCorr$ (\CvTsevTsessionVoneRawCorrPos/3) & $\CvTsevTsessionVoneZCorr$ (\CvTsevTsessionVoneZCorrPos/3) & $\CvTsevTsessionVoneRawEucl$ (\CvTsevTsessionVoneRawEuclPos/3) & $\CvTsevTsessionVoneZEucl$ (\CvTsevTsessionVoneZEuclPos/3) \\
THINGS-720 & LOC &  & T-raw & $\CvTsevTrawLocRawCorr$ (\CvTsevTrawLocRawCorrPos/3) & $\CvTsevTrawLocZCorr$ (\CvTsevTrawLocZCorrPos/3) & $\CvTsevTrawLocRawEucl$ (\CvTsevTrawLocRawEuclPos/3) & $\CvTsevTrawLocZEucl$ (\CvTsevTrawLocZEuclPos/3) \\
THINGS-720 & LOC &  & T-global & $\CvTsevTglobalLocRawCorr$ (\CvTsevTglobalLocRawCorrPos/3) & $\CvTsevTglobalLocZCorr$ (\CvTsevTglobalLocZCorrPos/3) & $\CvTsevTglobalLocRawEucl$ (\CvTsevTglobalLocRawEuclPos/3) & $\CvTsevTglobalLocZEucl$ (\CvTsevTglobalLocZEuclPos/3) \\
THINGS-720 & LOC &  & T-session & $\CvTsevTsessionLocRawCorr$ (\CvTsevTsessionLocRawCorrPos/3) & $\CvTsevTsessionLocZCorr$ (\CvTsevTsessionLocZCorrPos/3) & $\CvTsevTsessionLocRawEucl$ (\CvTsevTsessionLocRawEuclPos/3) & $\CvTsevTsessionLocZEucl$ (\CvTsevTsessionLocZEuclPos/3) \\
\midrule
THINGS test & V1 & 1 & T-raw & $\CvTtestTrawVoneRawCorrKOne$ (\CvTtestTrawVoneRawCorrKOnePos/3) & $\CvTtestTrawVoneZCorrKOne$ (\CvTtestTrawVoneZCorrKOnePos/3) & $\CvTtestTrawVoneRawEuclKOne$ (\CvTtestTrawVoneRawEuclKOnePos/3) & $\CvTtestTrawVoneZEuclKOne$ (\CvTtestTrawVoneZEuclKOnePos/3) \\
THINGS test & V1 & 1 & T-global & $\CvTtestTglobalVoneRawCorrKOne$ (\CvTtestTglobalVoneRawCorrKOnePos/3) & $\CvTtestTglobalVoneZCorrKOne$ (\CvTtestTglobalVoneZCorrKOnePos/3) & $\CvTtestTglobalVoneRawEuclKOne$ (\CvTtestTglobalVoneRawEuclKOnePos/3) & $\CvTtestTglobalVoneZEuclKOne$ (\CvTtestTglobalVoneZEuclKOnePos/3) \\
THINGS test & V1 & 1 & T-session & $\CvTtestTsessionVoneRawCorrKOne$ (\CvTtestTsessionVoneRawCorrKOnePos/3) & $\CvTtestTsessionVoneZCorrKOne$ (\CvTtestTsessionVoneZCorrKOnePos/3) & $\CvTtestTsessionVoneRawEuclKOne$ (\CvTtestTsessionVoneRawEuclKOnePos/3) & $\CvTtestTsessionVoneZEuclKOne$ (\CvTtestTsessionVoneZEuclKOnePos/3) \\
THINGS test & V1 & 12 & T-raw & $\CvTtestTrawVoneRawCorrKOneTwo$ (\CvTtestTrawVoneRawCorrKOneTwoPos/3) & $\CvTtestTrawVoneZCorrKOneTwo$ (\CvTtestTrawVoneZCorrKOneTwoPos/3) & $\CvTtestTrawVoneRawEuclKOneTwo$ (\CvTtestTrawVoneRawEuclKOneTwoPos/3) & $\CvTtestTrawVoneZEuclKOneTwo$ (\CvTtestTrawVoneZEuclKOneTwoPos/3) \\
THINGS test & V1 & 12 & T-global & $\CvTtestTglobalVoneRawCorrKOneTwo$ (\CvTtestTglobalVoneRawCorrKOneTwoPos/3) & $\CvTtestTglobalVoneZCorrKOneTwo$ (\CvTtestTglobalVoneZCorrKOneTwoPos/3) & $\CvTtestTglobalVoneRawEuclKOneTwo$ (\CvTtestTglobalVoneRawEuclKOneTwoPos/3) & $\CvTtestTglobalVoneZEuclKOneTwo$ (\CvTtestTglobalVoneZEuclKOneTwoPos/3) \\
THINGS test & V1 & 12 & T-session & $\CvTtestTsessionVoneRawCorrKOneTwo$ (\CvTtestTsessionVoneRawCorrKOneTwoPos/3) & $\CvTtestTsessionVoneZCorrKOneTwo$ (\CvTtestTsessionVoneZCorrKOneTwoPos/3) & $\CvTtestTsessionVoneRawEuclKOneTwo$ (\CvTtestTsessionVoneRawEuclKOneTwoPos/3) & $\CvTtestTsessionVoneZEuclKOneTwo$ (\CvTtestTsessionVoneZEuclKOneTwoPos/3) \\
THINGS test & LOC & 1 & T-raw & $\CvTtestTrawLocRawCorrKOne$ (\CvTtestTrawLocRawCorrKOnePos/3) & $\CvTtestTrawLocZCorrKOne$ (\CvTtestTrawLocZCorrKOnePos/3) & $\CvTtestTrawLocRawEuclKOne$ (\CvTtestTrawLocRawEuclKOnePos/3) & $\CvTtestTrawLocZEuclKOne$ (\CvTtestTrawLocZEuclKOnePos/3) \\
THINGS test & LOC & 1 & T-global & $\CvTtestTglobalLocRawCorrKOne$ (\CvTtestTglobalLocRawCorrKOnePos/3) & $\CvTtestTglobalLocZCorrKOne$ (\CvTtestTglobalLocZCorrKOnePos/3) & $\CvTtestTglobalLocRawEuclKOne$ (\CvTtestTglobalLocRawEuclKOnePos/3) & $\CvTtestTglobalLocZEuclKOne$ (\CvTtestTglobalLocZEuclKOnePos/3) \\
THINGS test & LOC & 1 & T-session & $\CvTtestTsessionLocRawCorrKOne$ (\CvTtestTsessionLocRawCorrKOnePos/3) & $\CvTtestTsessionLocZCorrKOne$ (\CvTtestTsessionLocZCorrKOnePos/3) & $\CvTtestTsessionLocRawEuclKOne$ (\CvTtestTsessionLocRawEuclKOnePos/3) & $\CvTtestTsessionLocZEuclKOne$ (\CvTtestTsessionLocZEuclKOnePos/3) \\
THINGS test & LOC & 12 & T-raw & $\CvTtestTrawLocRawCorrKOneTwo$ (\CvTtestTrawLocRawCorrKOneTwoPos/3) & $\CvTtestTrawLocZCorrKOneTwo$ (\CvTtestTrawLocZCorrKOneTwoPos/3) & $\CvTtestTrawLocRawEuclKOneTwo$ (\CvTtestTrawLocRawEuclKOneTwoPos/3) & $\CvTtestTrawLocZEuclKOneTwo$ (\CvTtestTrawLocZEuclKOneTwoPos/3) \\
THINGS test & LOC & 12 & T-global & $\CvTtestTglobalLocRawCorrKOneTwo$ (\CvTtestTglobalLocRawCorrKOneTwoPos/3) & $\CvTtestTglobalLocZCorrKOneTwo$ (\CvTtestTglobalLocZCorrKOneTwoPos/3) & $\CvTtestTglobalLocRawEuclKOneTwo$ (\CvTtestTglobalLocRawEuclKOneTwoPos/3) & $\CvTtestTglobalLocZEuclKOneTwo$ (\CvTtestTglobalLocZEuclKOneTwoPos/3) \\
THINGS test & LOC & 12 & T-session & $\CvTtestTsessionLocRawCorrKOneTwo$ (\CvTtestTsessionLocRawCorrKOneTwoPos/3) & $\CvTtestTsessionLocZCorrKOneTwo$ (\CvTtestTsessionLocZCorrKOneTwoPos/3) & $\CvTtestTsessionLocRawEuclKOneTwo$ (\CvTtestTsessionLocRawEuclKOneTwoPos/3) & $\CvTtestTsessionLocZEuclKOneTwo$ (\CvTtestTsessionLocZEuclKOneTwoPos/3) \\
\midrule
NSD & V1 & 1 & N-raw & $\CvNsdNrawVoneRawCorrKOne$ (\CvNsdNrawVoneRawCorrKOnePos/8) & $\CvNsdNrawVoneZCorrKOne$ (\CvNsdNrawVoneZCorrKOnePos/8) & $\CvNsdNrawVoneRawEuclKOne$ (\CvNsdNrawVoneRawEuclKOnePos/8) & $\CvNsdNrawVoneZEuclKOne$ (\CvNsdNrawVoneZEuclKOnePos/8) \\
NSD & V1 & 1 & N-global & $\CvNsdNglobalVoneRawCorrKOne$ (\CvNsdNglobalVoneRawCorrKOnePos/8) & $\CvNsdNglobalVoneZCorrKOne$ (\CvNsdNglobalVoneZCorrKOnePos/8) & $\CvNsdNglobalVoneRawEuclKOne$ (\CvNsdNglobalVoneRawEuclKOnePos/8) & $\CvNsdNglobalVoneZEuclKOne$ (\CvNsdNglobalVoneZEuclKOnePos/8) \\
NSD & V1 & 1 & N-session & $\CvNsdNsessionVoneRawCorrKOne$ (\CvNsdNsessionVoneRawCorrKOnePos/8) & $\CvNsdNsessionVoneZCorrKOne$ (\CvNsdNsessionVoneZCorrKOnePos/8) & $\CvNsdNsessionVoneRawEuclKOne$ (\CvNsdNsessionVoneRawEuclKOnePos/8) & $\CvNsdNsessionVoneZEuclKOne$ (\CvNsdNsessionVoneZEuclKOnePos/8) \\
NSD & V1 & 3 & N-raw & $\CvNsdNrawVoneRawCorrKThree$ (\CvNsdNrawVoneRawCorrKThreePos/8) & $\CvNsdNrawVoneZCorrKThree$ (\CvNsdNrawVoneZCorrKThreePos/8) & $\CvNsdNrawVoneRawEuclKThree$ (\CvNsdNrawVoneRawEuclKThreePos/8) & $\CvNsdNrawVoneZEuclKThree$ (\CvNsdNrawVoneZEuclKThreePos/8) \\
NSD & V1 & 3 & N-global & $\CvNsdNglobalVoneRawCorrKThree$ (\CvNsdNglobalVoneRawCorrKThreePos/8) & $\CvNsdNglobalVoneZCorrKThree$ (\CvNsdNglobalVoneZCorrKThreePos/8) & $\CvNsdNglobalVoneRawEuclKThree$ (\CvNsdNglobalVoneRawEuclKThreePos/8) & $\CvNsdNglobalVoneZEuclKThree$ (\CvNsdNglobalVoneZEuclKThreePos/8) \\
NSD & V1 & 3 & N-session & $\CvNsdNsessionVoneRawCorrKThree$ (\CvNsdNsessionVoneRawCorrKThreePos/8) & $\CvNsdNsessionVoneZCorrKThree$ (\CvNsdNsessionVoneZCorrKThreePos/8) & $\CvNsdNsessionVoneRawEuclKThree$ (\CvNsdNsessionVoneRawEuclKThreePos/8) & $\CvNsdNsessionVoneZEuclKThree$ (\CvNsdNsessionVoneZEuclKThreePos/8) \\
NSD & LO-obj & 1 & N-raw & $\CvNsdNrawLoobjRawCorrKOne$ (\CvNsdNrawLoobjRawCorrKOnePos/8) & $\CvNsdNrawLoobjZCorrKOne$ (\CvNsdNrawLoobjZCorrKOnePos/8) & $\CvNsdNrawLoobjRawEuclKOne$ (\CvNsdNrawLoobjRawEuclKOnePos/8) & $\CvNsdNrawLoobjZEuclKOne$ (\CvNsdNrawLoobjZEuclKOnePos/8) \\
NSD & LO-obj & 1 & N-global & $\CvNsdNglobalLoobjRawCorrKOne$ (\CvNsdNglobalLoobjRawCorrKOnePos/8) & $\CvNsdNglobalLoobjZCorrKOne$ (\CvNsdNglobalLoobjZCorrKOnePos/8) & $\CvNsdNglobalLoobjRawEuclKOne$ (\CvNsdNglobalLoobjRawEuclKOnePos/8) & $\CvNsdNglobalLoobjZEuclKOne$ (\CvNsdNglobalLoobjZEuclKOnePos/8) \\
NSD & LO-obj & 1 & N-session & $\CvNsdNsessionLoobjRawCorrKOne$ (\CvNsdNsessionLoobjRawCorrKOnePos/8) & $\CvNsdNsessionLoobjZCorrKOne$ (\CvNsdNsessionLoobjZCorrKOnePos/8) & $\CvNsdNsessionLoobjRawEuclKOne$ (\CvNsdNsessionLoobjRawEuclKOnePos/8) & $\CvNsdNsessionLoobjZEuclKOne$ (\CvNsdNsessionLoobjZEuclKOnePos/8) \\
NSD & LO-obj & 3 & N-raw & $\CvNsdNrawLoobjRawCorrKThree$ (\CvNsdNrawLoobjRawCorrKThreePos/8) & $\CvNsdNrawLoobjZCorrKThree$ (\CvNsdNrawLoobjZCorrKThreePos/8) & $\CvNsdNrawLoobjRawEuclKThree$ (\CvNsdNrawLoobjRawEuclKThreePos/8) & $\CvNsdNrawLoobjZEuclKThree$ (\CvNsdNrawLoobjZEuclKThreePos/8) \\
NSD & LO-obj & 3 & N-global & $\CvNsdNglobalLoobjRawCorrKThree$ (\CvNsdNglobalLoobjRawCorrKThreePos/8) & $\CvNsdNglobalLoobjZCorrKThree$ (\CvNsdNglobalLoobjZCorrKThreePos/8) & $\CvNsdNglobalLoobjRawEuclKThree$ (\CvNsdNglobalLoobjRawEuclKThreePos/8) & $\CvNsdNglobalLoobjZEuclKThree$ (\CvNsdNglobalLoobjZEuclKThreePos/8) \\
NSD & LO-obj & 3 & N-session & $\CvNsdNsessionLoobjRawCorrKThree$ (\CvNsdNsessionLoobjRawCorrKThreePos/8) & $\CvNsdNsessionLoobjZCorrKThree$ (\CvNsdNsessionLoobjZCorrKThreePos/8) & $\CvNsdNsessionLoobjRawEuclKThree$ (\CvNsdNsessionLoobjRawEuclKThreePos/8) & $\CvNsdNsessionLoobjZEuclKThree$ (\CvNsdNsessionLoobjZEuclKThreePos/8) \\
NSD & lateral & 1 & N-raw & $\CvNsdNrawLatRawCorrKOne$ (\CvNsdNrawLatRawCorrKOnePos/8) & $\CvNsdNrawLatZCorrKOne$ (\CvNsdNrawLatZCorrKOnePos/8) & $\CvNsdNrawLatRawEuclKOne$ (\CvNsdNrawLatRawEuclKOnePos/8) & $\CvNsdNrawLatZEuclKOne$ (\CvNsdNrawLatZEuclKOnePos/8) \\
NSD & lateral & 1 & N-global & $\CvNsdNglobalLatRawCorrKOne$ (\CvNsdNglobalLatRawCorrKOnePos/8) & $\CvNsdNglobalLatZCorrKOne$ (\CvNsdNglobalLatZCorrKOnePos/8) & $\CvNsdNglobalLatRawEuclKOne$ (\CvNsdNglobalLatRawEuclKOnePos/8) & $\CvNsdNglobalLatZEuclKOne$ (\CvNsdNglobalLatZEuclKOnePos/8) \\
NSD & lateral & 1 & N-session & $\CvNsdNsessionLatRawCorrKOne$ (\CvNsdNsessionLatRawCorrKOnePos/8) & $\CvNsdNsessionLatZCorrKOne$ (\CvNsdNsessionLatZCorrKOnePos/8) & $\CvNsdNsessionLatRawEuclKOne$ (\CvNsdNsessionLatRawEuclKOnePos/8) & $\CvNsdNsessionLatZEuclKOne$ (\CvNsdNsessionLatZEuclKOnePos/8) \\
NSD & lateral & 3 & N-raw & $\CvNsdNrawLatRawCorrKThree$ (\CvNsdNrawLatRawCorrKThreePos/8) & $\CvNsdNrawLatZCorrKThree$ (\CvNsdNrawLatZCorrKThreePos/8) & $\CvNsdNrawLatRawEuclKThree$ (\CvNsdNrawLatRawEuclKThreePos/8) & $\CvNsdNrawLatZEuclKThree$ (\CvNsdNrawLatZEuclKThreePos/8) \\
NSD & lateral & 3 & N-global & $\CvNsdNglobalLatRawCorrKThree$ (\CvNsdNglobalLatRawCorrKThreePos/8) & $\CvNsdNglobalLatZCorrKThree$ (\CvNsdNglobalLatZCorrKThreePos/8) & $\CvNsdNglobalLatRawEuclKThree$ (\CvNsdNglobalLatRawEuclKThreePos/8) & $\CvNsdNglobalLatZEuclKThree$ (\CvNsdNglobalLatZEuclKThreePos/8) \\
NSD & lateral & 3 & N-session & $\CvNsdNsessionLatRawCorrKThree$ (\CvNsdNsessionLatRawCorrKThreePos/8) & $\CvNsdNsessionLatZCorrKThree$ (\CvNsdNsessionLatZCorrKThreePos/8) & $\CvNsdNsessionLatRawEuclKThree$ (\CvNsdNsessionLatRawEuclKThreePos/8) & $\CvNsdNsessionLatZEuclKThree$ (\CvNsdNsessionLatZEuclKThreePos/8) \\
\bottomrule
\end{tabular}
\end{adjustbox}
\caption{All cells of step 48: $\Delta$ (untrained minus Backprop) under the three conventions per data set, ROI and $k$; in brackets the number of subjects with $\Delta > 0$.}
\label{tab:convfull}
\end{table}

\begin{table}[H]
\centering
\footnotesize
\setlength{\tabcolsep}{3pt}
\begin{adjustbox}{max width=\textwidth, max totalheight=0.82\textheight}
\begin{tabular}{@{}llrrrrrrrr@{}}
\toprule
Data & Convention & \multicolumn{4}{c}{brain $g$: Spearman with} & \multicolumn{2}{c}{$\rho(g, g^M)$} & \multicolumn{2}{c}{brain dominance} \\
\cmidrule(lr){3-6}\cmidrule(lr){7-8}\cmidrule(lr){9-10}
 & & $L$ & $C$ & CE & DD & untrained & Backprop & sum & diff. \\
\midrule
THINGS-720 V1 & T-raw & $\AmpTsevTrawVoneKOneL$ & $\AmpTsevTrawVoneKOneC$ & $\AmpTsevTrawVoneKOneCE$ & $\AmpTsevTrawVoneKOneDD$ & $\AmpGmTsevTrawVoneKOneRand$ & $\AmpGmTsevTrawVoneKOneBp$ & $\BDomTsevTrawVoneKOneSum$ & $\BDomTsevTrawVoneKOneDiff$ \\
THINGS-720 V1 & T-global & $\AmpTsevTglobalVoneKOneL$ & $\AmpTsevTglobalVoneKOneC$ & $\AmpTsevTglobalVoneKOneCE$ & $\AmpTsevTglobalVoneKOneDD$ & $\AmpGmTsevTglobalVoneKOneRand$ & $\AmpGmTsevTglobalVoneKOneBp$ & $\BDomTsevTglobalVoneKOneSum$ & $\BDomTsevTglobalVoneKOneDiff$ \\
THINGS-720 V1 & T-session & $\AmpTsevTsessionVoneKOneL$ & $\AmpTsevTsessionVoneKOneC$ & $\AmpTsevTsessionVoneKOneCE$ & $\AmpTsevTsessionVoneKOneDD$ & $\AmpGmTsevTsessionVoneKOneRand$ & $\AmpGmTsevTsessionVoneKOneBp$ & $\BDomTsevTsessionVoneKOneSum$ & $\BDomTsevTsessionVoneKOneDiff$ \\
\addlinespace
THINGS test V1, $k = 12$ & T-raw & $\AmpTestTrawVoneKOneTwoL$ & $\AmpTestTrawVoneKOneTwoC$ & $\AmpTestTrawVoneKOneTwoCE$ & $\AmpTestTrawVoneKOneTwoDD$ & $\AmpGmTestTrawVoneKOneTwoRand$ & $\AmpGmTestTrawVoneKOneTwoBp$ & $\BDomTestTrawVoneKOneTwoSum$ & $\BDomTestTrawVoneKOneTwoDiff$ \\
THINGS test V1, $k = 12$ & T-global & $\AmpTestTglobalVoneKOneTwoL$ & $\AmpTestTglobalVoneKOneTwoC$ & $\AmpTestTglobalVoneKOneTwoCE$ & $\AmpTestTglobalVoneKOneTwoDD$ & $\AmpGmTestTglobalVoneKOneTwoRand$ & $\AmpGmTestTglobalVoneKOneTwoBp$ & $\BDomTestTglobalVoneKOneTwoSum$ & $\BDomTestTglobalVoneKOneTwoDiff$ \\
THINGS test V1, $k = 12$ & T-session & $\AmpTestTsessionVoneKOneTwoL$ & $\AmpTestTsessionVoneKOneTwoC$ & $\AmpTestTsessionVoneKOneTwoCE$ & $\AmpTestTsessionVoneKOneTwoDD$ & $\AmpGmTestTsessionVoneKOneTwoRand$ & $\AmpGmTestTsessionVoneKOneTwoBp$ & $\BDomTestTsessionVoneKOneTwoSum$ & $\BDomTestTsessionVoneKOneTwoDiff$ \\
\addlinespace
NSD V1, $k = 1$ & N-raw & $\AmpNsdNrawVoneKOneL$ & $\AmpNsdNrawVoneKOneC$ & $\AmpNsdNrawVoneKOneCE$ & $\AmpNsdNrawVoneKOneDD$ & $\AmpGmNsdNrawVoneKOneRand$ & $\AmpGmNsdNrawVoneKOneBp$ & $\BDomNsdNrawVoneKOneSum$ & $\BDomNsdNrawVoneKOneDiff$ \\
NSD V1, $k = 1$ & N-global & $\AmpNsdNglobalVoneKOneL$ & $\AmpNsdNglobalVoneKOneC$ & $\AmpNsdNglobalVoneKOneCE$ & $\AmpNsdNglobalVoneKOneDD$ & $\AmpGmNsdNglobalVoneKOneRand$ & $\AmpGmNsdNglobalVoneKOneBp$ & $\BDomNsdNglobalVoneKOneSum$ & $\BDomNsdNglobalVoneKOneDiff$ \\
NSD V1, $k = 1$ & N-session & $\AmpNsdNsessionVoneKOneL$ & $\AmpNsdNsessionVoneKOneC$ & $\AmpNsdNsessionVoneKOneCE$ & $\AmpNsdNsessionVoneKOneDD$ & $\AmpGmNsdNsessionVoneKOneRand$ & $\AmpGmNsdNsessionVoneKOneBp$ & $\BDomNsdNsessionVoneKOneSum$ & $\BDomNsdNsessionVoneKOneDiff$ \\
\addlinespace
NSD V1, $k = 3$ & N-raw & $\AmpNsdNrawVoneKThreeL$ & $\AmpNsdNrawVoneKThreeC$ & $\AmpNsdNrawVoneKThreeCE$ & $\AmpNsdNrawVoneKThreeDD$ & $\AmpGmNsdNrawVoneKThreeRand$ & $\AmpGmNsdNrawVoneKThreeBp$ & $\BDomNsdNrawVoneKThreeSum$ & $\BDomNsdNrawVoneKThreeDiff$ \\
NSD V1, $k = 3$ & N-global & $\AmpNsdNglobalVoneKThreeL$ & $\AmpNsdNglobalVoneKThreeC$ & $\AmpNsdNglobalVoneKThreeCE$ & $\AmpNsdNglobalVoneKThreeDD$ & $\AmpGmNsdNglobalVoneKThreeRand$ & $\AmpGmNsdNglobalVoneKThreeBp$ & $\BDomNsdNglobalVoneKThreeSum$ & $\BDomNsdNglobalVoneKThreeDiff$ \\
NSD V1, $k = 3$ & N-session & $\AmpNsdNsessionVoneKThreeL$ & $\AmpNsdNsessionVoneKThreeC$ & $\AmpNsdNsessionVoneKThreeCE$ & $\AmpNsdNsessionVoneKThreeDD$ & $\AmpGmNsdNsessionVoneKThreeRand$ & $\AmpGmNsdNsessionVoneKThreeBp$ & $\BDomNsdNsessionVoneKThreeSum$ & $\BDomNsdNsessionVoneKThreeDiff$ \\
\midrule
Model $g^M$ & & \multicolumn{4}{c}{Spearman with $L$, $C$, CE, DD} & \multicolumn{2}{c}{model dominance} & \multicolumn{2}{c}{$\rho$(dominance, $m$)} \\
\midrule
THINGS-720, Conv1 & untrained & $\AmpModelTsevConvOneRandL$ & $\AmpModelTsevConvOneRandC$ & $\AmpModelTsevConvOneRandCE$ & $\AmpModelTsevConvOneRandDD$ & $\DomTsevConvOneSumRand$ & $\DomTsevConvOneDiffRand$ & $\DomMrTsevConvOneSum$ & $\DomMrTsevConvOneDiff$ \\
THINGS-720, Conv1 & Backprop & $\AmpModelTsevConvOneBpL$ & $\AmpModelTsevConvOneBpC$ & $\AmpModelTsevConvOneBpCE$ & $\AmpModelTsevConvOneBpDD$ & $\DomTsevConvOneSumBp$ & $\DomTsevConvOneDiffBp$ &  &  \\
THINGS test, Conv1 & untrained & $\AmpModelTestConvOneRandL$ & $\AmpModelTestConvOneRandC$ & $\AmpModelTestConvOneRandCE$ & $\AmpModelTestConvOneRandDD$ & $\DomTestConvOneSumRand$ & $\DomTestConvOneDiffRand$ & $\DomMrTestConvOneSum$ & $\DomMrTestConvOneDiff$ \\
THINGS test, Conv1 & Backprop & $\AmpModelTestConvOneBpL$ & $\AmpModelTestConvOneBpC$ & $\AmpModelTestConvOneBpCE$ & $\AmpModelTestConvOneBpDD$ & $\DomTestConvOneSumBp$ & $\DomTestConvOneDiffBp$ &  &  \\
NSD, Conv1 & untrained & $\AmpModelNsdConvOneRandL$ & $\AmpModelNsdConvOneRandC$ & $\AmpModelNsdConvOneRandCE$ & $\AmpModelNsdConvOneRandDD$ & $\DomNsdConvOneSumRand$ & $\DomNsdConvOneDiffRand$ & $\DomMrNsdConvOneSum$ & $\DomMrNsdConvOneDiff$ \\
NSD, Conv1 & Backprop & $\AmpModelNsdConvOneBpL$ & $\AmpModelNsdConvOneBpC$ & $\AmpModelNsdConvOneBpCE$ & $\AmpModelNsdConvOneBpDD$ & $\DomNsdConvOneSumBp$ & $\DomNsdConvOneDiffBp$ &  &  \\
\bottomrule
\end{tabular}
\end{adjustbox}
\caption{Response amplitude (steps 48, 48b). Brain $g$: cosine of each pattern with the mean pattern; Spearman with the image statistics (mean over subjects) and with the model $g^M$ (mean over seeds and subjects); brain dominance: Spearman(brain RDM, $A$) in the sum and the difference form. Model rows: Spearman of $g^M$ with the image statistics, model dominance Spearman(model RDM, $A^M$) (mean over seeds), and its Spearman with the moment ratio $m$ across the 10 checkpoints. DD is close to a mirror of $L$ (Section~\ref{sec:ampmethods}).}
\label{tab:amplitude}
\end{table}

\begin{table}[H]
\centering
\small
\begin{adjustbox}{max width=\textwidth, max totalheight=0.82\textheight}
\begin{tabular}{@{}llrrrrrr@{}}
\toprule
Data & Model & seed 0 & 1 & 2 & 3 & 4 & mean \\
\midrule
THINGS-720 & untrained & $\DomTsevConvOneSumRandSeedZero$ & $\DomTsevConvOneSumRandSeedOne$ & $\DomTsevConvOneSumRandSeedTwo$ & $\DomTsevConvOneSumRandSeedThree$ & $\DomTsevConvOneSumRandSeedFour$ & $\DomTsevConvOneSumRand$ \\
THINGS-720 & Backprop & $\DomTsevConvOneSumBpSeedZero$ & $\DomTsevConvOneSumBpSeedOne$ & $\DomTsevConvOneSumBpSeedTwo$ & $\DomTsevConvOneSumBpSeedThree$ & $\DomTsevConvOneSumBpSeedFour$ & $\DomTsevConvOneSumBp$ \\
THINGS test & untrained & $\DomTestConvOneSumRandSeedZero$ & $\DomTestConvOneSumRandSeedOne$ & $\DomTestConvOneSumRandSeedTwo$ & $\DomTestConvOneSumRandSeedThree$ & $\DomTestConvOneSumRandSeedFour$ & $\DomTestConvOneSumRand$ \\
THINGS test & Backprop & $\DomTestConvOneSumBpSeedZero$ & $\DomTestConvOneSumBpSeedOne$ & $\DomTestConvOneSumBpSeedTwo$ & $\DomTestConvOneSumBpSeedThree$ & $\DomTestConvOneSumBpSeedFour$ & $\DomTestConvOneSumBp$ \\
NSD & untrained & $\DomNsdConvOneSumRandSeedZero$ & $\DomNsdConvOneSumRandSeedOne$ & $\DomNsdConvOneSumRandSeedTwo$ & $\DomNsdConvOneSumRandSeedThree$ & $\DomNsdConvOneSumRandSeedFour$ & $\DomNsdConvOneSumRand$ \\
NSD & Backprop & $\DomNsdConvOneSumBpSeedZero$ & $\DomNsdConvOneSumBpSeedOne$ & $\DomNsdConvOneSumBpSeedTwo$ & $\DomNsdConvOneSumBpSeedThree$ & $\DomNsdConvOneSumBpSeedFour$ & $\DomNsdConvOneSumBp$ \\
\bottomrule
\end{tabular}
\end{adjustbox}
\caption{D1 (step 48b): Spearman(model RDM, $A^M$) per seed, sum form, Conv1.}
\label{tab:dom}
\end{table}

\begin{table}[H]
\centering
\footnotesize
\setlength{\tabcolsep}{3pt}
\begin{adjustbox}{max width=\textwidth, max totalheight=0.82\textheight}
\begin{tabular}{@{}lllrrrrrrr@{}}
\toprule
Data & ROI & $k$ & Convention & $\Delta$ & ($n > 0$) & two-sided partial, sum form & ($n > 0$) & ($n \leq 0.020$) & difference form \\
\midrule
THINGS-720 & V1 & 1 & T-raw & $\PbTsevTrawVoneKOneNone$ & (\PbTsevTrawVoneKOneNonePos) & $\PbTsevTrawVoneKOneSum$ & (\PbTsevTrawVoneKOneSumPos) & (\PbTsevTrawVoneKOneSumLe) & $\PbTsevTrawVoneKOneDiff$ \\
THINGS-720 & V1 & 1 & T-global & $\PbTsevTglobalVoneKOneNone$ & (\PbTsevTglobalVoneKOneNonePos) & $\PbTsevTglobalVoneKOneSum$ & (\PbTsevTglobalVoneKOneSumPos) & (\PbTsevTglobalVoneKOneSumLe) & $\PbTsevTglobalVoneKOneDiff$ \\
THINGS-720 & V1 & 1 & T-session & $\PbTsevTsessionVoneKOneNone$ & (\PbTsevTsessionVoneKOneNonePos) & $\PbTsevTsessionVoneKOneSum$ & (\PbTsevTsessionVoneKOneSumPos) & (\PbTsevTsessionVoneKOneSumLe) & $\PbTsevTsessionVoneKOneDiff$ \\
THINGS-720 & LOC & 1 & T-raw & $\PbTsevTrawLocKOneNone$ & (\PbTsevTrawLocKOneNonePos) & $\PbTsevTrawLocKOneSum$ & (\PbTsevTrawLocKOneSumPos) & (\PbTsevTrawLocKOneSumLe) & $\PbTsevTrawLocKOneDiff$ \\
THINGS-720 & LOC & 1 & T-global & $\PbTsevTglobalLocKOneNone$ & (\PbTsevTglobalLocKOneNonePos) & $\PbTsevTglobalLocKOneSum$ & (\PbTsevTglobalLocKOneSumPos) & (\PbTsevTglobalLocKOneSumLe) & $\PbTsevTglobalLocKOneDiff$ \\
THINGS-720 & LOC & 1 & T-session & $\PbTsevTsessionLocKOneNone$ & (\PbTsevTsessionLocKOneNonePos) & $\PbTsevTsessionLocKOneSum$ & (\PbTsevTsessionLocKOneSumPos) & (\PbTsevTsessionLocKOneSumLe) & $\PbTsevTsessionLocKOneDiff$ \\
\midrule
THINGS test & V1 & 1 & T-raw & $\PbTestTrawVoneKOneNone$ & (\PbTestTrawVoneKOneNonePos) & $\PbTestTrawVoneKOneSum$ & (\PbTestTrawVoneKOneSumPos) & (\PbTestTrawVoneKOneSumLe) & $\PbTestTrawVoneKOneDiff$ \\
THINGS test & V1 & 1 & T-global & $\PbTestTglobalVoneKOneNone$ & (\PbTestTglobalVoneKOneNonePos) & $\PbTestTglobalVoneKOneSum$ & (\PbTestTglobalVoneKOneSumPos) & (\PbTestTglobalVoneKOneSumLe) & $\PbTestTglobalVoneKOneDiff$ \\
THINGS test & V1 & 1 & T-session & $\PbTestTsessionVoneKOneNone$ & (\PbTestTsessionVoneKOneNonePos) & $\PbTestTsessionVoneKOneSum$ & (\PbTestTsessionVoneKOneSumPos) & (\PbTestTsessionVoneKOneSumLe) & $\PbTestTsessionVoneKOneDiff$ \\
THINGS test & V1 & 12 & T-raw & $\PbTestTrawVoneKOneTwoNone$ & (\PbTestTrawVoneKOneTwoNonePos) & $\PbTestTrawVoneKOneTwoSum$ & (\PbTestTrawVoneKOneTwoSumPos) & (\PbTestTrawVoneKOneTwoSumLe) & $\PbTestTrawVoneKOneTwoDiff$ \\
THINGS test & V1 & 12 & T-global & $\PbTestTglobalVoneKOneTwoNone$ & (\PbTestTglobalVoneKOneTwoNonePos) & $\PbTestTglobalVoneKOneTwoSum$ & (\PbTestTglobalVoneKOneTwoSumPos) & (\PbTestTglobalVoneKOneTwoSumLe) & $\PbTestTglobalVoneKOneTwoDiff$ \\
THINGS test & V1 & 12 & T-session & $\PbTestTsessionVoneKOneTwoNone$ & (\PbTestTsessionVoneKOneTwoNonePos) & $\PbTestTsessionVoneKOneTwoSum$ & (\PbTestTsessionVoneKOneTwoSumPos) & (\PbTestTsessionVoneKOneTwoSumLe) & $\PbTestTsessionVoneKOneTwoDiff$ \\
THINGS test & LOC & 1 & T-raw & $\PbTestTrawLocKOneNone$ & (\PbTestTrawLocKOneNonePos) & $\PbTestTrawLocKOneSum$ & (\PbTestTrawLocKOneSumPos) & (\PbTestTrawLocKOneSumLe) & $\PbTestTrawLocKOneDiff$ \\
THINGS test & LOC & 1 & T-global & $\PbTestTglobalLocKOneNone$ & (\PbTestTglobalLocKOneNonePos) & $\PbTestTglobalLocKOneSum$ & (\PbTestTglobalLocKOneSumPos) & (\PbTestTglobalLocKOneSumLe) & $\PbTestTglobalLocKOneDiff$ \\
THINGS test & LOC & 1 & T-session & $\PbTestTsessionLocKOneNone$ & (\PbTestTsessionLocKOneNonePos) & $\PbTestTsessionLocKOneSum$ & (\PbTestTsessionLocKOneSumPos) & (\PbTestTsessionLocKOneSumLe) & $\PbTestTsessionLocKOneDiff$ \\
THINGS test & LOC & 12 & T-raw & $\PbTestTrawLocKOneTwoNone$ & (\PbTestTrawLocKOneTwoNonePos) & $\PbTestTrawLocKOneTwoSum$ & (\PbTestTrawLocKOneTwoSumPos) & (\PbTestTrawLocKOneTwoSumLe) & $\PbTestTrawLocKOneTwoDiff$ \\
THINGS test & LOC & 12 & T-global & $\PbTestTglobalLocKOneTwoNone$ & (\PbTestTglobalLocKOneTwoNonePos) & $\PbTestTglobalLocKOneTwoSum$ & (\PbTestTglobalLocKOneTwoSumPos) & (\PbTestTglobalLocKOneTwoSumLe) & $\PbTestTglobalLocKOneTwoDiff$ \\
THINGS test & LOC & 12 & T-session & $\PbTestTsessionLocKOneTwoNone$ & (\PbTestTsessionLocKOneTwoNonePos) & $\PbTestTsessionLocKOneTwoSum$ & (\PbTestTsessionLocKOneTwoSumPos) & (\PbTestTsessionLocKOneTwoSumLe) & $\PbTestTsessionLocKOneTwoDiff$ \\
\midrule
NSD & V1 & 1 & N-raw & $\PbNsdNrawVoneKOneNone$ & (\PbNsdNrawVoneKOneNonePos) & $\PbNsdNrawVoneKOneSum$ & (\PbNsdNrawVoneKOneSumPos) & (\PbNsdNrawVoneKOneSumLe) & $\PbNsdNrawVoneKOneDiff$ \\
NSD & V1 & 1 & N-global & $\PbNsdNglobalVoneKOneNone$ & (\PbNsdNglobalVoneKOneNonePos) & $\PbNsdNglobalVoneKOneSum$ & (\PbNsdNglobalVoneKOneSumPos) & (\PbNsdNglobalVoneKOneSumLe) & $\PbNsdNglobalVoneKOneDiff$ \\
NSD & V1 & 1 & N-session & $\PbNsdNsessionVoneKOneNone$ & (\PbNsdNsessionVoneKOneNonePos) & $\PbNsdNsessionVoneKOneSum$ & (\PbNsdNsessionVoneKOneSumPos) & (\PbNsdNsessionVoneKOneSumLe) & $\PbNsdNsessionVoneKOneDiff$ \\
NSD & V1 & 3 & N-raw & $\PbNsdNrawVoneKThreeNone$ & (\PbNsdNrawVoneKThreeNonePos) & $\PbNsdNrawVoneKThreeSum$ & (\PbNsdNrawVoneKThreeSumPos) & (\PbNsdNrawVoneKThreeSumLe) & $\PbNsdNrawVoneKThreeDiff$ \\
NSD & V1 & 3 & N-global & $\PbNsdNglobalVoneKThreeNone$ & (\PbNsdNglobalVoneKThreeNonePos) & $\PbNsdNglobalVoneKThreeSum$ & (\PbNsdNglobalVoneKThreeSumPos) & (\PbNsdNglobalVoneKThreeSumLe) & $\PbNsdNglobalVoneKThreeDiff$ \\
NSD & V1 & 3 & N-session & $\PbNsdNsessionVoneKThreeNone$ & (\PbNsdNsessionVoneKThreeNonePos) & $\PbNsdNsessionVoneKThreeSum$ & (\PbNsdNsessionVoneKThreeSumPos) & (\PbNsdNsessionVoneKThreeSumLe) & $\PbNsdNsessionVoneKThreeDiff$ \\
NSD & LO-obj & 1 & N-raw & $\PbNsdNrawLoobjKOneNone$ & (\PbNsdNrawLoobjKOneNonePos) & $\PbNsdNrawLoobjKOneSum$ & (\PbNsdNrawLoobjKOneSumPos) & (\PbNsdNrawLoobjKOneSumLe) & $\PbNsdNrawLoobjKOneDiff$ \\
NSD & LO-obj & 1 & N-global & $\PbNsdNglobalLoobjKOneNone$ & (\PbNsdNglobalLoobjKOneNonePos) & $\PbNsdNglobalLoobjKOneSum$ & (\PbNsdNglobalLoobjKOneSumPos) & (\PbNsdNglobalLoobjKOneSumLe) & $\PbNsdNglobalLoobjKOneDiff$ \\
NSD & LO-obj & 1 & N-session & $\PbNsdNsessionLoobjKOneNone$ & (\PbNsdNsessionLoobjKOneNonePos) & $\PbNsdNsessionLoobjKOneSum$ & (\PbNsdNsessionLoobjKOneSumPos) & (\PbNsdNsessionLoobjKOneSumLe) & $\PbNsdNsessionLoobjKOneDiff$ \\
NSD & LO-obj & 3 & N-raw & $\PbNsdNrawLoobjKThreeNone$ & (\PbNsdNrawLoobjKThreeNonePos) & $\PbNsdNrawLoobjKThreeSum$ & (\PbNsdNrawLoobjKThreeSumPos) & (\PbNsdNrawLoobjKThreeSumLe) & $\PbNsdNrawLoobjKThreeDiff$ \\
NSD & LO-obj & 3 & N-global & $\PbNsdNglobalLoobjKThreeNone$ & (\PbNsdNglobalLoobjKThreeNonePos) & $\PbNsdNglobalLoobjKThreeSum$ & (\PbNsdNglobalLoobjKThreeSumPos) & (\PbNsdNglobalLoobjKThreeSumLe) & $\PbNsdNglobalLoobjKThreeDiff$ \\
NSD & LO-obj & 3 & N-session & $\PbNsdNsessionLoobjKThreeNone$ & (\PbNsdNsessionLoobjKThreeNonePos) & $\PbNsdNsessionLoobjKThreeSum$ & (\PbNsdNsessionLoobjKThreeSumPos) & (\PbNsdNsessionLoobjKThreeSumLe) & $\PbNsdNsessionLoobjKThreeDiff$ \\
NSD & lateral & 1 & N-raw & $\PbNsdNrawLatKOneNone$ & (\PbNsdNrawLatKOneNonePos) & $\PbNsdNrawLatKOneSum$ & (\PbNsdNrawLatKOneSumPos) & (\PbNsdNrawLatKOneSumLe) & $\PbNsdNrawLatKOneDiff$ \\
NSD & lateral & 1 & N-global & $\PbNsdNglobalLatKOneNone$ & (\PbNsdNglobalLatKOneNonePos) & $\PbNsdNglobalLatKOneSum$ & (\PbNsdNglobalLatKOneSumPos) & (\PbNsdNglobalLatKOneSumLe) & $\PbNsdNglobalLatKOneDiff$ \\
NSD & lateral & 1 & N-session & $\PbNsdNsessionLatKOneNone$ & (\PbNsdNsessionLatKOneNonePos) & $\PbNsdNsessionLatKOneSum$ & (\PbNsdNsessionLatKOneSumPos) & (\PbNsdNsessionLatKOneSumLe) & $\PbNsdNsessionLatKOneDiff$ \\
NSD & lateral & 3 & N-raw & $\PbNsdNrawLatKThreeNone$ & (\PbNsdNrawLatKThreeNonePos) & $\PbNsdNrawLatKThreeSum$ & (\PbNsdNrawLatKThreeSumPos) & (\PbNsdNrawLatKThreeSumLe) & $\PbNsdNrawLatKThreeDiff$ \\
NSD & lateral & 3 & N-global & $\PbNsdNglobalLatKThreeNone$ & (\PbNsdNglobalLatKThreeNonePos) & $\PbNsdNglobalLatKThreeSum$ & (\PbNsdNglobalLatKThreeSumPos) & (\PbNsdNglobalLatKThreeSumLe) & $\PbNsdNglobalLatKThreeDiff$ \\
NSD & lateral & 3 & N-session & $\PbNsdNsessionLatKThreeNone$ & (\PbNsdNsessionLatKThreeNonePos) & $\PbNsdNsessionLatKThreeSum$ & (\PbNsdNsessionLatKThreeSumPos) & (\PbNsdNsessionLatKThreeSumLe) & $\PbNsdNsessionLatKThreeDiff$ \\
\bottomrule
\end{tabular}
\end{adjustbox}
\caption{$\Delta$ (paper cell) before and after the two-sided partial (step 48b): the brain RDM is rank-residualised on its amplitude RDM $A$, the model RDM on its own $A^M$, and the residuals are correlated. Sum form $A_{ij} = 2 - g_i - g_j$ (the form of D1--D3); difference form $|g_i - g_j|$, reported for comparison.}
\label{tab:partial48b}
\end{table}

\end{document}